\documentclass[prd,twocolumn,superscriptaddress,floatfix,amsmath,amssymb]{revtex4-1}

\usepackage[english]{babel}
\usepackage{hyperref}	
\usepackage{graphicx}
\usepackage{dcolumn}
\usepackage{bm}
\usepackage{hyperref}
\usepackage{braket}
\usepackage{amsmath,amssymb,latexsym}
\usepackage{dsfont}
\usepackage{cancel}
\usepackage{bbold}
\usepackage[usenames, dvipsnames]{color}
\usepackage{amsfonts}
\usepackage{comment}
\usepackage{natbib}
\usepackage{mathrsfs}
\usepackage{soul}
\usepackage{verbatim}
\usepackage{textcomp}
\newcommand{\spliteq}[1]{\begin{equation}
\begin{split}
#1
\end{split}
\end{equation}
}

\DeclareMathOperator{\tr}{tr}

\DeclareMathOperator{\Ad}{Ad}

\begin{document}


\title{A Universal Training Algorithm for Quantum Deep Learning}

 \affiliation{Department of Applied Mathematics, University of Waterloo, Waterloo, Ontario, N2L 3G1, Canada
}

 \affiliation{Institute for Quantum Computing, University of Waterloo, Waterloo, Ontario, N2L 3G1, Canada
}
\affiliation{School of Computer Science, University of Waterloo, Waterloo, Ontario, N2L 3G1, Canada
}

 \affiliation{Perimeter Institute for Theoretical Physics, Waterloo, Ontario, N2L 2Y5, Canada
}

\author{Guillaume Verdon}
 \affiliation{Department of Applied Mathematics, University of Waterloo, Waterloo, Ontario, N2L 3G1, Canada
}
 \affiliation{Institute for Quantum Computing, University of Waterloo, Waterloo, Ontario, N2L 3G1, Canada
}
 \affiliation{Perimeter Institute for Theoretical Physics, Waterloo, Ontario, N2L 2Y5, Canada
}

\author{Jason Pye}
 \affiliation{Department of Applied Mathematics, University of Waterloo, Waterloo, Ontario, N2L 3G1, Canada
}
 \affiliation{Institute for Quantum Computing, University of Waterloo, Waterloo, Ontario, N2L 3G1, Canada
}
 \affiliation{Perimeter Institute for Theoretical Physics, Waterloo, Ontario, N2L 2Y5, Canada
}

\author{Michael Broughton}
 \affiliation{School of Computer Science, University of Waterloo, Waterloo, Ontario, N2L 3G1, Canada
}

\date{\today}

\begin{abstract}We introduce the Backwards Quantum Propagation of Phase errors (Baqprop) principle, a central theme upon which we construct multiple universal optimization heuristics for training both parametrized quantum circuits and classical deep neural networks on a quantum computer. Baqprop encodes error information in relative phases of a quantum wavefunction defined over the space of network parameters; it can be thought of as the unification of the phase kickback principle of quantum computation and of the backpropagation algorithm from classical deep learning. We propose two core heuristics which leverage Baqprop for quantum-enhanced optimization of network parameters: Quantum Dynamical Descent (QDD) and Momentum Measurement Gradient Descent (MoMGrad). QDD uses simulated quantum coherent dynamics for parameter optimization, allowing for quantum tunneling through the hypothesis space landscape. MoMGrad leverages Baqprop to estimate gradients and thereby perform gradient descent on the parameter landscape; it can be thought of as the quantum-classical analogue of QDD. In addition to these core optimization strategies, we propose various methods for parallelization, regularization, and meta-learning as augmentations to MoMGrad and QDD. We introduce several quantum-coherent adaptations of canonical classical feedforward neural networks, and study how Baqprop can be used to optimize such networks. We develop multiple applications of parametric circuit learning for quantum data, and show how to perform Baqprop in each case. One such application allows for the training of hybrid quantum-classical neural-circuit networks, via the seamless integration of Baqprop with classical backpropagation. Finally, for a representative subset of these proposed applications, we demonstrate the training of these networks via numerical simulations of implementations of QDD and MoMGrad.
\end{abstract}
\maketitle
\tableofcontents


\section{Introduction}

The field of classical deep learning \cite{Goodfellow-et-al-2016} has seen a rapid expansion in interest and number of breakthrough applications in recent years \cite{2014arXiv1412.6980K,2012arXiv1212.5701Z,2016arXiv160407316B,2014arXiv1409.3215S,2016arXiv160903499V,2013arXiv1301.3781M,2015arXiv151203385H,2013arXiv1312.5602M,2014arXiv1412.6980K,2013arXiv1312.6114K,2014arXiv1406.2661G}.

Deep learning consists of a class of algorithms within the broader class of machine learning algorithms, which are mostly employed either to identify patterns in a given dataset and/or generate new data mimicking such patterns.
At their core, many machine learning algorithms consist of three main components.
First is the model: a parametrized hypothesis class of functions, usually arranged in a network of layered compositions of simpler parametric functions.
Networks with multiple layers are called \emph{deep}, and are the subclass of models considered in deep learning.
Second is a cost function: a metric to evaluate how well specific hypotheses model the data.
The third and final key component is the optimizer: an algorithmic strategy used to search over the space of hypotheses in order to minimize the cost function to a sufficient degree.

A central concept in the optimization of such networks of compositions is the principle of \emph{backwards propagation of errors}, also known as the \emph{backpropagation algorithm}.
Typically the cost function (\emph{error}) of such a layered network is a function strictly of the output (final layer) of the network (or occasionally of the output of certain subsets of the network).
The backpropagation algorithm is a means for information about the gradient of the cost function (with respect to the network parameters) to spread efficiently throughout the network, beginning at the output and propagating backwards through the compositional layers.
Since the (negative) gradient provides the direction of steepest descent in the landscape of hypotheses, this propagation can be leveraged to optimize the network parameters in order to find a local minimum of the cost function.
Many, if not all, canonical network optimization methods employ the backpropagation principle in some manner \cite{Rumelhart1986,2012arXiv1212.5701Z}. 
It is often deemed that the recent resurgence and successes of classical deep learning can be traced back to the first demonstrations of implementations backpropagation algorithm \cite{Rumelhart1986,domingos2015master}. 

In this paper, we introduce a quantum-native backpropagation principle (Sec.~\ref{sec:phase-kick}), called the \emph{backwards quantum propagation of phase errors} (Baqprop).
This Baqprop principle allows for the efficient optimization of quantumly-parametrized networks on a quantum computer.
Previously, such quantum networks typically consisted of classically-parametrized quantum operations.
By considering versions of these networks using quantum parameters, we can exploit the quantum mechanical properties of the wavefunction over the hypothesis space to aid in the propagation of gradient information throughout the network.
More specifically, Baqprop employs the phase kickback principle of quantum computing to induce relative phases between different branches of the wavefunction in the superposition over hypothesis space.
These relative phases will contain the desired gradient information.
Baqprop will thus allow for quantum-enhanced optimization over multiple types of quantum parametric network hypothesis classes.
Note that the technique of leveraging phase kickback for gradients was originally pioneered by Jordan \cite{2005PhRvL..95e0501J}, and later improved upon in Ref.~\cite{2017arXiv171100465G}.
In our background section (Sec.~\ref{sec:bkgd}), we show how this gradient technique is related to phase estimation in the context of both continuous variable quantum information and qudits/qubits.
Therefore, in the context of training quantum-parametric networks, Baqprop provides a unified view of both classical backpropagation and quantum phase estimation.

Further included in this work is the introduction of two main Baqprop-based parameter optimization strategies (Sec.~\ref{sec:qdd} \& \ref{sec:momgrad}).
Both approaches leverage the cost function error signal encoded in the relative phases of the quantum parameter wavefunction, but provide different means of using this error signal to update the parameters during an iteration of the optimization.
The first of these strategies is a fully quantum-coherent method of optimization, called \emph{Quantum Dynamical Descent} (QDD).
This method is motivated by the recognition that these relative phases can be seen as induced by an effective potential acting on the parameters.
The QDD algorithm is then a descent of the parameter optimization landscape via quantum simulation of the Schr\"odinger dynamics under the influence of this effective potential.
The second method is a quantum-classical approach, which involves a quantum measurement of the Baqprop-induced relative phase shifts in the wavefunction of the parameters.
This allows for the estimation of the local gradient of the cost function in the parameter space, which can then be used in gradient descent to descend the cost landscape.
Since these relative phase shifts can be interpreted as kicks in the \emph{momenta} of the parameters, we call this approach \emph{Momentum Measurement Gradient Descent} (MoMGrad).

The broad aim of this work is to bridge classical and quantum deep learning theory within a unified framework for optimization on quantum computers.
Establishing this bridge between theories allows for an exchange of powerful tools across fields, as well as the possibility to mutually improve understanding of both topics.
In this spirit, in Section~\ref{sec:misc} we introduce multiple techniques as augmentations of the core optimization methods of Section~\ref{sec:opt} (QDD and MoMGrad), which are directly inspired from methods of classical deep learning \cite{Catanzaro2011,dropout,weightdecay}.
For example, we introduce methods for parallelization (Sec.~\ref{sec:CAMP}), regularization (Sec.~\ref{sec:reg}), and hyper-parameter optimization (meta-learning, Sec.~\ref{sec:meta}).
In addition to these various augmented optimization strategies, in Sections~\ref{sec:qnn} and \ref{sec:qdata_algs} we explore ways of leveraging Baqprop in numerous applications of quantum parametric transformation learning for classical and quantum data modelling.
In particular, for classical data learning we examine quantum-coherent analogues of traditional classical neural networks (Sec.~\ref{sec:qnn}), while for quantum data we discuss the training of a number of applications of Quantum Parametric Circuits (Sec.~\ref{sec:qdata_algs}).
We later test the efficacy of training some of these proposed applications with QDD and MoMGrad via numerical simulations of quantum computation in Section~\ref{sec:num}.

To provide further context for this work, let us briefly describe how it fits into the current state of quantum machine learning literature.
Inspired by classical machine learning, the field of quantum machine learning began as an exploration of the possibility of using quantum algorithms to identify patterns in either quantum or classical data using a quantum computer \cite{biamonte2017quantum}.
Early quantum machine learning work took a similar path as early classical machine learning;
before the advent of the connectionist approach (deep learning), the focus lied mostly on so-called analogizer-type algorithms \cite{domingos2015master}
.
Such early quantum algorithms include Quantum Principal Component Analysis \cite{lloyd2014quantum}, Quantum Support Vector Machines \cite{rebentrost2014quantum}, and other kernel methods \cite{schuld2018quantum,havlicek2018supervised}.
Many of these algorithms focused on the analysis of classical data embedded into a quantum wavefunction via a theoretical quantum computer component called a Quantum Random Access Memory \cite{giovannetti2008quantum}.
The goal of such an embedding was to exploit the exponential dimensionality of the Hilbert space to encode data in the probability amplitudes of a multi-qubit wavefunction, in order to potentially gain an exponential speedup over classical algorithms.
The feasibility and practicality of this data-loading scheme, with realistic noise conditions and error correction overheads taken into account, remains a debated topic to this day \cite{arunachalam2015robustness}.
Beyond the data loading issue, part of the quantum machine learning field has moved away from analogizer-type methods \cite{domingos2015master} towards parametric networks (resembling deep learning) for similar reasons to those responsible for the eventual dominance of classical deep learning over classical kernel-type methods. \cite{farhi2018classification,chen2018universal,grant2018hierarchical,peruzzo2014variational}
Namely, the reasoning being flexibility and modelling capacity: not all data is linearly separable (using SVMs), thus requiring a hand-picked kernel, and not all data is well-suited to a Principal Component Analysis.

Before we delve into the more recent literature on quantum parametric networks, we will first mention earlier work involving deep learning on quantum computers.
Similar to the progression of classical deep learning, the first forms of quantum neural networks to be studied were Boltzmann machines. 
In classical machine learning, some of the work first incorporating backpropagation was in the context of deep networks of coupled spin-like neurons called Deep Boltzmann Networks \cite{salakhutdinov2010efficient}.
On the quantum side, analog quantum computers allowed for a physical implementation of networks of qubits whose statistics mimic those of Boltzmann machines \cite{amin2018quantum,
adachi2015application,
neven2008image,
mohseni2016constructing}.
This general avenue of research focused on determining whether quantum computers can accelerate the training of classical neural network models. 
Due to the possibility of superpositions of the joint state of the neurons, and thereby of quantum tunneling in the energy landscape, it was hoped that Quantum Annealing could provide a speedup over classical annealing optimization methods for such neural network models.
Despite early claims of a speedup \cite{ronnow2014defining},
certain bottlenecks such as the embedding problem, qubit quality, and thermal noise \cite{benedetti2016estimation}
would obscure whether there could truly be a quantum advantage for Quantum Annealing, especially with the advent of quantum-inspired classical algorithms designed to compete with these machines \cite{katzgraber2015seeking}. 

The question thus remained: is there a way to leverage the quantum properties of superposition, entanglement, and tunneling in order to gain an optimization advantage for a classical neural network? 
Later work continued on this avenue of research, \cite{verdon2017quantum}
but most work pivoted to quantum parametric circuits, which we will return to below.

In this paper, we provide a comprehensive approach to training classical neural networks on a quantum computer for the purposes of classical data learning (Sec.~\ref{sec:qnn}).
All techniques make use of superposition and entanglement (Sec.~\ref{sec:opt}), and some
techniques employ tunneling directly (Sec.~\ref{sec:qdd}, \ref{sec:meta}).
We also provide an in-depth analysis of quantum backpropagation of the error signal in these quantum-coherent neural networks, thus explicitly relating quantum and classical backpropagation.
This bridging of the theories allows for further exchange of insights and techniques, as well as a merging of both the classical and quantum backpropagation principles (see Sec.~\ref{sec:hybrid}).
Furthermore, not only can the network parameters be optimized, but so can the network architecture and hyper-parameters in a quantum tunneling procedure in the space of trained networks, which we call Quantum Meta-Learning (QMetaL, Section~\ref{sec:meta}).

Although we do not directly claim a general speedup for training classical neural nets, in Section~\ref{sec:qdd} we explicitly relate Quantum Dynamical Descent (QDD) to the Quantum Approximate Optimization Algorithm (QAOA) \cite{farhi2014quantum,farhi2016quantum,hadfield2017quantum} and the Quantum Adiabatic Algorithm (QAA) \cite{farhi2000quantum,farhi2002quantum,
crosson2014different}.
QAA is the predecessor to Quantum Annealing, the latter of which is considered to be the open quantum system analogue of QAA.
The QAOA is akin to a variationally-optimized, temporally coarse-grained, approximate quantum simulation of the QAA.
Both the QAA and the QAOA have been shown to exhibit a quantum advantage in some optimization scenarios \cite{farhi2016quantum,crosson2016simulated}. 
As such, the possibility may be open to show a speedup for Quantum Dynamical Descent and/or Quantum Meta-Learning for certain types of networks and optimization scenarios.
We leave further analysis of such advantages for future work. 

More recent approaches to quantum deep learning have moved away from attempting to train classical models on a quantum computer, and have rather involved a quantum-native model called quantum parametric circuits (QPCs) \cite{farhi2018classification,chen2018universal,grant2018hierarchical,peruzzo2014variational}
.
As their name implies, QPCs consist of multiple parametric quantum operations arranged in a sequential, layered structure, in a similar fashion to a neural network.
In the literature, QPCs are sometimes called Quantum Variational Algorithms \cite{peruzzo2014variational} or Quantum Neural Networks \cite{farhi2018classification,chen2018universal}.
To avoid confusion with the quantum-coherent neural networks from Section~\ref{sec:qnn}, we will exclusively use the term Quantum Parametric Circuits.

QPCs can learn from either classical data or quantum data, and have been shown to be able to do so on near-term noisy quantum hardware \cite{zeng2017quantum,benedetti2018generative}, mainly through the use of classical-quantum hybrid optimization schemes \cite{mcclean2016theory}.
Such optimization schemes first execute multiple runs of a parametric circuit on a quantum processing unit for a certain set of parameters.
Through these runs the expectation value of certain observables at the output of the circuit are obtained and fed to a classical processing unit.
The classical computer is then tasked with the extremization of this expectation value subject to variations in the parameters, using the quantum computer as a black box.
Thus the classical and quantum computer work in tandem to variationally optimize over the space of parametric circuits, hence the name quantum-classical hybrid optimization.

Despite a recent rapid expansion of this body of work, the question remains open as to whether there can be a more efficient means to optimize over the space of quantum networks either in the short term (Noisy Intermediate Scale Quantum Devices \cite{preskill2018quantum} era) and long term (post Fault-Tolerance and Error Correction \cite{gottesman2010introduction}).
Furthermore, a backpropagation principle could provide a unified approach to the optimization of quantum networks, and provide insights as to which ansatze are efficiently trainable \cite{mcclean2018barren}. 

The work presented in this paper tackles these issues.
We show explicitly how to use Baqprop for a number of applications of Quantum Parametric Circuits in Section~\ref{sec:qdata_algs}.
A main draw using Baqprop is that it requires only on the order of one query (feedforward) per optimization step.
This can be compared to the above-mentioned classical finite-difference (quantum-classical) methods which usually require a number of queries which scales at least linearly with the number of parameters.
The applications featured in Section~\ref{sec:qdata_algs} either build upon previously existing work \cite{romero2017quantum,
johnson2017qvector}, 
or relate to works released during the writing of this paper \cite{lloyd2018quantum,farhi2018classification,chen2018universal}. 
In particular, we study the following tasks: quantum state learning (Sec.~\ref{sec:qstate_learn}), quantum unitary learning (Sec.~\ref{sec:qunitary_sup_learn} \& \ref{sec:qunitary_unsup_learn}), quantum channel learning (Sec.~\ref{sec:qchannel_sup_learn} \& \ref{sec:qchannel_unsup_learn}), quantum classification/regression (Sec.~\ref{sec:meas_learn}), quantum compression code (Sec.~\ref{sec:q_auto_learn} \& \ref{sec:q_noiseauto_learn}) and quantum error correcting code learning (Sec.~\ref{sec:qecc_learn}), quantum generative adversarial learning (Sec.~\ref{sec:gaq}), as well as parametric Hamiltonian optimization (Sec.~\ref{sec:hammer}).
Finally, we propose an application which combines both classical neural networks and quantum parametric circuits in a hybrid network (Sec.~\ref{sec:hybrid}).
We show how to leverage Baqprop to train these hybrid neural-circuit networks either exclusively on a quantum processing unit, or in a hybrid quantum-classical fashion, combining classical backpropagation with Baqprop, and allowing for the seamless backpropagation of error information through the classical-quantum interface.

As this paper is intended for a broad audience, we begin with an introduction of core background quantum computing concepts in Section~\ref{sec:bkgd}, including continuous-variable (CV) and discrete variable (DV) phase space, phase estimation, basic operations, and gradient estimation.
Although not essential to understanding this paper, a knowledge of standard deep learning may be useful to the reader who would like to compare classical versions of certain protocols to their respective quantum versions introduced in this paper.
We encourage the reader looking to fully connect concepts of classical and quantum deep learning to consult one of many possible references which cover the basics, such as gradient descent, stochastic gradient descent, minibatch gradient descent, and hyper-parameter optimization \cite{Goodfellow-et-al-2016}.

\section{Background}\label{sec:bkgd}

\subsection{Continuous Quantum Registers}\label{sec:CQR}

A quantum register that stores a real number is defined by an observable with a spectrum consisting of $\mathds{R}$, which we will denote here by $\hat{\Phi} := \int_{\mathds{R}} d\Phi \hspace{1mm} \Phi \ket{\Phi}\!\bra{\Phi}$.
The Hilbert space upon which this operator acts is $L_2(\mathds{R})$.
Shifts between eigenstates of the operator $\hat{\Phi}$ are generated by a conjugate momentum operator, denoted $\hat{\Pi}$, which satisfies $[ \hat{\Phi}, \hat{\Pi} ] = i$ (where throughout we set $\hbar = 1$).

Addition and multiplication of real numbers are common operations performed during a computation.
In order to implement these operations as unitaries on quantum registers, they need to be reversible.
This is typically achieved by retaining one or more of the inputs in the output of the computation.
For example, addition of two quantum real numbers in eigenstates $\ket{\Phi_1}$ and $\ket{\Phi_2}$ (respectively), can be achieved with
\begin{equation}
  e^{-i \hat{\Phi}_1 \hat{\Pi}_2} : \ket{\Phi_1,\Phi_2} \mapsto \ket{\Phi_1,\Phi_1+\Phi_2}.
\end{equation}
Here we have used one of the output registers to store one of the inputs, and the other to store the sum of the two input values.
Note that this implementation of addition can be thought of as a von Neumann measurement of the first register by the second.
Addition can also be achieved somewhat less efficiently by retaining both input values and storing the output in a third register (which is initialized to the state $\ket{\Phi_3 = 0}$):
\begin{equation}\label{eq:addition}
  e^{-i \hat{\Phi}_1 \hat{\Pi}_3} e^{-i \hat{\Phi}_2 \hat{\Pi}_3} : \ket{\Phi_1,\Phi_2,0} \mapsto \ket{\Phi_1,\Phi_2,\Phi_1+\Phi_2}.
\end{equation}
Enacting multiplication of two quantum real numbers typically involves using a third register for the output (initialized to $\ket{\Phi_3=0}$).
The unitary for multiplication is
\begin{equation}\label{eq:multiplication}
  e^{-i \hat{\Phi}_1 \hat{\Phi}_2 \hat{\Pi}_3} : \ket{\Phi_1,\Phi_2,0} \mapsto \ket{\Phi_1,\Phi_2,\Phi_1\Phi_2}.
\end{equation}
These basic operations will be used extensively in the paper.

\begin{figure}[h!]
 \begin{center}
\includegraphics[width=0.95\columnwidth]{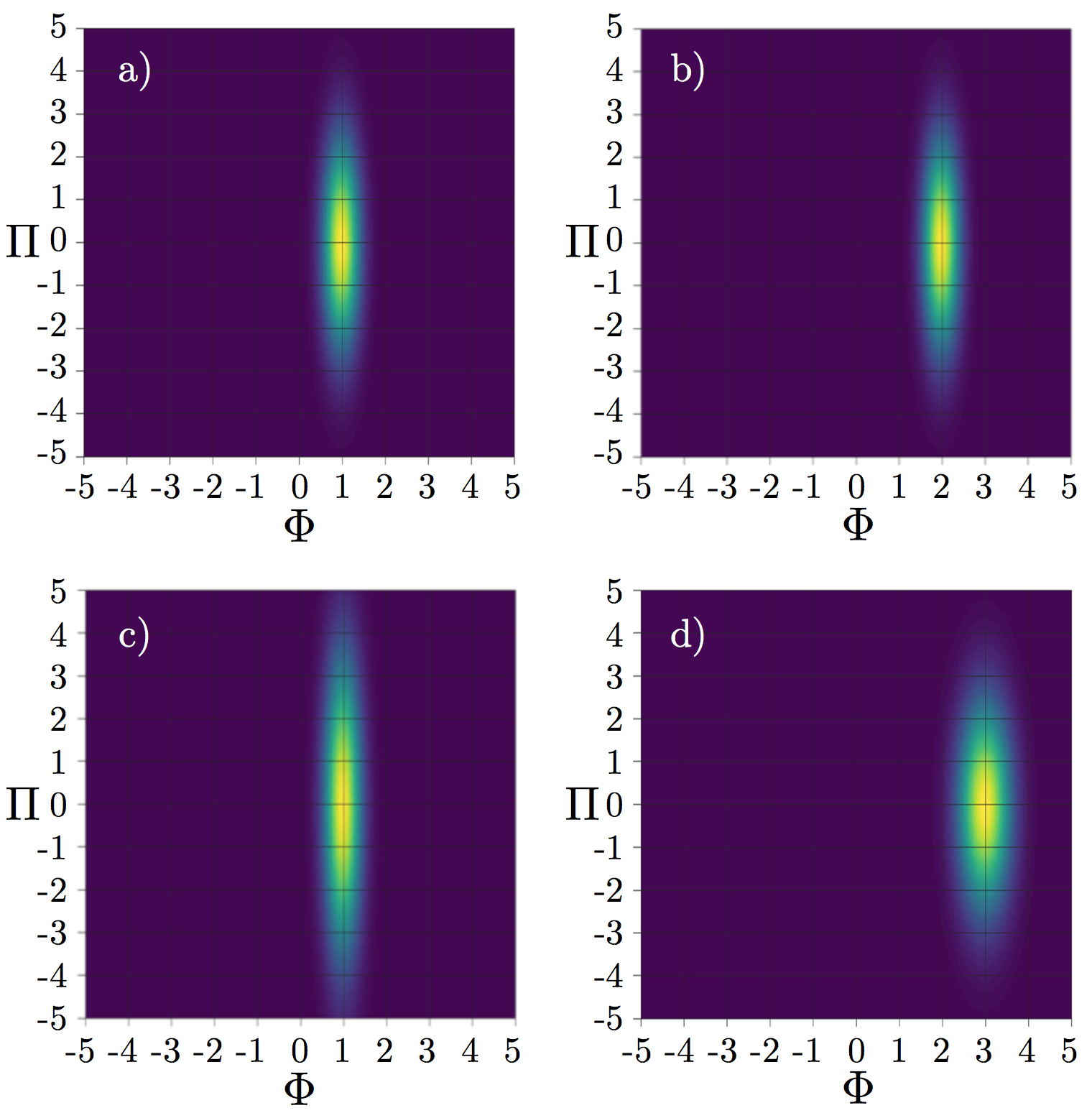}
\caption{CV adder. In plots (a) and (b) we have two initial Gaussian pointer states initialized with $\Phi$ centered at 1 and 2, respectively. In plots (c) and (d) we show these two registers after the addition, where plot (c) retains the input of the first register and plot (d) shows the output on the second register. We see that, as expected, the Gaussian in (d) is centered at the value 3. Here we have employed Gaussian states of finite variance, since $\hat{\Phi}$ eigenstates are unphysical, hence there is some noise in the input and output values for these registers. Also, in plot (c) we have that the Wigner function is broadened in the $\Pi$ direction due to phase kickback (which we will treat in Subsection~\ref{sec:qpk}). In plot (d), the variance in the $\Phi$ direction is increased due to uncertainty in $\Phi$ of the two registers before the addition.
} \label{fig:CV_add}
\end{center}
\end{figure}

An illustration of addition of two continuous registers is provided in Fig.~\ref{fig:CV_add}.
The figure consists of the plots of the Wigner functions before and after the first version of addition for two registers initialized to Gaussian states.
Here we will only employ Wigner functions for purposes of illustration, but for completeness, we have defined the Wigner function of a continuous register in state $\hat{\rho}$ to be:
\begin{equation}
  W_c(x,p) := \int \frac{dx' dp'}{2\pi} \tr[\hat{D}_c^\dagger(x',p') \hat{\rho}] e^{-i(x'p + p'x + p'x'/2)},
\end{equation}
where
\begin{equation}
  \hat{D}_c(x,p) := \frac{1}{\sqrt{2\pi}} e^{-ip \hat{\Phi}} e^{-ix \hat{\Pi}}.
\end{equation}

One of the simplest physical systems with a quantum real number as its degree of freedom is the quantum harmonic oscillator, defined by a Hamiltonian:
\begin{equation}
  \hat{H} = \frac{1}{2m} \hat{\Pi}^2 + \frac{m}{2} \omega^2 (\hat{\Phi} - \Phi_0)^2,
\end{equation}
where $m$ and $\omega$ are the oscillator mass and frequency (respectively), and $\Phi_0$ simply sets the location of the minimum of the potential term.
A collection of continuous quantum variables will often be arranged as a vector of operators, $\bm{\hat{\Phi}} := ( \hat{\Phi}_n )_{n=1}^N$.
(Throughout, we will be using the notation $\bm{a} = (a_n)_n$ to denote a vector whose $n^\text{th}$ component is $a_n$.)
A set of coupled quantum harmonic oscillators, with degrees of freedom $\bm{\hat{\Phi}}$, in the simplest case (of equal masses) are associated with a Hamiltonian
\begin{equation}
  H = \frac{1}{2m} \bm{\hat{\Pi}}^T \bm{\hat{\Pi}} + \frac{1}{2} (\bm{\hat{\Phi}} - \bm{\Phi}_0)^T \bm{K} (\bm{\hat{\Phi}} - \bm{\Phi}_0)
\end{equation}
where $\bm{K}$ is a positive-definite matrix which encodes the couplings.
It is a basic fact of such a system that its ground state, $\ket{0}$, is a Gaussian wavefunction when represented in the joint eigenbasis of $\bm{\hat{\Phi}}$:
\begin{equation}\label{eq:ground}
  \langle \bm{\Phi} | 0 \rangle = [\det(m\bm{W}/\pi)]^{1/4} e^{-\frac{m}{2} (\bm{\Phi}-\bm{\Phi}_0)^T \bm{W} (\bm{\Phi}-\bm{\Phi}_0)},
\end{equation}
where $\bm{W} := \sqrt{\tfrac{1}{m}\bm{K}}$ (recall, $\bm{K}$ is positive-definite).

\subsection{Discrete Simulation of Continuous Quantum Registers}

If one has access to a discrete system, for example, a collection of qubits on a quantum computer, then it is possible to approximate the behavior of a continuous register.
A register which stores a qudit of dimension $d$ is defined by the operator
\begin{equation}
  \hat{J}_d := \sum_{j=0}^{d-1} j \ket{j}\!\bra{j},
\end{equation}
acting on the Hilbert space $\mathcal{H} = \mathds{C}^d$.
If one has access to $N$ qubits, it is possible to construct $\hat{J}_d$ for $d = 2^N$ as
\begin{equation}
\begin{split}
  \hat{J}_{2^N} &= \sum_{n=1}^N 2^{n-2} ( \hat{I}_2^{(n)} - \hat{Z}_2^{(n)} ) \\
  &= \frac{2^N-1}{2} - \sum_{n=1}^N 2^{n-2} \hat{Z}_2^{(n)},
\end{split}
\end{equation}
where $\hat{I}_2^{(n)}$ and $\hat{Z}_2^{(n)}$ are the identity and the Pauli-Z operator (respectively) for the $n^\text{th}$ qubit.
Note that in the above equation, and throughout this paper, constant terms added to operators should be treated as proportional to the identity.

The operator $\hat{J}_d$ can be used to simulate a continuous operator $\hat{\Phi}$ on a finite interval, $[a,b] \subset \mathds{R}$, by identifying the eigenvalues of $\hat{J}_d$ with discrete samples on the interval.
Then one can write the simulated continuous variable on the interval $[a,b]$ as:
\begin{equation}\label{eq:qudit}
  \hat{\Phi}_d := \frac{(b-a)}{(d-1)} \hat{J}_d + a \hat{I}_d,
\end{equation}
where $\hat{I}_d$ is the identity operator for the qudit.
One means of defining a momentum operator is as the generator of shifts in the value of the continuous or discrete register.
Such an operator can be written as the Fourier transform (continuous or discrete, respectively) of the observable corresponding to the value of the register.
For a continuous register storing a quantum real number, the Fourier transform is:
\begin{equation}
  \hat{F}_c \ket{x} := \int_{\mathds{R}} \frac{dy}{\sqrt{2\pi}} e^{-ixy} \ket{y}.
\end{equation}
The momentum operator from the previous section can thus alternatively be defined as
\begin{equation}
  \hat{\Pi} = \hat{F}_c^\dagger \hat{\Phi} \hat{F}_c.
\end{equation}
From this definition, it is straightforward to show that $\hat{\Pi}$ generates shifts in the register: $e^{-i \alpha \hat{\Pi}} \ket{x} = \ket{x+\alpha}$, for any $\alpha \in \mathds{R}$.

In analogy with the continuum, one can define a discrete Fourier transform by
\begin{equation}
  \hat{F}_d \ket{j} := \frac{1}{\sqrt{d}} \sum_{k=0}^{d-1} \omega_d^{-jk} \ket{k},
\end{equation}
(where $\omega_d := e^{2\pi i/d}$), and an analogous discrete momentum operator by
\begin{equation}
  \hat{K}_d := \hat{F}_d^\dagger \hat{J}_d \hat{F}_d.
\end{equation}
It is easy to show that this operator also generates shifts in the eigenbasis of $\hat{J}_d$.
Explicitly, for some $a \in \mathds{Z}_d$,
\begin{equation}
  \omega_d^{-a \hat{K}_d} \ket{j} = \ket{(j+a) \mod d}.
\end{equation}

Although at times we will find it necessary to work with $\hat{J}_d$ and $\hat{K}_d$ directly, often it is more convenient to work with the exponentiated operators: $\hat{Z}_d := \omega_d^{-\hat{J}_d}$ and $\hat{X}_d := \omega_d^{-\hat{K}_d}$.
Notice that we could also write $\hat{X}_d = \hat{F}_d^\dagger \hat{Z}_d \hat{F}_d$.
These are the Heisenberg-Weyl operators which satisfy the relation, $\hat{Z}_d \hat{X}_d = \omega_d^{-1} \hat{X}_d \hat{Z}_d$, in analogy with the displacement operators used in the Weyl relations for continuous systems.

Simulation of a continuous momentum operator on the interval $[a,b]$ using a qudit can be achieved with:
\begin{equation}
  \hat{\Pi}_d := \frac{(d-1)}{(b-a)} \hat{K}_d.
\end{equation}
Note that this is not simply the discrete Fourier transform of $\hat{\Phi}_d$, since in the continuum we should have a Schr\"odinger representation of $\hat{\Pi}$ as $-i \partial / \partial \Phi$.
If $\hat{\Phi}$ has units of length, then $\hat{\Pi}$ should have units of inverse length, therefore the scaling of the two operators should be different.
Also, we do not have a constant offset for the momentum operator so that it is centered at zero momentum.

Now let us examine the exponentiated versions of these simulated position and momentum operators.
First, it will be useful to derive an expression for $\hat{Z}_d^\alpha$ and $\hat{X}_d^\alpha$ for arbitrary $\alpha \in \mathds{R}$ (note that the case where $\alpha \in \mathds{Z}_d$ is straightforward).
Because the map $z \mapsto z^\alpha$ is locally analytic, we can use the (Riesz) functional calculus to define $\hat{Z}_d^\alpha$ as
\begin{equation}
  \hat{Z}_d^\alpha := \sum_{j=0}^{d-1} \omega_d^{-\alpha j} \ket{j}\!\bra{j},
\end{equation}
and continue to write $\hat{X}_d^\alpha := \hat{F}_d^\dagger \hat{Z}_d^\alpha \hat{F}_d$.
Note that although we can write $\hat{Z}_d = \sum_{j \in \mathds{Z}_d} \omega_d^{-j} \ket{j}\!\bra{j}$, with a sum over $\mathds{Z}_d$, the sum for $\hat{Z}_d^\alpha$ is taken over the set $\{0, \dots, d-1\}$.
This is because, for arbitrary $\alpha \in \mathds{R}$, in order to determine $(\omega_d^{-j})^\alpha \equiv e^{\alpha \log(\omega_d^{-j})}$, we must choose a branch of the complex logarithm, which breaks the periodic structure of the phases, i.e., $\omega_d^{-\alpha(j+d)} \neq \omega_d^{-\alpha j}$ for $\alpha \not\in \mathds{Z}_d$.

There is an alternative form of the operator $\hat{Z}_d^\alpha$ which we will occasionally find convenient.
Due the local analyticity of the map $z \mapsto z^\alpha$, one can in principle write a local power series for $\hat{Z}_d^\alpha$ in terms of integer powers of $\hat{Z}_d$.
However, since we also have that $\hat{Z}_d^d = \hat{I}_d$, this series will collapse in order to give a representation of $\hat{Z}_d^\alpha$ in terms of a superposition of finite powers, $\hat{Z}_d^k$, with $k \in \{0,\dots,d-1\}$.
To obtain such a form explicitly, let us first define the following orthonormal basis for operators on $\mathds{C}^d$:
\begin{equation}
  \hat{D}_d(q,p) := \frac{1}{\sqrt{d}} \hat{Z}_d^p \hat{X}_d^q,
\end{equation}
where $q,p \in \mathds{Z}_d$.
These operators are orthonormal with respect to the Hilbert-Schmidt inner product, $\langle A, B \rangle := \tr( A^\dagger B )$.
To demonstrate that they are a basis, we note that
\begin{equation}
  \ket{j}\!\bra{k} = \frac{1}{\sqrt{d}} \sum_{p \in \mathds{Z}_d} \omega_d^{jp} \hat{D}_d(p,j-k),
\end{equation}
i.e., they can be used to represent an arbitrary matrix element of an operator acting on $\mathds{C}^d$.
Using this relation, it is simple to show that, for arbitrary $\alpha \in \mathds{R}$, one can write
\begin{eqnarray}\label{eq:QPE_filter}
  \hat{Z}_d^\alpha &=& \sum_{k \in \mathds{Z}_d} \Delta(\alpha - k) \hat{Z}_d^k, \\
  \hat{X}_d^\alpha &=& \sum_{k \in \mathds{Z}_d} \Delta(\alpha - k) \hat{X}_d^k,
\end{eqnarray}
where
\begin{equation}
\begin{split}
  \Delta(\gamma) &:= \frac{1}{d} \sum_{j=0}^{d-1} \omega_d^{\gamma j} \\
  &= \frac{1}{d} \omega_d^{(d-1)\gamma/2} \frac{\sin(\pi \gamma)}{\sin(\pi \gamma / d)}.
\end{split}
\end{equation}
Hence, $\hat{Z}_d^\alpha$ and $\hat{X}_d^\alpha$ can be represented as a superposition of phases and displacements (respectively).

Notice that $\Delta(\gamma)$ is peaked around $\gamma = 0$.
Therefore, for example, if we apply $\hat{X}_d^\alpha$ to a computational basis state, $\ket{j}$, then the value of the register is shifted to a digital approximation of $\alpha$, with errors if $\alpha \not\in \mathds{Z}_d$.
(We demonstrate this fact explicitly in Fig.~\ref{fig:QPE_diag} below in the context of phase estimation.)

With this in hand, let us consider the exponentiated versions of our simulated continuous operators $\hat{\Phi}_d$ and $\hat{\Pi}_d$.
Using the above, one can write
\begin{eqnarray}
  \omega_d^{-\alpha \hat{\Pi}_d} &= \sum_{k \in \mathds{Z}_d} \Delta \left( \alpha \left( \frac{d-1}{b-a} \right) - k \right) \hat{X}_d^k,\\
  \omega_d^{-\beta \hat{\Phi}_d} &= \sum_{k \in \mathds{Z}_d} \Delta \left( \beta \left( \frac{b-a}{d-1} \right) - k \right) \hat{Z}_d^k.
\end{eqnarray}
Consider, for example, beginning with the state $\ket{j}$.
In the simulated interval, this corresponds to an eigenvector of the simulated position operator, $\hat{\Phi}_d$, with eigenvalue $\left( \frac{b-a}{d-1} \right) j + a \in [a,b]$.
Now suppose we apply a simulated continuous displacement, $\omega_d^{-\alpha \hat{\Pi}_d}$, to this state.
Then we arrive at the state
\begin{equation}
  \omega_d^{-\alpha \hat{\Pi}_d} \ket{j} = \sum_{k \in \mathds{Z}_d} \Delta \left( \alpha \left( \frac{d-1}{b-a} \right) - k  \right) \ket{j+k}.
\end{equation}
This state is peaked around a digital approximation to the value $j + k \sim j + \alpha \left( \frac{d-1}{b-a} \right)$.
Thus, it is approximately an eigenvalue of the simulated position operator $\hat{\Phi}_d$ with eigenvalue $\left( \frac{b-a}{d-1} \right) j + a + \alpha$.
That is, the value of $\hat{\Phi}_d$ has been shifted by approximately $\alpha$, hence the operator $\omega_d^{-\alpha \hat{\Pi}_d}$ is a simulation of a displacement of the eigenstates of $\hat{\Phi}_d$.

Of course, the error in approximation of the shift in $\alpha$ is due to the fact that one can only get a certain precision for a finite $d$.
Thus, for a given $d$, ideally we would like the shift of the discrete register to be the closest integer $k$ to $\alpha \left( \frac{d-1}{b-a} \right)$.
However, here we have a probabilistic distribution over different integer values, so we will not obtain the closest integer approximation with certainty.
If using qubits, one technique for suppressing the probability of error is to use more qubits than the desired precision and ignore these extra qubits when performing a measurement or subsequent computations. 
It is a standard result (see, for example, Ref.~\cite{nielsen2002quantum}) that in order to obtain an approximation accurate to $n$ bits of precision with probability $1-\epsilon$, one must use at least $N = n + \lceil \log(2+1/2\epsilon) \rceil$ qubits for the simulation.

Other issues which can arise during a computation with the simulated continuous operators are \emph{overflow} and \emph{underflow} errors.
An overflow error is simply a shift, $\alpha$, of the register which goes beyond the range $[a,b]$.
An underflow error is a shift that is too small to be resolved by the discretization scale.

So far in this section, we have seen that it is possible to simulate the kinematic structure of a continuous quantum system using a sufficiently large digital system.
Insofar as the dynamics is concerned, in Ref.~\cite{somma2015quantum}, it was shown that it is possible to simulate the dynamics of a quantum harmonic oscillator using a Hamiltonian
\begin{equation}
  \hat{H}_d = \tfrac12 \hat{p}_d^2 + \tfrac12 \hat{x}_d^2,
\end{equation}
where in our notation these operators can be written
\begin{equation}
\begin{split}
  \hat{x}_d &= \sqrt{ \frac{2\pi}{d} } \left[ \left( \frac{d-1}{b-a} \right) ( \hat{\Phi}_d - a ) - \frac{d}{2} \right],\\
  \hat{p}_d &= - \sqrt{ \frac{2\pi}{d} } \left[ \left( \frac{b-a}{d-1} \right) \hat{\Pi}_d + \frac{d}{2} \right].
\end{split}
\end{equation}

\subsubsection{Quantum Phase Estimation}\label{sec:QPE}

The final tool we will review in this subsection is the phase estimation algorithm, which can be seen as a hybrid continuous-discrete variable operation.
As above, let $\hat{\Phi}$ and $\hat{\Pi}$ be the continuous variable and its conjugate momentum, and let $\hat{\Phi}_d$ and $\hat{\Pi}_d$ be corresponding simulated continuous operators.
The phase estimation algorithm is a von Neumann measurement of the continuous variable by the simulated discrete variable, and can be summarized by the operator:
\begin{equation}
  \omega_d^{- \hat{\Phi} \hat{\Pi}_d}.
\end{equation}
We see that this is a straightforward extension of the continuous shifts of the discrete registers from before, but now the magnitude of the shift is controlled by a continuous quantum register.
Often this algorithm is implemented by taking the discrete Fourier transforms out of the exponential and applying a controlled phase operator: $\omega_d^{- \hat{\Phi} \hat{\Pi}_d} = \hat{F}_d^\dagger \omega_d^{- \left( \frac{d-1}{b-a} \right)^2 \hat{\Phi} (\hat{\Phi}_d-a)} \hat{F}_d$.
This is due to the fact that it is straightforward to construct $\hat{\Phi}_d$ from Pauli-Z qubit operators for $d = 2^N$, and then the controlled-phase gate breaks into a product of 2-local controlled-phase gates:
\begin{equation}
  \omega_{2^N}^{- \hat{\Phi} \hat{\Pi}_{2^N}} = \omega_{2^N}^{- \frac{(2^N-1)^2}{2(b-a)} \hat{\Phi}} \hat{F}_{2^N}^\dagger \prod_{n=1}^N \omega_{2^N}^{ \left( \frac{2^N-1}{b-a} \right) 2^{n-2} \hat{\Phi} \hat{Z}_2^{(n)}} \hat{F}_{2^N}.
\end{equation}

Note that throughout this paper, all products of operators will be ordered as:
\begin{equation}
  \prod_{n=1}^N \hat{U}_n := \hat{U}_N \cdots \hat{U}_2 \hat{U}_1.
\end{equation}

\begin{figure}[h!]
 \begin{center}
\includegraphics[width=0.95\columnwidth]{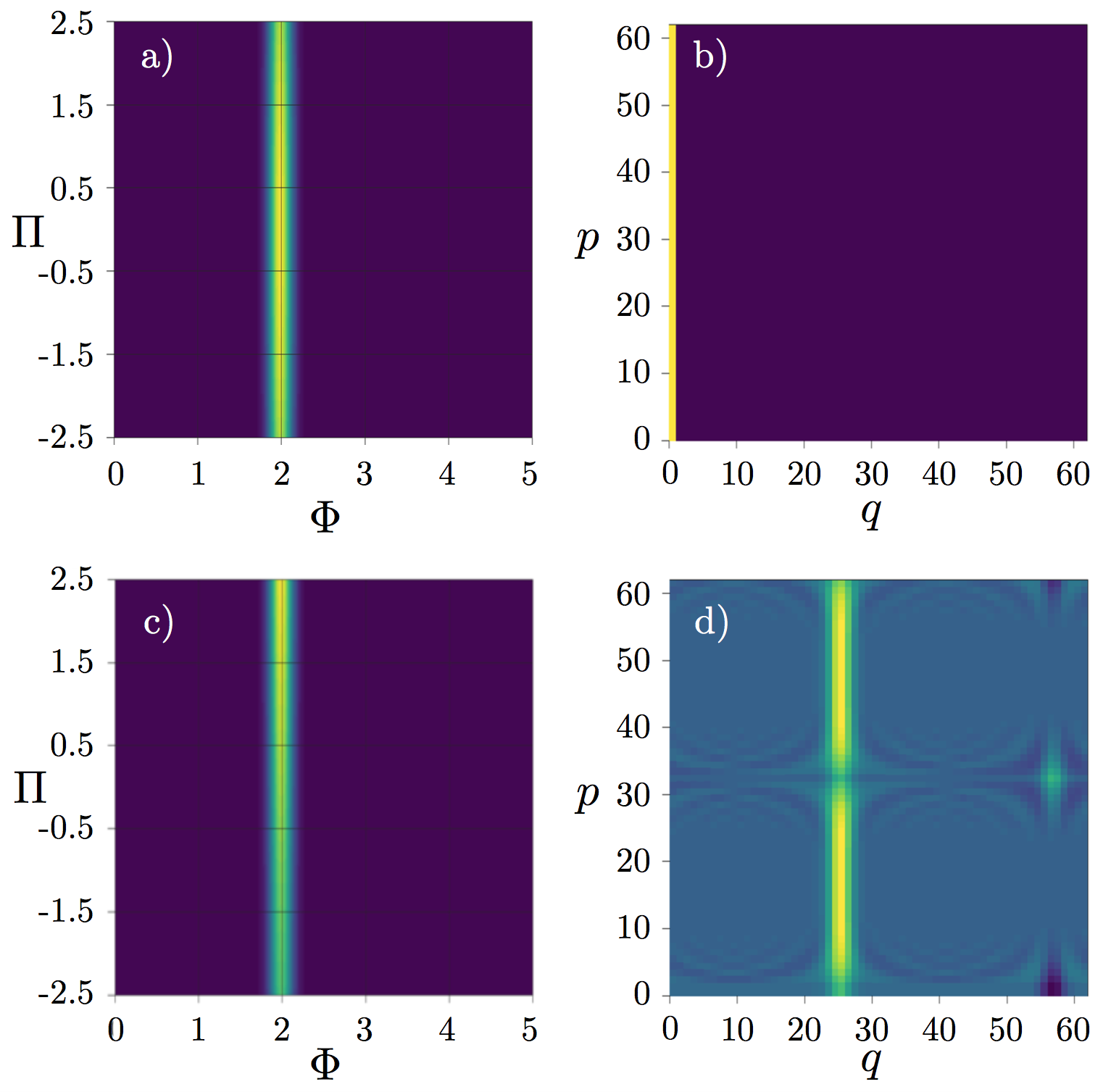}
\caption{Phase Estimation. In plots (a) and (b), we have the Wigner functions of the initial states of both the continuous and discrete registers (respectively). The discrete register was chosen to be a qudit of dimension $d = 63$. The continuous register is initialized to a highly squeezed Gaussian pointer state with $\Phi$ centered at 2. The discrete register is initialized to a null state. Plot (c) demonstrates the phase kickback on the momentum of the continuous register (discussed in the next subsection), while plot (d) is the result of the von Neumann measurement. Here we chose $a=0$ and $b=5$, so that the discrete-continuous conversion factor is $(d-1)/(b-a) = 12.4$. Hence, for a von Neumann measurement of a value of $\Phi = 2$, the discrete register is centered at $q = 25$, which is the closest integer approximation to $2 \times (d-1)/(b-a) = 24.8$.
} \label{fig:QPE_diag}
\end{center}
\end{figure}

An illustration of the phase estimation algorithm is provided in Fig.~\ref{fig:QPE_diag}.
The plots are of the Wigner functions of the continuous and discrete registers before and after the algorithm.
The continuous Wigner function is defined as in the previous subsection.
The definition of the discrete Wigner function used here (for odd $d$) is:
\begin{equation}
  W_d(q,p) := \frac{1}{d} \sum_{q',p' \in \mathds{Z}_d} \tr[ \hat{D}_d^\dagger(q',p') \hat{\rho} ] \omega_d^{-(q'p + p'q + 2^{-1} p'q')},
\end{equation}
where $2^{-1} := (d+1)/2$ is the multiplicative inverse of $2$ in $\mathds{Z}_d$ for odd $d$ (cf., Ref.~\cite{gross2006hudson}).

More generally, the shift could be controlled by an arbitrary observable, $\hat{A}$, from any type of register.
For example, this observable could be a Hamiltonian.
If this register is in an eigenstate of the observable, then the phase estimation algorithm provides a digital approximation to the eigenvalue of that observable.
Explicitly, suppose $\hat{A} \ket{\alpha} = \alpha \ket{\alpha}$, then
\begin{equation}
  \omega_d^{- \hat{A} \hat{\Pi}_d } \ket{\alpha,0} = \ket{\alpha} \otimes \sum_{k \in \mathds{Z}_d} \Delta \left( \alpha \left( \frac{d-1}{b-a} \right) - k  \right) \ket{k},
\end{equation}
where $a, b$ should be chosen so that the interval $[a,b]$ contains the spectrum of $\hat{A}$ (or at least the desired eigenvalue $\alpha$).

\subsection{Quantum Phase Kickback}\label{sec:qpk}

\textit{To every action, there is an equal and opposite reaction.}
Although this archaic dogma is no longer a central tenet of modern physics, remnants of Newton's third law make occasional appearances, particularly in quantum computing.
In more general terms, this law makes the point that generically a coupling between two physical systems causes them to react to one another.
Concretely this is due to the coupling inducing generalized forces appearing in the equations of motion of \emph{both} systems.
In quantum mechanics, these coupling terms appear in the generators of the unitary operators which evolve the interacting systems, although the effect on one or both systems may not be apparent.

For example, consider the case of a controlled-displacement on the position basis of two continuous registers,
\begin{equation}
  e^{-i \hat{\Phi}_\textsc{c} \hat{\Pi}_\textsc{t}} : \ket{\Phi_\textsc{c}, \Phi_\textsc{t}} \mapsto \ket{\Phi_\textsc{c}, \Phi_\textsc{t} + \Phi_\textsc{c}}.
\end{equation}
It would seem that the operation had no effect on the control register.
However, we can see this is not so by examining the action on the momentum basis,
\begin{equation}
  e^{-i \hat{\Phi}_\textsc{c} \hat{\Pi}_\textsc{t}} : \ket{\Pi_\textsc{c}, \Pi_\textsc{t}} \mapsto \ket{\Pi_\textsc{c}-\Pi_\textsc{t}, \Pi_\textsc{t}}.
\end{equation}
Hence, the control register is only left unchanged if in a position eigenstate, and the target register is only left unchanged if in a momentum eigenstate.

Often, this back-action is more easily visualized in the Heisenberg picture.
We see that under this controlled-displacement, the operators of the two registers evolve as:
\spliteq{
  &\Ad[ e^{i \hat{\Phi}_\textsc{c} \hat{\Pi}_\textsc{t}} ] ( \hat{\Phi}_\textsc{c} ) = \hat{\Phi}_\textsc{c} \\
  &\Ad[ e^{i \hat{\Phi}_\textsc{c} \hat{\Pi}_\textsc{t}} ] ( \hat{\Pi}_\textsc{c} ) = \hat{\Pi}_\textsc{c} - \hat{\Pi}_\textsc{t} \\
  &\Ad[ e^{i \hat{\Phi}_\textsc{c} \hat{\Pi}_\textsc{t}} ] ( \hat{\Phi}_\textsc{t} ) = \hat{\Phi}_\textsc{t} + \hat{\Phi}_\textsc{c} \\
  &\Ad[ e^{i \hat{\Phi}_\textsc{c} \hat{\Pi}_\textsc{t}} ] ( \hat{\Pi}_\textsc{t} ) = \hat{\Pi}_\textsc{t}
}
where $\Ad[\hat{U}](\hat{A}) := \hat{U} \hat{A} \hat{U}^\dagger$.

An analogous effect plays a prominent role in quantum computing, where it goes by the name of phase kickback.
Suppose we consider a controlled-NOT gate,
\begin{equation}
  \hat{C}_{NOT} := \ket{0}\!\bra{0}_\textsc{c} \otimes \hat{I}_\textsc{t} + \ket{1}\!\bra{1}_\textsc{c} \otimes \hat{X}_\textsc{t},
\end{equation}
acting on two qubits.
In the $Z$-basis, this acts as
\begin{equation}
  \hat{C}_{NOT} : \ket{z_\textsc{c},z_\textsc{t}} \mapsto \ket{z_\textsc{c}, z_\textsc{t} \oplus z_\textsc{c}},
\end{equation}
and in the $X$-basis,
\begin{equation}
  \hat{C}_{NOT} : \ket{x_\textsc{c},x_\textsc{t}} \mapsto \ket{x_\textsc{c} \oplus x_\textsc{t}, x_\textsc{t}},
\end{equation}
where the $X$ eigenstates are identified as $\ket{x=0} = \ket{+}$ and $\ket{x=1} = \ket{-}$.
In the Heisenberg picture, this looks like
\spliteq{
  &\Ad [\hat{C}_{NOT}] (\hat{Z}_\textsc{c}) = \hat{Z}_\textsc{c} \\
  &\Ad [\hat{C}_{NOT}] (\hat{X}_\textsc{c}) = \hat{X}_\textsc{c} \otimes \hat{X}_\textsc{t} \\
  &\Ad [\hat{C}_{NOT}] (\hat{Z}_\textsc{t}) = \hat{Z}_\textsc{c} \otimes \hat{Z}_\textsc{t} \\
  &\Ad [\hat{C}_{NOT}] (\hat{X}_\textsc{t}) = \hat{X}_\textsc{t}.
}
We see that the operation not only affects the target qubit, but also the control qubit.
More precisely, the operation changes the computational value of the target qubit ($z_\textsc{t}$) and the phase of the control qubit ($x_\textsc{c}$), hence the name \emph{phase kickback}.

The point here is that phase kickback may seem like odd quantum behavior, since it is not intuitive to think of a controlled operation affecting the state of the control register.
However, it is simply the fact that we are including the physics of the control register in our model, and that the back-action on the control register is simply due to a remnant of Newton's third law within quantum mechanics.
In particular, we are keeping track of the effects these operations have in the conjugate of the computational basis.
This back-action could also be seen with classical bit strings, however one does not typically consider the ``momentum'' of bit strings in classical computing.

However, that is not to say that we are only doing classical physics.
In the realm of quantum mechanics, these phases can interfere with each other to produce highly non-classical phenomena.
This effect is used throughout quantum computing, e.g., in Shor's algorithm.
In this paper, we will use controlled-unitaries of the form,
\begin{equation}
  \hat{U}(\hat{{\bm{\Phi}}}) := \sum_{\bm{\Phi}} \ket{{\bm{\Phi}}}\!\bra{{\bm{\Phi}}} \otimes \hat{U}({\bm{\Phi}}),
\end{equation}
which are generally more sophisticated than the controlled-displacement and controlled-NOT gates just described, but the phase kickback behaves similarly.
We will be using this kickback in the following to train machine learning algorithms parametrized by quantum registers, $\bm{\hat{\Phi}}$.

\subsubsection{Quantum Gradients}

An application of phase kickback that we will use throughout the paper is for computing the gradient of an oracle for a function, $f$, whose input is a continuous or simulated continuous register. This phase estimation of gradients technique for a black box oracle was first pioneered by Jordan \cite{2005PhRvL..95e0501J} and later improved upon by Wiebe et al. \cite{2017arXiv171100465G}. Consider a von Neumann measurement of the output of this function, described abstractly by:
\begin{equation}
  e^{-i f(\hat{\Phi}_1) \hat{\Pi}_2} : \ket{\Phi_1,\Phi_2} \mapsto \ket{\Phi_1, \Phi_2 + f(\Phi_1)}.
\end{equation}
We can think of this operation as computing the function $f$ on the first register and storing the result in the second.
In the Heisenberg picture, one can view this operation as
\begin{equation}
  \Ad [ e^{i f(\hat{\Phi}_1) \hat{\Pi}_2} ] (\hat{\Phi}_2) = \hat{\Phi}_2 + f(\hat{\Phi}_1).
\end{equation}
The phase kickback appears as a shift in the momentum of the first register,
\begin{equation}
  \Ad [ e^{i f(\hat{\Phi}_1) \hat{\Pi}_2} ] (\hat{\Pi}_1) = \hat{\Pi}_1 - f'(\hat{\Phi}_1) \hat{\Pi}_2,
\end{equation}
where $f'$ is the derivative of $f$.
Thus we see that if the second register begins in a state of small uncertainty in $\hat{\Pi}_2$, the momentum of the first register is shifted proportional to the gradient of $f$.
Of course, if there is large uncertainty in either of the quadratures of the first register, then this shift will provide little information.
However, we see that it will possible to extract some information about the gradient by measuring the momentum $\hat{\Pi}_1$.
Below we will show how to make use of this observation to implement backpropagation in a variety of contexts.

\section{Quantum Parametric Optimization}\label{sec:opt}

This section will be devoted to explaining abstractly the training algorithm used to accomplish quantum parameter optimization for general parametrized quantum algorithms.
Then, Sections~\ref{sec:qnn} and \ref{sec:qdata_algs} will examine more concretely how these techniques can be used to train quantum-coherent neural networks as well as parametrized quantum algorithms for performing various quantum information tasks.
 
\subsection{Basic Principles}

\subsubsection{Quantum Feedforward and Baqprop}\label{sec:phase-kick}

Machine learning consists of the task of finding a suitable algorithm among a parametrized class of algorithms.
On a quantum computer, an algorithm consists of a unitary operator acting on a collection of registers.
Naturally, one can consider parametrizing a quantum algorithm (unitary operator) with some collection of parameters, ${\bm{\Phi}} = ( \Phi_n )_n$.
Abstractly, we can denote this algorithm as $\hat{U}({\bm{\Phi}})$.
For example, the algorithm may consist of a set of single qubit rotations along with controlled-NOT gates, and the parameters could be taken as the Euler angles parametrizing each rotation.

In such considerations, the algorithm is quantum but the parameters remain externally-controlled and classical.
Here, we will extend this by using parameters which are quantized.
To this end, we introduce registers to store the parameters in addition to those upon which the algorithm is performing its computation.
Let us denote the full Hilbert space as $\mathcal{H}_{\bm{\Phi}} \otimes \mathcal{H}_\textsc{c}$, where $\mathcal{H}_{\bm{\Phi}}$ is the parameter Hilbert space and $\mathcal{H}_\textsc{c}$ is the computational Hilbert space.
The combined unitary operator will be denoted
\begin{equation}
  \hat{U}(\hat{{\bm{\Phi}}}) := \sum_{\bm{\Phi}} \ket{{\bm{\Phi}}}\!\bra{{\bm{\Phi}}} \otimes \hat{U}({\bm{\Phi}}).
\end{equation}
Note that the sum over ${\bm{\Phi}}$ is only formal; we also include the case where this consists of integrals over continuous variables.
Every fixed set of parameters ${\bm{\Phi}}$ applies a parametrized algorithm $\hat{U}({\bm{\Phi}})$.
Allowing for quantum parameters allows us to apply a superposition of quantum algorithms.
Of course, when the state of the parameters is such that the uncertainty in ${\bm{\Phi}}$ is small then we recover the case of a quantum algorithm controlled by classical parameters.

Including the parameters as a part of the system under consideration will allow us to analyze the computation of the class of algorithms and the training of the parameters as a closed dynamical system.
Furthermore, the quantum mechanical nature of the parameters can allow for some advantages in training, as we will explore throughout this paper.

Note that although the parameters have been promoted to quantum operators, in this paper we will only consider seeking a classical value for the parameters for the purposes of inference at the end of the training.
In one of the optimization strategies we will present below (Quantum Dynamical Descent, Subsection~\ref{sec:qdd}), the end of the training will result in a wavefunction over the set of parameters.
At this stage, one could perform a measurement of the parameters or multiple measurements to obtain an expectation value, and use the result as the classical value.
The second optimization strategy is semi-classical (Momentum Measurement Gradient Descent, Subsection~\ref{sec:momgrad}), and the end of the training will directly result in a set of classical parameters to be used for inference.

Once the parametrized class of algorithms is fixed (i.e., the hypothesis space), next is to provide an appraisal of each algorithm according to some metric, and then search for a set of parameters which optimizes this metric.
In machine learning, this parameter search is guided by training data.
The basic element of training is to feed a single example input into the algorithm, and evaluate a \emph{loss function} at the output.
Typically, the gradient (with respect to the parameters) of the loss function for multiple training examples are combined to update the values of the parameters using some variant of gradient descent.
If the algorithm is comprised of many parametrized components, as in deep learning, then the gradients of the loss function at the output need to be propagated back to the relevant component in order perform this update.
In this section, we will explain the use of quantum phase kickback to obtain a gradient of the loss function for a single training example.
The following sections will elaborate upon various schemes for making use of these gradients, as well as combining the phase kicks for multiple training examples.

The remainder of this section will be used to describe, abstractly, the Quantum Feedforward and Baqprop (QFB) algorithm, which evaluates the gradient of the loss function for a single training example and stores it in the momenta of the parameter registers via an effective phase kick.
To this end, let us begin by denoting $\ket{\xi} \in \mathcal{H}_\textsc{c}$ as the input associated with a single training example to the quantum algorithm $\hat{U}(\bm{\hat{\Phi}})$.
For example, this state could denote the encoding of a classical number (or set of numbers) in a continuous or discrete quantum register.
However, this could be any state in $\mathcal{H}_\textsc{c}$ for a general quantum algorithm.
Further discussion of the structure of input states will be provided below for particular applications.
Let us also suppose the parameters are initialized in an arbitrary state, $\ket{\Psi_0} \in \mathcal{H}_{\bm{\Phi}}$, expressed in the parameter eigenbasis as $\ket{\Psi_0} = \sum_{\bm{\Phi}} \Psi_0(\bm{\Phi}) \ket{\bm{\Phi}}$.
The algorithm then acts on this joint initial state to produce a superposition of parametrized algorithms on the example input state $\ket{\xi}$, and yields
\begin{equation}
  \sum_{\bm{\Phi}} \Psi_0(\bm{\Phi}) \ket{\bm{\Phi}} \hat{U}(\bm{\Phi}) \ket{\xi}.
\end{equation}
This will be called the \emph{feedforward} step of the QFB algorithm.

The next step in a machine learning training algorithm is to evaluate the performance of the algorithm, using a loss function, based on the output for a particular input.
In this case, the loss function will be an operator, which will be denoted $\hat{L}$, which acts on the computational Hilbert space, $\mathcal{H}_\textsc{c}$ (and acts as the identity on the parameters).
After the feedforward step of the QFB training algorithm, we apply the exponential of the loss function as a phase gate,
\begin{equation}
  \hat{I} \otimes e^{-i \eta \hat{L}},
\end{equation}
where $\eta$ is the \emph{phase kicking rate}, which will influence the learning rate of the algorithm.
Methods for exponentiating various loss functions will be described below when discussing particular applications.
Finally, after evaluating the loss function, we transmit the effect of the phase gate back to the parameters of the algorithm by performing a \emph{backpropagation} step, consisting of the application of the inverse of the feedforward step, namely, $\hat{U}^\dagger(\bm{\hat{\Phi}})$.
This backward quantum propagation of phase errors will be referred to as \emph{Baqprop}.

In all, the quantum feedforward and Baqprop (QFB) circuit is
\begin{equation}
\begin{split}
  \hat{U}_{\textsc{QFB}} &:= \hat{U}(\hat{\bm{\Phi}})^\dagger e^{-i \eta \hat{L}} \hat{U}(\hat{\bm{\Phi}}) \\
  &= e^{-i \eta \hat{L}(\bm{\hat{\Phi}}) },
\end{split}
\end{equation}
where $\hat{L}(\bm{\hat{\Phi}}) := \hat{U}(\hat{\bm{\Phi}})^\dagger \hat{L} \hat{U}(\hat{\bm{\Phi}})$ can be seen as the loss function operator evolved under the feedforward unitary $\hat{U}(\bm{\hat{\Phi}})$.
Applied to the joint initial state of the parameters and the training example input state, $\ket{\Psi_0} \otimes \ket{\xi}$, we get
\begin{equation}
  \hat{U}_{\textsc{QFB}} \ket{\Psi_0} \ket{\xi} = \sum_{\bm{\Phi}} \Psi_0(\bm{\Phi}) \ket{\bm{\Phi}} e^{-i \eta \hat{L}(\bm{\Phi})} \ket{\xi}.
\end{equation}
We can view this output state as a superposition of ancilla-assisted phase gates on the parameters by decomposing the operator $\hat{L}(\bm{\Phi})$ (for fixed $\bm{\Phi}$) into its eigenbasis, which we will denote as $\hat{L}(\bm{\Phi}) \ket{\lambda_{\bm{\Phi}}} = \lambda_{\bm{\Phi}} \ket{\lambda_{\bm{\Phi}}}$.
Then if we decompose the input state in this basis, $\ket{\xi} = \sum_{\lambda_{\bm{\Phi}}} \xi(\lambda_{\bm{\Phi}}) \ket{\lambda_{\bm{\Phi}}}$, we have
\begin{equation}
  \hat{U}_{\textsc{QFB}} \ket{\Psi_0} \ket{\xi} = \sum_{\bm{\Phi},\lambda_{\bm{\Phi}}} e^{-i \eta \lambda_{\bm{\Phi}}} \Psi_0(\bm{\Phi}) \xi(\lambda_{\bm{\Phi}}) \ket{\bm{\Phi}} \ket{\lambda_{\bm{\Phi}}}.
\end{equation}
We see that the QFB algorithm acts as a nonlinear phase gate for the parameters in each branch of $\lambda_{\bm{\Phi}}$.
Notice that if $\ket{\xi}$ is an eigenstate of the operator $\hat{L}(\bm{\hat{\Phi}})$ (in the sense that $\hat{L}(\bm{\hat{\Phi}}) \ket{\xi} = \xi \ket{\xi}$ for all $\bm{\Phi}$), then the QFB algorithm acts as a pure phase kick.
In particular, we will show in Section~\ref{sec:cl_data} that this generally occurs when training neural networks for classical data learning on a quantum computer.

In the generic case, the QFB algorithm, $\hat{U}_{\textsc{QFB}}$, causes the parameter and computational registers to become entangled.
Since the purpose of the training data is to play an auxiliary role to guide the training of the parameters, let us focus solely on the effect the QFB algorithm has on the parameters.
Not only will the momenta be shifted, but the entanglement between the parameters and the computational registers will cause the parameter wavefunction to decohere, as can be seen from the channel:
\begin{equation}
\begin{split}
  \hat{\rho}_{\bm{\Phi}}(\eta) &= \tr_C \left[ e^{-i \eta \hat{L}(\bm{\hat{\Phi}})} \hat{\rho}_{\bm{\Phi}}(0) \otimes \ket{\xi}\!\bra{\xi} e^{i \eta \hat{L}(\bm{\hat{\Phi}})} \right] \\
  &= \sum_i \hat{A}_i(\bm{\hat{\Phi}}) \hat{\rho}_{\bm{\Phi}}(0) \hat{A}_i^\dagger(\bm{\hat{\Phi}}),
\end{split}
\end{equation}
where $\hat{\rho}_{\bm{\Phi}}(0) = \ket{\Psi_0}\!\bra{\Psi_0}$, $\hat{A}_i(\bm{\hat{\Phi}}) := \bra{i} e^{-i \eta \hat{L}(\bm{\hat{\Phi}})} \ket{\xi}$ are the Kraus operators for the channel, and $\{ \ket{i} \}_i$ is an arbitrary basis for $\mathcal{H}_C$.
We see that the decoherence can be interpreted as due to a noisy measurement of the parameters by the computational registers which causes them to become entangled.
Generically this will have the effect of causing phase decoherence in the parameter eigenbasis and increasing the uncertainty in the values of the parameters.
To minimize the effect of this decoherence, one must train the algorithm slowly, i.e., tune the learning rate $\eta$ to be sufficiently small.
Then, if we expand the above channel perturbatively in $\eta$, we see that it is unitary to first order:
\begin{equation}
\begin{split}
  \hat{\rho}_{\bm{\Phi}}(\eta) &= \tr_C \left[ (1 - i \eta \hat{L}(\bm{\hat{\Phi}}) + \dots ) \hat{\rho}_{\bm{\Phi}}(0) \otimes \ket{\xi}\!\bra{\xi} \right.\\
  & \qquad \qquad \qquad \left. \times (1 + i \eta \hat{L}(\bm{\hat{\Phi}}) + \dots ) \right] \\
  &= \hat{\rho}_{\bm{\Phi}}(0) - i \eta [ \mathcal{L}(\bm{\hat{\Phi}}), \hat{\rho}_{\bm{\Phi}}(0) ] + \mathcal{O}(\eta^2),
\end{split}
\end{equation}
with an effective Hamiltonian,
\begin{equation}\label{eq:eff_phase}
  \mathcal{L}(\bm{\hat{\Phi}}) := \bra{\xi} \hat{L}(\bm{\hat{\Phi}}) \ket{\xi}.
\end{equation}
Therefore, we see that insofar as the parameters are concerned, the QFB algorithm acts as an effective unitary phase gate $e^{-i \eta \mathcal{L}(\bm{\hat{\Phi}})}$ (to first order in $\eta$).
Any decoherence does not occur until higher orders in $\eta$.
For the convenience of notation, in the following we will use $e^{-i \eta \mathcal{L}(\bm{\hat{\Phi}})}$ to denote the effect of the QFB algorithm on the parameters, and it should be understood that it is valid only to first order in $\eta$.

Now let us examine how the momenta of the parameters are affected by this effective phase gate,
\begin{equation}\label{eq:momm_update}
\begin{split}
  \bm{\hat{\Pi}} \quad \mapsto \quad & e^{i \eta \mathcal{L}(\bm{\hat{\Phi}})} \bm{\hat{\Pi}} e^{-i \eta \mathcal{L}(\bm{\hat{\Phi}})} + \mathcal{O}(\eta^2) \\
  &= \bm{\hat{\Pi}} - \eta \frac{\partial \mathcal{L} (\bm{\hat{\Phi}})}{\partial \bm{\hat{\Phi}}} + \mathcal{O}(\eta^2).
\end{split}
\end{equation}
We see that to first order in $\eta$, the momenta are kicked according to the gradient of the loss function.
This gradient update can be interpreted as an effective force on the parameters in the direction of decreasing values of the loss function.
We will elaborate upon this analogy in the Section~\ref{sec:eff_forces}.

We leave as future work a more careful analysis of the open systems nature of the parameter-data interactions.
In particular, of interest would be to elaborate upon the decoherence at higher orders in $\eta$ and frame the problem in terms of repeated interactions between the parameters with multiple data points.

\subsubsection{Full-batch Effective Phase Kicks}

Above we considered effective phase kicks for an abstract state, $\ket{\xi}$ (which is associated with an input example on the computational space), and an abstract loss function $\hat{L}$.
Now let us examine this again with some more emphasis on the machine learning aspects of these phase kicks, without yet examining the details of particular applications.
Specifically, we will consider how multiple loss function phase kicks can be batched over a dataset in order to induce a \emph{cost} function as an effective phase.
Here we will only consider batching the full dataset, whereas later (Section~\ref{sec:batch}) we will discuss more refined techniques for combining kicks from multiple data points.
We illustrate the concept for input-output pairs of data which would occur in supervised learning, but as we will show in later sections it can extend to many other cases, including unsupervised scenarios and Hamiltonian optimization.

A classical dataset for a supervised learning problem consists of a collection of input/output pairs, $\{(\bm{x}_j,\bm{y}_j)\}_{j \in \mathcal{B}}$, which we will assume to be real vectors.
In this setting, we consider the computational Hilbert space, $\mathcal{H}_\textsc{c}$, to be partitioned into an input space as well as an auxiliary work space, so that (respectively) $\mathcal{H}_\textsc{c} = \mathcal{H}_\textsc{i} \otimes \mathcal{H}_\textsc{w}$.
(If one is training via superpositions of classical data points, it would be necessary to assign a Hilbert space for the outputs as well.)
Before the QFB procedure is applied for a single data point, $(\bm{x}_j,\bm{y}_j)$, the input state on $\mathcal{H}_\textsc{i}$ must be prepared in a computational basis state corresponding to the input, $\bm{x}_j$.
For an initially blank input register, $\ket{\bm{0}}_\textsc{i}$, we can apply the classically-controlled unitary $\hat{U}_\textsc{i}(\bm{x}_j) = e^{-i \bm{x}_j \cdot \bm{\hat{p}}_\textsc{i}} : \ket{\bm{0}}_\textsc{i} \mapsto \ket{\bm{x}_j}_\textsc{i}$.
Once this state is prepared, we apply the QFB algorithm as above to the combined parameter and computational spaces $\mathcal{H}_{\bm{\Phi}} \otimes \mathcal{H}_\textsc{c}$, with the parameters initialized to some state $\ket{\Psi_0} = \sum_{\bm{\Phi}} \Psi_0(\bm{\Phi}) \ket{\bm{\Phi}}$.
Because in a supervised learning problem the loss function will depend on the output data point, $\bm{y}_j$, the exponentiated loss function occurring after the feedforward operation will generally be a classically-controlled unitary, where the classical control registers are those which store the desired output, $\bm{y}_j$.
We will label the classically-controlled loss function as $\hat{L}(\bm{y}_j)$.
After the uncomputation step, we can also uncompute the state preparation by acting $\hat{U}_\textsc{i}^\dagger(\bm{x}_j) = e^{+i \bm{x}_j \cdot \bm{\hat{p}}_\textsc{i}}$.
It turns out that this will indeed uncompute the state preparation because, as we mentioned above, in the case of an embedded classical machine learning problem, the computational registers are left unchanged at the end of the QFB algorithm, so we get a perfect unitary phase kick of the loss function and hence the parameter and computational registers are left unentangled.
Again, the details of this fact will be provided in \ref{sec:cl_data}.

\begin{figure}[h!]
 \begin{center}
\includegraphics[width=0.95\columnwidth]{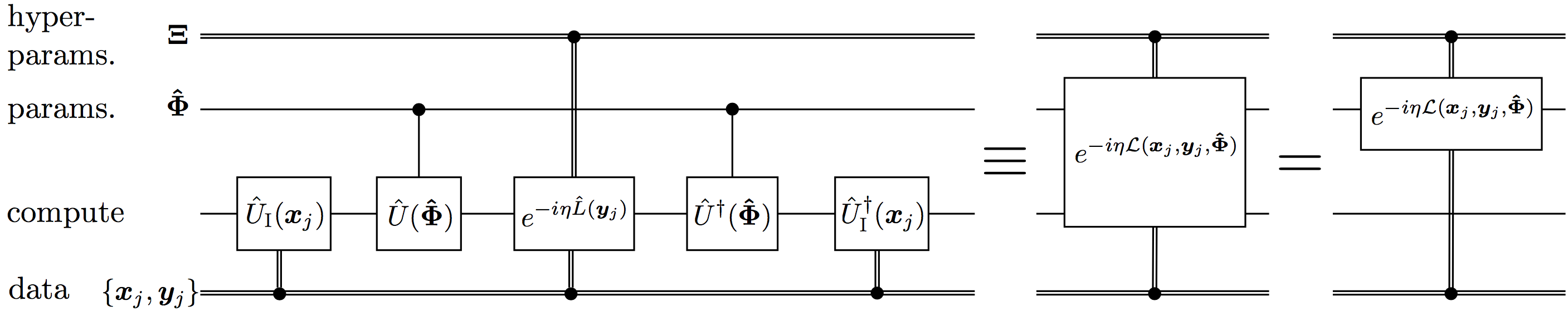}
\caption{Quantum Feedforward and Baqprop procedure for a backpropagating loss phase error of a single classical data point. The effective phase kick is an exact unitary on the parameter registers. Note that for this diagram and throughout this paper the \textit{rate} hyper-parameters will be labelled as $\bm{\Xi}$, in this case the \textit{phase kicking rate} $\eta\in\bm{\Xi}$ is considered a hyper-parameter.
} \label{fig:cdat_QFB}
\end{center}
\end{figure}

As a whole, this procedure applied onto the initial state $\ket{\Psi_0}_{\bm{\Phi}} \ket{\bm{0}}_\textsc{c}$ yields
\spliteq{
   \hat{U}^\dagger_\textsc{i}(\bm{x}_j) & \hat{U}^\dagger(\bm{\hat{\Phi}}) e^{-i\eta \hat{L}(\bm{y}_j)}  \hat{U}(\bm{\hat{\Phi}}) \hat{U}_\textsc{i}(\bm{x}_j) \ket{\Psi_0}_{\bm{\Phi}} \ket{\bm{0}}_\textsc{c}\\ 
   &= e^{-i\eta\mathcal{L}(\bm{x}_j,\bm{y}_j,\bm{\hat{\Phi}})} \ket{\Psi_0}_{\bm{\Phi}} \otimes \ket{\bm{0}}_\textsc{c},
}
where $\mathcal{L}(\bm{x}_j,\bm{y}_j,\bm{\hat{\Phi}}) := \hat{U}^\dagger_\textsc{i}(\bm{x}_j)  \hat{U}^\dagger(\bm{\hat{\Phi}}) \hat{L}(\bm{y}_j) \hat{U}(\bm{\hat{\Phi}}) \hat{U}_\textsc{i}(\bm{x}_j)$.
In Figure~\ref{fig:cdat_QFB} we represent this classical-data-induced phase kick for a single data point.
In the same figure, we represent pictorially how the effective phase kick amounts to an operation strictly on the parameters, using the computational registers effectively as an auxiliary space to assist the phase kick.

\onecolumngrid

\twocolumngrid

\begin{figure}[h!]
 \begin{center}
\includegraphics[width=1.0\columnwidth]{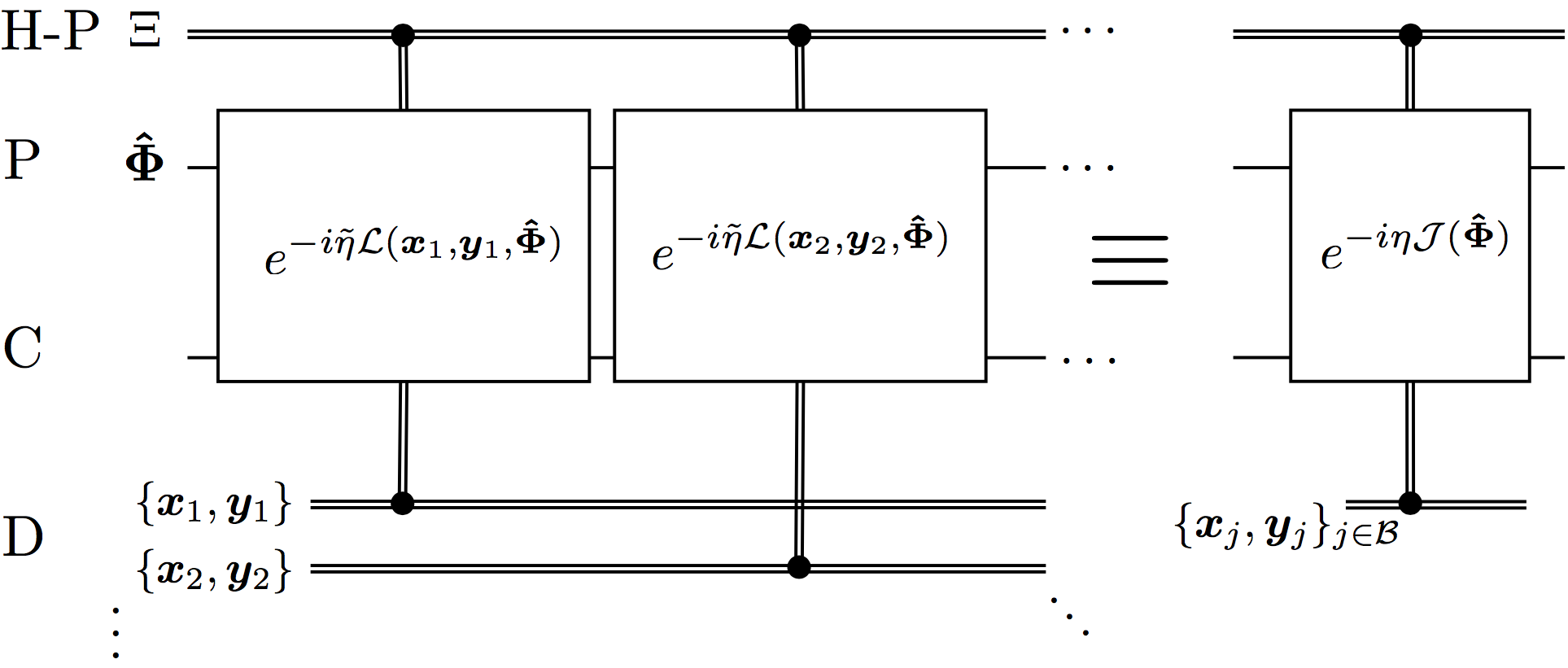}
\caption{Phase kick batching. By sequentially applying the Quantum Feedforward and Baqprop for multiple data points in the full batch $\mathcal{B}$, we can act an effective phase kick according to the cost function for the full data set by accumulating the loss exponentials. Here $\tilde{\eta}=\eta / |\mathcal{B}|$ is the phase kicking parameter normalized by the batch size. The registers labelled as H-P, P, C, and D, correspond to the hyper-parameters, the parameters, the compute, and the data registers respectively. Note that most diagrams in Sections~\ref{sec:opt} and \ref{sec:misc} will use classical data registers, but all protocols also work for quantum data kicking.} \label{fig:phase_batch}
\end{center}
\end{figure}

Repeating this procedure for subsequent data points is straightforward: first prepare the input state to a computational state representing the input data point $\bm{x}_{j+1}$, then apply the QFB algorithm with the classically-controlled loss function $\hat{L}(\bm{y}_{j+1})$, and finally uncompute the input state preparation.
As one proceeds through the dataset $\{ (\bm{x}_j,\bm{y}_j) \}_{j \in \mathcal{B}}$, the phase kicks on the parameter registers accumulate, resulting in a total phase kick,
\begin{equation}
  e^{-i \eta \mathcal{J}(\bm{\hat{\Phi}})} \ket{\Psi_0}_{\bm{\Phi}} \otimes \ket{\bm{0}}_\textsc{c},
\end{equation}
where we have defined the average \emph{cost} function for classical data,
\begin{equation}
  \mathcal{J}(\bm{\hat{\Phi}}) := \frac{1}{|\mathcal{B}|} \sum_{j \in \mathcal{B}} \mathcal{L}(\bm{x}_j,\bm{y}_j,\bm{\hat{\Phi}}).
\end{equation}
Here we have also redefined $\eta$ to be the total phase kicking rate of the batch, and $\tilde{\eta} := \eta / |\mathcal{B}|$ the phase kicking rate normalized by the batch size, which would appear in the exponentiated loss functions for the individual data points.
This accumulation of phase kicks for the dataset is illustrated in Figure~\ref{fig:phase_batch}.

For quantum data, the procedure is slightly different due to the fact that the state preparation at the input of the parametrized algorithm and the exponentiated loss function must be controlled from quantum data sources rather than classical.
Consider, for illustration, a case of supervised quantum data learning, where we are handed a set of input/output pairs of quantum states, $\{ (\hat{\rho}_j^{\textsc{i}},\hat{\rho}_j^{\textsc{o}}) \}_{j \in \mathcal{B}}$, which are states on respective Hilbert spaces $\mathcal{H}_{\textsc{i}_j} \otimes \mathcal{H}_{\textsc{o}_j}$.
We consider beginning with a blank computational register, $\ket{\bm{0}}_\textsc{c}$, and swap in an input state, $\hat{\rho}_j^\textsc{i}$, from the dataset using the input preparation unitary $\hat{U}_\textsc{i}(\hat{\rho}_j^\textsc{i})$.
Once again, we being with a certain state for the parameters, $\ket{\Psi_0} = \sum_{\bm{\Phi}} \Psi_0(\bm{\Phi}) \ket{\bm{\Phi}}$, and apply the parametric unitary  $\hat{U}(\bm{\hat{\Phi}})$ onto the compute and parameter Hilbert space jointly.
After this, a certain exponential of a loss operator dependent on the desired output state $\hat{L}_j(\hat{\rho}_j^{\textsc{o}})$ is applied.
In general, generating an exponentiated loss depending on the state $\hat{\rho}_j^{\textsc{o}}$ can consume multiple copies of $\hat{\rho}_j^{\textsc{o}}$.
We refer the reader to Section~\ref{sec:qdata_algs} for specific examples of quantum data learning problems, and to Section~\ref{sec:qse} for particular examples of state-dependent loss functions.
After the loss function is applied, the parametric unitary is uncomputed, and the quantum data preparation unitary is uncomputed in order to establish a fresh compute register for the next iteration.
The data registers $\mathcal{H}_j^{\textsc{i}} \otimes \mathcal{H}_j^{\textsc{o}}$ are also discarded after generating the effective phase kick.
In all, the algorithm schematically consists of the transformation,
\spliteq{
  \hat{U}^\dagger_{\textsc{i}}(\hat{\rho}_j^{\textsc{i}}) \hat{U}^\dagger(\bm{\hat{\Phi}}) e^{-i\eta \hat{L}_j(\hat{\rho}_j^{\textsc{o}})} & \hat{U}(\bm{\hat{\Phi}}) \hat{U}_\textsc{i}(\hat{\rho}_j^{\textsc{i}}) \ket{\Psi_0}_{\bm{\Phi}} \ket{\bm{0}}_\textsc{c} \\
  & \overset{\text{tr}_{\textsc{c}}}{\mapsto}  e^{-i\eta\mathcal{L}(\hat{\rho}_j^{\textsc{i}},\hat{\rho}_j^{\textsc{o}},\bm{\hat{\Phi}})}\ket{\Psi_0}_{\bm{\Phi}} + \mathcal{O}(\eta^2),
}
where
\begin{equation}
\begin{split}
  &e^{-i\eta\mathcal{L}(\hat{\rho}_j^{\textsc{i}},\hat{\rho}_j^{\textsc{o}},\bm{\hat{\Phi}})} :=\\
  &\qquad \bra{\bm{0}}_{\textsc{c}}\hat{U}^\dagger_{\textsc{i}}(\hat{\rho}_j^{\textsc{i}}) \hat{U}^\dagger(\bm{\hat{\Phi}})e^{-i\eta \hat{L}_j(\hat{\rho}_j^{\textsc{o}})}\hat{U}(\bm{\hat{\Phi}})\hat{U}_\textsc{i}(\hat{\rho}_j^{\textsc{i}})\ket{\bm{0}}_{\textsc{c}}.
\end{split}
\end{equation}

\pagebreak
\onecolumngrid

\begin{figure}[h!]
 \begin{center}
\includegraphics[width=0.95\columnwidth]{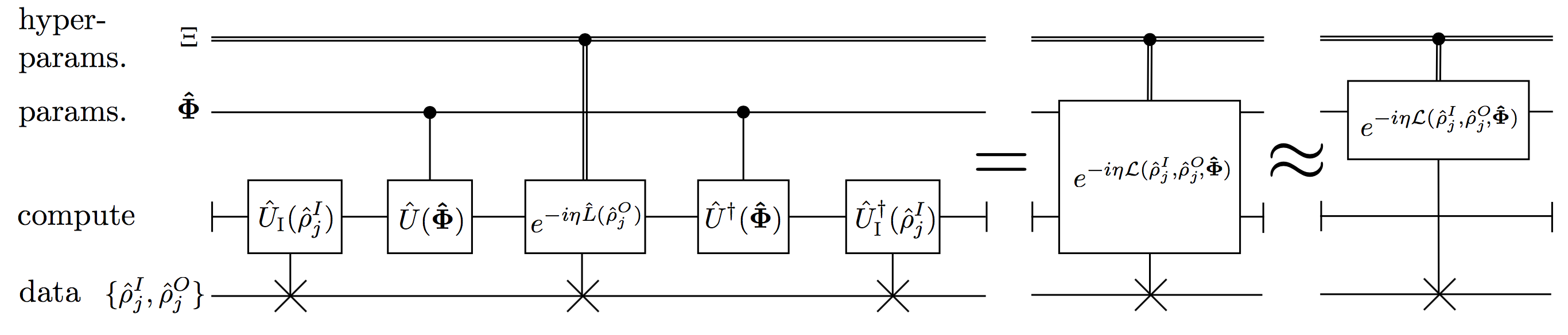}
\caption{Quantum Feedforward and Baqprop procedure for a backpropagating loss phase of a single quantum data point. The effective phase kick is effectively a unitary phase kick on the parameters to first order in $\eta$. We use a swap-control symbol for both the input and output, for the input the data is fully swapped onto the compute registers, whereas for application of an output-dependent phase kick the swap-control symbolizes the consumption of multiple copies from state exponentiation, which we treat in-depth in Subsection~\ref{sec:qse}.
} \label{fig:qdat_QFB}
\end{center}
\end{figure}

\twocolumngrid
This swapping out of the computational register is important for quantum data problems since the QFB procedure generally does not entirely disentangle the computational and parameter registers.
By swapping out and discarding the output of the computational register, we are tracing out this register, and as such, on average, to first order in $\eta$ the expected value of the phase kick is that of \eqref{eq:eff_phase}.
We present a quantum data QFB procedure pictorially in Figure~\ref{fig:qdat_QFB}.

With this procedure, it is again straightforward to accumulate phase kicks for multiple data points.
After applying the algorithm for the full data batch, $\{ (\hat{\rho}_j^{\textsc{i}},\hat{\rho}_j^{\textsc{o}}) \}_{j \in \mathcal{B}}$, we obtain a total effective phase kick,
\begin{equation}
  e^{-i \eta \mathcal{J}(\bm{\hat{\Phi}})} \ket{\Psi_0}_{\bm{\Phi}} \otimes \ket{\bm{0}}_\textsc{c},
\end{equation}
where we have defined the average effective cost function for quantum data as
\begin{equation}
  \mathcal{J}(\bm{\hat{\Phi}}) := \frac{1}{|\mathcal{B}|} \sum_{j \in \mathcal{B}} \mathcal{L}(\hat{\rho}_j^\textsc{i},\hat{\rho}_j^\textsc{o},\bm{\hat{\Phi}}),
\end{equation}
and where $\eta$ is the total phase kicking rate.
Again, we emphasize that this phase kick is only valid to first order in $\eta$.

Now that we have an abstract method for obtaining the gradient of the (effective) cost function for the parametrized quantum algorithm, next is to examine methods for making use of these gradients to update the weights.

\subsubsection{Effective Forces}\label{sec:eff_forces}

In the previous section, we saw that for any parametrized quantum programs, we can consider using a set of quantum registers for the parameters.
We can initiate the parameters in a superposition of values, and by applying a forward computation, a loss function phase-kick, and an uncomputation, we can push the wavefunction in the momentum basis according to the gradient of the (effective) loss function.

Let us consider how these momentum kicks can be interpreted from the perspective of classical Hamiltonian dynamics.
Consider the Hamiltonian $H(\bm{q},\bm{p})$ function, where $\bm{q}$ is a position variable, and $\bm{p}$ its conjugate momentum.
In the simplest cases, the Hamiltonian is composed of a sum of two terms, the \textit{kinetic} term $T$ and the \textit{potential} term $V$.
In many physical scenarios, the kinetic term is strictly a function of the momentum, e.g., $T(\bm{p})= \tfrac{\bm{p} \cdot \bm{p}}{2m}$ (where $m$ is the \textit{mass} parameter), while the potential term, $V(\bm{q})$, is strictly a function of position.
Hence, we can write $H(\bm{q},\bm{p}) = T(\bm{p}) + V(\bm{q})$.
In canonical Hamiltonian mechanics, the change in the momentum per unit time is given by $\dot{\bm{p}} = -\partial_{\bm{q}} H$, hence $\dot{\bm{p}} = -\partial_{\bm{q}} V(\bm{q})$ for the case above.
For a sufficiently small unit of time, the change in the momentum vector is proportional to the negative gradient of the potential $\Delta \bm{p} = -\nabla_{\bm{q}} V(\bm{q}) \, \Delta t$.
We can compare this to the finite-difference version of Newton's second law, where the change in momentum per unit time is defined as a \textit{force}, $\Delta \bm{p} = \bm{F} \Delta t$, and we see the force is then equal to the negative gradient of the potential $\bm{F} = -\nabla_{\bm{q}} V(\bm{q})$.
Notice that in \eqref{eq:momm_update}, the change in momentum for a single data point is proportional to the \textit{kicking rate} $\eta$ times the negative gradient of the effective loss function $\mathcal{L}$.
After batching multiple data points, the momentum kicks accumulate to change the momentum according to the negative gradient of the total effective cost function,
\begin{equation}
  \bm{\hat{\Pi}} \mapsto \bm{\hat{\Pi}} - \eta \frac{\partial \mathcal{J}(\bm{\hat{\Phi}})}{\partial \bm{\hat{\Phi}}} + \mathcal{O}(\eta^2).
\end{equation}
Thus, in our physics analogy, the cost function acts as a potential for the weights, with parameters $\bm{\Phi}$ and conjugate momenta $\bm{\Pi}$ playing the respective roles of position and momentum, $\bm{q}$ and $\bm{p}$.

Therefore, we see that the momentum shifts induced by the batched quantum feedforward and Baqprop procedure (QFB) can be seen as analogous to applying a force onto the parameters, with $\eta$ playing the role of the time step parameter and $\mathcal{J}$ the potential.
Note that a momentum kick is not a step of gradient descent, just as a force-induced momentum update does not alone cause a particle to move.
For this, one must also take into account the remaining Hamilton equations, which in canonical scenarios give the change in the position per unit time as $\dot{\bm{q}} = \partial_{\bm{p}} H$.
In the case above, where $H(\bm{q},\bm{p}) = T(\bm{p}) + V(\bm{q})$, this is given by $\dot{\bm{q}} = \partial_{\bm{p}} T(\bm{p}) = \frac{1}{m} \bm{p}$.
For a small time step $\Delta t$, the position is updated according to $\Delta \bm{q} = \frac{1}{m} \bm{p} \Delta t$.
Therefore, any update in the momentum is channelled to the position via the kinetic term.
We do not yet have an analogue of this kinetic term for training quantum machine learning programs.
In the next section we will introduce such terms in order to \emph{use} the force induced by the batched QFB algorithm to update the values of the parameters.

\subsection{Quantum Dynamical Descent}\label{sec:qdd}

The idea behind Quantum Dynamical Descent (QDD) is to simulate Schrodinger time-evolution of the weights as they would evolve if they were particles moving under the influence of the potential induced by the cost function.
Dynamical time-evolution of a quantum mechanical system is governed by a Hamiltonian \textit{operator}, similar to classical Hamiltonian mechanics, but instead of a scalar function, we have an operator for the Hamiltonian.
For a time-independent Hamiltonian, $\hat{H}$, the time evolution operator is simply $\hat{U}(\tau) = \exp{(-i\hat{H}\tau)}$, while for a time-dependent Hamiltonian $\hat{H}(\tau)$, with $\tau$ as the time parameter, the Schrodinger equation dictates that the time evolution operator is a time-ordered exponential of the form $\hat{U}(\tau) = \textstyle\mathcal{T}\!\exp(-i\int_0^\tau\! d\tau' \hat{H}(\tau')) $. 

Now, in the previous subsection, we established that our cost function $\mathcal{J}$ was analogous to a potential term $V$ in the Hamiltonian, as derived from the momentum update rule.
In fact, since the cost function, $\mathcal{J}(\hat{\bm{\Phi}})$, is an operator on the Hilbert space of the parameters, it is more akin to an operator-valued potential term which would appear in the Hamiltonian operator $\hat{H} = \hat{T} + \hat{V}$.
Thus for QDD, we want to introduce a kinetic term along with the cost function potential in order to construct a Hamiltonian operator under which we can evolve the parameter wavefunction.
Let us consider a time-dependent Hamiltonian,
\begin{equation}\label{eq:ham}
  \hat{H}(\tau) = \tfrac{1}{2m(\tau)}\hat{\bm{\Pi}}^2 + \mathcal{J}(\hat{\bm{\Phi}})
\end{equation}
where $\tau$ is the time parameter and $\hat{\bm{\Pi}}^2 \equiv \hat{\bm{\Pi}}^T \hat{\bm{\Pi}}$.
Notice that we consider a standard kinetic term strictly dependent on the momentum, but with a time-dependent mass parameter.
This is a standard Hamiltonian for Schrodinger dynamics of a single particle in $N$ spatial dimensions (apart from the time-dependent mass), where $N$ is the number of parameters.
Using a time-varying mass is less standard, but it will allow us to control the rate of descent of the potential landscape as the descent proceeds.
The optimization of how to initialize this parameter and its rate of change will fall into the category of hyper-parameter optimization tasks.
We will examine a few approaches for the design of the mass function, one inspired by the Quantum Approximate Optimization Algorithm, and another by the Quantum Adiabatic Algorithm.

\subsubsection{Core Algorithm}

The Quantum Dynamical Descent (QDD) algorithm consists of applying alternating pulses as in the Trotterization of the time evolution generated by a time-dependent Hamiltonian of the form $\hat{H} = \hat{T} + \hat{V}$.
In other words, the algorithm consists of applying operations which mimic a finite-time-step quantum simulation of the Hamiltonian in \eqref{eq:ham} which has the cost function as the potential and a time-variable mass kinetic term. 

More explicitly, the time-ordered exponential generated by the Hamiltonian \eqref{eq:ham} is approximated with a product of single time-step exponentials.
To describe this mathematically, first partition the time interval of interest, $\mathcal{I} \subset \mathds{R}$, into sub-intervals $\mathcal{I}_k := [\tau_k,\tau_{k+1}] \subset \mathcal{I}$ with $\tau_{k-1} < \tau_k < \tau_{k+1}$ for all $k$ and $\mathcal{I} = \cup_k \mathcal{I}_k$.
The evolution associated with each sub-interval will be called an \emph{epoch}, since it corresponds to a gradient and parameter update for the full batch of data.
(Alternative batching schemes will be discussed in Subsection~\ref{sec:batch}.)
With this partitioning of the time interval, we can approximate the time-evolution operator by decomposing it into a product of time-evolution operators generated by averaged operators on each sub-interval:
\begin{equation}
U(\tau)  = \mathcal{T}\!\exp\left(-i\int_\mathcal{I}\! d\tau \hat{H}(\tau)\right)\approx \prod_{k} e^{-i\Delta \tau_k \hat{H}_k},
\end{equation}
where $\Delta \tau_k := \tau_{k+1} - \tau_k$ is the length of sub-interval $\mathcal{I}_k$ and $\hat{H}_k := \int_{\mathcal{I}_k}\!\mathrm{d}\tau \hat{H}(\tau)$ is the averaged Hamiltonian for the $k^\text{th}$ time interval.
Note that the above expression is approximate since the time-dependent mass prevents the Hamiltonian from commuting with itself at different times.
We proceed by using the Lie-Suzuki-Trotter formula to divide the exponential of the Hamiltonian in each sub-interval into that of the cost potential and kinetic terms, 
\begin{equation}
  U(\tau) \approx  \prod_{k} e^{-i\Delta \tau_k \hat{H}_k} \approx \prod_{k} e^{-i\Delta \tau_k \bm{\hat{\Pi}}^2/2m_k}e^{-i\Delta \tau_k \mathcal{J}(\bm{\hat{\Phi}})}
\end{equation}
where $m_k := (\int_{\mathcal{I}_k}\!\mathrm{d}\tau m^{-1}(\tau) )^{-1}$ is the inverse averaged inverse mass.

In a following subsubsection, we argue that a small time step (fine temporal partition) with a slowly decreasing mass parameter can yield the minimum of $\mathcal{J}(\bm{\Phi})$ through the adiabatic theorem.
In general, it is up to the discretion of the practitioner to determine an applicable time step and mass parameter schedule.
In classical machine learning, this process of finding the optimal initializations and optimal learning rates are a part of a process called hyper-parameter optimization.
Algorithms for hyper-parameter optimization are often called \textit{meta-learning} algorithms.
In Subsubsection~\ref{sec:meta}, we offer a set of quantum meta-learning algorithms for Quantum Dynamical Descent.
Since we will generally optimize the parameters for each pulse, we can write the unitary corresponding to the quantum dynamical descent as 
\begin{equation}\label{eq:qdd}
    \hat{U}_{\textsc{qdd}} = \prod_k e^{-i\gamma_k\bm{\hat{\Pi}}^2}e^{-i\eta_k\mathcal{J}(\bm{\hat{\Phi}})}
\end{equation}
where we call each parameter $\eta_k$ the \textit{phase kicking rate} and $\gamma_k$ the \textit{kinetic rate} for epoch $k$.
We will argue in Subsubsection~\ref{sec:adiab} that a heuristic one may wish to use for choosing the phase kicking rate and kinetic rate is to begin with $\gamma_k \gg \eta_k$ for small $k$ (early time), and for large $k$ (late time) shift towards $\gamma_k \ll \eta_k$.
Later we will discuss how beginning with a large kinetic rate and transitioning to a (relatively) larger phase kicking rate aids in converging to a local minimum.

Recall that the cost function for each epoch is the loss function for each data point averaged over the entire batch (dataset).
That is, for a batch set $\mathcal{B}$,
\begin{equation}
   \mathcal{J}(\bm{\hat{\Phi}}) = \frac{1}{|\mathcal{B}|} \sum_{j \in \mathcal{B}} \mathcal{L}_j(\bm{\hat{\Phi}}).
\end{equation}
Above we wrote the loss function in supervised classical learning for a data point $(\bm{x}_j,\bm{y}_j)$ as $\mathcal{L}(\bm{x}_j,\bm{y}_j,\bm{\hat{\Phi}})$.
Similarly, for supervised quantum data learning, the loss function was denoted $\mathcal{L}(\hat{\rho}_j^\textsc{i}, \hat{\rho}_j^\textsc{o}, \bm{\hat{\Phi}})$ for a data point $(\hat{\rho}_j^\textsc{i}, \hat{\rho}_j^\textsc{o})$.
Here we will simply denote the loss function as $\mathcal{L}_j(\bm{\hat{\Phi}})$ for a data point (classical or quantum) indexed by points in the data batch $j \in \mathcal{B}$.

For each epoch, we can split the exponential of the cost function into a product of exponentiated loss functions for each data point in the batch,
\begin{equation}
    e^{-i \eta_k \mathcal{J}(\bm{\hat{\Phi}})} = \prod_{j \in \mathcal{B}} e^{-i \frac{\eta_k}{|\mathcal{B}|} \mathcal{L}_j(\bm{\hat{\Phi}})}.
\end{equation}
Recall that each of these exponentials in turn consist of applying an iteration of QFB (feedforward, classical- or quantum-controlled loss pulse, and backpropagation/uncomputation).
The above decomposition presumes an application of QFB for each data point sequentially.
Later, in Section~\ref{sec:batch}, we will examine methods for parallelizing this phase accumulation.

Now, for Quantum Dynamical Descent, as seen in \eqref{eq:qdd} we need to alternate between phase kicks and kinetic term exponentials.
Recall that $\bm{\hat{\Phi}}$ is a vector of parameters with $n^\text{th}$ component $\hat{\Phi}_n$, and the conjugate momentum $\bm{\hat{\Pi}}$ with components $\hat{\Pi}_n = \hat{F}^{(n)\dagger} \hat{\Phi}_n \hat{F}^{(n)}$, where we use $\hat{F}^{(n)}$ to denote the Fourier transform on the $n$th component.
For the kinetic term exponentials, because of the commutation relation $[\Phi_n,\Phi_{n'}] = 0$ (hence $[\Pi_n,\Pi_{n'}] = 0$), one can apply all kinetic terms in parallel on the different parameter registers, 
\begin{equation}
    e^{-i\gamma_k \bm{\hat{\Pi}}^2} = \bigotimes_n e^{-i\gamma_k (\hat{\Pi}_n)^2}  = \bigotimes_n \hat{F}^{(n)\dagger} e^{-i\gamma_k (\hat{\Phi}_n)^2} \hat{F}^{(n)}.
\end{equation}
Therefore, the depth of this part of the circuit is strictly dependent upon the qubital precision of the parameter registers in the case of simulated continuous parameters, and on the speed of the analog Fourier transform in the case of continuous parameters.

Recall that the Hilbert space upon which this algorithm is being run is a tensor product of the parameter and computational Hilbert spaces, $\mathcal{H}_{\bm{\Phi}} \otimes \mathcal{H}_\textsc{c}$.
The initial state on the computational space is the blank state $\ket{\bm{0}}_\textsc{c}$, whereas the parameters will generically be initialized to some Gaussian pointer state,
\begin{equation}\label{eq:qdd_init}
  \Psi_0(\bm{\Phi}) = \frac{1}{\det^{1/4}(2 \pi \bm{\Sigma}_0)} e^{+i \bm{\Pi}_0 \cdot \bm{\Phi}} e^{-\frac{1}{4} (\bm{\Phi}-\bm{\Phi}_0)^T \bm{\Sigma}_0^{-1} (\bm{\Phi}-\bm{\Phi}_0)},
\end{equation}
where $\bm{\Phi}_0$ is the initial mean for the wavefunction, $\bm{\Pi}_0$ its initial momentum, and $\bm{\Sigma}_0$ any initial correlations between the parameters (often they will be chosen as uncorrelated, hence $\bm{\Sigma}_0$ will be diagonal).
Some implications of different choices for the initial wavefunction parameters, $\{\bm{\Phi}_0, \bm{\Pi}_0, \bm{\Sigma}_0\}$, will be discussed below in the context of the adiabatic theorem.
We will find it useful when discussing hyper-parameter optimization to also include preparation of this initial parameter wavefunction as a part of the QDD algorithm.
Hence, we will write a unitary, $\hat{U}_\text{p}(\bm{\Theta})$, where $\bm{\Theta} := \{\bm{\Phi}_0, \bm{\Pi}_0, \bm{\Sigma}_0\}$, to denote the preparation of the parameter wavefunction $\Psi_0(\bm{\Phi})$ from some initial reference state.
This preparation unitary can be thought of as classically controlled by the hyper-parameters $\bm{\Theta}$.

\begin{figure}[h!]
 \begin{center}
\includegraphics[width=1.00\columnwidth]{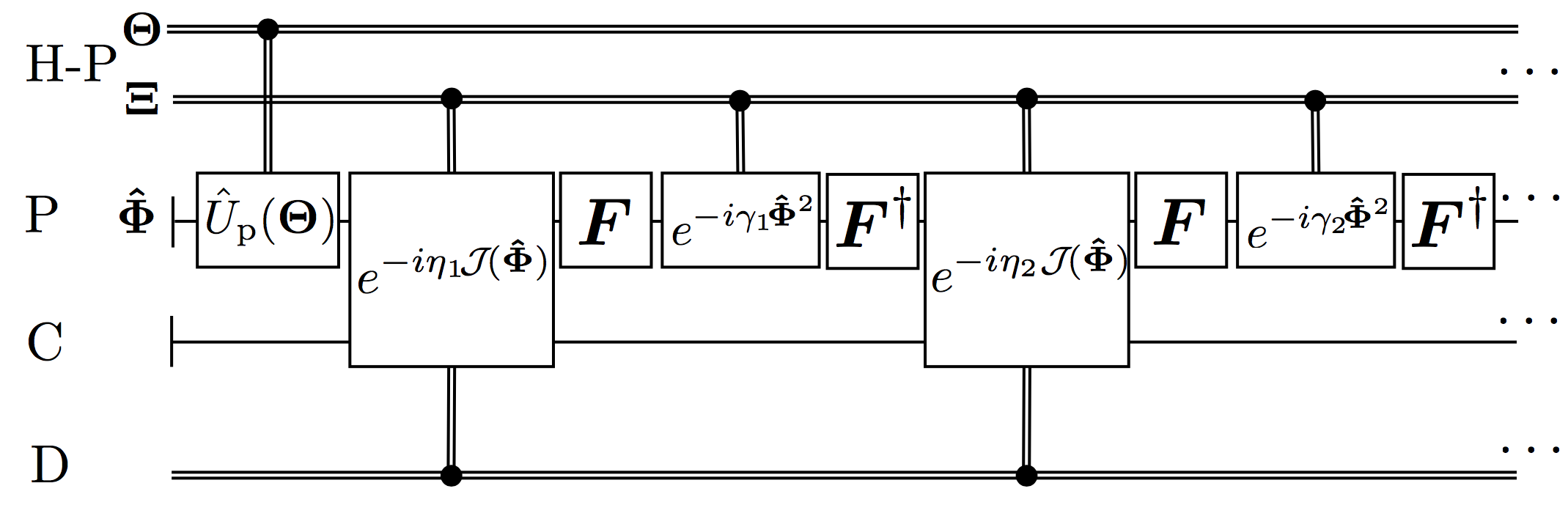}
\caption{Two iterations of Quantum Dynamical Descent. The descent rate hyper-parameters are represented by the classical vector $\bm{\Xi}= \{\bm{\gamma},\bm{\eta}\}$, while the preparation hyper-parameters are represented by $\bm{\Theta}= \{\bm{\Phi}_0, \bm{\Pi}_0, \bm{\Sigma}_0\}$. The unitary $\hat{U}_\text{p}(\bm{\Theta})$ prepares the pointer state of the parameters as in equation \eqref{eq:qdd_init} and is thus dependent on these preparation hyper-parameters. In this diagram and in most diagrams for Section~\ref{sec:opt}, we use classical registers to represent the data, but one may as well use quantum data registers, the latter of which might require a fresh swap-in after every phase kick. Additionally, we use H-P, P, C, and D to denote the hyper-parameter, parameter, compute, and data registers, respectively.} \label{fig:QDD}
\end{center}
\end{figure}

A circuit diagram for two iterations of QDD, including the initial state preparation, is provided in Figure~\ref{fig:QDD}.

\subsubsection{Heisenberg Picture Update rule}\label{sec:qdd_update}

We can now derive a Heisenberg picture update rule for single-epoch update for the \emph{parameters} under QDD.
Recall we derived the update rule for the momentum under conjugation by the cost exponential,
\begin{equation}\label{eq:momshift}
    e^{i\eta \mathcal{J}(\bm{\hat{\Phi}})}    \bm{\hat{\Pi}}e^{-i\eta \mathcal{J}(\bm{\hat{\Phi}})} = \bm{\hat{\Pi}}- \eta \bm{\nabla} \mathcal{J}(\bm{\hat{\Phi}}) +\mathcal{O}(\eta^2),
\end{equation}
i.e., the momentum is shifted by the negative gradient of the cost function, up to second order error in the kicking rate.
Conversely, the exponential kinetic term shifts the value of the parameter operator, 
\begin{equation}\label{eq:parshift}
    e^{i\gamma\bm{\hat{\Pi}}^2/2}   \, \bm{\hat{\Phi}} \,  
    e^{-i \gamma\bm{\hat{\Pi}}^2/2} =\bm{\hat{\Phi}} + \gamma \bm{\hat{\Pi}}.
\end{equation}
We see that the parameter operator is updated according to the momentum times the kinetic rate, $\gamma$.
The QDD algorithm applies \eqref{eq:momshift} followed by \eqref{eq:parshift}, hence we can derive how the parameters are updated according to the gradient, to first order in the kicking rate,
\spliteq{
       e^{i\eta \mathcal{J}(\bm{\hat{\Phi}})}  e^{i\gamma\bm{\hat{\Pi}}^2/2}   &\bm{\hat{\Phi}}e^{-i\gamma\bm{\hat{\Pi}}^2/2}  e^{-i\eta \mathcal{J}(\bm{\hat{\Phi}})}\\ &= \bm{\hat{\Phi}} +\gamma\bm{\hat{\Pi}}- \eta\gamma \bm{\nabla} \mathcal{J}(\bm{\hat{\Phi}}) +\mathcal{O}(\eta^2).
}
We thus see from the Heisenberg picture that the parameters get updated according to the momentum and gradient of the cost function at each epoch. 
 
Note that attempting to recursively conjugate the alternating operators of QDD analytically in the Heisenberg picture rapidly becomes intractable, unless the cost function can be approximated as a second-order polynomial.
This is due to the fact that the QFB unitary phase kick is a non-Gaussian operation, and keeping track of the operators as they evolve through such a set of non-Gaussian operations gets exponentially hard.
It is well-known that Gaussian operations are classically efficiently simulatable \cite{bartlett2002efficient}, whereas non-Gaussian operations generally are not.
Letting the parameters coherently descent the cost landscape leads to a state of complexity which can be seen as classically hard to simulate. As an additional inclination to suspect that Quantum Dynamical Descent is also classically hard to simulate, as we describe immediately below, it can be seen as a type of Quantum Approximate Optimization Algorithm (QAOA), and this class of algorithms has been proven to be classically hard to simulate \cite{farhi2016quantum}.

\subsubsection{Connections to QAOA}

Here we will relate the Quantum Dynamical Descent (QDD) approach to the Quantum Approximate Optimization Algorithm (QAOA) \cite{farhi2014quantum}, and its most general form the Quantum Alternating Operator Ansatz \cite{hadfield2017quantum}.
First we will review the QAOA algorithm.
QAOA is a quantum-classical hybrid algorithm in which a classical optimizer is used to optimize the hyper-parameters for a sequence of alternating operators generated by non-commuting Hamiltonians.
One usually defines a \textit{cost} Hamiltonian $\hat{H}_\textsc{c}$ and a \textit{mixer} Hamiltonian $\hat{H}_\textsc{m}$.
The goal of QAOA is to find states of low cost expectation value $\braket{\hat{H}_\textsc{c}}$.
The algorithm begins with a simple initial state, (e.g., the ground state of $\hat{H}_\textsc{m}$) and applying the sequence of alternating unitaries,
\begin{equation}\label{eq:qaoa}
    \hat{U}_{\textsc{qaoa}}(\bm{\eta},\bm{\gamma}) = \prod_{k=1}^P e^{-i\gamma_k \hat{H}_\textsc{m}}e^{-i\eta_k \hat{H}_{\textsc{c}}},
\end{equation}
where the number of pulses $P$ is fixed.
The set of parameters $\bm{\gamma} := ( \gamma_k )_{k=1}^P$ and $\bm{\eta} := ( \eta_k )_{k=1}^P$ are optimized classically to minimize the expectation $\braket{\hat{H}_\textsc{c}}_{\bm{\gamma},\bm{\eta}}$, which is usually estimated using multiple runs and measurements.
The optimization is done in a feedback loop between the quantum and classical computers; the quantum computer is seen as a black-box in which one inputs parameters and obtains a scalar corresponding to the expectation value.
As such, the classical optimizer is always using a finite-difference algorithm, whether it is using Nelder-Mead, gradient descent, or particle swarms.

Typically, the cost Hamiltonian is diagonal in the computational basis, and the mixer Hamiltonian is a generator of shifts of computational basis states.
For example, a typical cost Hamiltonian seen for QAOA with qubits \cite{farhi2014quantum} would be of the form $\hat{H}_{\textsc{c}} = -\sum_{jk}\alpha_{jk}\hat{Z}_j\hat{Z}_j -\sum_{j}\beta_j\hat{Z}_k$ which is a coupling of standard basis observables.
This choice of cost function is typical for any Quadratic Unconstrained Binary Optimization (QUBO) problem.
A typical choice of mixer Hamiltonian would be $H_{\textsc{m}} = \sum_j X_j$, i.e., an uncoupled sum of generators of shifts of the standard basis.
In our case, for machine learning, the parameters we are optimizing are often continuous (real) values, hence the choice of cost and mixer Hamiltonian has to be adapted, but we can repeat this same pattern of using a cost that is a polynomial of standard-basis diagonal operators, and a mixer which is the sum of generators of shifts of each register. 

Comparing \eqref{eq:qdd} to \eqref{eq:qaoa}, we see that the Quantum Dynamical Descent unitary is of the form of the Quantum Alternating Operator Ansatz, with cost Hamiltonian being the effective phase shift induced on the parameters by the QFB, $\hat{H}_\textsc{c} = \mathcal{J}(\bm{\hat{\Phi}})$, and the mixer Hamiltonian made up of generators of shifts of each register, $\hat{H}_{\textsc{m}} = \bm{\hat{\Pi}}^2$, i.e., the kinetic term we pulse to perform parameter shifts.

Since the QAOA algorithm consists both of applying the unitaries and also optimizing the pulse lengths, we see that we can consider the hyper-parameter optimization of Quantum Dynamical Descent to be a case of a QAOA problem, where optimizing the kinetic rates $\bm{\gamma} := ( \gamma_k )_{k=1}^P$ and phase kicking rates $\bm{\eta} := ( \eta_k )_{k=1}^P$ would count as a meta-learning.
We discuss in depth how to perform meta-learning in Section~\ref{sec:meta}.

As mentioned in the last subsubsection (\ref{sec:qdd_update}), in certain cases classically simulating the QDD for certain super-quadratic effective cost functions becomes rapidly intractable, since this would effectively be a task of simulating non-Gaussian operations.
As we have detailed above, QDD is like a continuous-variable QAOA problem, and QAOA itself has been shown to potentially demonstrate a quantum advantage for certain optimization problems \cite{farhi2016quantum}.
We leave as future work a formal proof of whether this quantum advantage result can be extended to QDD, and whether this is achieved only in the cases where the effective potential is non-Gaussian.

Finally, QAOA is technically inspired from the Quantum Adiabatic Algorithm, it can be considered as a temporally coarse-grained, variationally optimized quantum simulation of adiabatic evolution.
In the limit of many pulses, one can show that there exists a solution for the hyper-parameters which converges to a quantum simulation of adiabatic evolution, hence effectively a quantum simulated implementation of the Quantum Adiabatic Algorithm.
In the next subsection (\ref{sec:adiab}), we establish a similar limit for the QDD algorithm, i.e., a limit where many pulses are applied and show how it can be seen as a quantum simulation of a continuous-quantum-variable adiabatic evolution.
This will be useful to discuss convergence and to use the physical intuition to derive a heuristic for the initialization of hyper-parameters.

\subsubsection{Adiabatic Limit}\label{sec:adiab}

Before we can connect to the adiabatic theorem, let us briefly review the Quantum Adiabatic Algorithm (QAA).
Suppose we would like to find the ground state of a certain cost Hamiltonian $\hat{H}_\textsc{c}$.
We begin in a ground state, $\ket{g_0}$, of some simpler Hamiltonian, $\hat{H}_0$.
Then we evolve the system with the unitary generated by a time-dependent interpolating Hamiltonian of the form $\hat{H}(\tau) = (1-\tau/T)\hat{H}_0 + (\tau/T) \hat{H}_\textsc{c}$, where $\tau \in [0,T]$ is the time parameter.
Explicitly, this evolution operator is given by the time-ordered exponential $\textstyle\mathcal{T}\!\exp(-i\int_0^T\! d\tau \hat{H}(\tau))$.
In this scenario, the adiabatic theorem states that for sufficiently long time $T$, the evolution will track the instantaneous ground state $\ket{g(\tau)}$ for all times $\tau \in [0,T]$.
More precisely, the condition for $T$ is given by $T \gg 1/\Delta^2$, where $\Delta_g$ is the smallest ground state energy gap of the interpolating Hamiltonian: $\Delta_g := \min_\tau \Delta_g(\tau)$.
(There exist more general statements of the adiabatic theorem \cite{messiah2014quantum}.
We leave an examination of the connection of QDD to these more general versions to the zealous reader.)

We can consider a set of cases where our Quantum Dynamical Descent obeys the adiabatic theorem.
So far we have yet to specify which initial states are best for the quantum parameter wavefunction, however, to apply the adiabatic theorem we will have to pick an initial Hamiltonian, and hence a corresponding ground state.
We will work in the continuous-variable formalism but much transfers over to the qubital case through quantum simulation theory. 

We can pick an initial Hamiltonian for $N$ quantum harmonic oscillators (for $N$ parameters) with hyper-parameters $m_0$, $k_0$ and $\bm{\Phi}_0$
\begin{equation}
    \hat{H}_0 = \tfrac{1}{2m_0}\bm{\hat{\Pi}}^2 +  \tfrac{k_0 }{2}(\bm{\hat{\Phi}}-\bm{\Phi}_0)^2.
\end{equation}
The corresponding ground state is the Gaussian wavefunction given by
\begin{equation}\label{eq:adiab_init}
    \Psi_0(\bm{\Phi}) = \left(\tfrac{m_0 \omega_0}{\pi}\right)^{N/4} e^{-\frac{m_0 \omega_0}{2} (\bm{\Phi}-\bm{\Phi}_0)^2},
\end{equation}
where $\omega_0 := \sqrt{k_0/m_0}$.
We see that the variance in each of the parameters is $\sigma^2 = 1/2 m_0 \omega_0 = 1/2\sqrt{m_0 k_0}$.
For simplicity, we picked hyperparamters symmetrically over dimensions, but in general, one could have a different $m_0$ and $k_0$ for each dimension.
That is, one could use the Hamiltonian
\begin{equation}
       \hat{H}_0 = \sum_{n=1}^N \left[ \tfrac{1}{2m_{0,n}} (\hat{\Pi}_n)^2 +  \tfrac{k_{0,n}}{2} (\hat{\Phi}_n-\Phi_{0,n})^2 \right],
\end{equation}
and a corresponding ground state with different variances for each dimension. 

Note that, by the uncertainty principle, the wider the variance is in position space, the narrower it is in momentum space.
Concretely, if we look at the Fourier transform of \eqref{eq:adiab_init},
\begin{equation}
  \tilde{\Psi}_0(\bm{\Pi}) = \left( \tfrac{1}{\pi m_0 \omega_0} \right)^{N/4} e^{-i \bm{\Phi}_0 \cdot \bm{\Pi}} e^{-\frac{1}{2m_0 \omega_0} \bm{\Pi}^2},
\end{equation}
we see that the variance in each of the parameter momenta is inversely proportional to that of the corresponding parameter: $\tilde{\sigma}^2 = m_0 \omega_0 / 2 = \sqrt{m_0 k_0}/2 = 1/4 \sigma^2$.
In most cases, it will be advantageous to keep the variance in position space larger than that in momentum space, initially.
This is tied to our choice of initial Hamiltonian, since for $k_0 \rightarrow 0$ we have the position space variance go to infinity.

For the final Hamiltonian, we would like to find the optimum of the cost function $\mathcal{J}(\bm{\Phi})$.
In general, this will be achieved through minibatching (stochastic gradient descent), but for this discussion we will assume that this cost function includes all of the data in the batch so that the potential itself is not stochastically time-dependent.
In the continuous-variable setting, achieving an infinitely sharp wavefunction centered at the optimum would be physically impossible.
Therefore, instead we will aim to minimize a \textit{regularized} cost Hamiltonian, whose regularizer will be an added kinetic term.
That is, our cost Hamiltonian will be of the form:
\begin{equation}
    \hat{H}_{\textsc{c}}   = \tfrac{1}{2m_{\textsc{c}}}\bm{\hat{\Pi}}^2 +  \mathcal{J}(\bm{\hat{\Phi}}).
\end{equation}
We will choose the cost mass parameter to be much larger than the initial mass parameter, $m_\textsc{c} \gg m_0$.
The reasoning for this will become clear later.
For now, let us consider a time-ordered Trotterized exponentiation of our adiabatic time evolution.
That is, consider the evolution under the interpolating Hamiltonian, $\hat{H}(\tau) = (1-\tau/T)\hat{H}_0 + (\tau/T) \hat{H}_{\textsc{c}}$.
Let us partition the time interval $\tau \in [0,T]$ into sub-intervals $\mathcal{I}_k := [\tau_k, \tau_{k+1}]$ with $\tau_{k-1} < \tau_k < \tau_{k+1}$ for all $k$ and $\mathcal{I} = \cup_k \mathcal{I}_k$.
For convenience, let us also denote $\Delta \tau_k := \tau_{k+1} - \tau_k$, $\bar{\tau}_k := (\tau_{k+1} + \tau_k)/2$, and the time-averaged Hamiltonian on the sub-interval $\mathcal{I}_k$ as $\hat{H}_k := \int_{\mathcal{I}_k} d\tau \hat{H}(\tau)$.
Then the evolution operator over the interval $\tau \in [0,T]$ generated by the interpolating Hamiltonian is:
\begin{equation}
\begin{split}
  \hat{U}(T) &\approx \prod_k e^{-i \Delta \tau_k \hat{H}_k} \\
  &\approx \prod_k e^{-i \gamma_k \bm{\hat{\Pi}}^2} e^{-i \eta_k \mathcal{J}(\bm{\hat{\Phi}})} e^{-i \lambda_k (\bm{\hat{\Phi}} - \bm{\hat{\Phi}}_0)^2},
\end{split}
\end{equation}
where
\begin{eqnarray}
  \gamma_k &:=& \Delta \tau_k \left( \frac{(1-\bar{\tau}_k/T)}{2m_0} + \frac{\bar{\tau}_k/T}{2m_\textsc{c}} \right) \\
  \eta_k &:=& \Delta \tau_k (\bar{\tau}_k/T)\\
  \lambda_k &:=& \Delta \tau_k (1 - \bar{\tau}_k/T) \frac{k_0}{2}.
\end{eqnarray}
We see that the adiabatic evolution can be split up into a kinetic term, a (batched) QFB exponential, and an extra phase term, which can be interpreted as a fading parameter regularization term.
Therefore, we see that this evolution can be viewed as a regularized QDD algorithm.
(We will refrain from discussing regularization until we treat it in detail in Section~\ref{sec:reg}.)
We see that as $\tau \rightarrow T$, the coefficient of the kinetic term becomes that associated with the cost mass, while the weight decay term fades, and the QFB phase kicking rate dominates.
Thus starting with a small mass parameter $m_0$ and ending in a much larger mass $m_\textsc{c}$, we get an effective kinetic rate $\gamma_k$ that is decreasing, while we get an increasing QFB phase kicking rate.
Although in practical scenarios where one may want a local minimum rather than a global minimum, one may want to ignore the adiabatic condition.
The intuition of starting with a large uncertainty state, a large kinetic rate, and slowly increasing the phase kicking rate relative to the kinetic rate, can be very useful for practitioners manually optimizing hyper-parameters.

Now, let us examine the final cost Hamiltonian close to a local minimum $\bm{\Phi}^\ast$.
Let us assume that there is a local Taylor series expansion for the potential $\mathcal{J}(\bm{\Phi})$ around this point in parameter space.
Because it is a local optimum, the first order term, $\sim \nabla \mathcal{J}(\bm{\Phi}^\ast)$, should vanish, giving us
\begin{equation}\label{eq:adiab_taylor}
    \mathcal{J}(\bm{\hat{\Phi}}) \approx \mathcal{J}(\bm{\Phi}^\ast) + \frac{1}{2} (\bm{\hat{\Phi}}-\bm{\Phi}^\ast)^T \left( \bm{\nabla} \bm{\nabla}^T \mathcal{J}(\bm{\Phi}^\ast) \right) (\bm{\hat{\Phi}}-\bm{\Phi}^\ast).
\end{equation}
This is simply a quadratic potential.
Let us write $\bm{K} := \bm{\nabla} \bm{\nabla}^T \mathcal{J}(\bm{\Phi}^\ast)$.
Since $\bm{\Phi}^\ast$ is a local minimum, the Hessian $\bm{K}$ will have strictly positive eigenvalues, hence the cost Hamiltonian around the local minimum is simply that of a collection of quantum harmonic oscillators.
The eigenbasis of $\bm{K}$ is called the set of normal modes for the oscillators.
Let us write $\bm{O}$ to denote the orthogonal matrix which maps to the normal mode basis of $\bm{K}$, and let $\{ \kappa_n \}_{n=1}^N$ be the set of eigenvalues of $\bm{K}$.
Let us also write the quadratures in this basis as $\bm{\hat{q}} := \bm{O} \bm{\hat{\Phi}}$ (correspondingly, $\bm{q}^\ast := \bm{O} \bm{\Phi}^\ast$) and $\bm{\hat{p}} := \bm{O} \bm{\hat{\Pi}}$.
Then we can write the cost Hamiltonian as:
\begin{equation}
  \hat{H}_\textsc{c} = \sum_{n=1}^N \left[ \tfrac{1}{2m_\textsc{c}} \hat{p}_n^2 + \tfrac12 \kappa_n (\hat{q}_n - q_n^\ast)^2 \right].
\end{equation}
In the parameter Hilbert space, this Hamiltonian is diagonalized in the joint Fock bases of the normal modes.
The gap between adjacent Fock states of the $n$th mode is the frequency $\omega_n := \sqrt{\kappa_n/m_\textsc{c}}$.
Therefore, the gap of the entire Hamiltonian is the smallest of these gaps: $\Delta_\textsc{c} := \min_n \omega_n$.

\begin{figure}[h!]
 \begin{center}
\includegraphics[width=1.00\columnwidth]{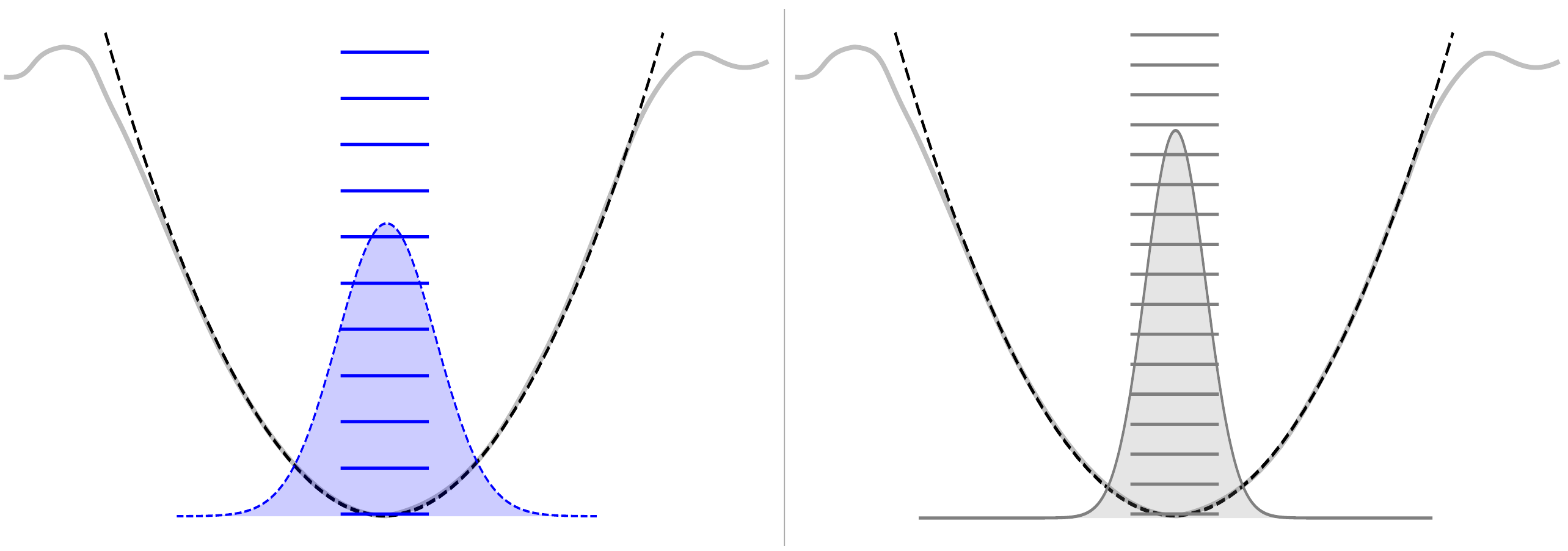}
\caption{Local ground state of at a local minimum of the cost function, for two distinct values of kinetic term. Represented above are the cost function (gray), its Taylor expansion near the local minimum (dashed), the spectrum of the effective Hamiltonians (horizontal lines) and the wavefunction in parameter space of the corresponding ground states, for a large  (left) and small (right) final kinetic rate relative to phase kicking rate of the simulated adiabatic evolution.
Taylor-expanded cost function potential can be locally approximated with a harmonic (second order) potential. Depending on the ratio of final kinetic vs. phase kicking rates of the simulated adiabatic evolution, the effective Hamiltonian near the local minimum admits a ground state of higher variance for a greater gap, or lower variance with smaller gap. Hence the slower the adiabatic evolution the sharper one can concentrate the wavefunction on the local minimum value.} \label{fig:ground}
\end{center}
\end{figure}

The ground state of the cost Hamiltonian written in the original variables is given by:
\begin{equation}
  \Psi_\textsc{c}(\bm{\Phi}) = [ \det(\sqrt{m_\textsc{c} \bm{K}}/\pi) ]^{1/4} e^{-\frac{\sqrt{m_\textsc{c}}}{2} (\bm{\Phi}-\bm{\Phi}^\ast)^T \sqrt{\bm{K}} (\bm{\Phi} - \bm{\Phi}^\ast)},
\end{equation}
which is a Gaussian with mean $\bm{\Phi}^\ast$ and covariance matrix $\bm{\Sigma} := \tfrac{1}{2\sqrt{m_\textsc{c}}} \bm{K}^{-1/2}$.
We see that having a higher mass parameter will concentrate the wavefunction around the optimum, but reduce the gap.
Thus, we can get a sharper estimate of the optimum value at the cost of having a longer runtime if we want to be sure to stay in the ground state and obey the adiabatic principle.
We illustrate this in Figure~\ref{fig:ground}.

We have discussed the gap of the final (cost) Hamiltonian (near a local minimum), but to obey the adiabatic condition (ensuring that the system stays in the instantaneous ground state at all time), we want to make sure to evolve the system adiabatically on a time scale greater than the gap of the \textit{interpolating} Hamiltonian.
To show that the interpolating Hamiltonian has a ground state gap will generally not be possible.
However, if the potential in the initial Hamiltonian is centered near the optimum of the cost potential (i.e., $\bm{\Phi}_0 \approx \bm{\Phi}^\ast$), then we can use the Taylor expansion \eqref{eq:adiab_taylor} of the cost potential and approximate the interpolating Hamiltonian as:
\begin{equation}
  \hat{H}(\tau) \approx \frac{\bm{\hat{\Pi}}^2}{2m(\tau)} + \tfrac12 ( \bm{\hat{\Phi}} - \bm{\Phi}_0(\tau) )^T \bm{K}(\tau) ( \bm{\hat{\Phi}} - \bm{\Phi}_0(\tau) ) + E_0(\tau),
\end{equation}
where 
\begin{eqnarray}
  m(\tau) &:=& \left[ \left( 1-\tfrac{\tau}{T} \right) \tfrac{1}{m_0} + \left( \tfrac{\tau}{T} \right) \tfrac{1}{m_\textsc{c}} \right]^{-1}, \\
  \bm{K}(\tau) &:=& \left( 1-\tfrac{\tau}{T} \right) k_0 + \left( \tfrac{\tau}{T} \right) \bm{K}, \\
  \bm{\Phi}_0(\tau) &:=& \bm{K}^{-1}(\tau) \left[ \left( 1-\tfrac{\tau}{T} \right) k_0 \bm{\Phi}_0 + \left( \tfrac{\tau}{T} \right) \bm{K} \bm{\Phi}^\ast \right], \\
  E_0(\tau) &:=& \left( \tfrac{\tau}{T} \right) \mathcal{J}(\bm{\Phi}^\ast) - \tfrac12 \bm{\Phi}_0^T(\tau) \bm{K}(\tau) \bm{\Phi}_0(\tau) \nonumber \\
  && \quad + \tfrac12 \left( 1-\tfrac{\tau}{T} \right) k_0 \bm{\Phi}_0^2 + \tfrac12 \left( \tfrac{\tau}{T} \right) \bm{\Phi}^{\ast T} \bm{K} \bm{\Phi}^\ast.
\end{eqnarray}
Notice that since the initial coupling matrix is proportional to the identity, $k_0 \hat{I}$, then the two terms in the expression for $\bm{K}(\tau)$ can be simultaneously diagonalized.
Hence, the eigenvalues of $\bm{K}(\tau)$ will be $\kappa_n(\tau) = (1-\tau/T) k_0 + (\tau/T) \kappa_n$.
Due to the fact that $k_0 > 0$ and the spectrum of $\bm{K}$ is positive, then the eigenvalues of $\bm{K}(\tau)$ are also positive as they consist of a convex combination of two positive numbers.
Similarly, we have that $m(\tau)^{-1} > 0$ as it is a convex combination of positive numbers (inverse masses of the initial and cost Hamiltonians).
Therefore, the gap of the interpolating Hamiltonian, for any $\tau$, which is the minimum eigenvalue of the matrix $\bm{W}(\tau) = \sqrt{\tfrac{1}{m(\tau)}\bm{K}(\tau)}$, will be positive.
That is, $\Delta_g(\tau) = \min_n \sqrt{\kappa_n(\tau)/m(\tau)} > 0$ for all $\tau \in [0,T]$.

\subsection{Momentum Measurement Gradient Descent}\label{sec:momgrad}

Momentum Measurement Gradient Descent (MoMGrad) is a way to perform a descent of the parameter landscape in a way leveraging a quantum advantage relative to finite-difference methods, \cite{2005PhRvL..95e0501J} but which requires far less coherence time than Quantum Dynamical Descent.
In many ways, it is very similar to typical gradient descent, but in this case rather than via classical backpropagation or automatic differentiation, the gradient is obtained via the phase kickback principle and by measurement of momentum perturbations.
As was outlined in a previous subsection, acting a phase kick with respect to the cost function and backpropagating this phase kick through the parametric circuit is akin to a finite time-step contribution to shifting the momentum of each parameter.
It is thus a ``force kick'', and by measuring the shift in momenta, we can estimate the gradient of the landscape, since we showed that the shift in momenta's expectation values will be proportional to the gradient of the cost function.
Rather than letting the Schrodinger dynamics naturally shift the parameter values after having their momenta kicked, we use classical computation to update the initialization of parameters for the next iteration. 

The algorithm consists of first preparing an appropriate momentum pointer state of the parameters, then applying the Quantum Feedforward and Baqprop (QFB) algorithm outlined above, followed my a measurement the momentum to obtain an estimate of the gradient.
In general, multiple runs and measurements are necessary to get a sharp estimate of the gradient.
We briefly analyze the tradeoffs between having a pointer state of high certainty in parameter value versus obtaining a high precision in the gradient.
We will also consider how the classical parameters for the momentum pointer state will interface with the quantum algorithm in the classical-quantum updating loop.

The type of initial pointer state of the parameters for MoMGrad can be a Gaussian state, similar to that used to initialize the parameters for Quantum Dynamical Descent.
For example, in the parameter eigenbasis, one could write the wavefunction in the form
\begin{equation}\label{eq:momgrad_init}
  \Psi_0(\bm{\Phi}) = \frac{1}{\left| 2 \pi \bm{\Sigma}_0 \right|^{1/4}} e^{+i \bm{\Pi}_0 \cdot \bm{\Phi}} e^{-\frac{1}{4} (\bm{\Phi}-\bm{\Phi}_0)^T \bm{\Sigma}_0^{-1} (\bm{\Phi}-\bm{\Phi}_0)},
\end{equation}
where $\bm{\Sigma}_0$ is the covariance matrix of the Gaussian, $\bm{\Phi}_0$ and $\bm{\Pi}_0$ are the parameter and momentum expectation values, respectively.
For the sake of generality, we include a preemptive momentum bias $\bm{\Pi}_0$, which is occasionally used in variants of the gradient descent algorithms in classical machine learning \cite{ruder2016overview}.
These sets of classical parameters can be considered as hyper-parameters which can be optimized using trial and error or using techniques from Section~\ref{sec:meta}.

We can contrast the spread of the wavefunction with that in the Fourier conjugate basis,
\begin{equation}
  \tilde{\Psi}_0(\bm{\Pi}) = \left| \frac{2 \bm{\Sigma}_0}{\pi} \right|^{1/4} e^{-i \bm{\Phi}_0 \cdot \bm{\Pi}} e^{- (\bm{\Pi}-\bm{\Pi}_0)^T \bm{\Sigma}_0 (\bm{\Pi}-\bm{\Pi}_0)}.
\end{equation}
(Ignoring global phases.)
The covariance matrix in momentum space is now inverted.
For example, in the case where this matrix is proportional to the identity, $\bm{\Sigma} = \sigma^2 I$, the variance in each of the parameters is $\sigma^2$, whereas the variance of the corresponding momentum is $1/4\sigma^2$.
Thus they are inversely related as dictated by the Heisenberg uncertainty principle.

Let us consider how the classical parametrization of the wavefunction, $\{\bm{\Phi}_0,\bm{\Pi}_0,\bm{\Sigma}_0\}$, gets updated at a given epoch, through the classical-quantum approach.
Once such an initial state has been prepared, the next step is to apply the Quantum Feedforward and Baqprop (QFB) circuit.
Recall that after running this over a full data batch, the momentum is updated according to
\begin{equation}
  e^{i \eta \mathcal{J}(\bm{\hat{\Phi}})} \bm{\Pi} e^{-i \eta \mathcal{J}(\bm{\hat{\Phi}})} = \bm{\hat{\Pi}} - \eta \bm{\nabla} \mathcal{J}(\bm{\hat{\Phi}}) + \mathcal{O}(\eta^2).
\end{equation}

We could also examine the effect this has on the wavefunction.
Eventually we want to perform a measurement of the momentum in order to obtain an estimate for the gradient.
By studying the statistics of each run in the Schr\"odinger picture, we can assess how many runs are necessary to obtain a particular precision for the gradient estimate.
In the Schr\"odinger picture, the QFB algorithm acts as a non-linear phase shift on the initial wavefunction:
\begin{equation}
  \ket{\Psi_0} \mapsto e^{-i \eta \mathcal{J}(\bm{\hat{\Phi}})} \ket{\Psi_0}.
\end{equation}
If the wavefunction is sufficiently localized in parameter space, around the mean $\bm{\Phi}_0$, then we can Taylor-expand the cost function,
\begin{equation}
  \mathcal{J}(\bm{\hat{\Phi}}) \approx \mathcal{J}(\bm{\Phi}_0) + (\bm{\hat{\Phi}} - \bm{\Phi}_0)^T \bm{\nabla} \mathcal{J}(\bm{\Phi}_0) + \mathcal{O} \left( (\bm{\hat{\Phi}} - \bm{\Phi}_0)^2 \right),
\end{equation}
and the QFB algorithm can be approximated as a linear phase shift according to the gradient of the cost function at the point $\bm{\Phi}_0$,
\begin{equation}
\begin{split}
  &\bra{\bm{\Phi}} e^{-i \eta \mathcal{J}(\bm{\hat{\Phi}})} \ket{\Psi_0} \\
  &\quad \approx \frac{1}{|2\pi \bm{\Sigma}_0|} e^{+i (\bm{\Pi}_0 - \eta \bm{\nabla} \mathcal{J}(\bm{\Phi}_0)) \cdot \bm{\Phi}} e^{-\frac{1}{4} (\bm{\Phi}-\bm{\Phi}_0)^T \bm{\Sigma}_0^{-1} (\bm{\Phi}-\bm{\Phi}_0)}
\end{split}
\end{equation}
(Ignoring global phases.)
Of course, this is equivalent to shifting the average of the momentum,
\begin{equation}
\begin{split}
  &\bra{\bm{\Pi}} e^{-i \eta \mathcal{J}(\bm{\hat{\Phi}})} \ket{\Psi_0} \\
  &\quad \approx \left| \frac{2 \bm{\Sigma}_0}{\pi} \right|^{1/4} e^{-i \bm{\Phi}_0 \cdot \bm{\Pi}} \\
  &\qquad \quad \times e^{- (\bm{\Pi}-\bm{\Pi}_0 + \eta \bm{\nabla} \mathcal{J}(\bm{\Phi}_0))^T \bm{\Sigma}_0 (\bm{\Pi}-\bm{\Pi}_0 + \eta \bm{\nabla} \mathcal{J}(\bm{\Phi}_0))}.
\end{split}
\end{equation}

Now that we have shown how the wavefunction gets shifted in momentum space, the next step in the algorithm is to measure the wavefunction in the momentum basis to get an estimate for the average gradient.
To perform this measurement, one simply applies the Fourier transform on each of the parameter registers, $\bm{\hat{F}} := \otimes_{n=1}^N \hat{F}^{(n)}$, and subsequently measures the state in the standard (parameter) basis.
From the Born rule, we see that this measurement will draw a sample from the probability distribution, $p(\bm{\Pi}) = | \bra{\bm{\Pi}} e^{-i \eta \mathcal{J}(\bm{\hat{\Phi}})} \ket{\Psi_0} |^2$.
For the case where the wavefunction is highly localized in parameter space, this distribution is approximately Gaussian:
\begin{equation}
\begin{split}
  &| \bra{\bm{\Pi}} e^{-i \eta \mathcal{J}(\bm{\hat{\Phi}})} \ket{\Psi_0} |^2 \\
  &\quad \approx \left| \frac{2\bm{\Sigma}_0}{\pi} \right|^{1/2} e^{-2 (\bm{\Pi} - \bm{\Pi}_0 + \eta \bm{\nabla} \mathcal{J}(\bm{\Phi}_0))^T \bm{\Sigma}_0 (\bm{\Pi} - \bm{\Pi}_0 + \eta \bm{\nabla} \mathcal{J}(\bm{\Phi}_0))},
\end{split}
\end{equation}
with mean, $\bm{\Pi}_0 - \eta \bm{\nabla} \mathcal{J}(\bm{\Phi}_0)$, and covariance, $\tfrac14 \bm{\Sigma}_0^{-1}$.
Supposing that we perform $r$ measurement runs, our point-estimate of the mean of the momentum can be seen as a sample from the normal distribution $\mathcal{N}( \bm{\Pi}_0 - \eta \bm{\nabla} \mathcal{J}(\bm{\Phi}_0), \tfrac14 \bm{\Sigma}_0^{-1} / r )$.
Thus, with $M$ runs, and initial parameter covariance, $\bm{\Sigma}_0 = \sigma^2 \hat{I}$, we can estimate the components of the average gradient each with a precision (standard deviation) $1/2\sigma \sqrt{r}$.
We see that increasing the uncertainty in the initial parameter space wavefunction allows for quadratically fewer runs to estimate the gradient within the same precision.

In the general case, without the simplifying assumption of the wavefunction being highly localized in the parameter eigenbasis, the outcomes of multiple measurement runs can be combined into a sample mean to generate an estimate of the average momentum.
Given that the original momentum expectation value, $\bm{\Pi}_0$, is known (since it is a parameter of the initial pointer state preparation), and the average momentum is updated as
\begin{equation}
  \bra{\Psi_0} \bm{\hat{\Pi}} \ket{\Psi_0} \mapsto \bm{\Pi}_0 - \eta \langle \bm{\nabla} \mathcal{J}(\bm{\hat{\Phi}}) \rangle_0 + \mathcal{O}(\eta^2),
\end{equation}
where we have written $\langle \bm{\nabla} \mathcal{J}(\bm{\hat{\Phi}}) \rangle_0 := \bra{\Psi_0} \bm{\nabla} \mathcal{J}(\bm{\hat{\Phi}}) \ket{\Psi_0}$.
From an estimate of the average momentum, we can extract an estimate for the average gradient (to first order in $\eta$), $\eta \langle \bm{\nabla} \mathcal{J}(\bm{\hat{\Phi}}) \rangle_0$.

Once we have estimated this average gradient to the desired precision, we use the result to reinitialize the wavefunction with a new set of parameters, $\{\bm{\Phi}_0,\bm{\Pi}_0,\bm{\Sigma}_0\}$, in order to repeat this process at the next epoch.
The momentum parameter is updated according to the shift of the average momentum induced by the cost potential phase kick, i.e.,
\begin{equation}\label{eq:mom_pi_upt}
  \bm{\Pi}_0 \mapsto \bm{\Pi}_0 - \eta \langle \bm{\nabla} \mathcal{J}(\bm{\hat{\Phi}}) \rangle_0.
\end{equation}
Then we use this new momentum to update the parameter value:
\begin{equation}\label{eq:mom_phi_upt}
  \bm{\Phi}_0 \mapsto \bm{\Phi}_0 + \gamma \left( \bm{\Pi}_0 - \eta \langle \bm{\nabla} \mathcal{J}(\bm{\hat{\Phi}}) \rangle_0 \right).
\end{equation}
Note the appearance of the hyper-parameters $\eta$ and $\gamma$.
These are in direct analogy to their counterparts in the Quantum Dynamical Descent algorithm from the previous subsection.
We will provide a visual comparison of the phase space behavior of both MoMGrad and QDD in the following subsection.

An alternative to the above update rule is to discard the momentum between updates, and always initialize the momentum of the wavefunction to zero, $\bm{\Pi}_0 = \bm{0}$.
In this case, we simply update the parameter values according to the rule,
\begin{equation}
  \bm{\Phi}_0 \mapsto \bm{\Phi}_0 - \gamma \eta \langle \bm{\nabla} \mathcal{J}(\bm{\hat{\Phi}}) \rangle_0.
\end{equation}
Then we recover the classical update rule for gradient descent (without momentum).
Here, the learning rate is a product of the phase kicking and kinetic rates.

Above we discussed the fact that increasing the uncertainty in the parameter wavefunction will allow for a more precise estimate of the average gradient using fewer measurement runs.
However, of course increasing the uncertainty means that the wavefunction is no longer well localized in parameter space.
Therefore, even if we obtain a precise estimate of the average gradient,
\begin{equation}
  \langle \bm{\nabla} \mathcal{J}(\bm{\hat{\Phi}}) \rangle_0 = \int d\bm{\Phi} | \Psi_0(\bm{\Phi}) |^2 \bm{\nabla} \mathcal{J}(\bm{\Phi}),
\end{equation}
the information we obtain is only a coarse-grained indication of the direction in which the parameters should be shifted in order to achieve the minimum of the cost function.
Ideally, as in the case of QDD, one would like to use the full operator $\bm{\nabla} \mathcal{J}(\bm{\hat{\Phi}})$ to update the wavefunction in each branch of the superposition over the parameters $\bm{\Phi}$.
In the case of MoMGrad, we only extract an expectation value for this operator and move the entire wavefunction to be centered at this updated parameter value.
Therefore, there is a trade-off which is forced upon us by the uncertainty principle, namely, that obtaining a more precise estimate of the gradient requires less uncertainty in momentum space (a sharper momentum pointer state), yet this estimate will only contain information about the gradient averaged over a larger region of parameter space.

\begin{figure}[h!]
 \begin{center}
\includegraphics[width=1.00\columnwidth]{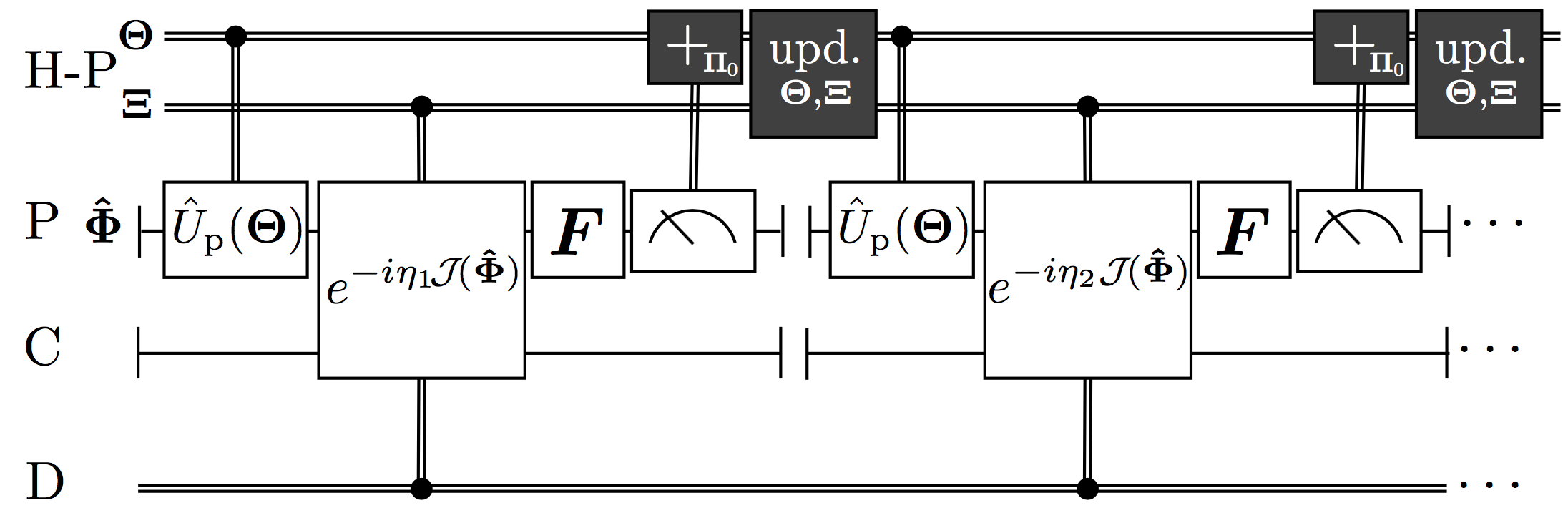}
\caption{Two iterations of Momentum Measurement Gradient Descent. We use the same graphical notation as in figure \ref{fig:QDD};  $\bm{\Xi}= \{\bm{\gamma},\bm{\eta}\}$ and $\bm{\Theta}= \{\bm{\Phi}_0, \bm{\Pi}_0, \bm{\Sigma}_0\}$ represent the descent rate and preparation hyper-parameters respectively. The measurement-controlled adder represents equation \eqref{eq:mom_pi_upt}, while the update of the hyper-parameters represents equation \eqref{eq:mom_phi_upt}, as well as the change in choice of preparation and rate hyper-parameters for the next epoch. The register labels are the same as in figure \ref{fig:QDD}. Once again we picture the procedure for classical data learning but the above is directly extendible to quantum data learning scenarios.} \label{fig:MoMGrad}
\end{center}
\end{figure}

\subsection{Phase Space Visualization}

\begin{figure}[h]
 \begin{center}
\includegraphics[width=0.95\columnwidth]{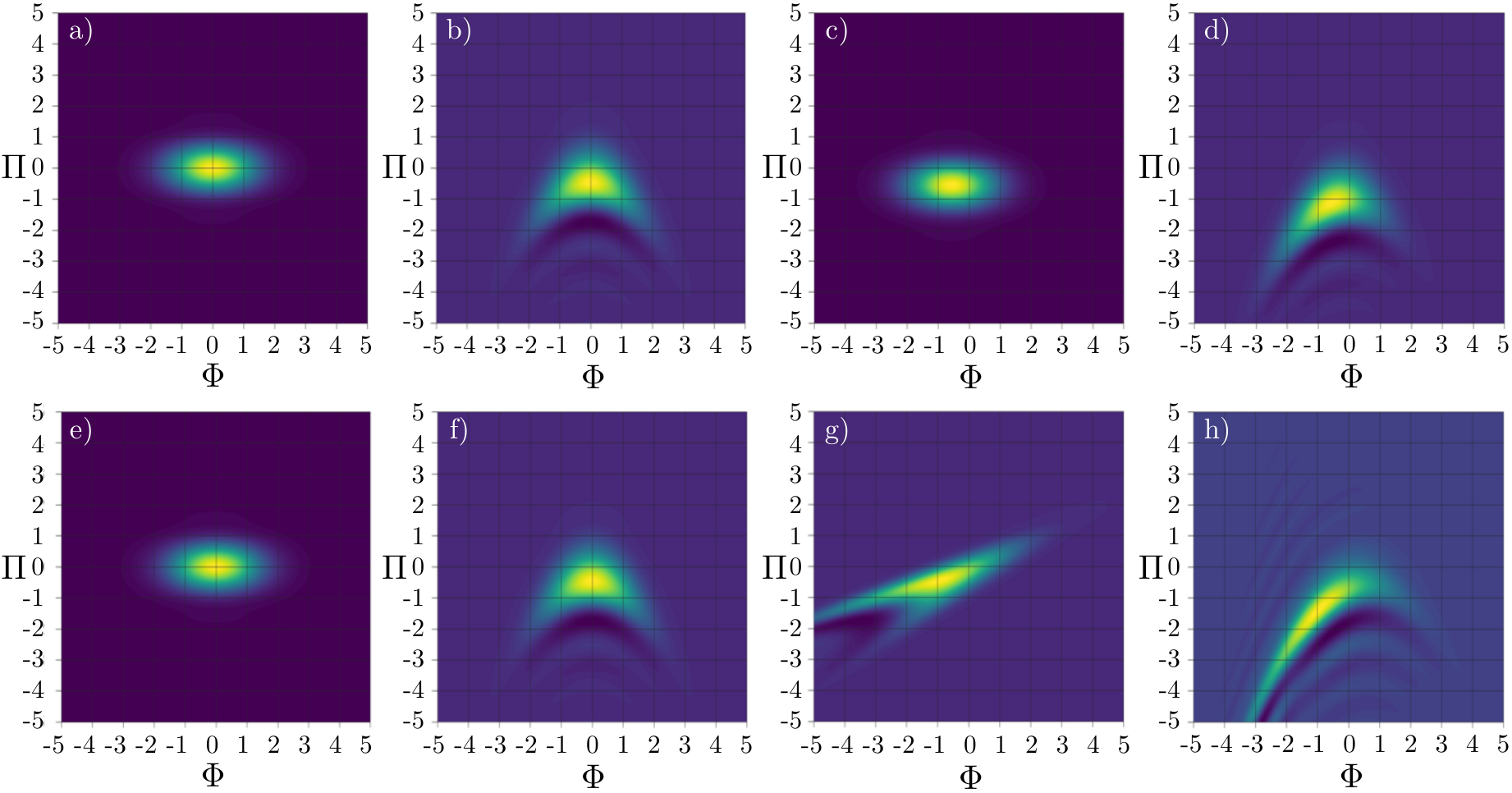}
\caption{Comparison of the Momentum Measurement Gradient Descent (MoMGrad) algorithm (a)-(d), and the Quantum Dynamical Descent (QDD) algorithm (e)-(h). Both algorithms are initialized to the same Gaussian state in (a) and (e), and undergo the same phase kick in (b) and (f). The momentum measurement and reinitialization of MoMGrad is shown in (c), whereas the result of the kinetic pulse of QDD is shown in (g). Plots (d) and (h) show the final phase kick for MoMGrad and QDD (respectively).
} \label{fig:flows}
\end{center}
\end{figure}

In Figure~\ref{fig:flows}, we present a visual side-by-side comparison of the Momentum Measurement Gradient Descent and Quantum Dynamical Descent algorithms.
For the purpose of simply illustrating the differences, the figure examines the evolution of the wavefunction for a single continuous variable (plotted as a Wigner function, defined in Subsection~\ref{sec:CQR}) under a simple cubic potential, $\mathcal{J}(\Phi) = \Phi^3 + 2 \Phi$.
In the figure, we show the initial wavefunctions, and then three steps of each of the QDD and MoMGrad algorithms: a phase kick, a kinetic pulse/measurement and reinitialization, and a second phase kick.

Both of the algorithms are intialized to the same Gaussian wavefunction, with zero mean position and momentum ($\Phi_0 = 0$ and $\Pi_0 = 0$ in our notation from the previous sections), and initial position uncertainty set to $\sigma_0^2 = 1$ (correspondingly, the initial momentum uncertainty is $1/4\sigma_0^2 = 1/4$).
The first phase kick according to the cubic potential is also the same for both algorithms, i.e., the momentum is updated by $\Pi \mapsto \Pi - \eta (3\Phi^2+2) + \mathcal{O}(\eta^2)$.
Note that in the above simulation, we use the full phase kick $e^{-i \eta \mathcal{J}(\Phi)}$, rather than only the approximation to first order in $\eta$.

The point at which QDD and MoMGrad differ is in the next step.
After the first phase kick, in QDD the next step is to apply a kinetic pulse in order to evolve the $\Phi$ quadrature of the wavefunction, so that the parameters are updated according to the momentum kick.
Under this operation, all of the branches of the wavefunction move independently of one another according to how they were shifted by the local gradient.
For example, the edges of the wavefunction (in the $\Phi$ direction) which are initialized in an area of higher slope of the potential are kicked with a greater force than pieces of the wavefunction near the origin.
This causes these pieces of the wavefunction to update more significantly after the application of the kinetic pulse.
Note that in the figure we have chosen relatively large kinetic and phase kicking rates in order to exaggerate the evolution so that one can see more clearly the differences between the two algorithms.

In MoMGrad, one takes the expectation value of the momentum after the phase kick.
Since we began with $\Pi_0 = 0$, this average gives us $-\eta \frac{ d \mathcal{J}(\Phi) }{ d \Phi }$.
Then the step after the phase kick is to reinitialize to a new Gaussian shifted according to this measurement (which we show in the third step of MoMGrad in the figure).
Therefore, as opposed to QDD, the branches of the wavefunction are not shifted independently, but they are all updated in the same manner.
Note also that since MoMGrad parameter updates are achieved using the average of the potential gradient, not the gradient of the average potential, therefore, for example, even if the wavefunction happened to be centered at a saddle point (and had zero initial momentum), the location would still be shifted after the parameter update.

The last step we show is a second phase kick after the kinetic pulse of QDD and the measurement and reinitialization of MoMGrad.
Because MoMGrad is reinitialized to a Gaussian state in place of applying the kinetic pulse of QDD, it does not retain the non-Gaussian features of the evolution induced by the generally non-quadratic potential $\mathcal{J}(\Phi)$.
This is particularly apparent in the rightmost plots of the figure, where we see that although both methods roughly track the same evolution, in QDD these non-Gaussianities accumulate, whereas in MoMGrad they are periodically removed.
\pagebreak
\onecolumngrid

\twocolumngrid

\section{Further Quantum Descent Methods}\label{sec:misc}

In this section we discuss a collection of variations upon elements of the algorithms presented above, as well as possible augmentations one can use in conjunction with some of these procedures.

\subsection{Batching \& Parallelization}\label{sec:batch}

In Subsections~\ref{sec:qdd} and \ref{sec:momgrad}, we outlined basic cases of the Quantum Dynamical Descent (QDD) and Momentum Measurement Gradient Descent (MoMGrad) algorithms.
During this previous analysis, we assumed \textit{full batching} of the data at every parameter update.
That is, we assumed the cost function was the averaged loss over every data point in the dataset.
Here, we will make the distinction between an \emph{iteration}, corresponding to an update of the parameters, and an \emph{epoch}, which is a sweep through the entire dataset.
In the above discussion, each iteration was a full epoch; now we will examine alternatives.
As QDD and MoMGrad have numerous connections with classical deep learning, we can draw inspiration from these techniques to engineer such alternatives.

\subsubsection{Quantum Stochastic Descent}

One of the core techniques in classical deep learning is stochastic gradient descent (SGD) \cite{Goodfellow-et-al-2016}.
In this approach, one uses the loss function of a single data point as the cost function for each parameter update (iteration).

We will first describe how to apply this to Quantum Dynamical Descent.
Let us denote a dataset (classical or quantum) by $\mathcal{D}$.
For example, this could be a collection of input/output pairs of classical vectors or quantum states.
The data points in this set will be indexed by $j \in \mathcal{B}$.
For each data point (classical or quantum), the corresponding effective loss function arising from the Quantum Feedforward and Baqprop algorithm will be written as $\mathcal{L}_j(\bm{\hat{\Phi}})$.
Stochastic Quantum Dynamical Descent then consists of applying the unitary
\begin{equation}\label{eq:qsdd}
    \hat{U}_{\textsc{sqdd}} = \prod_{j\in \mathcal{B}} e^{-i\gamma_j\bm{\hat{\Pi}}^2}e^{-i\eta_j\mathcal{L}_j(\bm{\hat{\Phi}})}.
\end{equation}
That is, for each data point we have the exponentiated effective loss function from QFB (which holds to first order in $\eta_j$), as well as a kinetic pulse.
At the $j^\text{th}$ iteration, the parameters get shifted by the gradient of the loss function for the $j^\text{th}$ data point, to first order in the kicking rate, 
\spliteq{\label{eq:sqdd_update}
       e^{i\eta_j \mathcal{L}_j(\bm{\hat{\Phi}})} & e^{i\gamma_j\bm{\hat{\Pi}}^2/2}  \, \bm{\hat{\Phi}}\,e^{-i\gamma_j\bm{\hat{\Pi}}^2/2}  e^{-i\eta_j \mathcal{L}_j(\bm{\hat{\Phi}})}\\ &\!\!\!\!= \bm{\hat{\Phi}} +\gamma_j\bm{\hat{\Pi}}- \eta_j\gamma_j \bm{\nabla} \mathcal{L}_j(\bm{\hat{\Phi}}) +\mathcal{O}(\gamma_j^2,\eta_j^2).
}
We get the gradient update rule, similar to classical stochastic gradient descent.
Note the unitary \eqref{eq:qsdd} is for a single epoch (sweep over the entire dataset); it can be repeated for multiple epochs.

For the SGD variant of Momentum Measurement Gradient Descent, we similarly update the parameters after kicking with the exponential of the loss of each data point.
Before the $j^\text{th}$ parameter update the parameter space wavefunction can be reinitialized to a Gaussian state,
\begin{equation}
\begin{split}
\label{eq:par_rep}
    \Psi^{(j)}(\bm{\Phi}) &= \frac{1}{|2\pi \bm{\Sigma}^{(j)}|^{1/4}}  e^{+i\bm{\Pi}^{(j)} \bm{\Phi}} \\
    &\qquad \quad \times e^{- \tfrac14 (\bm{\Phi}-\bm{\Phi}^{(j)})^{T} (\bm{\Sigma}^{(j)})^{-1} (\bm{\Phi}-\bm{\Phi}^{(j)})},
\end{split}
\end{equation}
where $\bm{\Pi}^{(j)}$ is the expectation value of the momentum from the previous measurements, $\bm{\Phi}^{(j)}$ is the expectation value of the parameter vector, and $\bm{\Sigma}^{(j)}$ is the covariance matrix for the $j^\text{th}$ update round (the latter being considered as a classical hyper-parameter).
Now, by applying the QFB circuit using the loss function of the $j^\text{th}$ data point,
\begin{equation}
    \ket{\Psi^{(j)}} \mapsto e^{-i\eta_j \mathcal{L}_j(\bm{\hat{\Phi}})} \ket{\Psi^{(j)}},
\end{equation}
and then applying the Fourier transform on each parameter register and measuring the output (i.e., measuring the momentum), we obtain the updated average momentum as the expectation value
\begin{equation}
\begin{split}
    \bm{\Pi}^{(j+1)} &:= \braket{\Psi^{(j)}|e^{i\eta_j \mathcal{L}_j(\bm{\hat{\Phi}})} \bm{\hat{\Pi}} e^{-i\eta_j \mathcal{L}_j(\bm{\hat{\Phi}})} |\Psi^{(j)}} \\
    &=\bm{\Pi}^{(j)} - \eta_j \bm{\nabla} \mathcal{L}_j(\bm{\Phi}^{(j)}) + \mathcal{O}(\eta^2).
\end{split}
\end{equation}
We then classically update the parameter expectation value for the next round as
\begin{equation}
    \bm{\Phi}^{(j+1)} := \bm{\Phi}^{(j)} + \gamma_j \bm{\Pi}^{(j+1)}.
\end{equation}
This is the parameter iteration for the the $j^\text{th}$ data point.
We then sweep over the data set for a full epoch, updating both the momentum and parameter expectation values at each step.
This can be repeated for multiple sweeps over the dataset (epochs) as necessary.

In both stochastic QDD and stochastic MoMGrad, we have the hyper-parameters $\{\eta_j,\gamma_j\}_{j\in \mathcal{B}}$, which are the phase kicking and kinetic rates for each update.
To optimize these hyper-parameters there are various classical heuristics from which we can draw inspiration \cite{zeiler2012adadelta,kingma2014adam}.
Note that in our case the learning rate is a product of both the phase kicking rate and kinetic rates.

As in the classical case, stochastic descent has some perks; it tends to regularize the landscape and avoid overfitting \cite{bottou2010large}.
However, this comes with a tradeoff of being noisy and hence unstable for high learning rates.

\subsubsection{Sequential Mini-Batching}

In classical machine learning, a common practice is to partition the training data into \textit{mini-batches} of data.
That is, we can partition our dataset as $\mathcal{D} = \cup_{k \in \mathcal{S}} \mathcal{M}_k$, where $\mathcal{S}$ is an index set over the mini-batches.
In turn, each mini-batch, $\mathcal{M}_k$, consists of a number of data points indexed by $j \in \mathcal{B}_k$.

For the purposes of generating a cost function, sequential mini-batching will consist of consecutive applications of the phase kicks for each data point in a mini-batch, before either acting the kinetic term or classically shifting the mean parameter vector, in the cases of QDD and MoMGrad, respectively.
By sequentially applying the phase kicks for every data point in the mini-batch, there is a summation of the contributions to the shifts in the momentum of the parameters.
Therefore, the average momentum shift over the mini-batch can be used in the parameter update.
In the next subsubsection, we will explore an alternative where the average momentum shift over a mini-batch is produced through an accumulation of phase kicks applied in parallel rather than sequentially.

For the present case, we can write the explicit unitary corresponding to a sequentially mini-batched version of Quantum Dynamical Descent,
\spliteq{\label{eq:smqdd}
    \hat{U}_{\textsc{smqdd}} &= \prod_{k\in \mathcal{S}} \Big( e^{-i\gamma_k\bm{\hat{\Pi}}^2} \prod_{j\in \mathcal{B}_k}e^{-i\bar{\eta}_k \mathcal{L}_j(\bm{\hat{\Phi}})} \Big),\\
}
where we have the loss function $\mathcal{L}_j(\hat{\bm{\Phi}})$ for each data point, $j \in \mathcal{B}_k$.
We have also denoted a modified kicking rate $\bar{\eta}_k := \eta_k /|\mathcal{M}_k|$, which is normalized by the size of the mini-batch.
If we consider the the cost function for mini-batch $k$ to be the average loss over the mini-batch, 
\begin{equation}
   \mathcal{J}_k(\bm{\hat{\Phi}}) = \tfrac{1}{|\mathcal{M}_k|}\textstyle\sum_{j\in \mathcal{B}_k}\mathcal{L}_j(\bm{\hat{\Phi}}),
\end{equation}
then we see that sequentially applying the QFB losses for each data point in the mini-batch is the same as applying the mini-batch cost exponential,
\spliteq{
   \hat{U}_{\textsc{mqdd}} &= \prod_{k\in \mathcal{B}} e^{-i\gamma_k\bm{\hat{\Pi}}^2}e^{-i\bar{\eta}_k \sum_{j\in \mathcal{B}_k}\mathcal{L}_j(\bm{\hat{\Phi}})}\\
    &= \prod_{k\in \mathcal{B}} e^{-i\gamma_k\bm{\hat{\Pi}}^2}e^{-i\eta_k \mathcal{J}_k(\bm{\hat{\Phi}})}.
}
Note that for each application of $e^{-i\bar{\eta}_k\mathcal{L}_j(\bm{\hat{\Phi}})}$, one executes the QFB circuit for the $j^\text{th}$ data point, and in the above each data point's exponential loss is applied in sequence over the minibatch index before a kinetic pulse is applied.
Using the above expression, we can derive the update rule for the parameters for each minibatch,
\spliteq{
    e^{i\eta_k \mathcal{J}_k(\bm{\hat{\Phi}})} & e^{i\gamma_k\bm{\hat{\Pi}}^2/2}  \, \bm{\hat{\Phi}}\,e^{-i\gamma_k\bm{\hat{\Pi}}^2/2}  e^{-i\eta_k \mathcal{J}_k(\bm{\hat{\Phi}})}\\ &\!\!\!\!= \bm{\hat{\Phi}} +\gamma_k\bm{\hat{\Pi}}- \eta_k\gamma_k\nabla \mathcal{J}_k(\bm{\hat{\Phi}}) +\mathcal{O}(\gamma_k^2, \eta_k^2).
}
We see that we have the same update rule as in the case of SQDD \eqref{eq:sqdd_update}, but now for the averaged gradient over the minibatch.

For the minibatched Momentum Measurement Gradient descent, there are two options: accumulate momentum kicks sequentially in a quantum coherent fashion before measuring the momenta, or measuring the momentum shift for each data point and classically summing up the contributions for the parameter iteration.
Although the latter approach requires less coherence, more runs are necessary to get an accurate estimate of the momentum expectation value as compared to the former.
Let us consider the coherent momentum accumulation first, we start with a pointer state $\ket{\Psi^{(k)}}$ for the $k^\text{th}$ iteration, which we assume has a parameter space representation,
\begin{equation}
\begin{split}
\label{eq:form}
    \Psi^{(k)}(\bm{\Phi}) &= \frac{1}{|2\pi \bm{\Sigma}^{(k)}|^{1/4}}  e^{+i\bm{\Pi}^{(k)} \bm{\Phi}} \\
    &\qquad \quad \times e^{- \tfrac14 (\bm{\Phi}-\bm{\Phi}^{(k)})^{T} (\bm{\Sigma}^{(k)})^{-1} (\bm{\Phi}-\bm{\Phi}^{(k)})}.
\end{split}
\end{equation}
The coherently sequentially minibatched momentum measurement gradient descent approach then consists of applying all the QFB circuits for the loss function exponential of each data point in sequence:
\begin{equation}
    \ket{\Psi^{(k)}} \mapsto  \Big(\prod_{j\in \mathcal{B}_k}e^{-i\bar{\eta}_k \mathcal{L}_j(\bm{\hat{\Phi}})}\Big) \ket{\Psi^{(k)}} = e^{-i\eta_k \mathcal{J}_k(\bm{\hat{\Phi}})} \ket{\Psi^{(k)}}.
\end{equation}
Note that same notation as in SMQDD above was used.
If we then measure in momentum space (by applying the Fourier transform and measuring all the parameter registers), we can update the momentum expectation value
\begin{equation}\label{eq:momdate}
\begin{split}
    \bm{\Pi}^{(k+1)} &:= \braket{\Psi^{(k)}|\bm{\hat{\Pi}}|\Psi^{(k)}}\\
    &= \bm{\Pi}^{(k)} - \eta_k \braket{\Psi^{(k)}|\bm{\nabla} \mathcal{J}_k(\bm{\Phi})|\Psi^{(k)}} + \mathcal{O}(\eta_k^2).
\end{split}
\end{equation}
If the covariance matrix of the parameters is chosen to be diagonal with entries $(\sigma_n^{(k)})^2$ (where $n$ indexes the parameters), then the $n^\text{th}$ component of the gradient can be estimated to a precision (standard deviation) $1/2\sigma_n^{(k)} \sqrt{r}$ with $r$ runs of preparation and measurement.

In contrast, the classically accumulated minibatched momentum measurement gradient descent proceeeds by preparing a copy of the pointer state for each data point, i.e., $\ket{\Psi^{(k)}} = \bigotimes_{j\in \mathcal{B}_k}\ket{\Psi^{(k,j)}}$.
(Here we use the tensor product notation, but one could also consider measuring and resetting sequentially).
Assuming the pointer states $\ket{\Psi^{(k,j)}}$ are all identical copies of the form \eqref{eq:form}, by applying sequentially the exponential loss of each data point in the minibatch on the different copies
\begin{equation}
\begin{split}
    \ket{\Psi^{(k)}} \mapsto& \Big( \bigotimes_{j\in \mathcal{B}_k}e^{-i\bar{\eta}_k \mathcal{L}_j(\bm{\hat{\Phi}})} \Big)\ket{\Psi^{(k)}} \\
    &= \bigotimes_{j\in \mathcal{B}_k} e^{-i\bar{\eta}_k \mathcal{L}_j(\bm{\hat{\Phi}})} \ket{\Psi^{(k,j)}},
\end{split}
\end{equation}
measuring each copy's momenta, and classically summing up the results yields the expectation value 
\begin{equation}
\begin{split}
  \tfrac{1}{|\mathcal{M}_k|}\textstyle\sum_{j\in \mathcal{B}_k} & \braket{\Psi^{(k)}|\bm{\hat{\Pi}}^{(j)}|\Psi^{(k)}} \\  
  &= \bm{\Pi}^{(k)} - \eta_k \braket{\Psi^{(k)}|\bm{\nabla} \mathcal{J}_k(\bm{\Phi})|\Psi^{(k)}} + \mathcal{O}(\eta_k^2)
\end{split}
\end{equation}
Note that if the covariance matrix for all data points is diagonal with entries $(\sigma_n^{(k)})^2$, then with $r$ runs for each data point in the minibatch $\mathcal{M}_k$, the expectation value of the $n^\text{th}$ component of the gradient can be estimated to a precision $\sqrt{|\mathcal{M}_k|}/2 \sigma_n^{(k)} \sqrt{r}$.
Hence, an option is to increase the variance of the pointer states (in position) or to perform more runs.
Given a minimal variance of pointer states (e.g., limited number of qubits per parameter in the DV case, or limited squeezing levels in the CV case), then the coherent accumulation of momenta yields an advantage in terms of runs needed for a certain precision.
In any case, one then updates the average momentum parameter vector by the expectation value above,
\begin{equation}
    \bm{\Pi}^{(k+1)} := \bm{\Pi}^{(k)} - \eta_k \braket{\Psi^{(k)}|\bm{\nabla} \mathcal{J}_k(\bm{\Phi})|\Psi^{(k)}} + \mathcal{O}(\eta_k^2).
\end{equation}
We get the same result as with the coherently accumulated version, but possibly with a different precision as noted above. 

In both cases, once the momentum was updated, we can classically update the parameters expectation for the next round as
\begin{equation}\label{eq:pardate}
    \bm{\Phi}^{(k+1)} := \bm{\Phi}^{(k)} + \gamma_k \bm{\Pi}^{(k+1)}.
\end{equation}
Thus concludes an iteration of sequentially minibatched gradient descent, with either coherent or classical momentum accumulation. We can now consider how to parallelize the gradient accumulation over a minibatch.

\subsubsection{Coherently Accumulating Momentum Parallelization}\label{sec:CAMP}

An important development in the deployment of large neural networks is the possibility to parallelize the training over minibatch elements. Classically, this would be achieved by feeding forward the information of each data point and computing the gradient contribution on different replicas of the network, each replica running on different registers in spatially parallel fashion, either on different cores or different processors. Once the gradient contributions are computed, the replicas must communicate the gradient values to update their parameters synchronously. The same parallelization will be possible for Baqprop, but instead of adding gradients, we will accumulate momenta coherently. The non-coherent classical version of parallel accumulation of momenta adapts trivially from the above sequential version, hence we will focus on the coherent accumulation of momenta. We call this approach Coherently Accumulating Momentum Parallelization (CAMP).

The parallelism of CAMP relies on leveraging GHZ-like entanglement \cite{horodecki2009quantum} of the weights of the replicas. A central \textit{quantum parameter server} keeps in quantum coherent memory the weights at the beginning of an epoch, then by coherently adding the parameter values onto different replicas, acting the QFB circuit for each data point replica, and uncomputing the coherent addition, the central parameter server will have accumulated all the momenta contributions from the various replicas. This can be seen as extending the phase backpropagation through the computation that is the distribution and recollection of the parameter values via adder gates.

\begin{figure}[h!]
 \begin{center}
\includegraphics[width=0.4\columnwidth]{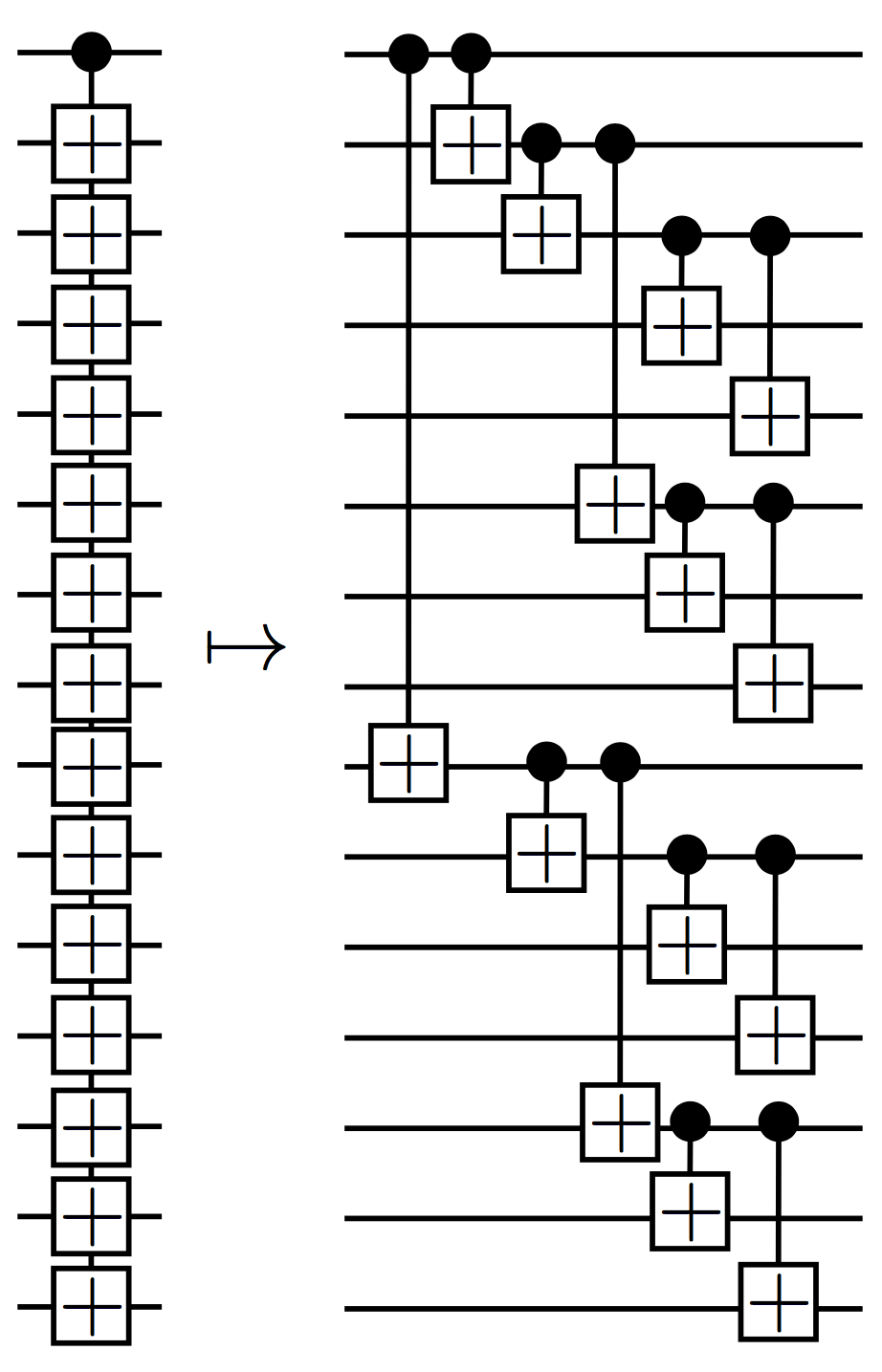}
\caption{Example of Tree Entangler (TENT) unitary with a binary tree structure. Left is a compact graphical representation for the TENT, and on the right is its expanded form. Each CV adder is as in equation \eqref{eq:addition}. For a complete $n$-ary tree structure of adders, we can create a GHZ state on $r+1$ registers in a depth $\mathcal{O} (n\log_n(r))$.
} \label{fig:TENT}
\end{center}
\end{figure}

Consider the parameter eigenstates of the various replicas to be labelled as $\ket{\bm{\Phi}}_{[c]}$ where $c$ is the index of the replica, with $c\!=\!0$ being the index of the central parameter server. For minibatch parallelization, we have a number of replicas equal to the minibatch size, as such, for the $k^\text{th}$ minibatch, $c\in\mathcal{B}_k$, we begin with the state of the central parameter server at iteration $k$, with the replica's parameters set to null (squeezed state for CV or null position state for qudit), i.e.,
\begin{equation}
    \ket{\Psi^{(k)}}_{[0]}\bigotimes_{c\in\mathcal{B}_k}\ket{\bm{0}}_{[c]} = \sum_{\bm{\Phi}} \Psi^{(k)}(\bm{\Phi}) \ket{\bm{\Phi}}_{[0]}\bigotimes_{c\in\mathcal{B}_k}\ket{\bm{0}}_{[c]}.
\end{equation}
Now, using adder gates (either CV or DV depending on the architecture), we transform this state to
\begin{equation}
    \ket{\Psi^{(k)}}_{[0]}\bigotimes_{c\in\mathcal{B}_k}\ket{\bm{0}}_{[c]} \mapsto \sum_{\bm{\Phi}} \Psi^{(k)}(\bm{\Phi}) \ket{\bm{\Phi}}_{[0]}\bigotimes_{c\in\mathcal{B}_k}\ket{\bm{\Phi}}_{[c]}.
\end{equation}
Effectively, we are applying a parameter replication unitary, which we will call the Tree Entangler (TENT), which adds the parameter values to the replicas, as in
\begin{equation}\label{eq:tent}
\hat{U}_{\textsc{tent}}=\prod_{c\in\mathcal{B}_k}e^{-i\bm{\hat{\Phi}}_{[0]}\otimes \bm{\hat{\Pi}}_{[c]}}.
\end{equation}
Rather than applying this addition sequentially as above, this addition can be achieved in logarithmic depth in the size of the minibatch, by using adders in a sequence shaped like a $n$-ary tree. This is a depth-optimal way to add standard basis values onto multiple target registers by using adders recursively, as to form a perfect (or complete) $n$-ary tree of adders. For a complete $n$-ary tree structure, we can create a GHZ state on $r+1$ registers in a depth $\mathcal{O} (n\log_n(r))$. Practically, the depth will be determined by the interconnectivity of the different sets of registers, as it is the case classically where the bottleneck of data parallelization is highly dependent on the interconnect between chips \cite{dean2012large}.
Figure~\ref{fig:TENT} provides a circuit diagram illustrating the Tree Entangler.

\begin{figure}[h!]
 \begin{center}
\includegraphics[width=1\columnwidth]{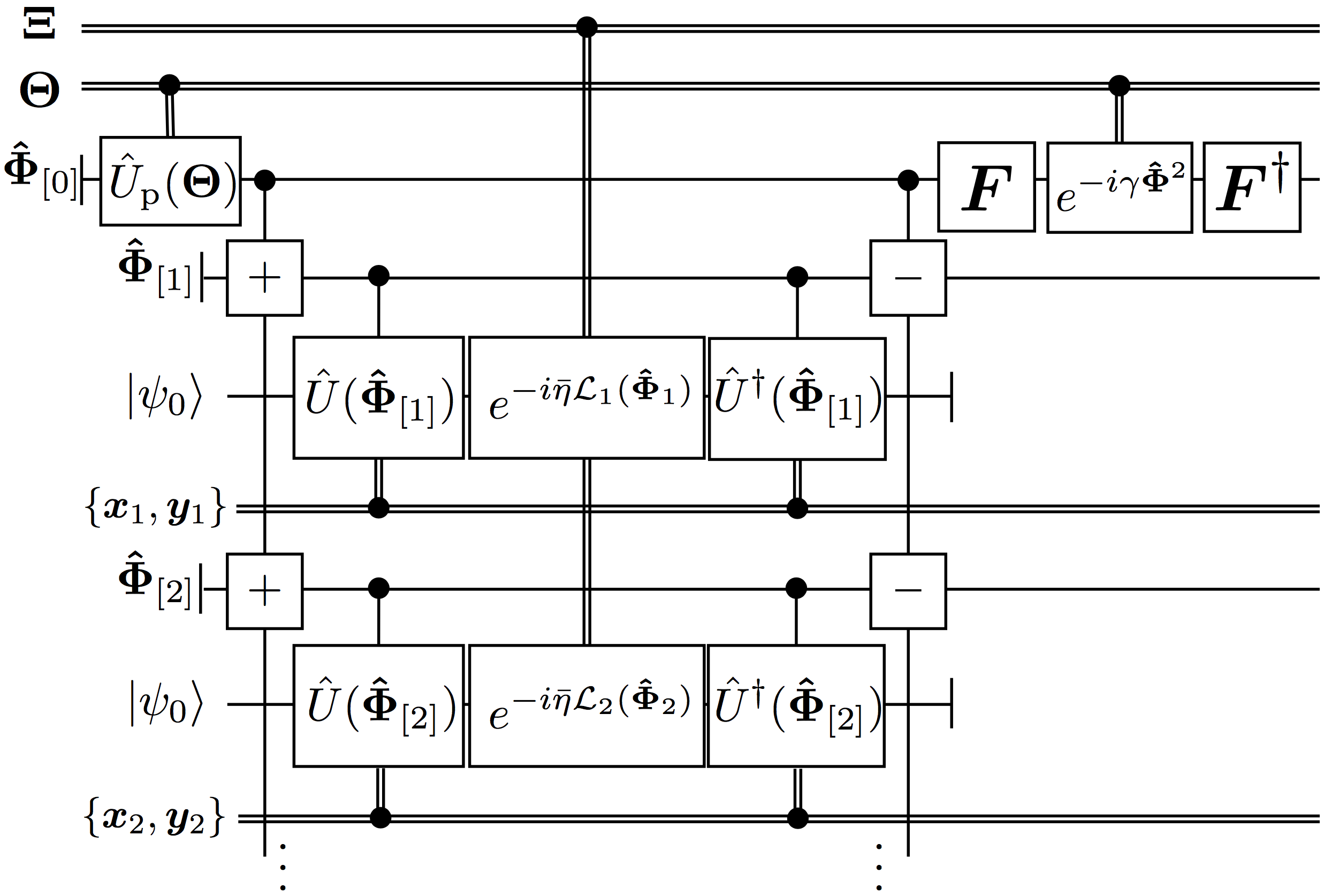}
\caption{Example of a first iteration of a Coherent Accumulation of Momentum Parallelization protocol (CAMP), applied to a Quantum Dynamical Descent (QDD), for classical data. The replica $\bm{\hat{\Phi}}_{[0]}$ is the quantum parameter server, the initial pointer state of the parameters is prepared in this register via the unitary $\hat{U}_\text{p}(\bm{\Theta})$, the TENT unitary (see Fig. \ref{fig:TENT}) is then applied to entangle the parameter server with the replicas $\bm{\hat{\Phi}}_{[j]}$. Following this, the QFB circuit is applied for a certain data point in each replica in a parallel fashion, then the TENT unitary is uncomputed (inverse TENT represented with [-] boxes). Finally the kinetic term exponential is applied on the parameter server.
} \label{fig:QDD_CAMP-og}
\end{center}
\end{figure}

Coherently Accumulating Momentum Parallelization (CAMP) consists of applying the TENT unitary, then applying the QFB circuit for each corresponding data point in the minibatch on the different replicas, and finally uncomputing the TENT. This is illustrated in Figure~\ref{fig:QDD_CAMP-og}.

For the minibatch of index $k$, the CAMP unitary would consist of:
\begin{equation}
      \hat{U}_{\textsc{camp},k} =  \hat{U}^\dagger_{\textsc{tent}}\Big(\bigotimes_{j\in \mathcal{B}_k}e^{-i\bar{\eta}_k \mathcal{L}_j(\bm{\hat{\Phi}}_{[j]})}\Big)\hat{U}_{\textsc{tent}}.
\end{equation}
We can compute the effect of this conjugation as
\spliteq{
      \hat{U}_{\textsc{camp},k} =  \prod_{j\in \mathcal{B}_k}e^{-i\bar{\eta}_k \mathcal{L}_j(\bm{\hat{\Phi}}_{[0]}+\bm{\hat{\Phi}}_{[j]})}.
}
We see that we get a simultaneous exponential loss function on both the parameter server and the replica, but since the replicas are initialized in null-parameter value pointer states, the effective unitary on the parameter server is the minibatch loss function 
\spliteq{
      \hat{U}_{\textsc{camp},k}\bigotimes_{c\in\mathcal{B}_k}\ket{\bm{0}}_{[c]} &=  \prod_{j\in \mathcal{B}_k}e^{-i\bar{\eta}_k \mathcal{L}_j(\bm{\hat{\Phi}}_{[0]})}\bigotimes_{c\in\mathcal{B}_k}\ket{\bm{0}}_{[c]}\\
      &= e^{-i\eta_k \mathcal{J}_k(\bm{\hat{\Phi}})}\bigotimes_{c\in\mathcal{B}_k}\ket{\bm{0}}_{[c]}.
}
 Hence we can consider CAMP as simply an ancilla-assisted way to enact the exponential loss function over the minibatch in a parallelized fashion,
 \begin{equation}
     e^{-i\eta_k \mathcal{J}_k(\bm{\hat{\Phi}})}.
 \end{equation}
 One main draw of this method is that there is a speedup to estimate the minibatch momentum update when done coherently as compared to a classically accumulated momentum.
 Assuming perfect null eigenstates, $\hat{\bm{\Phi}}\ket{\bm{0}}= 0$, initially in the replica's parameter registers, then inducing the GHZ-like entanglement, acting the phase kicks on each replica, and undoing the GHZ entanglement as described above, we can resolve the $n^\text{th}$ component of momentum within a standard deviation $1/2\sigma_n^{(k)}\sqrt{r}$ with $r$ runs, while classically accumulating momenta in different replicas one would get a $\sqrt{|\mathcal{M}_k|}/2\sigma_n^{(k)}\sqrt{r}$ standard deviation in the same number of runs. 
 
 Using our results from sequential minibatching from above, it is then straightforward to see how one applies CAMP to minibatched Quantum Dynamical Descent and minibatched Momentum Measurement Gradient Descent. For example, to perform paralellized Quantum Dynamical Descent, one applies the CAMP for each minibatch, using the initially null parameter replicas, interlaced with the kinetic pulses in the parameter server to update the parameter values: 
\spliteq{\label{eq:PQDD}
      \hat{U}_{\textsc{pqdd}} &= \prod_{k\in \mathcal{B}} e^{-i\gamma_k\bm{\hat{\Pi}}_{[0]}^2}U^\dagger_{\textsc{tent}}\Big(\bigotimes_{j\in \mathcal{B}_k}e^{-i\bar{\eta}_k \mathcal{L}_j(\bm{\hat{\Phi}}_{[j]})}\Big)\hat{U}_{\textsc{tent}}\\
      &= \prod_{k\in \mathcal{B}} e^{-i\gamma_k\bm{\hat{\Pi}}_{[0]}^2} \hat{U}_{\textsc{camp},k}.
}
For Momentum Measurement Gradient Descent, one can apply CAMP on the $k^\text{th}$ iteration pointer state
\begin{equation}
   \hat{U}_{\textsc{camp},k}  \ket{\Psi^{(k)}}_{[0]}\bigotimes_{c\in\mathcal{B}_k}\ket{\bm{0}}_{[c]}
\end{equation}
and follow the same steps and classical updates of the coherently sequentially minibatched MoMGrad, as in \eqref{eq:momdate} and \eqref{eq:pardate}.

The technique of employing GHZ-entanglement to get a sensitivity speedup for phase sensing is well known and widely used in the quantum sensing literature \cite{degen2017quantum}. In this literature, this speedup is usually achieved in the context of discrete qubit/qudit pointers.

\subsubsection{Quantum Random Access Memory Mini-batching}\label{sec:qramm}

One may consider attempting to train the algorithm using multiple data points of a mini-batch in superposition.
A possible means for doing mini-batching with a superposition of data is to use Quantum Random Access Memory (QRAM) to generate a sum of input states (along with an address register) for every data point in the mini-batch.
The general form of the entangled address-input state is:
\begin{equation}
  \ket{\bar{\xi}}_{\textsc{ac}} := \tfrac{1}{\sqrt{|\mathcal{B}_k|}} \sum_{j\in\mathcal{B}_k} \ket{j}_\textsc{a} \ket{\xi_{j}}_{\textsc{c}},
\end{equation}
where $j \in \mathcal{B}_k$ is an address for a data point in the mini-batch, and $\ket{\xi_j}_\textsc{c}$ is an input state on the computational Hilbert space, $\mathcal{H}_\textsc{c}$, associated with the corresponding data point.

In this approach, the parametric unitary, $\hat{U}(\bm{\hat{\Phi}})$, acting on $\mathcal{H}_{\bm{\Phi}} \otimes \mathcal{H}_\textsc{c}$ remains unchanged.
The loss exponential becomes a controlled-loss exponential, with the address register as the control,
\begin{equation}
  \sum_{j\in\mathcal{B}_k} \ket{j} \!\bra{j}_A \otimes e^{-i\eta \hat{L}_j},
\end{equation}
where $\hat{L}_j$ is the loss function associated with data point $j$.
The full QFB circuit in this case is, 
\spliteq{
  &\hat{U}^\dagger(\bm{\hat{\Phi}}) \Big( \sum_{j\in\mathcal{B}_k} \ket{j} \!\bra{j}_A \otimes e^{-i\eta \hat{L}_j} \Big) \hat{U}(\bm{\hat{\Phi}})  \ket{\bar{\xi}}_{AC}\\
  &\qquad \quad = \tfrac{1}{\sqrt{|\mathcal{B}_k|}} \sum_{j\in\mathcal{B}_k} \ket{j}_A \otimes e^{-i\eta \hat{L}_j(\bm{\hat{\Phi}}) }\ket{\xi_{j}}_C,
}
where $\hat{L}_j(\bm{\hat{\Phi}}) := \hat{U}(\bm{\hat{\Phi}})^\dagger  \hat{L}_j \hat{U}(\bm{\hat{\Phi}})$.
To examine the effect this has on training the parameters, we can compute the effective phase operator, $e^{-i\eta \mathcal{J}_k(\bm{\hat{\Phi}})}$, (again which holds on average, to first order in $\eta$), by tracing out both the address and compute registers ($\textsc{a}$ and $\textsc{c}$):
\begin{equation}
    \mathcal{J}_k(\bm{\hat{\Phi}}) = \tfrac{1}{|\mathcal{B}_k|}\sum_{j\in\mathcal{B}_k}\bra{\xi_{j}}\hat{L}_j(\bm{\hat{\Phi}})\ket{\xi_{j}}_\textsc{c}+\mathcal{O}(\eta^2).
\end{equation}

Thus we see that in the end we obtain the same effective phase as in either sequential mini-batching or CAMP.
Let us briefly compare the CAMP and QRAM approaches to mini-batching with a rough analysis of the errors in the effective cost function.
For CAMP, the error is $\mathcal{O}(|\mathcal{B}_k| \times \eta^2/|\mathcal{B}_k|^2)=\mathcal{O}( \eta^2/|\mathcal{B}_k|)$, since we have $|\mathcal{B}_k|$ copies each with error $\mathcal{O}(\eta^2/|\mathcal{B}_k|^2)$.
For QRAM, the error is $\mathcal{O}(\eta^2)$, since we have a single copy of QFB applying a phase kick with kicking rate $\eta$.
The sum in QRAM is obtained from averaging over multiple runs.
Thus, in QRAM, we are applying larger kicks and the cost function is obtained stochastically over the data points, whereas in CAMP we add logarithmic depth to the circuit in order to run a separate instance of QFB for each data point in parallel, and the phase kicks are accumulated coherently.
Even though in CAMP one has to add this logarithmic depth to the circuit, it seems that the error in the effective cost function is suppressed by the size of the mini-batch, and one also would not have the difficulty of building a QRAM \cite{giovannetti2008quantum}.

Furthermore, for classical data, the CAMP procedure for each data point is single-shot, because the computational register is left in a computational eigenstate at the end of the QFB procedure.
For quantum data, one would need to run the QFB procedure multiple times in order to obtain the average over the computational registers in the effective cost function.

Thus, it would seem that CAMP may be advantageous to QRAM in certain contexts.
We leave a more careful analysis of the overhead needed in both approaches for future work.

\subsection{Discrete Parametric Optimization}\label{sec:discrete}

In some cases, instead of optimizing over a continuous space of parameters, one may want to restrict the optimization to a discrete subspace of parameters. In this subsection, we consider how to perform both QDD and MoMGrad in order to optimize discretely-parametrized quantum circuits. Furthermore, we propose a way to embed a given discrete parametric optimization into the continuous-variable versions of the QDD and MoMGrad protocols via a specific choice of regularizing potential. Finally, we show how to approximate continuum gradients as a finite-difference using single qubits for each of the parameters.

Let us first formally define what we consider to be discrete parametric optimization. If $\mathcal{P}$ is the index set of parameters, then in previous considerations in this paper, the space of parameters was $\mathds{R}^{|\mathcal{P}|}$. Using $n$-qubit precision simulated continuous registers, whose parameter values form a lattice isomorphic to $\mathds{Z}_{2^n}^{|\mathcal{P}|}$ we considered approximating the set of possible parameters on some interval of $\mathds{R}^{|\mathcal{P}|}$, as described in sec. \ref{sec:bkgd}). We consider discrete optimization to be the case where a subset of parameters, $\{\Phi_j\}_{j\in \mathcal{S}}$, for some index subset $\mathcal{S}$, are discrete.
The optimization is then over $\mathds{R}^{|\mathcal{C}|}\times \mathds{Z}_{d}^{|\mathcal{S}|}$, where $\mathcal{C}:=\mathcal{P}\setminus \mathcal{S}$. In this subsection, we will mainly focus on having a hybrid set of parameters where some parameters are binary and some are continuous, i.e., hybrid discrete-continuous parametric optimization over $\mathds{R}^{|\mathcal{C}|}\times \mathds{Z}_{2}^{|\mathcal{S}|}$.

\subsubsection{Kicking Hybrid Discrete-Continuous Parameters}
Consider the case where we would like a subset of the parameters to be binary. In a nutshell, the strategy will be to replace the quadratures of the continuous parameters $\{\hat{\Phi}_j,\hat{\Pi}_j\}_{j\in\mathcal{S}}$ for Pauli operators $ \{\hat{Z}_j,\hat{X}_j\}_{j\in\mathcal{S}}$, in the various instances where the quadratures play a role in the optimization procedures. 

For this discrete continuous-binary hybrid parametric optimization, we start by going from a a continuous parametric feedforward unitary $\hat{U}(\bm{\hat{\Phi}})$ to a hybrid continuous-discrete-parametric feedforward unitary $\hat{U}(\bm{\hat{\Phi}},\bm{\hat{Z}})$ of the form
\begin{equation}\label{eq:hybrid_u}
  \hat{U}(\bm{\hat{\Phi}},\bm{\hat{Z}}) =  \!\!\! \sum_{\bm{\Phi}\in \mathds{R}^{|\mathcal{C}|}} \sum_{\bm{b}\in \mathds{Z}_2^{|\mathcal{S}|}}\!\! \ket{\bm{\Phi}}\!\bra{\bm{\Phi}} \otimes \ket{\bm{b}}\!\bra{\bm{b}} \otimes \hat{U}(\bm{\Phi},\bm{b}),
\end{equation} where $\ket{b_j}$ are the eigenstates of $\hat{Z}_j$ of eigenvalues $( -1)^{b_j}$, and $\ket{\bm{b}}\equiv \bigotimes_{j\in \mathcal{S}}\ket{b_j}$. 

For above hybrid-parametric unitary, given a batch of loss operators $\{\hat{L}_k\}_{k\in\mathcal{B}}$, the quantum feedforward and Baqprop circuit for the datum of index $k\in\mathcal{B}$ would have an induced effective phase kick of the form
\begin{equation}\label{eq:kickit}
  e^{-i\bar{\eta} \mathcal{L}_k(\bm{\hat{\Phi}},\bm{\hat{Z}})} = \braket{ \hat{U}^\dagger(\bm{\hat{\Phi}},\bm{\hat{Z}}) e^{-i\bar{\eta} \hat{L}_k}   \hat{U}(\bm{\hat{\Phi}},\bm{\hat{Z}})}_\textsc{c}
\end{equation}
where $\bm{\hat{Z}}$ is a vector of Pauli operators, $\bm{\hat{Z}} = (\hat{Z}_j)_{j\in\mathcal{S}}$, and the expectation value is taken over the initial state of the compute register as in \eqref{eq:eff_phase}. Here the rate $\bar{\eta} = \eta/|\mathcal{B}|$ is the phase kicking rate $\eta$ divided by the size of the batch. By concatenating multiple of these effective phase kicks in sequence or in parallel (see sec. \ref{sec:batch} for techniques to do so), we can get an effective phase kick for the full batch cost function
\begin{equation}\label{eq:phase_disc}
  e^{-i \eta \mathcal{J}(\bm{\hat{\Phi}},\bm{\hat{Z}})} = \prod_{k \in \mathcal{B}} e^{-i \bar{\eta} \mathcal{L}_k(\bm{\hat{\Phi}},\bm{\hat{Z}})}.
\end{equation}
We now have an effective phase kick on both the discrete and continuous parameters for the full cost function. 

Before we derive an update rule induced by this effective phase kick, let us define some notation and formalism to treat functions of Paulis.
Starting with a single qubit, any function, $f$, of the operator $\hat{Z}$ can be written as
\begin{equation}\label{eq:discrete_gradient}
\begin{split}
  f(\hat{Z}) &= f(1) \ket{0}\bra{0} + f(-1) \ket{1}\bra{1} \\
  &= \tfrac12 [ f(1) + f(-1) ] \hat{I} + \tfrac12 [ f(1) - f(-1) ] \hat{Z}.
\end{split}
\end{equation}
Then we define the derivative of $f$ with respect to the Pauli-Z operator as the coefficient of $\hat{Z}$ in the above expression, i.e.,
\begin{equation}
    \partial_{\hat{Z}} f(\hat{Z}) = \tfrac12 [ f(1) - f(-1) ].
\end{equation}
Note that since this derivative is simply a coefficient, as an operator it is taken to be proportional to the identity.
Hence we see that the commutator of a function of $\hat{Z}$ with $\hat{X}$ will give
\begin{equation}
    [ f(\hat{Z}), \hat{X} ] = 2 i \partial_{\hat{Z}} f(\hat{Z}) \hat{Y}.
\end{equation}
In general one can have a function of Pauli operators of multiple registers, $\bm{\hat{Z}} = \{\hat{Z}_j\}_{j=1}^M$,
which in general has a decomposition as a polynomial of Paulis,
\begin{equation}
    f(\bm{\hat{Z}}) = \sum_{\bm{b}\in\mathds{Z}_2^M} \alpha_{\bm{b}}\bm{\hat{Z}}^{\bm{b}},\quad\bm{\hat{Z}}^{\bm{b}}:=\Big(\textstyle\bigotimes_{j=1}^M\hat{Z}^{b_{j}}_j\Big), 
\end{equation}
where $\alpha_k\in \mathds{R}, b_{j} \in\mathds{Z}_2$ for all $j$ and $k$. For such multi-operator function, we define the partial derivative as
\begin{equation}\label{eq:comm_disc}
 \partial_{\hat{Z}_k}\!f(\bm{\hat{Z}}):= \sum_{\bm{b}\in\mathds{Z}_2^M}\alpha_{\bm{b}}\,\delta_{1b_k }  \Big(\textstyle\bigotimes_{j\neq k}\hat{Z}^{b_{j}}_j\Big)
\end{equation}
which is the sum over terms which had a non-null power of $\hat{Z}_k$, and for such terms the power of $\hat{Z}_k$ is removed. Finally notice that if we look at the commutator of $f(\bm{\hat{Z}})$ with $\hat{X}_k$, we get 
\begin{equation}\label{eq:disc_diff}
      [ f(\bm{\hat{Z}}), \hat{X}_k ] = 2 i \partial_{\hat{Z}_k} f(\bm{\hat{Z}})\otimes \hat{Y}_k  .
\end{equation}

Now, using this formalism, we can derive an update rule due to an effective phase kick of the form featured in equation \eqref{eq:kickit},
\spliteq{\label{eq:disc_update}
 & \braket{ e^{i\eta \mathcal{L}(\bm{\hat{\Phi}},\bm{\hat{Z}})} \hat{X}_k e^{-i\eta \mathcal{L}(\bm{\hat{\Phi}},\bm{\hat{Z}})}} \\ &= \braket{ e^{i\eta\, \text{ad}[\mathcal{L}(\bm{\hat{\Phi}},\bm{\hat{Z}})]} (\hat{X}_k)}\\&= \braket{ \cos(2\eta\partial_{\hat{Z}_k}\mathcal{L}(\bm{\hat{\Phi}},\bm{\hat{Z}}))\otimes \hat{X}_k} \\
 &\qquad \qquad \quad- \braket{ \sin(2\eta\partial_{\hat{Z}_k}\mathcal{L}(\bm{\hat{\Phi}},\bm{\hat{Z}}))\otimes \hat{Y}_k}\\
 &= \braket{\hat{X}_k} -2\eta\braket{  \partial_{\hat{Z}_k}\mathcal{L}(\bm{\hat{\Phi}},\bm{\hat{Z}})\otimes \hat{Y}_k} +\mathcal{O}(\eta^2).
}
Now that we have derived an update rule from the phase kick induced by the hybrid-parametric QFB, we can leverage this to perform hybrid-parametric variants of both Quantum Dynamical Descent (QDD) and Momentum Measurement Gradient Descent (MoMGrad).

\subsubsection{Continuous-Discrete Hybrid QDD}

We begin with the hybrid-parametric variant of QDD. As established in Section \ref{sec:qdd}, QDD with continuous quantum parameters is a form of Quantum Alternating Operator Ansatz (QAOA). Thus, naturally, a continuous-discrete variable hybrid QDD should be analogous to both the continuous-variable QDD and a discrete variable QAOA \cite{farhi2014quantum}. 

Our task is to optimize the cost function $\mathcal{J}(\bm{\hat{\Phi}},\bm{\hat{Z}})$, via alternating exponentials of operators. In order to construct such an optimization scheme, we draw inspiration from discrete QAOA. In typical discrete QAOA for qubits, the cost Hamiltonian is a function of the standard basis operators $\bm{\hat{Z}} = \{\hat{Z}_j\}_j$, i.e.
\begin{equation}
    \hat{H}_{\textsc{c}}=f(\bm{\hat{Z}}).
\end{equation}

For such a cost Hamiltonian, the traditional choice of mixer Hamiltonian is one of the form 
\begin{equation}
   \hat{H}_{\textsc{m}}= \sum_j \hat{X}_j,
\end{equation}
i.e. the sum of Pauli $\hat{X}$ operators. 

To some extent, the Pauli $\hat{Z}$ and $\hat{X}$ operators are to a qubit what $\hat{\Phi}$ and $\hat{\Pi}$ are to a harmonic oscillator (continuous quantum register). The Hadamard gate is the qubit's analogue of a discrete Fourier transform, and conjugation of $\hat{Z}$ by Hadamards gives $\hat{H}\hat{Z}\hat{H} =\hat{X}$. Recalling the formalism from section \ref{sec:bkgd}, the $\hat{\Pi}$ quadrature for a qudit is the Fourier conjugate to the position operator $\hat{\Phi}$. As such, in the context of this analogy, the mapping $\{\bm{\hat{\Phi}},\bm{\hat{\Pi}}\}\mapsto \{\bm{\hat{Z}},\bm{\hat{X}}\}$ of continuous to discrete operators is sensible. 

Although there is some analogy between $\hat{\Pi}$ and $\hat{X}$, this analogy has its limits as there are a few differences to keep in mind. Looking at the update rule in equation \eqref{eq:disc_diff}, we can contrast this with an analogous formula for functions $f(\bm{\hat{\Phi}})$ of the continuous parameter variables
\begin{equation}
     [ f(\bm{\hat{\Phi}}), \hat{\Pi}_k ] = i \partial_{\hat{\Phi}_k}\!f(\bm{\hat{\Phi}}) .
\end{equation}
We again get a derivative of the function, but in contrast to \eqref{eq:disc_diff}, there is no remaining operator with support on the $k^\text{th}$ register which tensored with the derivative, we have the identity on this register instead. Furthermore, for continuous QDD (see sec. \ref{sec:qdd}), the mixer Hamiltonian is $\sim\bm{\hat{\Pi}}^2$, but Pauli operators are involutory $\hat{X}^2=\hat{I}$, hence we will have to use $\hat{X}_j$ as the mixer Hamiltonian for each discrete parameter. 

Now that we have built some further intuition, we can proceed to building the continuous-discrete hybrid QDD procedure. For this hybrid QDD, the analogue to the cost Hamiltonian is the full-batch the effective phase operator from \eqref{eq:phase_disc}, 
\begin{equation}
    \hat{H}_\textsc{c} = \mathcal{J}(\bm{\hat{\Phi}},\bm{\hat{Z}})
\end{equation}
Since this effective phase
function is dependent on both the continuous parameter operators $\bm{\hat{\Phi}}$ and on the Pauli operators $\bm{\hat{Z}}$, we need to add mixer terms of both the continuous and discrete type. Thus, we chose the mixer Hamiltonian
\begin{equation}
    \hat{H}_{\textsc{m}}=\beta \sum_{k\in\mathcal{S}} \hat{X}_k + \sum_{j\in\mathcal{C}}\hat{\Pi}_j^2
\end{equation}
where $\beta$ is a hyperparameter which can serve as a modifier of the ratio between kinetic rates of the discrete and continuous parameters. Now that we have defined can defined our cost and mixer Hamiltonians, we can write out the whole hybrid-parametric QDD unitary for multiple epochs as
\begin{equation}\label{eq:hqdd}
    \hat{U}_{\textsc{hqdd}} = \prod_j  (e^{-i\gamma_j\bm{\hat{\Pi}}^2} \otimes e^{-i\beta\gamma_j\bm{\hat{X}}})e^{-i\eta_j \mathcal{J}(\bm{\hat{\Phi}},\bm{\hat{Z}})},
\end{equation}
where we use the following notation for the discrete mixer exponential
\begin{equation}
    e^{-i\beta\gamma_j\bm{\hat{X}}} := e^{-i\beta\gamma_j\sum_{k\in\mathcal{S}}\hat{X}_k}= \bigotimes_{k\in\mathcal{S}} e^{-i\beta\gamma_j\hat{X}_k}.
\end{equation}
In general, one could vary $\beta$ for each epoch, i.e. $\beta\mapsto \beta_j$ in \eqref{eq:hqdd}. This would allow for an independently variable kinetic rate specific to the discrete registers, as it could then be chosen as different from the continuous parameters' kinetic rate at each epoch. In general, one could go as far as having specific kinetic and kicking rates for each parameter, at the cost of having to optimize more hyper-parameters.

Now, we see in equation \eqref{eq:hqdd} that the QDD unitary becomes a hybrid continuous-discrete QAOA-type unitary. For a discrete QAOA \cite{farhi2014quantum}, the hyperparameters must usually be optimized in order to minimize the cost function. In this case, our hybrid QAOA's cost function is the QFB-induced phase  $\mathcal{J}(\bm{\hat{\Phi}},\bm{\hat{Z}})$, which is a function of both the continuous and discrete parameters. By optimizing the the hyperparameters $\{\gamma_j, {\beta}_j, \eta_j\}$ of the above Hybrid QDD, we can expect to minimize the cost function. 

We have established above that the hybrid QDD is like a QAOA problem, with $\mathcal{J}(\bm{\hat{\Phi}},\bm{\hat{Z}})$ as the QAOA cost function (Hamiltonian). From previous literature on discrete QAOA \cite{farhi2014quantum}, we know we can find an approximate minimum of the cost function by optimizing the hyper-parameters such as to minimize the cost function. These hyper-parameters are themselves continuous parameters and thus could be optimized via Meta-QDD or Meta-MoMGrad approach (covered in the Quantum Meta-Learning Section~\ref{sec:meta}) or via quantum-classical methods. After the QDD, as discussed in section \ref{sec:qdd}, the continuous parameters should concentrate near a local minimum in the continuous landscape, and from discrete QAOA \cite{farhi2014quantum} we expect a superposition of bitstrings in the discrete parameters which statistically favors bitstrings of low cost function value. Thus by jointly measuring both the continuous parameters with the discrete parameters, after applying the optimized hybrid QDD unitary, one should then obtain a set of discrete and continuous parameters for which the cost function is relatively low value with high probability. Once these measurements yield a set of classical parameters deemed sufficiently optimal, one can then perform inference with a classical-parametric variant of the feedforward circuit, as described in section \ref{sec:phase-kick}.

\subsubsection{Continuous-Discrete Hybrid Momentum Measurement Gradient Descent}

Now that we have derived a hybrid continuous-discrete variant of QDD, using the update rule derived in \eqref{eq:disc_update}, we can derive a hybrid variant of Momentum Measurement Gradient Descent (MoMGrad). 

To add to the intuition provided by the operator update rule in \eqref{eq:disc_update}, we can understand the full-batch effective phase kick as a conditional rotation of each qubit, conditioned on other parameters. To see this let us first define the notation $\bm{z}_{\bm{b}}\equiv (-1)^{\bm{b}}$ as the vector of eigenvalues for the bitstring $\bm{b}$. We then can rewrite cost-phase gate induced by the QFB circuit as a controlled-rotation of the qubit $k$ about its $Z$ axis, for any $k\in \mathcal{S}$,
\begin{equation}
    e^{-i\eta \mathcal{J}(\hat{\bm{\Phi}},\bm{\hat{Z}})} = \ket{\bm{\Phi}}\!\bra{\bm{\Phi}}\!\!\!\!\bigotimes_{j\in \mathcal{S}\setminus\{k\}}\!\!\!\!\ket{b_j}\!\bra{b_j}\, \otimes e^{-i\eta\hat{Z}_k\partial_k \mathcal{J}(\bm{\Phi},\bm{z}_{\bm{b}})}.
\end{equation}
We see that, conditioned on all parameters other than that of qubit $k$, the gate is effectively a rotation of the form $e^{i\varphi_k \hat{Z}_k}$ where the the angle $\varphi_k$ of rotation is $\varphi_k= -\eta\partial_k \mathcal{J}(\bm{\Phi},\bm{z}_{\bm{b}})$, i.e., proportional to the negative gradient. This controlled-rotation interpretation further provides intuition for the update rule derived in \eqref{eq:disc_update}.

\begin{figure}[h!]
 \begin{center}
\includegraphics[width=0.6\columnwidth]{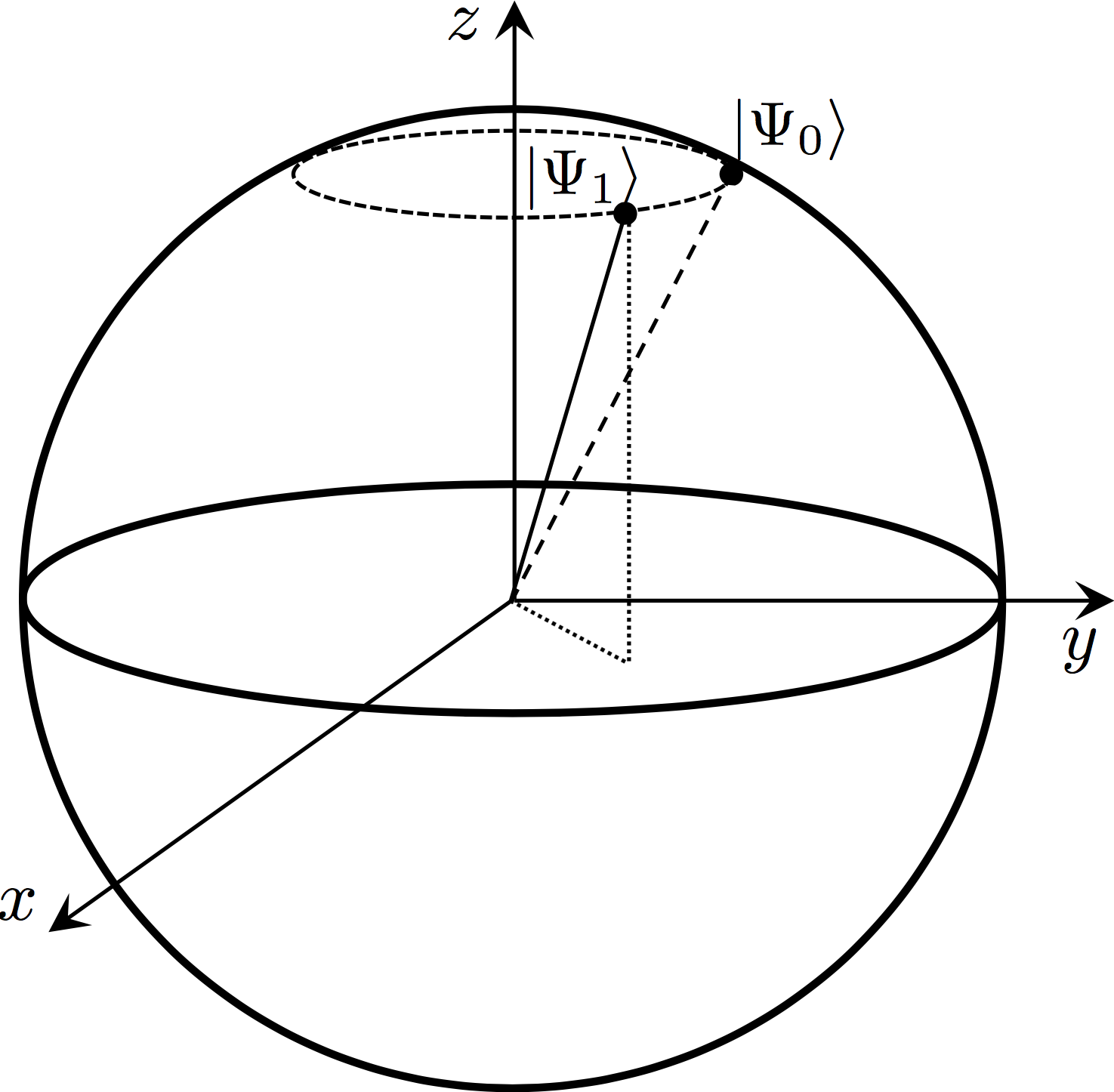}
\caption{Bloch sphere diagram of single-qubit gradient estimation. $\ket{\Psi_1} = e^{i\eta \mathcal{J}(\bm{\hat{\Phi}},\bm{\hat{Z}})}\ket{\Psi_0}$ the longitudinal angle of rotation relative to the $z$-$y$ plane is given by $2\eta\partial_{\hat{Z}_k}\mathcal{J}(\bm{\hat{\Phi}},\bm{\hat{Z}})$, hence allowing for estimation of the gradient of the cost function via multiple runs and measurements of the expectation value of the Pauli $\hat{X}$.
} \label{fig:rot}
\end{center}
\end{figure}

Now, using the update rule \eqref{eq:disc_update}, we can see that given an initial state of qubit $k$ in the $z$-$y$ plane of the Bloch sphere, we will be able to read off the gradient given small rotations. That is, given an initial state of the form
\begin{equation}\label{eq:disc_point}
    \ket{S_0} = \bigotimes_{j\in \mathcal{S}}\left(\cos(\theta_j)\ket{0}_j + i \sin(\theta_j)\ket{1}_j\right)
\end{equation}
which all have $\braket{\hat{X}_k}=0$, and
where the $\theta_j$ are hyper-parameters, which we consider as classical parameters for the time being. By applying the QFB phase kick, each qubit is effectively rotated about the $z$ axis, hence the state travels laterally at a certain lattitude in the Bloch sphere, as depicted in Figure \ref{fig:rot}. This rotation then converts initially null $x$ component of the state to one depending on the initial latitude on the Bloch sphere, and on the gradient of the cost function, as depicted in Figure \ref{fig:rot}. By measuring the expectation value $\braket{\hat{X}_k}$ after the QFB phase kicks from \eqref{eq:phase_disc}, for pointer states of the form of \eqref{eq:disc_point}, we can get the gradient as follows,  
\spliteq{
    \bra{S_0}& e^{i\eta \mathcal{J}(\bm{\hat{\Phi}},\bm{\hat{Z}})} \hat{X}_k e^{-i\eta \mathcal{J}(\bm{\hat{\Phi}},\bm{\hat{Z}})}\ket{S_0}\\ &= -2\eta\sin(2\theta_k) \braket{  \partial_{\hat{Z}_k}\mathcal{J}(\bm{\hat{\Phi}},\bm{\hat{Z}})} +\mathcal{O}(\eta^2).
}
We can then use this measurement of the gradient to update the parameters $\theta_k\mapsto \theta_k -\eta \gamma\braket{  \partial_{\hat{Z}_k}\mathcal{J}(\bm{\hat{\Phi}},\bm{\hat{Z}})}$ for all $k$, where $\gamma$ is a hyperparameter, while simultaneously performing the MoMGrad for the continuous parameters using update rules discussed in \ref{sec:momgrad}. 
Thus, one can perform a continuous-discrete hybrid MoMGrad. Once a sufficient number of iterations has been performed, since the $\bm{\theta}$ parameters are classical parameters which are known, one can round each of the binary parameters to the most probable value ($b^\ast_k =\lfloor\sin^2(\theta_k)\rceil$). One can then perform further MoMGrad on the continuous parameters $\bm{\Phi}$ for the hybrid cost function with the classical binary parameters $\bm{b}^\ast$ in the feedforward, i.e., perform continuous MoMGrad with the cost function $\mathcal{J}(\bm{\hat{\Phi}},\bm{z}_{\bm{b}^\ast})$. Finally one obtains sufficiently optimal continuous parameters $\bm{\Phi}^\ast$ which can then allow for inference with the hybrid discrete-continuous classically parametrized operation $\hat{U}(\bm{\Phi}^\ast,\bm{z}_{\bm{b}^\ast})$.

\subsubsection{Continuum-Embedded Discrete Optimization}
In this section we discuss the possibility of performing optimization over a discrete subset of hypotheses by embedding the problem into the continuum and adding a regularizing potential which forces the weights to converge onto the closest discrete value.

As was discussed above one can use discrete variable quantum registers, such as a qubits, in order to perform such a discrete optimization. These discrete optimization approaches came with caveats, namely, the hybrid MoMGrad necessitates multiple QFB runs and measurements to get an accurate estimate of the gradient, while the hybrid QDD requires hyperparameter optimization to yields approximately optimal bitstrings. By embedding the discrete parametric optimization into a continuous parametric optimization, we can directly leverage the continuous parameter variants of MoMGrad and QDD from section \ref{sec:opt}, rather than have to modify the latter to accommodate discrete parameters.
The approach which will be outlined in this subsection will be consist of adding a regularization potential to the continuous parameters which effectively turns these into a simulated qubit (or qudit). 

If we have some subset of parameters $\{\Phi_j\}_{j\in \mathcal{S}}$ (where $\mathcal{S}$ is the subset of indices) which we would like to be either $0$ or $\lambda_j$, then, for either quantum dynamical descent or momentum measurement gradient descent, we can add a regularizing potential to the cost: $\mathcal{J}(\bm{\hat{\Phi}}) \mapsto \mathcal{J}(\bm{\hat{\Phi}}) +V_{\mathcal{S}}(\bm{\hat{\Phi}})$, where the regularizing potential is of the kind
\begin{equation}\label{eq:quartic}
      V_{\mathcal{S}}(\bm{\hat{\Phi}}) \equiv \sum_{j\in \mathcal{S}}\tfrac{\omega_j^2}{2}\Big(\left(\hat{\Phi}_j-
    \tfrac{\lambda_j}{2}\right)^2 - \left(\tfrac{\lambda_j}{2}\right)^2\Big)^2 
\end{equation}
which is a sum of double-well potentials, each with wells centered at $0$ and $\lambda_j$.
To simulate QDD with this potential, one would need to apply the unitary
\begin{equation}\label{eq:rqdd}
    \hat{U}_{\textsc{qdd}} = \prod_j e^{-i\gamma_j\bm{\hat{\Pi}}^2}e^{-i\eta_j\mathcal{J}(\bm{\hat{\Phi}})}e^{-i\eta_jV_{\mathcal{S}}(\bm{\hat{\Phi}}) },
\end{equation}
with the quartic potential exponential being applicable through the use of multiple qubic phase gates with CV techniques \cite{marshall2015repeat,lau2017quantum}, or in the DV case using $\mathcal{O}((\log d)^4)$ 4-local Pauli terms (where $d$ is the range of the qudit parameter register, see section \ref{sec:bkgd}), similarly to how exponentials of $\hat{\Phi}^2$ are implemented, see section \ref{sec:reg} for details.

Although an extensive analysis of the dynamics of quartic oscillators is beyond the scope of this paper, there has been extensive physics literature on the subject \cite{jelic2012double}, hence we will stick to a qualitative understanding of its dynamics for our discussion. Notice that Taylor expanding the above potentials at points $\Phi_j=\{0,\lambda\}$, we see that each well is locally like a harmonic oscillator of harmonic frequency $\lambda_j \omega_j$. The ground states of each parameter are then approximately equal to a symmetric superposition of the ground states of each of the wells. Thus the added potential induced by a cost function $\mathcal{J}(\bm{\hat{\Phi}})$ can bias the effective potential on the parameters and break this energetic degeneracy, a specific well will then be favoured as having lower energy. Thus naturally a metaheuristic like Quantum Dynamical Descent or MoMGrad would nudge the parameter wavefunction into the well which minimizes $V_\mathcal{S} + \mathcal{J}$. Once the parameter wavefunction is concentrated into one of the wells, assuming it approximates the ground state of the well, its expectation value can be estimated with few measurements (the determinant of the covariance should be around $\sim\prod_{j\in\mathcal{S}}(\omega_j\lambda_j)^{-\frac{1}{2}}$ due to the locally harmonic dynamics of the well). One can then round to the closest value of each parameter on the lattice.

A near-term implementation of this approach might favour using an analog flux qubit for simplicity, but in this case the poor readout capabilities will limit precision and thus speed of execution of the optimization. Simulating a similar system at a high-level of precision at the logical level will allow for smoother state transitions, the ability to resolve a continuum of momenta and parameter values, and generally will allow for finer-grained dynamics by having the parameter query a continuum of possibilities. This is thus the long-term favourable approach.

\subsubsection{Estimating Continuum Gradients with Single Qubits}
Before we move on from this subsection, now that we have treated how we can embed a discrete optimization into continuous parameters, let us very briefly mention how we can use a low-dimensional discrete system such as a qubit to perform apprximate MoMGrad (\ref{sec:momgrad}). This is a straightforward application of the fairly general formalism developed in Section~\ref{sec:bkgd}. A reason this is worth mentioning is because this single-qubit readout might be relevant in the short-term for Noisy Intermediate Scale Quantum devices \cite{preskill2018quantum}, where the number of quantum degrees of freedom and level of control is limited.
Large qudits for the parameters should be more relevant in the long-term post Quantum Error Correction and Fault-Tolerance era of Quantum Computation.

We can use a single qubit for each parameter in MoMGrad, at the cost of having to perform more runs in order to get an accurate gradient estimate. That is, we can use a set of parameters for parametric controlled-unitaries whose parameters are a mixture of classical offset and a qubit's standard basis operator, $\hat{\Phi} \mapsto \Phi_0 \hat{I} + \epsilon\hat{Z}$, where $\epsilon>0$ is relatively small, i.e., $\Phi_0 \gg \epsilon$. 
This is akin to equation \eqref{eq:qudit} from the background, for a qubit instead of a qudit. Note a small enough $\epsilon$ is necessary for the discrete gradient to estimate the continuous gradient.
That is, if we have some unitary parametrized with $\Phi_0 \hat{I} + \epsilon \hat{Z}$, then by definition of the discrete gradient \eqref{eq:discrete_gradient}, we have
\begin{equation}
  \partial_{\hat{Z}} \hat{U}(\Phi_0 \hat{I} + \epsilon \hat{Z}) = \tfrac12 [ \hat{U}(\Phi_0 + \epsilon) - \hat{U}(\Phi_0 - \epsilon)].
\end{equation}
In the case of $\epsilon \ll \Phi_0$, the expression $\tfrac{1}{\epsilon} \partial_{\hat{Z}} \hat{U}(\Phi_0 \hat{I} + \epsilon \hat{Z})$ will approach a notion of a continuous derivative.

As the goal of the quantum parameters is to sense the phase kickback induced by Baqprop, one can use a detector which has a very low-resolution readout. Using a single qubit as the qudit from that equation \eqref{eq:qudit}, then we get a very rough estimate of the kickback on the parameters; this is equivalent to a single-qubit phase estimation; from the nature of phase space we only get a single-bit readout of the gradient. This then takes multiple runs so that the continuous gradient can be estimated.

\subsection{Regularization \& Variants}\label{sec:reg-var}

Regularization methods are useful tools for training deep neural networks in the classical literature \cite{dropout,weightdecay}. The role of regularization is to ensure a smooth training process and avoid overfitting to a particular dataset. Such techniques are thus indispensable when training very large networks whose training dynamics can be somewhat unwieldy, and ensures that the network is able to generalize beyond the given dataset.

In this section we will mainly focus on techniques which restrain the dynamics of the weights in a certain way, to either avoid weight blowup, or avoid over-reliance on certain connections in the network. hyper-parameter optimization, which is another important tool to avoid under/overfitting, will be treated in Subsection~\ref{sec:meta}.

\subsubsection{Parameter/Weight Decay}\label{sec:reg}
A technique from classical machine learning which allows one to dynamically bound the norm of the weights/parameters is \textit{weight decay}, or more generally parameter decay. The trick to weight decay is to add a regularization term to the cost function, the canonical choice being a simple quadratic penalty. For our quantum weights, we can also add a quadratic potential, 
\begin{equation}
    \mathcal{J}(\bm{\hat{\Phi}})\mapsto \mathcal{J}(\bm{\hat{\Phi}}) + \lambda\bm{\hat{\Phi}}^2
\end{equation}
where we use the notation $\bm{\hat{\Phi}}^T\bm{\hat{\Phi}} \equiv \bm{\hat{\Phi}}^2$. To see how this influences the execution of Quantum Dynamical descent, we can write
\begin{equation}\label{eq:qdd_reg}
    \hat{U}_{\textsc{qdd}} = \prod_j e^{-i\gamma_j\bm{\hat{\Pi}}^2}e^{-i\eta_j\mathcal{J}(\bm{\hat{\Phi}})}e^{-i\eta_j\lambda\bm{\hat{\Phi}}^2}.
\end{equation}
Note that the $e^{-i\eta_j\mathcal{J}(\bm{\hat{\Phi}})}$ and the $e^{-i\eta_j\lambda\bm{\hat{\Phi}}^2}$ terms can be executed simultaneously, partly due to the fact that these operations on the parameters commute, but also because the QFB effective phase kick involves many steps not involving every parameter.
If some parameter were used in every gate of the QFB algorithm, then the regularizing potential could not be applied simultaneously.
However, since the QFB algorithm involves an exponentiated loss operator which acts as the identity on the parameters, then the regularizing potential on the parameters could be applied at this stage.
That is, we could enact the above unitary as
\begin{equation}
   e^{-i\eta_j\mathcal{L}(\bm{\hat{\Phi}})}e^{-i\eta_j\lambda\bm{\hat{\Phi}}^2}\!\!=\! \hat{U}(\hat{\bm{\Phi}})^\dagger( e^{-i\eta_j\lambda\bm{\hat{\Phi}}^2}\otimes e^{-i \eta_j \hat{L}} )\hat{U}(\hat{\bm{\Phi}}).
\end{equation}
Note that in the above $\hat{U}(\hat{\bm{\Phi}})$ is the feedforward operation, and the Hilbert space factorization is  $\mathcal{H}_{\bm{\Phi}} \otimes \mathcal{H}_C$, where $\mathcal{H}_{\bm{\Phi}}$ and $\mathcal{H}_\textsc{c}$ are the parameter and computational Hilbert spaces respectively.
Alternatively, one could apply the regularizing potential phase shift for a given parameter at any other point in the QFB circuit where it is not being used to implement a controlled operation.

As we have seen previously, both Quantum Dynamical Descent and Momentum Measurement Gradient Descent have the parameters experiencing a force which is induced by the cost function acting as a potential. In the above case, adding a quadratic regularizing potential to each parameter register is effectively like adding a harmonic oscillator potential. For illustration, consider performing quantum dynamical descent for a \textit{null} cost function, i.e., $\mathcal{J}(\bm{\hat{\Phi}})=0$, then QDD becomes 
\begin{equation}
    \left.\hat{U}_{\textsc{qdd}}\right|_{\mathcal{J}=0}  =\prod_{j} e^{-i\gamma_j\bm{\hat{\Pi}}^2}e^{-i\eta_j \lambda \bm{\hat{\Phi}}^2}.
\end{equation}
For a certain set of hyper-parameters can be viewed as a simulation of Trotterized evolution under a free Hamiltonian of the form $\hat{H} \sim \sum_n ( \hat{\Pi}^2_n + \omega_n \hat{\Phi}^2_n )$, which represents a set of free (uncoupled) Harmonic oscillators.
The cost function $\mathcal{J}(\bm{\hat{\Phi}})$ is responsible for the coupling between the parameters during QDD.
This allows the parameters to entangle and dynamically influence one another.

\subsubsection{Meta-networked Interacting Swarm Optimization}\label{sec:miso}

An option for simultaneous parallelization and regularization is to have multiple replicas, similar to the parallelized minibatch method from \ref{sec:CAMP} above, which have their parameters coupled to each other with an attractive potential.
Such a potential can be used to correlate the dynamics of the replicas equilibrium point, while still allowing for some degree of independent dynamics. We call this approach Meta-networked Interacting Swarm Optimization (MISO).

To precisely describe how to couple replicas, first consider a \textit{meta-network} of replicas, which is a graph $\mathcal{G}=\{\mathcal{V},\mathcal{E}\}$, where each vertex is a replica with parameters $\bm{\hat{\Phi}}_{[c]},$ for  $c\in \mathcal{V}$. The edges $\mathcal{E}$ of the graph $\mathcal{G}$ represent couplings between replicas.
Each edge's weight will represent the coupling strength. The way this coupling will be introduced into the parameter dynamics is via an added potential. 

Let $\bm{\hat{\bar{\Phi}}} := ( \bm{\hat{\Phi}}_{[c]} )_{c\in \mathcal{V}}$ be the operator-valued meta-vector of all parameter vectors.
The global added potential is the sum of the coupling potentials,
\begin{equation}
    V(\bm{\hat{\bar{\Phi}}}) \!=\! \sum_{\{j,k\}\in \mathcal{E}} V_{j,k}(\bm{\hat{\Phi}}_{[j]},\bm{\hat{\Phi}}_{[k]}).
    \end{equation}
For example, we could choose each of these coupling potentials of the form,
\spliteq{\label{eq:couple}
  V_{j,k}(\bm{\hat{\Phi}}_{[j]},\bm{\hat{\Phi}}_{[k]}) &=    \lambda_{jk}\,(\bm{\hat{\Phi}}_{[j]}-\bm{\hat{\Phi}}_{[k]})^T(\bm{\hat{\Phi}}_{[j]}-\bm{\hat{\Phi}}_{[k]})\\
  &= \lambda_{jk} \big( \bm{\hat{\Phi}}_{[j]}^2 + \bm{\hat{\Phi}}_{[k]}^2 - 2 \bm{\hat{\Phi}}_{[j]} \cdot \bm{\hat{\Phi}}_{[k]} \big),
}
with all coupling scalars non-negative reals, $\lambda_{jk}\in \mathds{R}_{\geq 0}$, $\forall \{j,k\}\in\mathcal{E}$.
We see that each parameter in a given replica is coupled to its corresponding parameters from neighboring replicas, with locality of couplings determined by the graph of the replica network. We also see that the first two terms act as parameter decay terms to keep the norm of the parameters contained. Note that due to the fact that $\lambda_{jk}\geq 0$, coupled parameters will naturally want to correlate to minimize the third term, which is the interaction term.

\begin{figure}[h!]
 \begin{center}
\includegraphics[width=0.9\columnwidth]{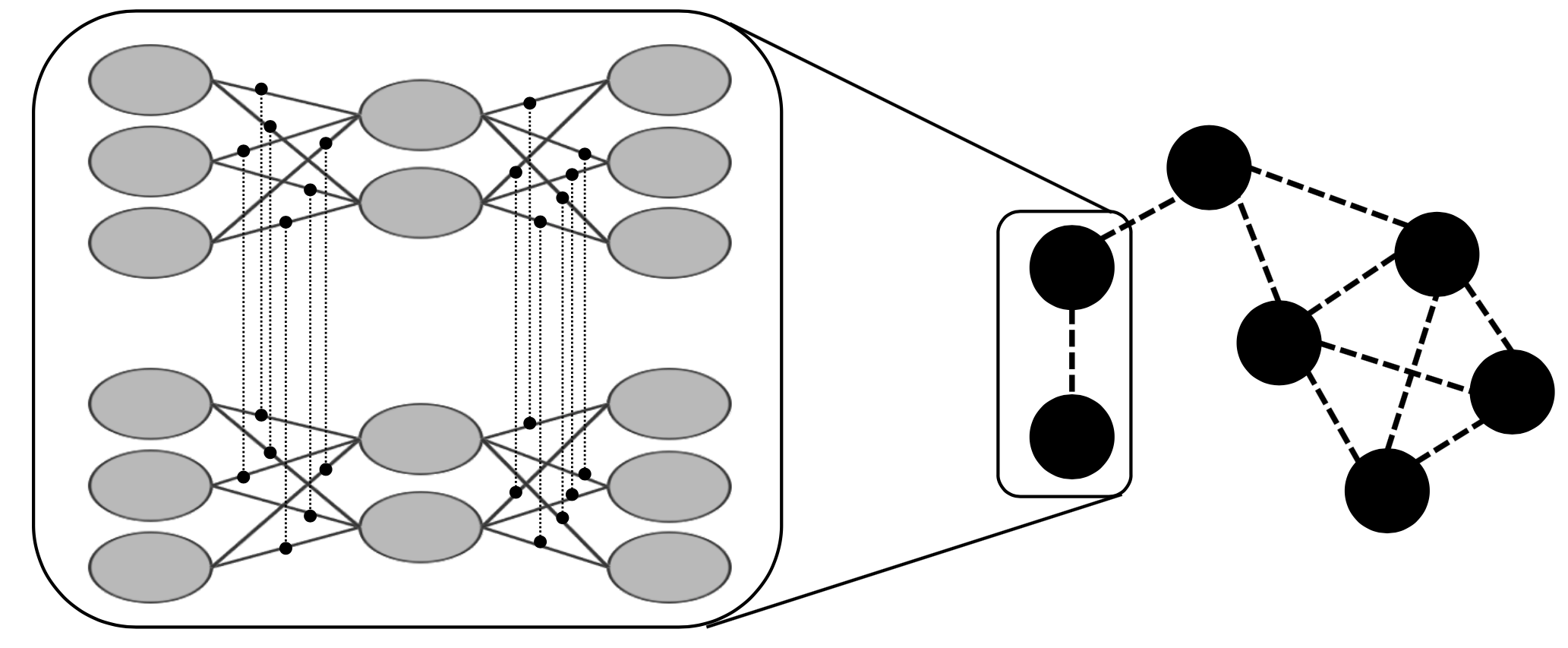}
\caption{Example of a graph $\mathcal{G}$ for meta-network of replicas (right), with the weight of each edge $\{j,k\}\in \mathcal{E}$ corresponding to a coupling strength $\lambda_{jk}$, for couplings of the form \eqref{eq:couple}. The replicas' parameter registers are coupled one-to-one with their corresponding parameter in coupled neighboring replicas (left), all according to the same inter-replica coupling. Here we give a simple neural network with synaptic weights as the parameters which are coupled between replicas.} \label{fig:metanet}
\end{center}
\end{figure}

Depending on the topology of the meta-network, different joint dynamics will be induced on the set of parameters of the replicas. The interesting advantage of this approach is that it is highly adaptable to the given interconnect capabilities of different networks of quantum processors; one can restrict the meta-network to be tailored to the natural topology of the given implementation of the algorithm. We represent pictorially the meta-network of replicas in Figure~\ref{fig:metanet}. 

In the previous subsection \ref{sec:CAMP} on Coherently Accumulating Momentum Parallelization, different replicas were assigned to different data points of a minibatch for each iteration of the parameter optimization (i.e., each momentum measurement or application of the kinetic pulse).
In this variant, we can consider assigning a different arrangement of minibatches $\mathcal{D}_{[j]} := \{\mathcal{M}_{k}^{[j]}\}_{k}$ to each replica $j\in \mathcal{V}$.
That is, we can consider each replica to have effectively a different dataset, in cases where data is scarce then one may simply shuffle the data points and create new partitions for the different replicas. For each replica, we can then execute the minibatch accumulation of momenta via the serialized or parallelized variants. In the parallelized variant as in subsection \ref{sec:CAMP}, each vertex of the meta-network would represent a parameter server node, which itself is connected to its sub-replicas with which it executes CAMP. In a sense, these sub-replicas are used simply as ancillas to accumulate on the parameter server node a phase shift equal to the cost function of a minibatch in a parallelized fashion. We do not count sub-replicas executing CAMP as part of the meta-network, since the dynamics of CAMP are akin to a single replica, that of the server.

Let us consider how to compile and execute this set of transformations including the added rgularizing potential. Let us denote the cost function for the minibatch $\mathcal{M}_k^{[j]}$ assigned to the $j^\text{th}$ replica's $k^\text{th}$ parameter iteration as 
\begin{equation}
    \mathcal{J}_k^{[j]} (\bm{\hat{\Phi}}_{[j]})\equiv \tfrac{1}{|\mathcal{M}_k^{[j]}|}\textstyle\sum_{x\in \mathcal{B}_k^{[j]}}\mathcal{L}_x(\bm{\hat{\Phi}}),
\end{equation}
where $\mathcal{B}_k^{[j]}$ is the index set of $\mathcal{M}_k^{[j]}$. We can write the the global replica meta-network cost function for the $k^\text{th}$ iteration as the sum of all the costs over the replicas in the meta-network
\begin{equation}
    \mathcal{J}_k(\bm{\hat{\bar{\Phi}}}) = \sum_{j\in \mathcal{V}} \mathcal{J}_k^{[j]} (\bm{\hat{\Phi}}_{[j]}).
\end{equation}
 
Consider the multi-iteration QDD unitary with Meta-networked Interacting Swarm Optimization (MISO), for an iteration index $k\in\mathcal{S}$.
For now we assume uniformity across replicas of the kicking and kinetic rates,
\spliteq{\label{eq:miso}
      \hat{U}_{\textsc{qdd.miso}} &= \prod_{k\in\mathcal{S}} e^{-i\gamma_k\bm{\hat{\bar{\Pi}}}^2}e^{-i\eta_k \left(\mathcal{J}_k(\bm{\hat{\bar{\Phi}}}) + V(\bm{\hat{\bar{\Phi}}}) \right)}. \\
}
The kinetic terms can be applied in parallel across replicas and parameters,
\spliteq{
e^{-i\gamma_k\bm{\hat{\bar{\Pi}}}^2} = \bigotimes_{j\in\mathcal{V}} e^{-i\gamma_k\bm{\hat{\Pi}}_{[j]}^2},
}
while the exponentials of the regularizing potential can be executed during any part of the feedforward, phase kick, and backpropagation operation, as long as a given parameter is not currently executing a controlled-operation which only happens during the feedforward and uncomputation portions of QFB.
This can be done in any order since the phase shifts commute.
As long as the kinetic term is not executed, there is freedom to choose exactly how to compile the operation.
A simple way is sequentially adding the potential pulse before or after the replica-parallelized QFB,
\spliteq{\label{eq:miso_phase}
  &e^{-i\eta_k \left(\mathcal{J}_k(\bm{\hat{\bar{\Phi}}}) + V(\bm{\hat{\bar{\Phi}}}) \right)} \\
  &= \Big( \bigotimes_{j\in\mathcal{V}} e^{-i\eta_k\mathcal{J}_k^{[j]} (\bm{\hat{\Phi}}_{[j]})}e^{-i\eta_k\tilde{\lambda}_{[j]}\bm{\hat{\Phi}}_{[j]}^2}  \Big)\!\! \prod_{\{m,l\}\in\mathcal{E}}\!\!\! e^{i2\eta_k\lambda_{lm}\bm{\hat{\Phi}}_{[l]}^T\bm{\hat{\Phi}}_{[m]}} ,
}
where we denoted the coupling strength averaged over all edges incident to a meta-network vertex as $\tilde{\lambda}_{[j]} := \sum_{k\in\mathcal{V}} \lambda_{jk}$.
Again, since all the above exponentials are commuting, there is opportunity to combine the execution of all these terms in the potential in a more efficient manner than serially.
Note that to execute a MoMGrad optimization with MISO, one simply prepare a pointer state of choice in all parameters of all replicas, as in \eqref{eq:momgrad_init}.
Then, one applies the above MISO phase unitary from \eqref{eq:miso_phase}, measures the momentum of all parameters and updates them according to the regular MoMGrad update rule from \eqref{eq:momdate}.

An option for the swarm approach is that one can have multiple networks with the same architecture (hence the name replica), but with different hyper-parameters, i.e., different initializations and/or kicking and kinetic rates at different iterations.
This would mean a modification of the above formulas to having replica-specific rates, i.e., $\{\gamma_k,\eta_k\} \mapsto \{\gamma_{k}^{[j]},\eta_{k}^{[j]}\}$, as well as replica-specific initializations (mean and variance) for the weights.
This can allow a sort of effective majority voting of where to go in the parameter landscape, which may possibly kick replicas with poor initializations out of a local well, but also might perturb a replica performing well in terms of cost optimization to get kicked off of its trajectory to a low cost function value. As this is a strategy which will increase the training set error to possibly improve the test set error, we consider it as a regularization technique.

As Quantum Dynamical Descent is effectively a QAOA approach to finding low energy states of a Hamiltonian, we can see that MISO is effectively like trying to find the ground state of a swarm of interacting particles.
Assuming uniform descent hyper-parameters, and considering full batch cost function, $\mathcal{J}(\bm{\hat{\bar{\Phi}}}) := \sum_k \mathcal{J}_k(\bm{\hat{\bar{\Phi}}})$, we can write down this effective Hamiltonian to be of the form
\begin{equation}
      \hat{\bar{H}} = \tfrac{1}{2}\bm{\hat{\bar{\Pi}}}^2 + \tfrac{\omega^2}{2}\mathcal{J}(\bm{\hat{\bar{\Phi}}}) + \tfrac{\omega^2}{2} V(\bm{\hat{\bar{\Phi}}}),  \\
\end{equation}
which resembles a lattice of oscillators with an added non-linear potential proportional to $\mathcal{J}$.
Theoretically one could expand $\mathcal{J}(\bm{\hat{\bar{\Phi}}})$ about its minimum value to second order and obtain a quadratic potential.
The approximate ground state would then be given by a Gaussian ground state of the form \eqref{eq:ground}. The important takeaway is that the joint system of parameters is like a coupled network of oscillators, with the intra-replica coupling induced by the cost function and the inter-replica couplings due to the meta-network's topology. There are many way to modify the approach described above, in terms of how to manage data, how to modify hyper-parameters, etc.. In the next subsection \ref{sec:meta}, we discuss how to leverage the quantum phase backpropagation of errors and quantum dynamical descent to optimize all these possible hyper-parameters via quantum dynamical descent.

\subsubsection{Dropout}\label{sec:drop}

The method of \textit{dropout} in classical machine learning encompasses a set of techniques which add noise to the training process in order to regularize the learning. The addition of noise to the neural information processing effectively forces the network to learn to process information in a redundant, robust manner. In a sense, adding errors forces the neurons to not over-rely on a specific neural pathway, and thus to split signals into multiple pathways, thereby spreading the computation over neural elements in order to add noise resistance. Traditional dropout consists of adding classical erasure noise to the neural information processing, this consists of effectively blocking the path of the information flowing forward by stochastically \emph{dropping out} certain neural elements. Modern techniques for dropout also include Gaussian multiplicative noise, or Gaussian additive noise \cite{neelakantan2015adding} for neural networks. In this section we focus on techniques to use quantum registers as stochastic classical variables which control whether certain subsets of parametric operations are applied. Note that we will reuse much of the machinery developed in this subsection in our subsection on network architecture optimization via Quantum Meta-Learning (section \ref{sec:NAO}), where instead of simply using the quantum registers as a source of stochastic noise, we can optimize over superpositions of network architectures via a quantum meta-learning optimization loop.

As our parameters naturally have Gaussian noise in both the gradient and parameter value due to our optimization approach outlined in Section~\ref{sec:opt} using Gaussian pointer state, the Gaussian multiplicative noise dropout comes for free for our schemes. In a sense the Quantum uncertainty of the wavefunction serves as natural regularizing noise. For Gaussian additive noise dropout, refer to Section~\ref{sec:qnn} where we describe quantum parametric circuits for neural networks. In this section, the computational registers are initialized in null-position qudit or qumode eigenstates $\ket{0}$. It would be straightforward to use computational registers which have some added Gaussian noise to their position value, i.e., are in a simulated squeezed state rather than a perfect position eigenstate initially. Because these types of dropout are straightforward to implement with our schemes, we focus on \textit{operation dropout}: stochastically removing certain subsets of parametric operations.

The goal of operation dropout is to probabilistically create a blockage of information flow in the feedforward computational graph. Furthermore, another important aspect of dropout is the ability to backpropagate errors with knowledge of this erasure error. As our backpropagation approach relies on the ability to backpropagate error signals through the quantum computational graph via uncomputation after the feedforward operation and phase kick, we will need to keep in memory the register which controls the erasure. We use a quantum state's computational basis statistics as the source of classical stochasticity in this section for notational convenience, but note that could equivalently replace these qubits with classical random Bernoulli variables of equivalent statistics.

Now, let us develop some formalism to characterize how to leverage ancillary quantum registers in order to stochastically control which architecture is used in the quantum feedforward and Baqprop. Whether it is a quantum parametric circuit, as those discussed in Section~\ref{sec:qdata_algs}, or a neural network embedded into a set of quantum parametric circuits, as discussed in Section~\ref{sec:qnn}, we can assume the parametric circuit ansatz can be written as a layered circuit of unitaries, i.e.
\begin{equation}\label{eq:full_par_circ}
    \hat{U}(\bm{\hat{\Phi}}) = \prod_{\ell=1}^\mathscr{L}\hat{U}^{(\ell)}(\bm{\hat{\Phi}}^{(\ell)}), \quad \hat{U}^{(\ell)}(\bm{\hat{\Phi}}^{(\ell)})\equiv \bigotimes_{j_\ell\in \mathcal{I}_\ell} \hat{U}_{j_\ell }(\hat{\Phi}_{j_\ell})
\end{equation}
where $\hat{U}^{(\ell)}(\bm{\hat{\Phi}}^{(\ell)})$ is the multi-parameter unitary corresponding to the $\ell^\text{th}$ layer, which can itself be composed of multiple parametric unitaries $\{\hat{U}_{j_\ell}(\hat{\Phi}_{j_\ell})\}_{j_\ell}$, and where $\mathcal{I} = \cup_{\ell=1}^\mathscr{L}\mathcal{I}_{\ell}$ is the partition of parameter indices for the parameters of each layer.

Now, if we would like to parametrize whether a number $N$ of certain subsets of parametric unitaries are applied or not, we need to first index which unitaries are controlled by the same variable. For this index, consider a partition of the indices $\mathcal{I}= \cup_{j=0}^N\mathcal{A}_j$, where $\mathcal{A}_j\subset \mathcal{I}\ \forall j$. For notational convenience, let us reserve the subset $\mathcal{A}_0$ as the set of unitaries over which we would not like to be stochastically controlled, i.e., we want to implement these with absolute certainty, the reason for this notation will be apparent below. To quantumly control the application of these subsets of unitaries, we will need a set of $N$ ancillary qubits which index the architecture, say we label these as $A_j$ where the $\ket{1}_{A_j}$ indicates we are applying the unitaries in subset $\mathcal{A}_j$. For notational convenience, consider the following operator-valued function, which takes indices of operations from $\mathcal{I}$ and maps them to operators on the architecture ancillas' Hilbert space $\mathcal{H}_A =\bigotimes_{j=1}^N\mathcal{H}_{A_j}$; $ \hat{C}: \mathcal{I} \rightarrow \mathcal{B}(\mathcal{H}_A)$, 
\begin{equation}\label{eq:ctrl}
    \hat{C}(j) = \bigotimes_{k=1}^N \ket{1}\!\bra{1}_{A_k}^{\bm{1}_{\mathcal{A}_k}(j)} 
\end{equation}
where we denote $\bm{1}_{\mathcal{A}_k}(j)$ as the indicator function for the set $\mathcal{A}_k\subset \mathcal{I}$, and $\ket{1}\!\bra{1}^0 = I$. This operator can serve a the control operator for a given index; essentially, given an index of an operation, it is a projector onto $\ket{1}_{A_k}$ for the ancilla whose index corresponds to that of the partition in which $j$ belongs. Also note that the above operator is a function of Pauli $\hat{Z}$'s of the architecture ancillas, hence it is only dependent on the vector of Paulis $\bm{\hat{Z}}_A = \{\hat{Z}_{A_k}\}_k$.

We can consider the architecture index to be a stochastically-determined hyper-parameter. We can then modify our parametric unitary to become a hyper-parametric unitary, which acts on both the Hilbert space of architecture indices (used as controls) and the joint computational and parameters' Hilbert spaces,
\begin{equation}\label{eq:AHP}
    \hat{\bar{U}}(\bm{\hat{\Phi}},\bm{\hat{Z}}_{A}) = \prod_{\ell=1}^\mathscr{L} \prod_{j_\ell\in \mathcal{I}_\ell} \hat{C}_A(j_\ell)\otimes \hat{U}_{j_\ell }(\hat{\Phi}_{j_\ell})
\end{equation}
note this is essentially the same unitary as previously \eqref{eq:full_par_circ}, except now each unitary in $\mathcal{I}\setminus \mathcal{A}_0$ is a controlled-unitary, and the control qubit for each index $j$ is that which corresponds to the partition of inidices $\mathcal{A}_k$ such that $j\in\mathcal{A}_k$.

Although the above operation may seem complex, the circuit to execute the above may be compiled efficiently, simply by adding a control to each operation. For example, each parametric unitary (see \ref{sec:par}) is of the form \begin{equation}
  \hat{U}_{j_\ell }(\hat{\Phi}_{j_\ell})= \sum_{\Phi_{j_\ell}}\ket{\Phi_{j_\ell}}\!\bra{\Phi_{j_\ell}}\otimes  \hat{U}_{j_\ell }(\Phi_{j_\ell})
\end{equation}
now assuming each unitary is generated by a certain Hamiltonian, i.e., $ \hat{U}_{j_\ell }(\Phi_{j_\ell})= e^{-i\Phi_{j_\ell}\hat{h}_{j_\ell }}$ then the above becomes \begin{equation}
  \hat{U}_{j_\ell }(\hat{\Phi}_{j_\ell})=   e^{-i\hat{\Phi}_{j_\ell}\otimes \hat{h}_{j_\ell }},
\end{equation}
which we see is an exponential with a generator $\hat{\Phi}_{j_\ell}\otimes \hat{h}_{j_\ell }$.
In order to convert a certain parametric unitary of index $k\in\mathcal{A}_j$ to have an added qubit control register, one simply has to exponentiate the modified generator $\ket{1}\!\bra{1}_{A_j}\otimes \hat{\Phi}_{k}\otimes \hat{h}_{k }$, i.e.
\begin{equation}
    \sum_{b_j \in\mathds{Z}_2}\ket{b_j}\!\bra{b_j}_{A_j} \otimes \hat{U}^{b_j}_{k}(\bm{\hat{\Phi}}) = e^{-i\ket{1}\!\bra{1}_{A_j}\otimes \hat{\Phi}_{k}\otimes \hat{h}_{k }}
\end{equation}
which can be synthesized into a product of $(v_k+2)$-local exponentials of Paulis, where $v_k$ is the locality of $ \hat{h}_{k}$.

Now that we have covered how to modify the parameteric circuit ansatz to include quantum controls, we can now describe how to modify the Quantum Feedforward and Baqprop (QFB) to include dropout. 
Suppose we would like to perform the QFB for a certain iteration where we have the loss operator $\hat{L}_j$, the usual QFB operation would consist of applying
\begin{equation}
    \hat{U}^\dagger(\bm{\hat{\Phi}})e^{-i\eta \hat{L}_j}\hat{U}(\bm{\hat{\Phi}})
\end{equation}
onto a computational register state $\ket{\xi_j}_C$, the effective phase on the parameters (see sec. \ref{sec:phase-kick}) would then be
\begin{equation}
    \mathcal{L}_j(\bm{\hat{\Phi}}) =  \bra{\xi_j} \hat{U}^\dagger(\bm{\hat{\Phi}}) \hat{L}_j\hat{U}(\bm{\hat{\Phi}})\ket{\xi_j}_C.
\end{equation}
Now to modify QFB to include dropout, we simply modify the regular feedforward parameteric unitary to be the controlled unitary from \eqref{eq:AHP}. 
\begin{equation}
        \hat{\bar{U}}^\dagger(\bm{\hat{\Phi}},\bm{\hat{Z}}_{A})e^{-i\eta \hat{L}_j} \hat{\bar{U}}(\bm{\hat{\Phi}},\bm{\hat{Z}}_{A}),
\end{equation}
and we act this hyper-parametric unitary on the same computational registers' state $\ket{\xi_j}_C$ and an initial state $\ket{\alpha_0}_A$ of our architecture qubits:
\begin{equation}\label{eq:drop_state}
   \ket{\alpha_0}_A = \bigotimes_{k=1}^N (\sin(\theta_k)\ket{0}_{A_k} + \cos(\theta_k)\ket{1}_{A_k}\!)
\end{equation}
where the $\bm{\theta} = \{\theta_k\}_{k=1}^N$ are hyper-parameters which will control the probability of dropout; the probability of applying a set of operations of index $\mathcal{A}_k$ will be given by $\cos^2(\theta_k)$. Tracing out the computational and architecture registers, the resulting effective phase on the parameters will be
\begin{equation}\label{eq:drop_eff_phase}
        \mathcal{L}_j(\bm{\hat{\Phi}}) =  \bra{\xi_j}\!\bra{\alpha_0} \hat{\bar{U}}^\dagger(\bm{\hat{\Phi}},\bm{\hat{Z}}_{A})\hat{L}_j \hat{\bar{U}}(\bm{\hat{\Phi}},\bm{\hat{Z}}_{A})\ket{\xi_j}_C\!\ket{\alpha_0}_A.
\end{equation}
Thus, we get the average cost function phase kick, averaged over the possible architectures. 

To see this more explicitly, we can expand the above expression, to do so it will be convenient to define some more notation. Let us begin with  
\begin{equation} \lambda_{\bm{a}} \equiv \prod_{k=1}^N (\sin(\theta_k))^{1-a_k}(\cos(\theta_k))^{a_k},
\end{equation}
hence the state from \eqref{eq:drop_state} could be written as $\ket{\alpha_0} = \sum_{\bm{a}\in\mathds{Z}_2^N} \lambda_{\bm{a}}\ket{\bm{a}}_A$.
Additionally, let $ \hat{\bar{U}}_{\bm{a}}(\bm{\hat{\Phi}})  = \bra{\bm{a}}\hat{\bar{U}}(\bm{\hat{\Phi}},\bm{\hat{Z}}_{A})\ket{\bm{a}}_A$ be the unitary corresponding to the parametric circuit of the following architecture
\begin{equation}
    \hat{\bar{U}}_{\bm{a}}(\bm{\hat{\Phi}}) \equiv \prod_{\ell=1}^\mathscr{L} \prod_{j_\ell\in \mathcal{I}_\ell\cap \mathcal{Z}_{\bm{a}}}  \hat{U}^{a_{j_\ell}}_{j_\ell }(\hat{\Phi}_{j_\ell}).
\end{equation}
where $\mathcal{Z}_{\bm{a}} = \{\mathcal{A}_k: 0\leq k\leq N, a_k=1\}$. This corresponds to acting all unitaries of index $j$ for which $j\in\mathcal{A}_k$ for some $k$ such that $a_k$ is nonzero. Now, we can use the above to expand equation \eqref{eq:drop_eff_phase}
\spliteq{
\mathcal{L}_j(\bm{\hat{\Phi}})  = \sum_{\bm{a}\in \mathds{Z}_2^N } \lambda_{\bm{a}}\bra{\xi_j} \hat{\bar{U}}^\dagger_{\bm{a}}(\bm{\hat{\Phi}})\hat{L}_j \hat{\bar{U}}_{\bm{a}}(\bm{\hat{\Phi}})\ket{\xi_j}_C,
}
and we see that we get, on average, the expectation over architectures of the effective phase kick. We can then use tools from Section \ref{sec:opt} to leverage this effective phase signal for optimization of the parameters via MoMGrad or QDD.

Note that, just like the regular effective phase kicking for quantum data, the registers other than the parameters must be reinitialized (refreshed) after each QFB run, in order to get the averaged behaviour. If we were to keep the same quantum ancillas indexing the architecture for multiple runs of QDD, then the parameters would entangle with the superposition of architectures, such as to optimize the cost function for each architecture in each individual branch of the superposition rather than optimizing for the mixture of architectures. We harness this very property of training different network architectures in superposition for meta-learning and architecture optimization in section \ref{sec:NAO}.

As dropout is integral to training classical neural networks, the above technique is useful to have for training classical neural networks on a quantum computer. This ``operation dropout'' may also be useful for robust parametric circuit learning in certain settings. Since dropout will emulate faulty execution of gates, this would force the parametric circuit to not rely too much on a single gate for a large change to the state, each parametric operation would stay not too far from the identity; each continuously parametrized gate would then have a small angle, hence keeping $\lVert\bm{\Phi}\rVert^2$ small. This becomes effectively similar to the parameter decay described in section \ref{sec:reg}. In general one would expect dropout will have other effects on the parameters that simple weight decay alone cannot emulate.

For general quantum parametric circuits, one could consider adding additional parametric unitaries which stochastically \textit{drop in}, potentially to emulate various forms of noise. For example, we could consider adding controlled-$X$ and controlled-$Z$ as additional operations in a given parametric circuit ansatz. Using two qubits for controls, one could then apply stochastically apply $X$ and/or $Z$ at each site using techniques from above. One could thus emulate a depolarizing channel for example. Optionally, one could stochastically swap out or swap in computational registers, again controlled by architecture binary hyper-parameters. This would simulate a form of erasure noise. Generally, one could use this technique to add a great variety of types of noise. There are many ways to add noise to a system, but dropout is used to regularize the training networks. It is not yet clear whether quantum parametric circuits need dropout for better training, nor what kind of noise map would be best, at this stage in the development of the field.

\subsection{Quantum Meta-Learning}\label{sec:meta}

\subsubsection{Overview}
In practical machine learning scenarios, it is often better to rapidly find a local minimum rather than a global optimum (which has a cost of longer runtime). This is where the low-depth limit becomes interesting. Rather than having many pulses in order to minimize the Suzuki-Trotter error approximating the adiabatic path, it will often be better to have a higher phase kicking and kinetic rate, and to variationally optimize these hyper-parameters. This variational optimization is done by training the model with a certain set of hyper-parameters, and by checking the value of the cost function with respect to a subset of data called the \textit{test set}. Oftentimes this is done via trial and error and careful hand-tuning, but there exists ways to automate this process. Automation of this hyper-parameter optimization is called meta-learning \cite{vilalta2002perspective}. 

Instead of using a classical optimizer which would involve finite-difference optimization, we can use the Quantum Dynamical Descent method at the hyper-parameter level.
Hyper-parameter optimization methods commonly used in classical deep learning, namely, grid search, random search, manual search, or even Bayesian optimization, come with multiple training cycles for optimization, often scaling exponentially in overhead with the number of hyper-parameters \cite{bergstra2012random}.
This problem of meta-training, training the hyper-parameters, is what these techniques address.
Meta-learning has been used to boost the learning speed (decrease training set error in less iterations) \cite{hochreiter2001learning,andrychowicz2016learning}, has allowed for better test set error and generalization error, and has been used to learn how to rapidly adapt a network trained for a given task to perform a new one, an approach known as transfer learning \cite{MAML}.

A recent approach to meta-learning has been to use gradient descent on the hyper-parameters, often with an additional neural network relating the choices of hyper-parameters between different iterations \cite{andrychowicz2016learning}. 
The optimization of this hyper-parameter network is done via a backpropagation of errors up the computational graph, which traces back the influence of the hyper-parameters on the output loss function.

The following techniques we will describe below are analogous in a sense to this hyper-parameter gradient descent.
Each hyper-parameter influences either the initialization, the descent rates, or even the architecture of the network.
In the rest of this section we will explore how to move from what we hitherto have considered to be classical (fixed) hyper-parameters, to quantum (continuous or discrete) parameters. By considering how to perform the feedforward and backpropagation with quantum hyper-parameters, we will then be able to perform \textit{meta-Baqprop}, once again using the quantum backpropagation of phases principle. We will then be able to apply either quantum dynamical descent or momentum measurement gradient descent on the hyper-parameters, and do so in an efficient manner, as Baqprop does not require knowledge of analytic derivatives of each part of the computation. 

Finally, note that the Quantum Meta-Learning approach relies heavily on the possibility of entanglement between the quantum hyper-parameters and the parameters/compute registers. Given a superposition of hyper-parameters, one can consider each branch of the wavefunction of these hyper-parameters. As the hyper-parameters influence the training of the network via Quantum Dynamical Descent, each value of the joint set of hyper-parameters will lead to a different trained network. Since the whole training process is kept quantum coherent, the result is an entangled superposition of hyper-parameters and their corresponding fully trained networks. At this point, applying a cost function exponential of choice for the network tags the different branches of the wavefunction with relative phases, and unitarily uncomputing the training allows for a backpropagation of errors all the way up to the hyperparameters. Thereby allowing for their optimization via a Meta-QDD or Meta-MoMGrad approach.

\subsubsection{Quantum hyper-parameter Descent}\label{sec:QHD}

In previous discussions of Quantum Dynamical Descent and Momentum Measurement Gradient Descent, given a fixed network architecture ansatz, there were sets of classical hyper-parameters for the preparation of the parameter's pointer states, denoted $\bm{\Theta}= \{\bm{\Phi}_0, \bm{\Pi}_0, \bm{\Sigma}_0\}$, and some for the choice of kicking and kinetic rates for each iteration, which were denoted $\bm{\Xi}= \{\bm{\gamma},\bm{\eta}\}$. We can then consider a parameter pointer state preparation unitary as a classically parametrized unitary $\hat{U}_p(\bm{\Theta})$, and similarly, the entire Quantum Dynamical Descent unitary, as featured in equation \eqref{eq:qdd}, can be seen as a unitary parametrized by classical hyper-parameters $\bm{\Xi}$, which acts both on computational registers and the parameter registers, $\hat{U}_{\textsc{qdd}}(\bm{\Xi})$. The key to our meta-learning problem will be to view the combination of the preparation and quantum dynamical descent unitaries as \textit{hyper-parametric} circuits to be optimized.

The task of meta-learning usually involves optimizing the initialization and execution of the training process in order to minimize some cost function which assesses either generalization or optimization performance. This loss function can be the same as the training loss/cost function, using the training data, or it can be some different cost function than that of the training, either through the use of the same loss applied to different data, or some different loss function altogether. In cases where the learning comes from data, either classical or quantum, the subset of data reserved for the hyper-parameter training is called the \textit{test set}, while the subset of data reserved for the training of the parameters is called the training set or \textit{development set} (dev set). In any case, there is a cost function which we want to optimize, whose effective phase we will call $\mathcal{J}_{\textsc{m}}$ subject to variations in the hyper-parameter vectors $\{\bm{\Theta},\bm{\Xi}\} $. 

Using the same approach as our parameter optimization for regular learning, we can quantize the hyper-parameters $\{\bm{\Theta},\bm{\Xi}\} \mapsto \{\bm{\hat{\Theta}},\bm{\hat{\Xi}}\}$, and using either Quantum Dynamical Descent or Momentum Measurement Gradient Descent for quantum-enhanced optimization of these hyper-parameters. We will refer to these approaches as Meta-QDD and Meta-MoMGrad, respectively. We regard the meta-feedforward hyper-parametric unitary to be 

\begin{equation}\label{eq:meta-feed}
    \hat{U}_{\textsc{meta}}(\bm{\hat{\Theta}},\bm{\hat{\Xi}})\equiv \hat{U}_{\textsc{qdd}}(\bm{\hat{\Xi}})\hat{U}_p(\bm{\hat{\Theta}}),
\end{equation}
i.e., the parameter state preparation unitary followed by the Quantum Dynamical Descent unitary. 

Before we proceed with how to leverage such a unitary, let us examine how exactly this upgraded quantum-hyperparametric unitary can be synthesized into elementary gates. Note the Quantum Dynamical descent unitary is now of the form 
\begin{equation}
  \hat{U}_{\textsc{qdd}}(\bm{\hat{\Xi}})=  \prod_j e^{-i\hat{\gamma}_j\otimes\bm{\hat{\Pi}}^2}e^{-i\hat{\eta}_j\otimes \mathcal{J}(\bm{\hat{\Phi}})},
\end{equation}
where the exponentials are now quantum-controlled. The synthesis of the kinetic exponential is straightforward, taking $\mathcal{O}(\log^3 d)$ 3-local exponentials of qubit Paulis to enact, where $d$ is the qudit dimension of our parameter and hyper-parameter registers. For the hyper-parametric effective phase, one can apply the regular feedforward unitary, but the exponential of the loss function now being quantum-parametric, i.e., apply
\begin{equation}
    e^{-i\hat{\eta}_j\otimes \hat{L}_j(\bm{\hat{\Phi}})}= \hat{U}^\dagger(\bm{\hat{\Phi}}) e^{-i\hat{\eta}_j\otimes \hat{L}_j}\hat{U}(\bm{\hat{\Phi}})
\end{equation}
and the expectation of the above for an input computational state will give the quantum-parametric effective phase. How to synthesize this exponential of the loss function will vary. In general for a compilation of $e^{-i\hat{\eta}_j\otimes \hat{L}_j}$ down to Clifford gates and $Z$-rotations gates of the form $e^{i\beta \eta \hat{Z}}$ for some constants $\beta$, we can then modify the classically parametric rotations to be quantum-hyper-parametric $e^{i\beta \eta  \hat{Z}} \mapsto e^{i\beta \hat{\eta}\otimes   \hat{Z}}$ which themselves can each be broken down into $\mathcal{O}(\log d)$ exponentials. For more details on parametric circuit synthesis see section \ref{sec:qpar_ans}. If the loss function is based on $\eta$-parametric exponential-swap, as we will treat in \ref{sec:qdata_algs}, we provide compilation of these into Fredkin and $Z$-rotation, hence can be quantum-hyperparametrized straightforwardly. Finally, for the preparation unitary, upgrading the hyper-parameters to quantum is straightforward, since for Gaussian state preparation we can have quantum-parametrized simulated continuous-variable displacement and squeezing operators to quantum parametrize the first and second moments of the Gaussian wavefunctions.

Now that we have covered how to synthesize the hyper-parametric unitary, we can now proceed to leveraging this unitary to perform the Quantum Feedforward and Phase Kick Backpropagation procedure at the meta-level. The cost function we are trying to optimize can be the loss over some minibatch which corresponds to the test data. Let $\hat{U}(\bm{\hat{\Phi}})$ be the parametric unitary acting on the compute and parameter registers, the exponential of the loss function for the meta-learning is the exponential loss 
\begin{equation}\label{eq:test_exp}
    e^{-i\mu \mathcal{J}_{\textsc{m}}(\bm{\hat{\Phi}})} = \prod_{j\in\mathcal{B}_{\textsc{t}}}e^{-i\tilde{\mu} \mathcal{L}(\bm{\hat{\Phi}})}
\end{equation}
where $\mathcal{B}_{\textsc{t}}$ is the test set batch index and $\mu$ is the phase kicking rate, $\tilde{\mu}\equiv \mu/|\mathcal{B}_{\textsc{t}}|$ is the same rate divided by the test batch size. Recall that to enact each of the loss function effective phase shifts exponentials, this entails applying the QFB procedure for the parameter circuit $\hat{U}(\bm{\hat{\Phi}})$, i.e., 
\begin{equation}\label{eq:recall_kick}
 e^{-i\tilde{\mu}\hat{L}_j(\bm{\hat{\Phi}})}= \hat{U}^\dagger(\bm{\hat{\Phi}}) e^{-i\tilde{\mu}\hat{L}_j}\hat{U}(\bm{\hat{\Phi}})
\end{equation}
and the expectation value of the above when the computational register is traced out is the effective phase, as in equation \eqref{eq:eff_phase}, hence to enact \eqref{eq:test_exp}, multiple applications of \eqref{eq:recall_kick} must be applied, using multiple ancillas that are swapped in and out of the compute register in the general case of quantum data training. In the case of training classical neural networks on a quantum computer, as described in section \ref{sec:qnn}, since the compute registers are in an eigenstate of the QFB circuit, we can simply concatenate the phase kicks without the need for swapout, simply need to flip the input registers to the right input after each round, which is done unitarily. For quantum data training, if the phase kicking rates are kept small during training, even if the compute register ancillas are tossed away, as was shown in section \ref{sec:opt}, the dynamics of the weights are effectively unitary to first order in $\eta$.

\begin{figure}[h!]
 \begin{center}
\includegraphics[width=1.00\columnwidth]{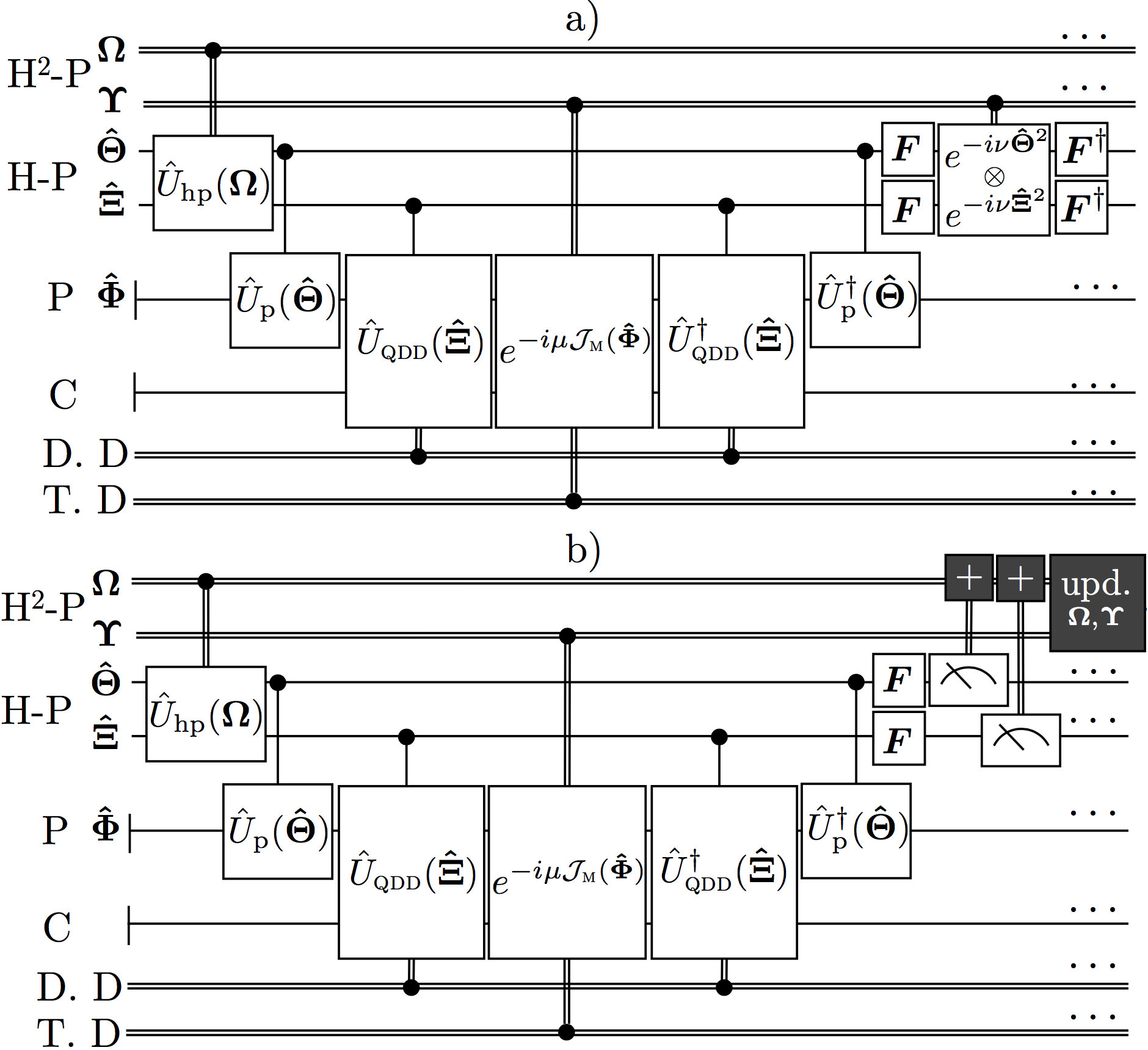}
\caption{Examples of the first iteration of both optimization strategies for quantum-enhanced hyper-parameter optimization via Quantum Meta-Learning (QMetaL). Represented in (a) is the Meta-QDD protocol, while in (b) is Meta-MoMGrad protocol. Note that in the above, $\text{H}^2\text{-P}$, H-P, and P are denote the hyper-hyper-parameters, the hyper-parameters, and the regular parameters, respectively. C is the compute register, while D.D and T.D denote the development (training) data, and the test data, respectively. The process begins with the preparation of the hyper-parameter pointer state using a unitary $\hat{U}_{\text{hp}}(\bm{\Omega})$, the meta-feedforward is then applied (eq. \eqref{eq:meta-feed}), the phase kick according to the test set error is then applied (see eq. \eqref{eq:test_exp}), and the meta-feedforward is then uncompute. Finally, in the case of QDD, a kinetic pulse is applied on the hyperparameters, whereas for MoMGrad the gradient of the hyper-parameters is measured and the hyper-hyper-parameters are updated for the next meta-iteration.  } \label{fig:EML}
\end{center}
\end{figure}

Now, we have defined the hyper-parametric unitary and the exponential loss function to be applied, we can consider applying the Meta-QFB (Quantum Feedforward and Phase Kick Backpropagation) procedure, for an iteration of such a phase kick, one must apply
\begin{equation}  \label{eq:meta_kick}
\hat{U}^\dagger_{\textsc{meta}}(\bm{\hat{\Theta}},\bm{\hat{\Xi}})  e^{-i\mu \mathcal{J}_{\textsc{m}}(\bm{\hat{\Phi}})}\hat{U}_{\textsc{meta}}(\bm{\hat{\Theta}},\bm{\hat{\Xi}}) 
\end{equation}
we can consider the effective phase function induced by the kickback from this meta-QFB. Let the compute and parameters' initial state be labelled as $\ket{\chi_0}_{CP}$, then the effective phase of the meta-QFB on the hyper-parameters can be labelled as 
\begin{equation}\label{eq:meta_phase}
\begin{split}
    &e^{-i\mu \mathcal{K}(\bm{\hat{\Theta}},\bm{\hat{\Xi}})}\\
    &\qquad \!\approx\! \bra{\chi_0}\hat{U}^\dagger_{\textsc{meta}}(\bm{\hat{\Theta}},\bm{\hat{\Xi}})  e^{-i\mu \mathcal{J}_{\textsc{m}}(\bm{\hat{\Phi}})}\hat{U}_{\textsc{meta}}(\bm{\hat{\Theta}},\bm{\hat{\Xi}})\ket{\chi_0}_{CP} 
\end{split}
\end{equation}
which is true to first order in $\mu$. Now, we have reduced the problem of hyper-parameter optimization to that of optimizing an effective exponential phase, as was the case before for the base case of QDD and MoMGrad. It is then straightforward to extend previous techniques to hyper-parameter optimization. First, for Quantum Dynamical descent, suppose we have a set of preparation hyper-hyper-parameters $\bm{\Omega}$, i.e., classical parameters which control how the initial quantum pointer states of the quantum hyper-parameter are initialized, in a sense the hyper-parameter analogue of $\bm{\Theta}$. Let $\hat{U}_{\text{hp}}(\bm{\Omega})$ be the hyper-parameter state preparation unitary. Let $\bm{\Upsilon}$ act as the classical hyper-hyper-parameters representing the meta-QDD or meta-MoMGrad kicking and kinetic rates, i.e., the hyper-hyper-parameters $\bm{\Upsilon}=\{\bm{\mu},\bm{\nu}\} $ are analogues of the hyper-parameters $\bm{\Xi} =\{\bm{\eta},\bm{\gamma}\}$ for the meta-optimization. The Meta-QDD algorithm, pictured in figure \ref{fig:EML}, can be summarized as applying the hyper-parameter preparation unitary, followed by the sequence

\begin{equation}\label{eq:mqdd}
    \hat{U}_{\textsc{m-qdd}} = \prod_{j\in \mathcal{B}_{\textsc{t}}} \bm{\hat{F}}^{\dagger}_h e^{-i\nu_j(\bm{\hat{\Omega}}^2+ \bm{\hat{\Xi}}^2)}\bm{\hat{F}}_{h} e^{-i\mu_j \mathcal{K}(\bm{\hat{\Theta}},\bm{\hat{\Xi}})}
\end{equation}
where the $\bm{\hat{F}}_{h}$ is the component-wise Quantum Fourier transform for all hyper-parameter registers.

For meta-MoMGrad, similarly, we can begin by preparing the quantum pointer states of the hyper-parameters, using a parametric unitary which itself is dependent on preparation hyper-hyper-parameters $\bm{\Omega}$, i.e., $\hat{U}_{\text{hp}}(\bm{\Omega})$, following this we can apply the meta-QFB circuit from equation \eqref{eq:meta_phase} in order to apply the effective phase kick $e^{-i \mu \mathcal{K}(\bm{\hat{\Theta}},\bm{\hat{\Xi}})}$. To complete Meta-MoMGrad, we can then apply the component-wise Fourier transform $\bm{\hat{F}}_h$ on the hyper-parameter registers, and then measure these registers in their computational bases. From this phase kick, the shift in expectation value will be proportional to the negative gradient of the effective phase, we can see this by looking at the Heisenberg picture, 
\spliteq{
    \text{Ad}[e^{i \mu \mathcal{K}(\bm{\hat{\Theta}},\bm{\hat{\Xi}})}]&(\bm{\hat{F}}^\dagger\bm{\hat{\Theta}}\bm{\hat{F}} )\\
    &= \bm{\hat{F}}^\dagger\bm{\hat{\Theta}}\bm{\hat{F}} -\mu \nabla_{\bm{\hat{\Theta}}} \mathcal{K}(\bm{\hat{\Theta}},\bm{\hat{\Xi}})+ \mathcal{O}(\mu^2)\\
    \text{Ad}[e^{i \mu \mathcal{K}(\bm{\hat{\Theta}},\bm{\hat{\Xi}})}]&(\bm{\hat{F}}^\dagger\bm{\hat{\Xi}}\bm{\hat{F}})\\
    &= \bm{\hat{F}}^\dagger\bm{\hat{\Xi}}\bm{\hat{F}} -\mu \nabla_{\bm{\hat{\Xi}}} \mathcal{K}(\bm{\hat{\Theta}},\bm{\hat{\Xi}})+ \mathcal{O}(\mu^2)
}
the computational basis hyper-parameter observables, after an Fourier transform, is shifted by the negative gradient multiplied by the hyper-parameter phase kicking rate. Similar to MoMGrad for the regular parameters (see \eqref{eq:mom_pi_upt}, \eqref{eq:mom_phi_upt}), one can then update the preparation hyper-parameters for the next iteration, i.e., update the initial expectation value of position and momentum, which are hyper-hyper-parameters in the vector $\bm{\Omega}$, akin to the $\bm{\sigma}, \bm{\Phi}_0$, and $\bm{\Pi}_0$ but for the hyper-parameters; the first and second moments of the pointer states. The rate at which the mean hyper-parameter value is updated can be multiplied by some constant $\nu$, akin to \eqref{eq:mom_phi_upt} but with $\nu$ replacing $\gamma$. This hyper-hyper-parameter $\nu$ can be considered the effective kinetic rate for the Meta-MoMGrad. We represent an iteration of Meta-MoMGrad in Figure~\ref{fig:EML}.

Finally, note that the hyper-hyper-parameters $\{\bm{\Upsilon},\bm{\Omega}\}$ remain to be optimized. Theoretically, just as we have shown above that one can optimize the hyper-parameters via MoMGrad/QDD if the parameters are being optimized by QDD, we could consider performing a quantum parameter descent on the hyper-hyper-parameters. To do so, one could consider quantizing the hyper-hyper-parameters $\{\bm{\Upsilon},\bm{\Omega}\}\mapsto \{\bm{\hat{\Upsilon}},\bm{\hat{\Omega}}\}$, and applying a meta-meta-optimization on these using MoMGrad or Quantum Dynamical Descent, with Meta-QDD taking the role of QDD. In a sense, QDD is self-concatenable to as many meta-levels of optimization as is desired. Practically, each additional level adds a nested QFB loop of optimization, which grows the overhead of execution exponentially. Additionally, the number of hyper-parameters increases with the number of meta-optimizations, since the parameter preparation hyper-parametric unitary has multiple hyper-parameters per parameter. For meta-levels of concatenation to be useful (in the sense of achieving a lower expectation value of the cost function of choice), one would need to consider a choice of hyper-parameters which reduces the ratio of hyper-parameters per parameter for each level of meta-optimization. Perhaps one could take inspiration from classical machine learning techniques \cite{andrychowicz2016learning} 
, where a recurrent neural network is used to relate the different rates of descent at subsequent iterations, thus providing an educated ansatz for how these should relate to each other, and thus reducing the number of unique degrees of freedom in the hyper-parameters. To incorporate such a technique into QDD would require adding a hyper-hyper-parametric circuit/neural network to relate the hyper-parameters between iterations, which would require a modification of the QDD approach, we leave this for future work.

\subsubsection{Network Architecture Optimization}\label{sec:NAO}

Another application of the Quantum Meta-Learning principle is for Quantum Network Architecture Optimization.
In some instances, one may want to optimize over various network architectures in order to improve performance, e.g., one may optimize whether a certain set of parametric circuit elements should be applied, or optimize over a space of possible neural network connectomes (topology of connections). This problem can be seen as a meta-learning problem, as the goal is to pick the network architecture which performs best once each network is trained. As such, the optimization must be done over a space of \textit{trained} networks, and this space of architectures is generally discretely parametrized. To enact this optimization, we can adapt techniques of quantum discrete parametric optimization from section \ref{sec:discrete}, and combine it with some of the machinery from our treatment of dropout \ref{sec:drop}, along with the principles of Meta-QDD or Meta-MoMGrad which were just discussed above. 

The key to network architecture optimization will be to have ancillary quantum registers to index the architecture. Luckily, we have already developed a formalism for this in section \ref{sec:drop}, as such, we will use the same notation in this section. Recall our general decomposition of the parametric unitary from \eqref{eq:full_par_circ}, 
\begin{equation}\label{eq:full_par_circ-prime}
      \hat{U}(\bm{\hat{\Phi}}) = \prod_{\ell=1}^\mathscr{L}\hat{U}^{(\ell)}(\bm{\hat{\Phi}}^{(\ell)}), \quad \hat{U}^{(\ell)}(\bm{\hat{\Phi}}^{(\ell)})\equiv \bigotimes_{j_\ell\in \mathcal{I}_\ell} \hat{U}_{j_\ell }(\hat{\Phi}_{j_\ell}).
\end{equation} We can add a set of $N$ control qubits ancillas (of Hilbert space $\mathcal{H}_A$), which will each control whether a certain subset of parametric unitaries is applied. These qubtits can be seen as hyper-parameters, and as such we can construct the hyper-parametric unitary (same as eq. \eqref{eq:AHP}) of the form
\begin{equation}\label{eq:AHP-prime}
        \hat{\bar{U}}(\bm{\hat{\Phi}},\bm{\hat{Z}}_{A}) = \prod_{\ell=1}^\mathscr{L} \prod_{j_\ell\in \mathcal{I}_\ell} \hat{C}(j_\ell)\otimes \hat{U}_{j_\ell }(\hat{\Phi}_{j_\ell})
\end{equation}
where $\mathcal{I}= \cup_{j=0}^N\mathcal{A}_j$ is the partition of indices which groups operations for which we wish to share the same control parameter (e.g. multiple parametric operations of a neuron). The subset of indices $\mathcal{A}_0$ corresponds to the set of indices of unitaries over which we would not like to optimize. The operator $\hat{C}(j)$ is defined in equation \eqref{eq:ctrl}, it is simply a way to index which control qubit each unitaries is assigned. For further details on how to compile this unitary refer to section \ref{sec:drop}.

Now, we have outlined how to convert a given parametric circuit to a hyper-parametric circuit with architecture index qubits, we can simply apply techniques from the Meta-MoMGrad/Meta-QDD from subsection \ref{sec:QHD}, combined with the adaptations of MoMGrad/QDD for discrete optimization from section \ref{sec:discrete}. The key is to replace the parametric unitary $\hat{U}(\bm{\hat{\Phi}})$ from \eqref{eq:full_par_circ-prime}, with the architecture-hyper-parametric unitary  $\hat{\bar{U}}(\bm{\hat{\Phi}},\bm{\hat{Z}}_{A})$ from \eqref{eq:AHP-prime}.

For architecture Meta-Learning MoMGrad, we can start in a pointer state of the architecture hyper-parameter registers as in \eqref{eq:disc_update}
\begin{equation}
    \ket{\alpha_0}_A = \bigotimes_{k=1}^N (\cos(\theta_k)\ket{0}_{A_k} + i\sin(\theta_k)\ket{1}_{A_k}\!)
\end{equation}
and a pointer state of choice for the regular parameters $\ket{\Psi_0}$ (see sec. \ref{sec:momgrad}). 
Onto this joint pointer state of choice, we can then apply the modified feedforward hyper-parametric unitary from $\hat{\bar{U}}(\bm{\hat{\Phi}},\bm{\hat{Z}}_{A})$ \eqref{eq:AHP}, then the usual loss function exponential, and then uncompute the feedforward in order to complete the effective hyper-parametric phase kick unitary; e.g.
\begin{equation}\label{eq:anotherkick}
 e^{-i\eta\hat{L}_j(\bm{\hat{\Phi}},\bm{\hat{Z}}_{A})}= \hat{\bar{U}}^\dagger(\bm{\hat{\Phi}},\bm{\hat{Z}}_{A}) e^{-i\eta\hat{L}_j}\hat{\bar{U}}(\bm{\hat{\Phi}},\bm{\hat{Z}}_{A})
\end{equation}
and by batching multiple kicks like the above into a minibatch $\mathcal{B}_k$, we can enact an effective phase kick on the joint system of hyper-parameters and parameters,
\begin{equation}
   e^{-i\eta_k \mathcal{J}(\bm{\hat{\Phi}},\bm{\hat{Z}}_{A})}= \prod_{j\in \mathcal{B}_k}e^{-i\bar{\eta}_k \mathcal{L}_j(\bm{\hat{\Phi}},\bm{\hat{Z}}_{A})}.
\end{equation}
This exponential effective cost function can then be used for a minibatched architecture-dependent Quantum Dynamical Descent, by interlacing some kinetic pulses on the parameters just as in regular QDD,
    \spliteq{\label{eq:AQDD}
    \hat{U}_{\textsc{aqdd}}(\bm{\hat{Z}}_{A}) &= \prod_{k\in \mathcal{B}} e^{-i\gamma_k\bm{\hat{\Pi}}^2 } e^{-i\eta_k \mathcal{J}(\bm{\hat{\Phi}},\bm{\hat{Z}}_{A})}.\\
}
Note that as opposed to our method in dropout (sec. \ref{sec:drop}), in this meta-learning approach the state for the architecture qubits $\ket{\alpha_0}$ is kept in quantum memory between QFB runs and for multiple QDD iterations rather than being reinitialized every run.
Now, similar to \eqref{eq:meta_phase} and \eqref{eq:meta_kick}, one can perform a meta-QFB, using $\hat{U}_{\textsc{AQDD}}$ as the meta-feedforward;
\begin{equation}\label{eq:hmm}
\hat{U}^\dagger_{\textsc{aqdd}}(\bm{\hat{Z}}_{A})  e^{-i\mu \mathcal{J}_{\textsc{m}}(\bm{\hat{\Phi}},\bm{\hat{Z}}_{A})}\hat{U}_{\textsc{aqdd}}(\bm{\hat{Z}}_{A}) 
\end{equation}
where $e^{-i\mu \mathcal{J}_{\textsc{m}}(\bm{\hat{\Phi}},\bm{\hat{Z}}_{A})}$ is a cost function hyper-parametric phase kick for the test set of the data. We can let $e^{-i\mu \mathcal{K}(\bm{\hat{Z}}_{A})}$ be the the effective phase induced on the architecture hyper-parameters by the operation in \eqref{eq:hmm}. 

To optimize the architecture hyper-parameters, we can then either apply a discrete Meta-QDD, 
\begin{equation}
   \prod_j   e^{-i\nu_j\bm{\hat{X}}_A}e^{-i\eta_j \mathcal{K}(\bm{\hat{Z}}_{A})}
\end{equation}
where the hyper-hyper-parameters will need to be optimized. One option being combining this architecture meta-optimization and the regular meta-learning from the previous subsection into one meta-optimization loop. We leave this as exercise to the reader.

Finally, another option is to perform a discrete Meta-MoMGrad by measuring $\braket{\bm{\hat{X}}_A}$ after phase kicks $e^{-i\eta_j \mathcal{K}(\bm{\hat{Z}}_{A})}$ and updating the angles $\theta_j$ according to the estimated gradient as prescribed in \ref{sec:discrete}. Another option which might be beneficial in this case would be to use continuum-embedding for the discrete parameters, since the gradient estimation can be much more fine-grained.

\section{Quantum Neural Network Learning}\label{sec:qnn}

In this section, we will elaborate upon the use of the ideas presented in the previous section for the purpose of quantumly training deep neural networks on a quantum computer, to solve machine learning problems involving classical data.
Here we will present a quantum neural network architecture which encodes classical neural networks as quantum parametric circuits, along with an in-depth analysis of the phase kick backpropagation procedure from the previous section, and how the error signals backpropagate through the quantum neural network.

\subsection{Quantum-Coherent Neural Networks}

In this subsection we show how to encode a classical feedforward neural network into a quantum computation and how to leverage the Quantum Feedforward and Baqprop as well as optimization techniques introduced in Sections~\ref{sec:opt} and \ref{sec:misc} for the training of such a network.

\subsubsection{Classical-to-Quantum Computational Embedding}

A central principle employed in this section is the ability to encode a classical computation into a quantum computation \cite{haner2018optimizing}. In general, for an $n$-bit input, $\bm{x}\in\mathds{Z}_2^n$, and a computable function from $n$ bits to $m$ bits, $f:\mathds{Z}_2^n \rightarrow \mathds{Z}_2^m$, we can embed the classical computation as a unitary quantum computation $\hat{U}_f:\mathcal{H}_{\textsc{a}}\otimes \mathcal{H}_{\textsc{b}}$ acting on $n+m$ qubits \cite{nielsen2002quantum}. This unitary takes computational basis state equivalent of the input, $\ket{\bm{x}}$, an maps it as follows:
\begin{equation}
  \hat{U}_f  \ket{\bm{x}}\ket{\bm{0}} =  \ket{\bm{x}}\ket{f(\bm{x})}.
\end{equation}
Thus this quantum-converted classical function maps computational basis states to computational basis states. Note that, trivially by the linearity of quantum operators, superpositions of computational basis states get mapped to entangled superpositons between the possible inputs and their corresponding standard basis outputs,
\begin{equation}
      \hat{U}_f  (\textstyle\sum_j\psi_j\ket{\bm{x}_j})\ket{\bm{0}} =  \sum_j\psi_j\ket{\bm{x}_j}\ket{f(\bm{x}_j)}.
\end{equation}
Of course, since the evaluation is unitary, the above computation is fully reversible:
\begin{equation}
        \hat{U}^\dagger_f    \left(\textstyle\sum_j\psi_j\ket{\bm{x}_j}\ket{f(\bm{x}_j)}\right) =(\textstyle\sum_j\psi_j\ket{\bm{x}_j})\ket{\bm{0}}.
\end{equation}

Notice that for such functions the probability amplitudes are unaffected during the evaluation.
That is, each branch of the wavefunction labelled by the $\bm{x}_j$'s evolves independently of the others.
This property will be harnessed during the computation and uncomputation stages of the Quantum Feedforward and Backwards Quantum Propagation of Phase errors (QFB).
That is, we use the ability to query classical functions in superposition in order to tag the output with relative phase shifts, and follow this with by uncomputation.
The combination of all three of these steps causes appropriate momentum kicks for the parameters which can be leveraged for optimization using techniques from Sections~\ref{sec:opt} and \ref{sec:misc}.

Simply by the nature of the embedding of a classical computation into a quantum computation, by the requirement of reversibility, we are forced to store the computational graph of the classical computation in quantum memory.
For a Directed Acyclic Graph representing the flow of classical variables being transformed by a composing multivariate functions, such as is the case for neural networks, this so-called computational graph \cite{rall1981automatic} then has to be embedded into an entangled set of quantum registers which hold the history of the computation.
The encoding of computation into multiple quantum registers can be seen, in a sense, as embedding the classical computational graph in quantum memory, it is then natural that one can backpropagate a phase error signal through the computational graph via uncomputation, which we know carries gradient information.
This generalized backpropagation through a general computational graph is called \textit{Automatic Differentiation} (AD), the specialization of AD to Neural Networks is what is considered to be the error backpropagation algorithm.
In subsection \ref{sec:deep_baq} we analyse in-depth how the phase signal is carried through during the uncomputation, and how one can rederive the classical neural network backpropagation principle from it.
Although we do not explicitly do so, this analysis could then easily be extendable to a general computational graph, thereby providing a demonstration of emergence of automatic differentiation through quantum phase backpropagation in a general setting.

Although recent progress has been made to perform common classical operations efficiently on a quantum computer \cite{haner2018optimizing}, in general, synthesizing quantum circuits for quantum-embedded classical computations may not always be efficient.
On the other hand, our focus is on training neural networks, which only require certain types of operations, namely multiplication, addition, and the evaluation of activation functions for continuous values.
In this section we will thus cover how to addition and multiplication using machinery introduced in the background section \ref{sec:CQR}.
Later in this section (see \ref{sec:zoo}), we cover possible implementations of activation functions commonly used in classical machine learning.

\subsubsection{Classical Data Phase Kicking}
\label{sec:cl_data}

First, let us begin by detailing how exactly to enact the phase kicking according to a classical loss function.
For purposes of demonstrating the key concepts, we will consider the employment of the QFB algorithm for an classical supervised learning, although it could also be used in other contexts.
In classical supervised learning, the goal is to build a model for a function $\bm{f} : \bm{x} \mapsto \bm{y}$ based on training data $\{ (\bm{x}_j,\bm{y}_j) \}_j$.
The model consists of a parametrized ansatz $\bm{f}(\bm{\Phi},\bm{x})$ with parameters $\bm{\Phi}$.
Every set of parameters gives a prediction denoted $\bm{\tilde{y}} = \bm{f}(\bm{\Phi},\bm{x})$.
As above, the Hilbert space for the parameters will be denoted $\mathcal{H}_{\bm{\Phi}}$.
The Hilbert space for the computation requires registers for the inputs and the prediction: $\mathcal{H}_C = \mathcal{H}_{\bm{x}} \otimes \mathcal{H}_{\bm{\tilde{y}}}$.
Note that in the case of training with a superposition of data, one would also requires a set of registers for the output, $\mathcal{H}_{\bm{y}}$.
For the moment we will only consider using a single data point at a time, so each $\bm{y}_i$ will only enter the loss functions as a classical parameter, although the extension to encoding the outputs in a quantum register is straightforward, depending on the loss function.

For now, consider a single input-output pair $(\bm{x},\bm{y})$.
The parametrized algorithm for classical training is a unitary, $\hat{U}_{\bm{f}}(\hat{\bm{\Phi}})$, that computes $f(\bm{\Phi},\bm{x})$, i.e.,
\begin{equation}
  \hat{U}_{\bm{f}}(\hat{\bm{\Phi}}) : \ket{\bm{\Phi}, \bm{x}, \bm{0}} \mapsto \ket{\bm{\Phi}, \bm{x}, \bm{f}(\bm{\Phi},\bm{x}) }.
\end{equation}
Later in this section, we will be constructing explicit circuits (quantum-coherent neural networks) that implement $\hat{U}_{\bm{f}}(\hat{\bm{\Phi}})$.
For now, we will write this unitary somewhat abstractly as
\begin{equation}
\begin{split}
  \hat{U}_{\bm{f}}(\hat{\bm{\Phi}}) &= \sum_{\bm{\Phi}, \bm{x}}  \ket{\bm{\Phi}}\!\bra{\bm{\Phi}} \otimes \ket{\bm{x}}\!\bra{\bm{x}} \otimes e^{-i \bm{f}(\bm{\Phi}, \bm{x}) \hat{\bm{p}}_{\bm{\tilde{y}}} } \\
  &= e^{-i \bm{f}(\hat{\bm{\Phi}},\hat{\bm{x}}) \cdot \hat{\bm{p}}_{\bm{\tilde{y}}} },
\end{split}
\end{equation}
where $\hat{\bm{p}}_{\bm{\tilde{y}}}$ is the generator of shifts in the prediction register.
Our notation will be suggestive of continuous registers, although this can be achieved for discrete registers as well. 

Now we apply a loss function which compares the output to the prediction:
\begin{equation}
  e^{-i \eta L(\hat{\bm{\tilde{y}}}, \bm{y}) }.
\end{equation}
For example, the loss function could be
\begin{equation}
    L(\hat{\bm{\tilde{y}}},\bm{y}) = \| \hat{\bm{\tilde{y}}} - \bm{y} \|_2^2,
\end{equation} note that this is still an operator which acts on $\mathcal{H}_{\bm{\tilde{y}}}$. The loss exponential of such a mean-squared error loss is efficiently compilable into a tensor product of second-order phase exponentials of each register. In principle, this loss function could be any classical computable function which maps the output to the set of reals. 

After uncomputing with $\hat{U}_f(\hat{\bm{\Phi}})^\dagger$, the entire QFB circuit is
\spliteq{\label{eq:cl_qfb}
    \hat{U}_{\textsc{QFB}} &= e^{i \bm{f}(\hat{\bm{\Phi}},\hat{\bm{x}}) \cdot \hat{\bm{p}}_{\bm{\tilde{y}}}} e^{-i \eta L(\hat{\bm{\tilde{y}}},\bm{y})} e^{-i \bm{f}(\hat{{\bm{\Phi}}},\hat{\bm{x}}) \cdot \hat{\bm{p}}_{\bm{\tilde{y}}}} \\
    & = e^{-i \eta L(\hat{\bm{\tilde{y}}} + \bm{f}(\hat{{\bm{\Phi}}},\hat{\bm{x}}),\bm{y})}
}
Applied to the momenta of the parameters, we have that Eq.~\eqref{eq:grad_rule} gives
\begin{equation}\label{eq:grad_rule}
  \hat{U}_{\textsc{QFB}}^\dagger \hat{\Pi}_k \hat{U}_{\textsc{QFB}} = \hat{\Pi}_k - \eta \frac{\partial}{\partial \hat{\Phi}_k} L(\hat{\bm{\tilde{y}}} + \bm{f}(\hat{\bm{\Phi}},\hat{\bm{x}}),\bm{y}).
\end{equation}
In particular, we see that all of the higher order terms vanish since all the operators in $L$ commute and hence $L$ commutes with its partial derivatives.
For classical data, all of the terms in the above update rules for the momentum truncate at first order in $\eta$.
Thus, for training a classical machine, the momentum of each parameter gets shifted by an amount equal to the partial derivative of the loss function.

We also have the important fact that the parameter and computational registers are not entangled at the end of the QFB circuit, hence the parameters will experience no decoherence due to these operations.
Of course, this is assuming perfect position eigenstates for the computational registers $\mathcal{H}_{\bm{x}}$ and $\mathcal{H}_{\bm{\tilde{y}}}$; the parameters will experience some decoherence if this is not the case (e.g., if one is using finitely squeezed continuous variable pointer states).

Notice that $\hat{U}_{\textsc{QFB}}$ applied to the initial state yields:
\spliteq{
  \hat{U}_{\textsc{QFB}} :& \sum_{\bm{\Phi}} \psi(\bm{\Phi}) \ket{\bm{\Phi}} \otimes \ket{\bm{x},0} \\&\mapsto \sum_{\bm{\Phi}} e^{-i \eta L( \bm{f}(\bm{\Phi},\bm{x}), \bm{y} ) } \psi(\bm{\Phi}) \ket{\bm{\Phi}} \otimes \ket{\bm{x},0}.
}

 Since $\hat{U}_{\textsc{QFB}}$ leaves $\ket{\bm{x},0} \in \mathcal{H}_C$ invariant, then the QFB circuit simply tags different values of $\bm{\Phi}$ with a phase depending on the corresponding output of the circuit.
 In this case, we get a true phase kickback.
 Because $\hat{U}_{\textsc{QFB}}$ acting on this initial state does not generate entanglement between the parameter and the computational registers, then for further data points, it is simple to coherently re-initialize $\ket{\bm{x},0}$ to input a new data point $\ket{\bm{x'},0}$.

For multiple data points in a minibatch, $\{ (\bm{x}_j, \bm{y}_j) \}_{j \in \mathcal{B}}$, we begin in a state
\begin{equation}
  \sum_{\bm{\Phi}} \psi(\bm{\Phi}) \ket{\bm{\Phi}} \otimes \ket{\bm{0},\bm{0}} \in \mathcal{H}_{\bm{\Phi}} \otimes \mathcal{H}_{\bm{x}} \otimes \mathcal{H}_{\bm{\tilde{y}}}.
\end{equation}
For each data point, we can first shift the input register to the appropriate $\bm{x}_j$, apply the QFB circuit with the appropriate output $\bm{y}_j$ in the loss function, and then shift the input register back to zero and repeat for all of the data points in the minibatch.
Explicitly, the algorithm is:
\begin{equation}
\begin{split}
  \prod_{j \in \mathcal{B}} e^{i \bm{x}_j \cdot \hat{\bm{p}}_{\bm{x}}} & e^{-i \eta L(\hat{\bm{\tilde{y}}} + \bm{f}(\hat{\bm{\Phi}},\hat{\bm{x}}), \bm{y}_j)} e^{-i \bm{x}_j \cdot \hat{\bm{p}}_{\bm{x}}} \\
  &= \prod_{j \in \mathcal{B}} e^{-i \eta L(\hat{\bm{\tilde{y}}} + \bm{f}(\hat{\bm{\Phi}},\hat{\bm{x}} + \bm{x}_j), \bm{y}_j)}.
\end{split}
\end{equation}
This maps the parameter momenta and the state to (respectively):
\begin{equation}
  \hat{\Pi}_k \mapsto \hat{\Pi}_k - \eta \sum_{j \in \mathcal{B}} \frac{\partial}{\partial \hat{\Phi}_k} L(\hat{\bm{\tilde{y}}} + \bm{f}(\hat{\bm{\Phi}},\hat{\bm{x}}+\bm{x}_j), \bm{y}_j),
\end{equation}
and
\begin{equation}
\begin{split}
  \sum_{\bm{\Phi}} &\psi(\bm{\Phi}) \ket{\bm{\Phi}} \otimes \ket{\bm{0},\bm{0}} \\
  &\mapsto \sum_{\bm{\Phi}} e^{-i \eta \sum_{j \in \mathcal{B}} L(\bm{f}(\bm{\Phi},\bm{x}_j),\bm{y}_j) } \psi(\bm{\Phi}) \ket{\bm{\Phi}} \otimes \ket{\bm{0},\bm{0}}.
\end{split}
\end{equation}
We see that the momenta of the parameters and the phase induced in the final state is according to the cost function
\begin{equation}
  J = \frac{1}{|\mathcal{B}|} \sum_{j \in \mathcal{B}} L(\bm{f}(\bm{\Phi},\bm{x}_j),\bm{y}_j),
\end{equation}
and accumulated kicking rate $\eta |\mathcal{B}|$.

Note that this discussion also applies if $\bm{f}$ is comprised of multiple layers:
\begin{equation}
  \bm{f}(\bm{\hat{\Phi}},\bm{\hat{x}}) = \bm{f}_N( \bm{\hat{\Phi}}_N, \cdots \bm{f}_2( \bm{\hat{\Phi}}_2, \bm{f}_1( \bm{\hat{\Phi}}_1, \bm{\hat{x}}) ) ),
\end{equation}
with parameters are divided up among these layers as $\bm{\hat{\Phi}} = \bm{\hat{\Phi}}_N \oplus \cdots \oplus \bm{\hat{\Phi}}_2 \oplus \bm{\hat{\Phi}}_1$.
The update rule for the momenta of the parameters that we derived above also holds in this special case, so abstractly we can conclude that the uncomputation step of the QFB algorithm indeed propagates gradient information back through the computational graph.
In Section~\ref{sec:deep_baq} we will explore in-depth how this mechanism behaves in the context of backpropagation of error in neural-network type computations, as a special case of the above analysis.

A final note about implementation of the loss function for classical data problems.
Occasionally it may be more practical to use an auxiliary register to store the computation of the loss function, rather than to exponentiate the loss function as a phase kick.
That is, suppose we added another register $\mathcal{H}_L$ and appended a computation $\hat{U}_L(\bm{y})$ to the feedforward operation:
\begin{equation}
\begin{split}
  \hat{U}_L(\bm{y}) \circ \hat{U}_{\bm{f}}(\bm{\hat{\Phi}}) :& \ket{\bm{\Phi},\bm{x},\bm{0},0} \\
  &\quad \mapsto \ket{\bm{\Phi},\bm{x},\bm{f}(\bm{\Phi},\bm{x}),L(\bm{f}(\bm{\Phi},\bm{x}),\bm{y})}
\end{split}
\end{equation}
which we can denote abstractly as $\hat{U}_L(\bm{y}) = e^{-i L(\bm{\hat{\tilde{y}}},\bm{y}) \hat{p}_L }$.
Then instead of exponentiating the loss function as a phase kick, we simply apply a linear phase shift, $e^{-i \eta \hat{x}_L}$, to this new register before uncomputing the modified feedforward circuit.
In all, this modified QFB algorithm is
\begin{equation}
\begin{split}
  &\hat{U}_{\textsc{qfb}+\textsc{l}} \\
  &\quad = e^{i \bm{f}(\bm{\hat{\Phi}},\bm{\hat{x}}) \cdot \bm{\hat{p}}_{\bm{\tilde{y}}}} e^{i L(\bm{\hat{\tilde{y}}},\bm{y}) \hat{p}_L } e^{-i \eta x_L} e^{-i L(\bm{\hat{\tilde{y}}},\bm{y}) \hat{p}_L } e^{-i \bm{f}(\bm{\hat{\Phi}},\bm{\hat{x}}) \cdot \bm{\hat{p}}_{\bm{\tilde{y}}}}\\
  &\quad = e^{i \bm{f}(\bm{\hat{\Phi}},\bm{\hat{x}}) \cdot \bm{\hat{p}}_{\bm{\tilde{y}}}}  e^{-i \eta (x_L + L(\bm{\hat{\tilde{y}}},\bm{y}) )} e^{-i \bm{f}(\bm{\hat{\Phi}},\bm{\hat{x}}) \cdot \bm{\hat{p}}_{\bm{\tilde{y}}}}\\
  &\quad = e^{-i \eta \left( x_L + L(\bm{\hat{\tilde{y}}} + \bm{f}(\bm{\hat{\Phi}},\bm{\hat{x}}), \bm{y}) \right)}.
\end{split}
\end{equation}
We see that if we initialize the new register $\mathcal{H}_L$ to $\ket{x_L = 0}$, then this is equivalent to \eqref{eq:cl_qfb}.

Now we will proceed to discuss the construction of the circuit $\hat{U}_f(\hat{\bm{\Phi}})$ which computes the output to a neural network with quantum parameters.
We will also discuss in detail the feedforward and backpropagation mechanisms in this setting in order to make some of the previous discussions more concrete.

\subsubsection{Abstract Quantum Neuron}

Classical neurons usually act by taking as input a collection of signals, adding up these contributions in a weighted fashion, and applying a nonlinearity to this sum to finally generate an output.
A simple example, given a vector of inputs $\bm{x}$, weights $\bm{w}$, and bias $b$, the mapping corresponding to the neuron is given by $\bm{x} \mapsto \sigma( \bm{w} \cdot \bm{x} + b )$ where $\sigma : \mathds{R} \rightarrow \mathds{R}$ is a nonlinear activation function.

\begin{figure}[h!]
\includegraphics[width=1\columnwidth]{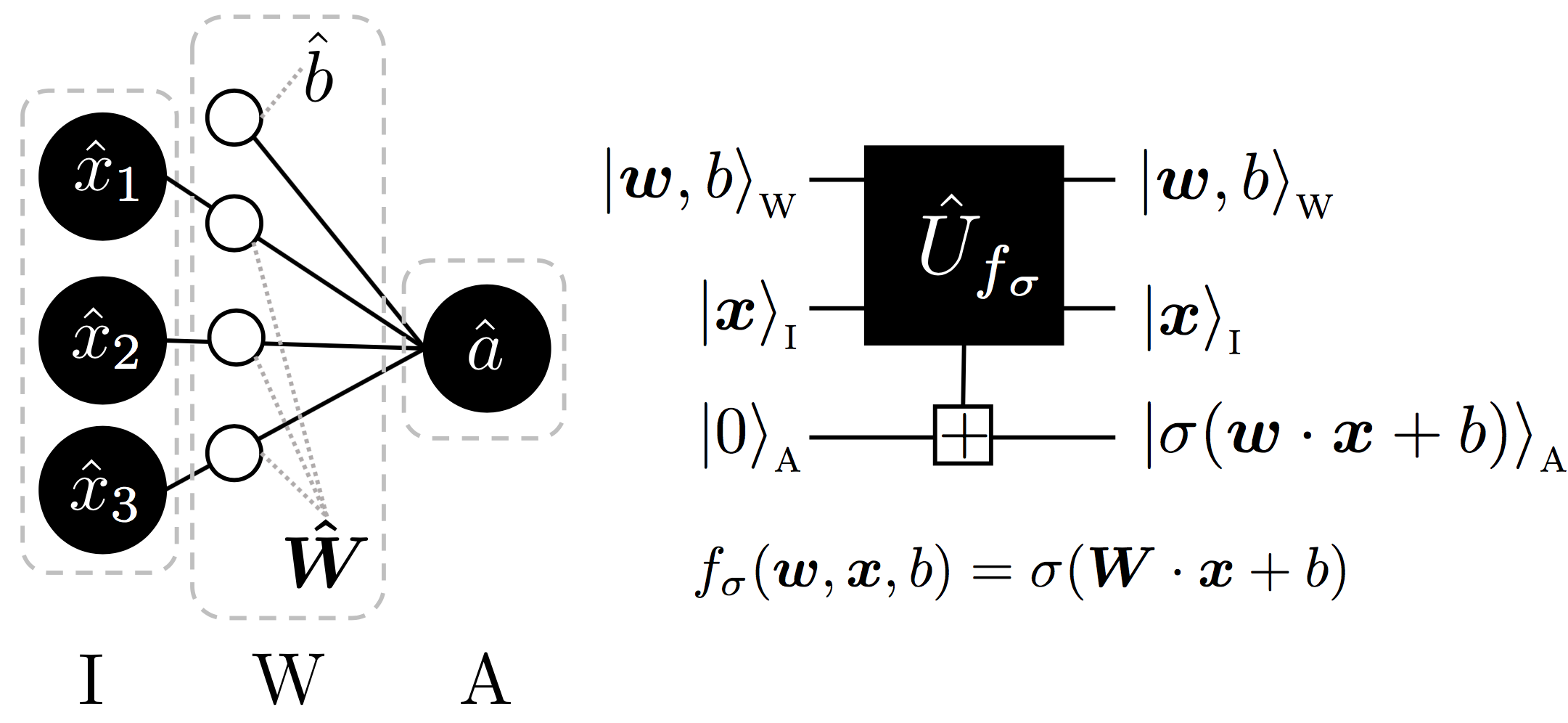}
\caption{Diagram of abstract quantum neuron model. Left is a representation of the neuron itself, right is an abstract quantum circuit representing its feedforward operation.
} \label{fig:abstract}
\end{figure}

To embed this mapping into a quantum computation, we need to make the whole process reversible, as quantum computation ultimately has to be enacted by unitary, hence invertible, operations.
To do this, we can assume that the weights, the inputs, and the outputs (activation), are all quantum number registers (either continuous variable or a discrete variable binary approximation thereof).
The quantum neuron should ideally map
\begin{equation}
  \ket{\bm{x}}_{\textsc{i}} \ket{\bm{w},b}_\textsc{w} \ket{0}_{\textsc{a}} \mapsto \ket{\bm{x}}_{\textsc{i}} \ket{\bm{w},b}_{\textsc{w}} \ket{\sigma( \bm{w} \cdot \bm{x} + b )}_{\textsc{a}}
\end{equation}
which could be implemented via an idealized unitary which enacts the above map:
\begin{equation}\label{eq:ideal_ff}
  e^{-i \sigma( \hat{\bm{w}} \cdot \hat{\bm{x}} + b ) \hat{p}_a},
\end{equation}
where $\hat{a}$ is the position quadrature of the activation register, and $\hat{p}_a$ is its canonical conjugate: $[ \hat{a}, \hat{p}_a ] = i$. In Figure~\ref{fig:abstract} we picture such an abstract neuron and a corresponding abstract quantum circuit.

\begin{figure}[h!]
\includegraphics[width=0.75\columnwidth]{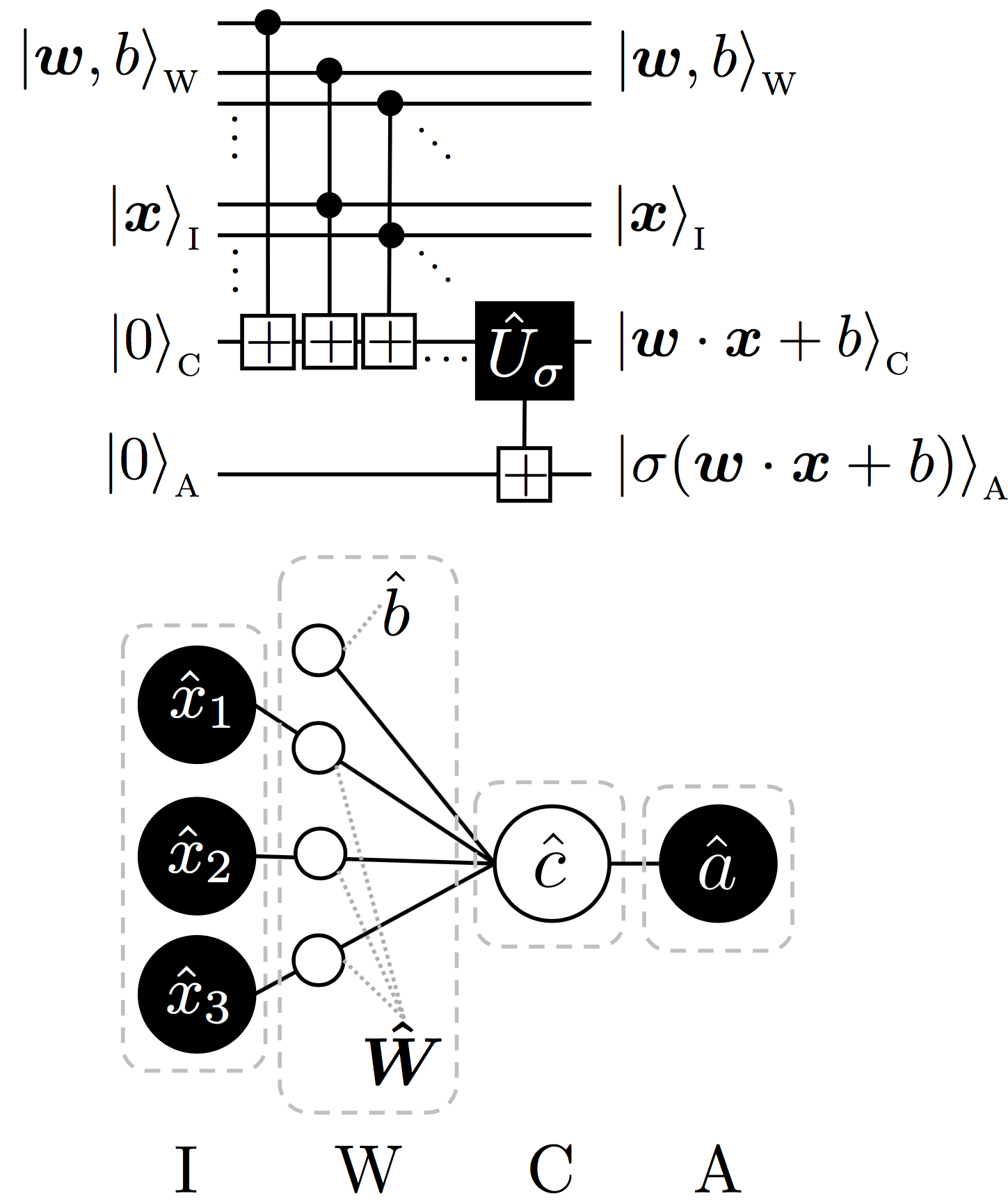}
\caption{Diagram of a neuron with collector, and its corresponding feedforward quantum circuit. The controlled-adders are of the form \eqref{eq:addition}, and the control-control adder are of the form \eqref{eq:multiplication}. The operator involving $\hat{U}_\sigma$ computes the non-linear activation function $\sigma$ on the collector register and stores the result in the activation.
} \label{fig:collect}
\end{figure}

As the above form is quite abstract, let us briefly describe how we could unpack the execution of the above feedforward operation, while remaining fairly abstract.
In section \ref{sec:zoo}, we outline various circuits and physical scenarios which would practically enact such a mapping, either using nonlinear optics, or using ancilla registers and phase estimation. In both cases, the weighted contributions of the input are first accumulated in \emph{collector} register, $\textsc{c}$, as such:
 $\ket{\bm{x}}_\textsc{i} \ket{\bm{w},b}_\textsc{w} \ket{0}_\textsc{c} \mapsto \ket{\bm{x}}_\textsc{i} \ket{\bm{w},b}_\textsc{w} \ket{z}_\textsc{c}$, where $z = \bm{w} \cdot \bm{x} + b$.
The remaining operation is to take the stored in the collector register, and synthesis the computation of the activation function $\sigma:\mathds{R}\rightarrow \mathds{R}$ as a quantum circuit, $\ket{z}_\textsc{c} \ket{0}_\textsc{a} \mapsto \ket{z}_\textsc{c} \ket{\sigma(z)}_\textsc{a}$, where $\textsc{a}$ is the label for the (continuous) activation register. For a general classically computable activation function for which we know a classical circuit, one could directly convert it to a quantum circuit using Toffoli gates \cite{nielsen2002quantum}, or more efficient coherent implementations of basic classical functions \cite{haner2018optimizing}, although this may end up being somewhat inefficient. For more details see section \ref{sec:zoo} for various examples of low-overhead implementations of certain activation functions. In figure \ref{fig:collect} we represent a neuron with a collector degree of freedom and the corresponding two-stage abstract circuit.

Until we reach \ref{sec:zoo}, for sake of compactness and generality of our analysis, we will use the ideal form of the feedforward unitary from equation \eqref{eq:ideal_ff}.

\subsubsection{Quantum Neural Network Feedforward \& Baqprop}\label{sec:simple_baq}

A typical feedforward neural network is comprised of a collection of neurons organized into layers.
Each layer can be thought of as a unit which takes the activations of the neurons of the previous layer and produces a collection of output activations based on some simple nonlinear function of the input activations.
The composition of many layers produces an output activation for the entire network, which overall can be seen as a nonlinear function of the input decomposed into the simpler functions of each layer.
The output of the network produces a prediction for the supervised learning problem.
The parameters of the function consist of the weights and biases of the collection of neurons in the network.

In our above notation, the feedforward step of computing the prediction of the neural network (on a quantum computer) is the computation of the unitary
\begin{equation}
\label{eqn:qnn_oracle}
  \hat{U}_f(\hat{\bm{\Phi}}) : \ket{\bm{\Phi},\bm{x},0} \mapsto \ket{\bm{\Phi},\bm{x},f(\bm{\Phi},\bm{x})}.
\end{equation}
Recall for each data point, the quantum phase kick backpropagation consists of three steps: the feedforward, cost-function phase kick, and backpropagation.

The quantum neural network described here will consist of a set of quantum number registers for the input to the network, as well as the weights, biases, and output activations for each neuron in the network.
Let the neurons in a single layer $\ell$ be indexed by $n_\ell = 1, \dots, N_\ell$, and let the layer index run from $\ell = 1, \dots, \mathscr{L}$.
Recall that ideally the neuron $n_\ell$ enacts
\begin{equation}
\begin{split}
  &\ket{\bm{a}_{\ell-1}}_{A_{\ell-1}} \ket{\bm{w}_{n_\ell},b_{n_\ell}}_{W_{n_\ell}} \ket{0}_{A_{n_\ell}} \mapsto \\
  &\quad \ket{\bm{a}_{\ell-1}}_{A_{\ell-1}} \ket{\bm{w}_{n_\ell},b_{n_\ell}}_{W_{n_\ell}} \ket{ \sigma( \bm{w}_{n_\ell} \cdot \bm{a}_{\ell-1} + b_{n_\ell} ) }_{A_{n_\ell}}
\end{split}
\end{equation}
using a unitary
\begin{equation}
  e^{ -i \sigma( \hat{\bm{w}}_{n_\ell} \cdot \hat{\bm{a}}_{\ell-1} + \hat{b}_{n_\ell} ) \hat{p}_{a_{n_\ell}} }
\end{equation}
which acts on the Hilbert space of the activations of the layer $\ell-1$ as well as those of the weights and activation of neuron $n_\ell$, i.e., $\mathcal{H}_{A_{\ell-1}} \otimes \mathcal{H}_{W_{n_\ell}} \otimes \mathcal{H}_{A_{n_\ell}}$.
Of course, our notation implies $\mathcal{H}_{A_\ell} := \bigotimes_{n_\ell=1}^{N_\ell} \mathcal{H}_{A_{n_\ell}}$, so, for example, $\hat{\bm{a}}_\ell := ( \hat{a}_{n_\ell} )_{n_\ell}$ (i.e., $\hat{\bm{a}}_\ell$ is a vector of operators whose $n_\ell^\text{th}$ component is $\hat{a}_{n_\ell}$).

Combining the action of all the neurons in layer $\ell$, we have
\begin{equation}
\begin{split}
  &\ket{\bm{a}_{\ell-1}}_{A_{\ell-1}} \bigotimes_{n_\ell=1}^{N_\ell} \ket{\bm{w}_{n_\ell},b_{n_\ell}}_{W_{n_\ell}} \ket{0}_{A_{n_\ell}} \mapsto \\
  & \ket{\bm{a}_{\ell-1}}_{A_{\ell-1}} \bigotimes_{n_\ell=1}^{N_\ell} \ket{\bm{w}_{n_\ell},b_{n_\ell}}_{W_{n_\ell}} \ket{\sigma(\bm{w}_{n_\ell} \cdot \bm{a}_{\ell-1} + b_{n_\ell})}_{A_{n_\ell}}.
\end{split}
\end{equation}
To compress the notation a little, let us write
\begin{eqnarray*}
  \bm{W}_\ell &:= ( \bm{w}_{n_\ell}^T )_{n_\ell} \\
  \bm{b}_\ell &:= ( \bm{b}_{n_\ell} )_{n_\ell}
\end{eqnarray*}
\vspace{-1cm}
\begin{equation*}
  \bm{\sigma}( \bm{W}_\ell \bm{a}_{\ell-1} + \bm{b}_\ell ) := ( \sigma( \bm{w}_{n_\ell} \cdot \bm{a}_{\ell-1} + b_{n_\ell} ) )_{n_\ell}
\end{equation*}
(note that $\bm{W}_\ell$ is a matrix with rows $\bm{w}_{n_\ell}^T$, and $\bm{\sigma}(\cdot)$ acts $\sigma(\cdot)$ componentwise on elements of the vector $\bm{W}_\ell \bm{a}_{\ell-1} + \bm{b}_\ell$), and
\begin{eqnarray*}
  \ket{\bm{W}_\ell,\bm{b}_\ell}_{W_\ell} & := \bigotimes_{n_\ell=1}^{N_\ell} \ket{\bm{w}_{n_\ell},b_{n_\ell}}_{W_{n_\ell}} \\
  \ket{\bm{a}_\ell}_{A_\ell} & := \bigotimes_{n_\ell=1}^{N_\ell} \ket{a_{n_\ell}}_{A_{n_\ell}}.
\end{eqnarray*}
The previous equation for the action of layer $\ell$ in this notation is then
\begin{equation}
\begin{split}
  &\ket{\bm{a}_{\ell-1}}_{A_{\ell-1}} \ket{\bm{W}_\ell,\bm{b}_\ell}_{W_\ell} \ket{\bm{0}}_{A_\ell} \mapsto \\
  &\quad \ket{\bm{a}_{\ell-1}}_{A_{\ell-1}} \ket{\bm{W}_\ell,\bm{b}_\ell}_{W_\ell} \ket{\bm{\sigma}( \bm{W}_\ell \bm{a}_{\ell-1} + \bm{b}_\ell )}_{A_\ell},
\end{split}
\end{equation}
under the unitary
\begin{equation}
  e^{-i \bm{\sigma}(\hat{\bm{W}}_\ell \hat{\bm{a}}_{\ell-1} + \hat{\bm{b}}_\ell ) \cdot \hat{\bm{p}}_{a_\ell}}.
\end{equation}

The feedforward for the entire network consists of a concatenation of these unitaries:
\begin{equation}
\begin{split}
  \hat{U}_{\textsc{ff}} &:= e^{-i \bm{\sigma} ( \hat{\bm{W}}_\mathscr{L} \hat{\bm{a}}_{\mathscr{L}-1} + \hat{\bm{b}}_\mathscr{L} ) \cdot \hat{\bm{p}}_{a_\mathscr{L}} } \times \\
  &\qquad \cdots e^{-i \bm{\sigma} ( \hat{\bm{W}}_2 \hat{\bm{a}}_1 + \hat{\bm{b}}_2 ) \cdot \hat{\bm{p}}_{a_2} } e^{-i \bm{\sigma} ( \hat{\bm{W}}_1 \hat{\bm{x}} + \hat{\bm{b}}_1 ) \cdot \hat{\bm{p}}_{a_1} }\\
  &= \prod_{\ell=1}^\mathscr{L} e^{-i \bm{\sigma} ( \hat{\bm{W}}_\ell \hat{\bm{a}}_{\ell-1} + \hat{\bm{b}}_\ell ) \cdot \hat{\bm{p}}_{a_\ell} }
\end{split}
\end{equation}
where $\hat{\bm{a}}_0 := \hat{\bm{x}}$ (i.e., the input to the network).
The feedforward unitary maps
\begin{equation}
\begin{split}
  \ket{\bm{x}}_I &\bigotimes_{\ell=1}^\mathscr{L} \ket{\bm{W}_\ell,\bm{b}_\ell}_{W_\ell} \ket{\bm{0}}_{A_\ell} \\
  &\mapsto \ket{\bm{x}}_I \bigotimes_{\ell=1}^\mathscr{L} \ket{\bm{W}_\ell,\bm{b}_\ell}_{W_\ell} \ket{\bm{a}_\ell}_{A_\ell},
\end{split}
\end{equation}
where the $\bm{a}_l$'s satisfy the recursion relation,
\begin{equation}
\begin{split}
  \bm{a}_\ell &= \bm{\sigma}( \bm{W}_\ell \bm{a}_{\ell-1} + \bm{b}_\ell ) \\
  \bm{a}_0 &:= \bm{x}.
\end{split}
\end{equation}
Of course, the output of the network is $\bm{a}_\mathscr{L}$ (satisfying the above recursion), which corresponds to the prediction of the network upon input $\bm{x}$.
Notice that this is indeed of the form of \eqref{eqn:qnn_oracle}, with the input register $\ket{\bm{x}}_I$, parameter registers $\otimes_{\ell=1}^\mathscr{L} \ket{ \bm{W}_\ell, \bm{b}_\ell }_{W_\ell}$, and prediction register $\ket{\bm{a}_\mathscr{L}}_{A_\mathscr{L}}$, along with auxiliary registers for the intermediate activations $\otimes_{\ell=1}^{\mathscr{L}-1} \ket{\bm{a}_\ell}_{A_\ell}$.

At the end of the feedforward, the phase kick for data point $(\bm{x},\bm{y})$ is generated by
\begin{equation}
  e^{-i L(\hat{\bm{a}}_\mathscr{L}, \bm{y})} : \ket{\bm{a}_\mathscr{L}}_{A_\mathscr{L}} \mapsto e^{-i L(\bm{a}_\mathscr{L}, \bm{y})} \ket{\bm{a}_\mathscr{L}}_{A_\mathscr{L}}
\end{equation}
The state after the phase kick is
\begin{equation}
  e^{-i L(\bm{a}_\mathscr{L}, \bm{y})} \ket{\bm{x}}_I \bigotimes_{l=1}^\mathscr{L} \ket{\bm{W}_l,\bm{b}_l}_{W_l} \ket{\bm{a}_l}_{A_l}
\end{equation}

After the phase kick, $\hat{U}_{\textsc{ff}}^\dagger$ is employed for backpropagation.
The phase is not affected by the backpropagation; $\hat{U}_{\textsc{ff}}^\dagger$ simply uncomputes the activations of each layer in the reverse order of the feedforward.
For example, the first uncompute is:
\begin{widetext}
\begin{equation}
\begin{split}
  e^{i \bm{\sigma} ( \hat{\bm{W}}_\mathscr{L} \hat{\bm{a}}_{\mathscr{L}-1} + \hat{\bm{b}}_\mathscr{L} ) \cdot \hat{\bm{p}}_{A_\mathscr{L}} } &: e^{-i L(\bm{a}_\mathscr{L}, \bm{y})} \ket{\bm{x}}_I \ket{\bm{W}_\mathscr{L},\bm{b}_\mathscr{L}}_{W_\mathscr{L}} \ket{\bm{a}_\mathscr{L}}_{A_\mathscr{L}} \bigotimes_{\ell=1}^{\mathscr{L}-1} \ket{\bm{W}_\ell,\bm{b}_\ell}_{W_\ell} \ket{\bm{a}_\ell}_{A_\ell} \\
  &\mapsto e^{-i L(\bm{a}_\mathscr{L}, \bm{y})} \ket{\bm{x}}_I \ket{\bm{W}_\mathscr{L},\bm{b}_\mathscr{L}}_{W_\mathscr{L}} \ket{\bm{0}}_{A_\mathscr{L}} \bigotimes_{\ell=1}^{\mathscr{L}-1} \ket{\bm{W}_\ell,\bm{b}_\ell}_{W_\ell} \ket{\bm{a}_\ell}_{A_\ell}.
\end{split}
\end{equation}
\clearpage
\end{widetext}

This continues until we arrive back at the initial state along with a phase:
\begin{equation}
  e^{-i L(\bm{a}_\mathscr{L},\bm{y})} \ket{\bm{x}}_I \bigotimes_{\ell=1}^\mathscr{L} \ket{\bm{W}_\ell,\bm{b}_\ell}_{W_\ell} \ket{\bm{0}}_{A_\ell},
\end{equation}
where in the phase we still have
\begin{equation}
\begin{split}
  \bm{a}_\ell &= \bm{\sigma}( \bm{W}_\ell \bm{a}_{\ell-1} + \bm{b}_\ell ), \\
  \bm{a}_0 &:= \bm{x}.
\end{split}
\end{equation}
This demonstrates more concretely how the quantum phase kick backpropagation algorithm can be performed for a quantum neural network.
In the next section we examine in further detail the quantum mechanism behind the backpropagation of the error signal during the uncomputation step, and explicitly show how the Quantum Phase Error Backpropagation relates to the classical Backpropagation of Errors.

\subsection{Quantum Phase Error Backpropagation: Layerwise Analysis}\label{sec:deep_baq}

Although we have already examined the trajectory of the state of the entire network under the QFB circuit and have a general form for the parameter momentum updates, here we want to examine the internal mechanisms of the feedforward and backpropagation more concretely.
For instance, we showed before that under the full QFB circuit, $\hat{U}_{\textsc{qfb}} = \hat{U}_{\textsc{ff}}^\dagger e^{-i \eta L(\hat{\bm{a}}_\mathscr{L}, \bm{y}) } \hat{U}_{\textsc{ff}}$, the momenta of the parameters are shifted by $\hat{U}_{\textsc{qfb}}^\dagger \hat{\bm{\Pi}} \hat{U}_{\textsc{qfb}} = \hat{\bm{\Pi}} - \eta \partial L( \hat{U}_{\textsc{ff}}^\dagger \hat{\bm{\tilde{y}}} \hat{U}_{\textsc{ff}}, \bm{y} ) / \partial \hat{\bm{\Phi}}$.
In this case, we have $\hat{\bm{\tilde{y}}} = \hat{\bm{a}}_\mathscr{L}$ and $\hat{\bm{\Phi}}$ abstractly represents the collection of weights and biases $\{ \hat{\bm{W}}_\ell, \hat{\bm{b}}_\ell \}_{\ell=1}^\mathscr{L}$.
However, here the purpose is to examine the propagation of impulses in the network, layer-by-layer, to better understand the behavior of the network during training algorithm.

A key observation of this section will be that during the training, the activations of a layer, $\hat{\bm{a}}_\ell$, are always influenced only by the activations of previous layers, whereas the momenta of the activations, weights, and biases ($\hat{\bm{p}}_{a_\ell}$, $\hat{\bm{p}}_{W_\ell}$, and $\hat{\bm{p}}_{b_\ell}$, respectively), are directly affected by activations of previous layers along with momenta of later layers.
Ultimately, this is what allows the feedforward operation to propagate signals forward in the network and the uncomputation to propagate momentum updates backward through the network.

For convenience, we will write the feedforward unitary for layer $\ell$ as
\begin{equation}
  \hat{U}_\ell := e^{-i \bm{\sigma}( \hat{\bm{W}}_\ell \hat{\bm{a}}_{\ell-1} + \hat{\bm{b}}_\ell ) \cdot \hat{\bm{p}}_{a_\ell}},
\end{equation}
and the feedforward from layer $\ell'$ to $\ell$ (with $\ell' < \ell$) as
$\ell > \ell'$
\begin{equation}
  \hat{U}_{\textsc{ff}}^{(\ell,\ell')} := \hat{U}_\ell \hat{U}_{\ell-1} \cdots \hat{U}_{\ell'+1} \hat{U}_{\ell'}.
\end{equation}
Of course, the feedforward for the entire network is $\hat{U}_{\textsc{ff}} = \hat{U}_{\textsc{ff}}^{(\mathscr{L},1)}$.

Notice that each of the operators $\hat{\bm{a}}_\ell$, $\hat{\bm{p}}_{a_\ell}$, $\hat{\bm{p}}_{W_\ell}$, and $\hat{\bm{p}}_{b_\ell}$ are only directly affected by one of these unitaries.
Of course, they can depend indirectly on the others.
For example, the activation for layer $\ell$, $\hat{\bm{a}}_\ell$, is only affected by $\hat{U}_\ell$ on the forward pass and $\hat{U}_\ell^\dagger$ on the backward pass, since these are the only operators in the QFB circuit containing the conjugate operator, $\hat{\bm{p}}_{a_\ell}$.

The feedforward unitary for layer $\ell$ changes the activation by
\begin{equation}
  \hat{U}_\ell^\dagger \hat{\bm{a}}_\ell \hat{U}_\ell = \hat{\bm{a}}_\ell + \bm{\sigma}( \hat{\bm{W}}_\ell \hat{\bm{a}}_{\ell-1} + \hat{\bm{b}}_\ell ).
\end{equation}
The operation in the backpropagation, $\hat{U}_\ell^\dagger$, just changes the sign of the shift
\begin{equation}
  \hat{U}_\ell^\dagger \hat{\bm{a}}_\ell \hat{U}_\ell = \hat{\bm{a}}_\ell - \bm{\sigma}( \hat{\bm{W}}_\ell \hat{\bm{a}}_{\ell-1} + \hat{\bm{b}}_\ell ).
\end{equation}
We see that the activation in one layer depends on the activation of the previous layer as well as the weights in the current layer.
Of course in the full feedforward circuit, the activations in the previous layer will also depend on the preceeding layers, so we get a recursion:
\begin{equation}
  \hat{U}_{\textsc{ff}}^{(\ell,1)\dagger}  \hat{\bm{a}}_\ell  \hat{U}_{\textsc{ff}}^{(\ell,1)} = \hat{\bm{a}}_\ell + \bm{\sigma}( \hat{\bm{W}}_\ell \hat{U}_{\textsc{ff}}^{(\ell-1,1)\dagger} \hat{\bm{a}}_{\ell-1} \hat{U}_{\textsc{ff}}^{(\ell-1,1)} + \hat{\bm{b}}_\ell ),
\end{equation}
which ends with
\begin{equation}
  \hat{U}_1^\dagger \hat{\bm{a}}_1 \hat{U}_1 = \hat{\bm{a}}_1 + \bm{\sigma}( \hat{\bm{W}}_1 \hat{\bm{x}} + \hat{\bm{b}}_1 ).
\end{equation}
Of course these expressions are unaffected by the remaining feedforward operations, $\hat{U}_{\textsc{ff}}^{(\mathscr{L},\ell+1)}$, so we could also write
\begin{equation}
  \hat{U}_{\textsc{ff}}^\dagger \hat{\bm{a}}_\ell \hat{U}_{\textsc{ff}} = \hat{\bm{a}}_\ell + \bm{\sigma}( \hat{\bm{W}}_\ell \hat{U}_{\textsc{ff}}^\dagger \hat{\bm{a}}_{\ell-1} \hat{U}_{\textsc{ff}} + \hat{\bm{b}}_\ell ).
\end{equation}

In the backpropagation steps, we see that the activation is still only affected by the activations of the previous layers:
\begin{equation}
\begin{split}
  \hat{U}_{\textsc{ff}} \hat{\bm{a}}_\ell \hat{U}_{\textsc{ff}}^\dagger &= \hat{U}_{\textsc{ff}}^{(\mathscr{L},\ell)}  \hat{\bm{a}}_\ell  \hat{U}_{\textsc{ff}}^{(\mathscr{L},\ell)\dagger} \\
  &= \hat{\bm{a}}_\ell - \bm{\sigma}( \hat{\bm{W}}_\ell \hat{\bm{a}}_{\ell-1} + \hat{\bm{b}}_\ell ).
\end{split}
\end{equation}
Therefore, we see clearly that the domain of influence of the activations consists only of activations (and weights) of the preceeding layers.
Furthermore, since the activations are not directly affected by the phase kick at the output, $e^{ -i \eta L(\hat{\bm{a}}_\mathscr{L}, \bm{y}) }$, the entire QFB circuit simply computes and then uncomputes the activations:
\begin{equation}
\begin{split}
  &\hat{U}_{\textsc{qfb}}^\dagger \hat{\bm{a}}_\ell \hat{U}_{\textsc{qfb}} \\
  &= \hat{U}_{\textsc{ff}}^\dagger e^{ i \eta L(\hat{\bm{a}}_\mathscr{L}, \bm{y}) } \hat{U}_{\textsc{ff}} \hat{\bm{a}}_\ell \hat{U}_{\textsc{ff}}^\dagger e^{ -i \eta L(\hat{\bm{a}}_\mathscr{L}, \bm{y}) } \hat{U}_{\textsc{ff}} \\
  &= \hat{U}_{\textsc{ff}}^\dagger e^{ i \eta L(\hat{\bm{a}}_\mathscr{L}, \bm{y}) } \left( \hat{\bm{a}}_\ell - \bm{\sigma}( \hat{\bm{W}}_\ell \hat{\bm{a}}_{\ell-1} + \hat{\bm{b}}_\ell ) \right) e^{ -i \eta L(\hat{\bm{a}}_\mathscr{L}, \bm{y}) } \hat{U}_{\textsc{ff}} \\
  &= \hat{U}_{\textsc{ff}}^\dagger \left( \hat{\bm{a}}_\ell - \bm{\sigma}( \hat{\bm{W}}_\ell \hat{\bm{a}}_{\ell-1} + \hat{\bm{b}}_\ell ) \right) \hat{U}_{\textsc{ff}} \\
  &= \hat{\bm{a}}_\ell + \bm{\sigma}( \hat{\bm{W}}_\ell \hat{U}_{\textsc{ff}}^\dagger \hat{\bm{a}}_{\ell-1} \hat{U}_{\textsc{ff}} + \hat{\bm{b}}_\ell )\\
  &\qquad \qquad - \bm{\sigma}( \hat{\bm{W}}_\ell \hat{U}_{\textsc{ff}}^\dagger \hat{\bm{a}}_{\ell-1} \hat{U}_{\textsc{ff}} + \hat{\bm{b}}_\ell ) \\
  &= \hat{\bm{a}}_\ell.
\end{split}
\end{equation}
This fact could have been deduced more easily by writing, $\hat{U}_{\textsc{qfb}} = \hat{U}_{\textsc{ff}}^\dagger e^{-i \eta  L( \hat{\bm{a}}_\mathscr{L}, \bm{y} )} \hat{U}_{\textsc{ff}} = e^{-i \eta L( \hat{U}_{\textsc{ff}}^\dagger \hat{\bm{a}}_\mathscr{L} \hat{U}_{\textsc{ff}}, \bm{y} )}$, and noticing that in the above recursion relation for $\hat{U}_{\textsc{ff}}^\dagger \hat{\bm{a}}_\mathscr{L} \hat{U}_{\textsc{ff}}$ that no $\hat{\bm{p}}_{a_\ell}$'s appear.
However, the purpose of this exercise was to demonstrate that the activations only depend on the values of the activations and weights in the previous layers.
Hence, insofar as the activations are concerned, there is only a forward propagation of information in the network.

Now we will discuss the momenta of the activations and the weights/biases.
We will see that these will be affected by both earlier and later layers in the network.
In particular, these respond to the activations in previous layers and momentum kicks in succeeding layers.
Therefore, to propagate information forward in the network, we have to act on the activations, and to propagate backwards we have to act on the momenta of the activations.

The momenta of the activations in layer $\ell$, $\hat{\bm{p}}_{a_\ell}$, are only affected by the unitary $\hat{U}_{\ell+1}$.
The single exception is $\hat{\bm{p}}_{a_\mathscr{L}}$, which is affected only by the phase kick $e^{-i \eta L(\hat{\bm{a}}_\mathscr{L},\bm{y})}$.
For the final layer, we get
\begin{equation}
  e^{i \eta L( \hat{\bm{a}}_\mathscr{L}, \bm{y} )} \hat{\bm{p}}_{a_\mathscr{L}} e^{-i \eta L( \hat{\bm{a}}_\mathscr{L}, \bm{y}) } = \hat{\bm{p}}_{a_\mathscr{L}} - \eta \partial L( \hat{\bm{a}}_\mathscr{L}, \bm{y} ) / \partial \hat{\bm{a}}_\mathscr{L},
\end{equation}
and for $\ell < \mathscr{L}$, we have
\begin{equation}
  \hat{U}_{\ell+1}^\dagger \hat{\bm{p}}_{a_\ell} \hat{U}_{\ell+1} = \hat{\bm{p}}_{a_\ell} - \hat{\bm{W}}_{\ell+1}^T [ \bm{\sigma}'( \hat{\bm{W}}_{\ell+1} \hat{\bm{a}}_\ell + \hat{\bm{b}}_{\ell+1} ) \odot \hat{\bm{p}}_{a_{\ell+1}} ]
\end{equation}
where $\bm{\sigma}'$ is the derivative of the nonlinear activation function acting on components of the vectorial argument, and $\odot$ denotes componentwise multiplication, i.e.,
\begin{equation}
  \begin{bmatrix} x_1 \\ y_1 \end{bmatrix} \odot \begin{bmatrix} x_2 \\ y_2 \end{bmatrix} = \begin{bmatrix} x_1 x_2 \\ y_1 y_2 \end{bmatrix}.
\end{equation}
Also, note that $\hat{U}_{\ell+1} \hat{\bm{p}}_{a_\ell} \hat{U}_{\ell+1}^\dagger$ is the same expression with the opposite sign for the shift.
We see explicitly that the shift in the momentum of the activation for a layer depends on the activation of that layer as well as the momentum of the activation (and the values of the weights/biases) of the following layer.
For the full feedforward circuit, we get
\begin{equation}
\begin{split}
  &\hat{U}_{\textsc{ff}}^\dagger \hat{\bm{p}}_{a_\ell} \hat{U}_{\textsc{ff}} \\
  &= \hat{U}_{\textsc{ff}}^{(\ell+1,1)\dagger} \hat{\bm{p}}_{a_\ell} \hat{U}_{\textsc{ff}}^{(\ell+1,1)} \\
  &= \hat{\bm{p}}_{a_\ell} \\
  &\qquad - \hat{\bm{W}}_{\ell+1}^T [ \bm{\sigma}'( \hat{\bm{W}}_{\ell+1} ( \hat{U}_{\textsc{ff}}^{(\ell,1)\dagger} \hat{\bm{a}}_\ell \hat{U}_{\textsc{ff}}^{(\ell,1)} ) + \hat{\bm{b}}_{\ell+1} ) \odot \hat{\bm{p}}_{a_{\ell+1}} ]
\end{split}
\end{equation}

Note that the momenta get kicked on the forward pass (not just in the backpropagation) since the shift depends on the current activation, which in turn implicitly depends on activations and weights/biases earlier in the network due to the feedforward $\hat{U}_{\textsc{ff}}$.

Now, if we look at the backpropagation in isolation (without the preceding feedforward and phase kick), we get
\begin{equation}
\begin{split}
  &\hat{U}_{\textsc{ff}} \hat{\bm{p}}_{a_\ell} \hat{U}_{\textsc{ff}}^\dagger \\
  &= \hat{U}_{\textsc{ff}}^{(\mathscr{L},\ell+1)\dagger} \hat{\bm{p}}_{a_\ell} \hat{U}_{\textsc{ff}}^{(\mathscr{L},\ell+1)} \\
  &= \hat{\bm{p}}_{a_\ell} + \hat{\bm{W}}_{\ell+1}^T [ \bm{\sigma}'( \hat{\bm{W}}_{\ell+1} \hat{\bm{a}}_\ell + \hat{\bm{b}}_{\ell+1} ) \\
  &\qquad \qquad \qquad \qquad \qquad \odot \hat{U}_{\textsc{ff}}^{(\mathscr{L},\ell+2)} \hat{\bm{p}}_{a_{\ell+1}} \hat{U}_{\textsc{ff}}^{(\mathscr{L},\ell+2)\dagger} ]
\end{split}
\end{equation}
Which, as before, is shifted according to the activation in the current layer as well as momenta and weights/biases in the following layer.
However, the full backpropagation also carries influences from later in the network through $\hat{U}_{\textsc{ff}}^{(\mathscr{L},\ell+2)} \hat{\bm{p}}_{a_{\ell+1}} \hat{U}_{\textsc{ff}}^{(\mathscr{L},\ell+2)\dagger}$.

In summary, we see that, roughly, the activations carry information forward through the network (via the feedforward operations), and the momenta of the activations carry information backward through the network (via the uncomputation operations).
Therefore, for the entire QFB circuit, we feedforward the activations to make the prediction, kick the momentum of the output activation, and then this momentum kick propagates back to the remaining activation momenta and returns the activations to their original state.
Explicitly, for $\ell < \mathscr{L}$,

\begin{widetext}
\begin{equation}
\begin{split}
  \hat{U}_{\textsc{qfb}}^\dagger \hat{\bm{p}}_{a_\ell} \hat{U}_{\textsc{qfb}} &= \hat{U}_{\textsc{ff}}^\dagger e^{i \eta L(\hat{\bm{a}}_\mathscr{L}, \bm{y}) } \hat{U}_{\textsc{ff}} \hat{\bm{p}}_{a_\ell} \hat{U}_{\textsc{ff}}^\dagger e^{-i \eta L(\hat{\bm{a}}_\mathscr{L}, \bm{y})} \hat{U}_{\textsc{ff}} \\
  &= \hat{U}_{\textsc{ff}}^\dagger e^{i \eta L(\hat{\bm{a}}_\mathscr{L}, \bm{y}) } \left( \hat{\bm{p}}_{a_\ell} + \hat{\bm{W}}_{\ell+1}^T [ \bm{\sigma}'( \hat{\bm{W}}_{\ell+1} \hat{\bm{a}}_\ell + \hat{\bm{b}}_{\ell+1} ) \odot \hat{U}_{\textsc{ff}}^{(\mathscr{L},\ell+2)} \hat{\bm{p}}_{a_{\ell+1}} \hat{U}_{\textsc{ff}}^{(\mathscr{L},\ell+2)\dagger} ] \right) e^{-i \eta L(\hat{\bm{a}}_\mathscr{L}, \bm{y})} \hat{U}_{\textsc{ff}} \\
  &= \hat{U}_{\textsc{ff}}^\dagger \left( \hat{\bm{p}}_{a_\ell} + \hat{\bm{W}}_{\ell+1}^T [ \bm{\sigma}'( \hat{\bm{W}}_{\ell+1} \hat{\bm{a}}_\ell + \hat{\bm{b}}_{\ell+1} ) \odot e^{i \eta L(\hat{\bm{a}}_\mathscr{L}, \bm{y}) } \hat{U}_{\textsc{ff}}^{(\mathscr{L},\ell+2)} \hat{\bm{p}}_{a_{\ell+1}} \hat{U}_{\textsc{ff}}^{(\mathscr{L},\ell+2)\dagger} e^{-i \eta L(\hat{\bm{a}}_\mathscr{L}, \bm{y}) } ] \right)  \hat{U}_{\textsc{ff}} \\
  &= \hat{\bm{p}}_{a_\ell} - \hat{\bm{W}}_{\ell+1}^T [ \bm{\sigma}'( \hat{\bm{W}}_{\ell+1} ( \hat{U}_{\textsc{ff}}^{(\ell,1)\dagger} \hat{\bm{a}}_\ell \hat{U}_{\textsc{ff}}^{(\ell,1)} ) + \hat{\bm{b}}_{\ell+1} ) \odot \hat{\bm{p}}_{a_{\ell+1}} ] \\
  & \qquad + \hat{\bm{W}}_{\ell+1}^T [ \bm{\sigma}'( \hat{\bm{W}}_{\ell+1} \hat{U}_{\textsc{ff}}^\dagger \hat{\bm{a}}_\ell \hat{U}_{\textsc{ff}} + \hat{\bm{b}}_{\ell+1} ) \odot \hat{U}_{\textsc{ff}}^\dagger e^{i \eta L(\hat{\bm{a}}_\mathscr{L}, \bm{y}) } \hat{U}_{\textsc{ff}}^{(\mathscr{L},\ell+2)} \hat{\bm{p}}_{a_{\ell+1}} \hat{U}_{\textsc{ff}}^{(\mathscr{L},\ell+2)\dagger} e^{-i \eta L(\hat{\bm{a}}_\mathscr{L}, \bm{y}) } ] \hat{U}_{\textsc{ff}} \\
  &= \hat{\bm{p}}_{a_\ell} - \hat{\bm{W}}_{\ell+1}^T [ \bm{\sigma}'( \hat{\bm{W}}_{\ell+1} ( \hat{U}_{\textsc{ff}}^\dagger \hat{\bm{a}}_\ell \hat{U}_{\textsc{ff}} ) + \hat{\bm{b}}_{\ell+1} ) \odot \hat{\bm{p}}_{a_{\ell+1}} ] \\
  & \qquad + \hat{\bm{W}}_{\ell+1}^T [ \bm{\sigma}'( \hat{\bm{W}}_{\ell+1} \hat{U}_{\textsc{ff}}^\dagger \hat{\bm{a}}_\ell \hat{U}_{\textsc{ff}} + \hat{\bm{b}}_{\ell+1} ) \odot \hat{U}_{\textsc{qfb}}^\dagger \hat{\bm{p}}_{a_{\ell+1}} \hat{U}_{\textsc{qfb}} ]. \\
\end{split}
\end{equation}
\clearpage
\end{widetext}

We have a shift which is a sum of the shift on the forward pass and the backward pass.
A more illustrative way to look at this is in terms of differences:
\begin{equation}\label{eqn:activation_backprop}
\begin{split}
  \Big( \hat{U}_{\textsc{qfb}}^\dagger & \hat{\bm{p}}_{a_\ell} \hat{U}_{\textsc{qfb}} - \hat{\bm{p}}_{a_\ell} \Big) \\
  &= \hat{\bm{W}}_{\ell+1}^T \left[ \bm{\sigma}'( \hat{\bm{W}}_{\ell+1} \hat{U}_{\textsc{ff}}^\dagger \hat{\bm{a}}_\ell \hat{U}_{\textsc{ff}} + \hat{\bm{b}}_{\ell+1} ) \right. \\
  &\qquad \qquad \qquad \left. \odot \left( \hat{U}_{\textsc{qfb}}^\dagger \hat{\bm{p}}_{a_{\ell+1}} \hat{U}_{\textsc{qfb}} - \hat{\bm{p}}_{a_{\ell+1}} \right) \right]. 
\end{split}
\end{equation}
This shows that the differences of the activation momenta before and after the QFB circuit propagate back recursively, i.e., here from $\Delta \hat{\bm{p}}_{a_{\ell+1}}$ to $\Delta \hat{\bm{p}}_{a_\ell}$.
Of course, the recursion ends with the output of the network, where we apply the loss function to kick the output activation momentum,
\begin{equation}
  \hat{U}_{\textsc{qfb}}^\dagger \hat{\bm{p}}_{a_\mathscr{L}} \hat{U}_{\textsc{qfb}} - \hat{\bm{p}}_{a_\mathscr{L}} = - \eta \bm{\nabla} L ( \hat{U}_{\textsc{ff}}^\dagger \hat{\bm{a}}_\mathscr{L} \hat{U}_{\textsc{ff}}, \bm{y} ),
\end{equation}
where the derivative on the loss function is understood to be with respect to the first argument.
This propagates back via the above recursion to kick the momenta of the activations throughout the network.

These activation momentum updates in turn affect the momenta of the weights and biases, which are the shifts that we are actually interested in for the training.
The calculation is similar to that for $\hat{\bm{p}}_{a_\ell}$.
It is simple to show that for a single feedforward step,
\begin{eqnarray}
  \hat{U}_\ell^\dagger \hat{\bm{p}}_{W_\ell} \hat{U}_\ell &=& \hat{\bm{p}}_{W_\ell} - [ \bm{\sigma}'( \hat{\bm{W}}_\ell \hat{\bm{a}}_{\ell-1} + \hat{\bm{b}}_\ell ) \odot \hat{\bm{p}}_{a_\ell} ] \hat{\bm{a}}_{\ell-1}^T,\\
  \hat{U}_\ell^\dagger \hat{\bm{p}}_{b_\ell} \hat{U}_\ell &=& \hat{\bm{p}}_{b_\ell} -  \bm{\sigma}'( \hat{\bm{W}}_\ell \hat{\bm{a}}_{\ell-1} + \hat{\bm{b}}_\ell ) \odot \hat{\bm{p}}_{a_\ell},
\end{eqnarray}
where we note that here we have, respectively, a matrix and a vector of operators.
For the full feedforward and uncomputation, we get (respectively),
\begin{equation}
\begin{split}
  &\hat{U}_{\textsc{ff}}^\dagger \hat{\bm{p}}_{W_\ell} \hat{U}_{\textsc{ff}} \\
  &\quad= \hat{\bm{p}}_{W_\ell} - [ \bm{\sigma}'( \hat{\bm{W}}_\ell \hat{U}_{\textsc{ff}}^\dagger \hat{\bm{a}}_{\ell-1} \hat{U}_{\textsc{ff}} + \hat{\bm{b}}_\ell ) \odot \hat{\bm{p}}_{a_\ell} ] \hat{U}_{\textsc{ff}}^\dagger \hat{\bm{a}}_{\ell-1}^T \hat{U}_{\textsc{ff}},
\end{split}
\end{equation}
\begin{equation}
  \hat{U}_{\textsc{ff}}^\dagger \hat{\bm{p}}_{b_\ell} \hat{U}_{\textsc{ff}} = \hat{\bm{p}}_{b_\ell} - \bm{\sigma}'( \hat{\bm{W}}_\ell \hat{U}_{\textsc{ff}}^\dagger \hat{\bm{a}}_{\ell-1} \hat{U}_{\textsc{ff}} + \hat{\bm{b}}_\ell ) \odot \hat{\bm{p}}_{a_\ell},
\end{equation}
and
\begin{equation}
  \hat{U}_{\textsc{ff}} \hat{\bm{p}}_{W_\ell} \hat{U}_{\textsc{ff}}^\dagger = \hat{\bm{p}}_{W_\ell} + [ \bm{\sigma}'( \hat{\bm{W}}_\ell \hat{\bm{a}}_{\ell-1} + \hat{\bm{b}}_\ell ) \odot \hat{U}_{\textsc{ff}} \hat{\bm{p}}_{a_\ell} \hat{U}_{\textsc{ff}}^\dagger ] \hat{\bm{a}}_{\ell-1}^T,
\end{equation}
\begin{equation}
  \hat{U}_{\textsc{ff}} \hat{\bm{p}}_{b_\ell} \hat{U}_{\textsc{ff}}^\dagger = \hat{\bm{p}}_{b_\ell} + \bm{\sigma}'( \hat{\bm{W}}_\ell \hat{\bm{a}}_{\ell-1} + \hat{\bm{b}}_\ell ) \odot \hat{U}_{\textsc{ff}} \hat{\bm{p}}_{a_\ell} \hat{U}_{\textsc{ff}}^\dagger.
\end{equation}
Using these, one obtains for the full algorithm that
\begin{equation}
\begin{split}
  \Big( \hat{U}_{\textsc{qfb}}^\dagger & \hat{\bm{p}}_{W_\ell} \hat{U}_{\textsc{qfb}} - \hat{\bm{p}}_{W_\ell} \Big) \\
  &= \left[ \bm{\sigma}'( \hat{\bm{W}}_\ell \hat{U}_{\textsc{ff}}^\dagger \hat{\bm{a}}_{\ell-1} \hat{U}_{\textsc{ff}} + \hat{\bm{b}}_\ell ) \right. \\
  &\qquad \qquad \left. \odot \left( \hat{U}_{\textsc{qfb}}^\dagger \hat{\bm{p}}_{a_\ell} \hat{U}_{\textsc{qfb}} - \hat{\bm{p}}_{a_\ell} \right) \right] \hat{U}_{\textsc{ff}}^\dagger \hat{\bm{a}}_{\ell-1}^T \hat{U}_{\textsc{ff}}, \label{eqn:weight_backprop}
\end{split}
\end{equation}
\begin{equation}
\begin{split}
  \Big( \hat{U}_{\textsc{qfb}}^\dagger & \hat{\bm{p}}_{b_\ell} \hat{U}_{\textsc{qfb}} - \hat{\bm{p}}_{b_\ell} \Big) \\
  &= \bm{\sigma}'( \hat{\bm{W}}_\ell \hat{U}_{\textsc{ff}}^\dagger \hat{\bm{a}}_{\ell-1} \hat{U}_{\textsc{ff}} + \hat{\bm{b}}_\ell )  \\
  &\qquad \qquad \qquad \odot \left( \hat{U}_{\textsc{qfb}}^\dagger \hat{\bm{p}}_{a_\ell} \hat{U}_{\textsc{qfb}} - \hat{\bm{p}}_{a_\ell} \right). \label{eqn:bias_backprop}
\end{split}
\end{equation}
Therefore, the update for the momentum of the weights is directly related to the update of the momentum of the activation of the same layer.
With the formula we derived before for the update of the activation momentum, the kick in the activation momentum of this layer depends on the updates of the following layers back to the kick at the output of the network.
Together, the equations \eqref{eqn:activation_backprop}, \eqref{eqn:weight_backprop}, and \eqref{eqn:bias_backprop} provide the key insight into the physics of the backpropagation of errors in the quantum neural network.

\subsubsection{Operator Chain Rule}

There is yet another way of viewing backpropagation in the Heisenberg picture, by directly applying the chain rule to the loss function.
This perspective of the backpropagation of errors is not as vivid as in the previous section, but is more closely related to classical backpropagation, which would be written schematically as
\begin{equation}
  \frac{\partial L}{\partial \bm{W}_\ell} = \frac{\partial L}{\partial \bm{a}_\mathscr{L}} \cdot \frac{\partial \bm{a}_\mathscr{L}}{\partial \bm{a}_{\mathscr{L}-1}} \cdots \frac{\partial \bm{a}_\ell}{\partial \bm{W}_\ell}.
\end{equation}

Recall that from above we have the QFB circuit for the neural network as $\hat{U}_\textsc{qfb} = e^{-i \eta L( \hat{U}_\textsc{ff}^\dagger \bm{\hat{a}}_\mathscr{L} \hat{U}_\textsc{ff}, \bm{y} )}$.
Therefore, we can write
\begin{equation}
\begin{split}
  \hat{U}_{\textsc{qfb}}^\dagger (\hat{\bm{p}}_{a_\ell})^i &\hat{U}_{\textsc{qfb}} - (\hat{\bm{p}}_{a_\ell})^i \\
  &= i \eta \bm{\nabla}^T L( \hat{U}_{\textsc{ff}}^\dagger \hat{\bm{a}}_\mathscr{L} \hat{U}_{\textsc{ff}}, \bm{y} ) [ \hat{U}_{\textsc{ff}}^\dagger \hat{\bm{a}}_\mathscr{L} \hat{U}_{\textsc{ff}}, (\hat{\bm{p}}_{a_\ell})^i ],
\end{split}
\end{equation}
where $(\hat{\bm{p}}_{a_\ell})^i$ denotes the $i$th component of the vector $\hat{\bm{p}}_{a_\ell}$.
It is straightforward to write a similar expression with $(\hat{\bm{p}}_{a_\ell})^i$ replaced by $(\hat{\bm{p}}_{W_\ell})^{ij}$ or $(\hat{\bm{p}}_{b_\ell})^i$.
Note that the term on the right-hand side is analogous to writing
\begin{equation}
  \frac{\partial L}{\partial \bm{a}_\ell} = \frac{\partial L}{\partial \bm{a}_\mathscr{L}} \cdot \frac{\partial \bm{a}_\mathscr{L}}{\partial \bm{a}_\ell},
\end{equation}
which is akin to forward mode accumulation of automatic differentiation.
One typically continues with backpropagation by iterating this procedure of using the chain rule.
In our operator picture, this proceeds by successively using the following identity,
\begin{equation}
\begin{split}
  [ \hat{U}_{\textsc{ff}}^\dagger \hat{\bm{a}}_\ell \hat{U}_{\textsc{ff}}, (\hat{\bm{p}}_{a_{\ell'}})^i ] =& \bm{\sigma}'(\hat{\bm{W}}_\ell \hat{U}_{\textsc{ff}}^\dagger \hat{\bm{a}}_{\ell-1} \hat{U}_{\textsc{ff}} + \hat{\bm{b}}_\ell ) \\
  &\odot \hat{\bm{W}}_\ell [ \hat{U}_{\textsc{ff}}^\dagger \hat{\bm{a}}_{\ell-1} \hat{U}_{\textsc{ff}}, (\hat{\bm{p}}_{a_{\ell'}})^i ],
\end{split}
\end{equation}
which holds for $\ell > \ell'$.
One can check that this commutator vanishes for the cases where $\ell < \ell'$.
Hence this backpropagation procedure terminates at $\ell = \ell'$, where one can show that
\begin{widetext}
\begin{eqnarray}
  \left[ \hat{U}_{\textsc{ff}}^\dagger (\hat{\bm{a}}_\ell)^k \hat{U}_{\textsc{ff}}, (\hat{\bm{p}}_{a_\ell})^{i} \right] &=& i \delta^{ki}\\
  \left[ \hat{U}_{\textsc{ff}}^\dagger (\hat{\bm{a}}_\ell)^k \hat{U}_{\textsc{ff}}, (\hat{\bm{p}}_{W_\ell})^{ij} \right] &=& i \delta^{ki} [ \bm{e}_k \cdot \bm{\sigma}' ( W_\ell \hat{U}_{\textsc{ff}}^\dagger \hat{\bm{a}}_{\ell-1} \hat{U}_{\textsc{ff}} + \hat{\bm{b}}_\ell ) ] \hat{U}_{\textsc{ff}}^\dagger (\hat{\bm{a}}_{\ell-1})^j \hat{U}_{\textsc{ff}}\\
  \left[ \hat{U}_{\textsc{ff}}^\dagger (\hat{\bm{a}}_\ell)^k \hat{U}_{\textsc{ff}}, (\hat{\bm{p}}_{b_\ell})^i \right] &=& i \delta^{ki} [ \bm{e}_k \cdot \bm{\sigma}' ( W_\ell \hat{U}_{\textsc{ff}}^\dagger \hat{\bm{a}}_{\ell-1} \hat{U}_{\textsc{ff}} + \hat{\bm{b}}_\ell ) ]
\end{eqnarray}
\end{widetext}
\vspace*{-25px}
where $\bm{e}_k$ denotes the $k$th standard basis vector.
It is simple to show that these expressions can be used to derive the equations \eqref{eqn:activation_backprop}, \eqref{eqn:weight_backprop}, and \eqref{eqn:bias_backprop} from the previous section.

\subsection{Implementations of Quantum Coherent Neurons }\label{sec:zoo}

Recall that the idealized neuron takes a vector of inputs, combines it with a vector of weights and scalar bias, and outputs a scalar activation as a nonlinear function of this combination:
\begin{equation}
\begin{split}
  e^{-i \sigma( \hat{\bm{w}} \cdot \hat{\bm{x}} + b ) \hat{p}_a} &: \ket{\bm{x}}_I \ket{\bm{w},b}_W \ket{0}_A \\
  &\mapsto \ket{\bm{x}}_I \ket{\bm{w},b}_W \ket{\sigma(\bm{w} \cdot \bm{x} + b)}_A.
\end{split}
\end{equation}
The linear operations can be implemented in a straightforward manner, using digital or continuous adders (generalized CNOTs), and the multiplications from two registers into a third can be done using generalized CCNOTs, all of which were outlined in Section~\ref{sec:bkgd}.

The step which is less palpable is the application of the nonlinear activation function $\sigma$.
Two activation functions which are commonly used in classical neural networks are the sigmoid function,
\begin{equation}
  \sigma_\beta(z) = \frac{1}{1+e^{-\beta z}}, \quad \mbox{where } \beta \in \mathds{R},
\end{equation}
and the rectified linear unit (ReLU),
\begin{equation}
  \sigma(z) = \max(0,z).
\end{equation}
(The parameter $\beta$ in the sigmoid function controls the sharpness of the transition.
In the limit $\beta \to \infty$, it becomes a step function.)

In this section, we will discuss means of approximating the sigmoid and ReLU activation functions with quantum circuits, for the purpose of implementing the quantum neural network described above.
We will first examine an implementation using a hybrid discrete-continuous variable (or simulated continuous variable) system, which is based on the phase estimation algorithm.
Although using phase estimation requires some overhead of gates, projection onto the positive subspace of the input can be done easily.
The second method is a fully continuous variable (or simulated continuous variable) implementation.
In this case, projection onto the positive subspace of the input requires some overhead to enact a non-linear phase gate and also requires squeezed states for precision.
Of course this second issue can be overcome if using simulated continuous variables on a sufficiently large quantum computer.

In both cases of hybrid CV-DV neurons and CV-only, we will separate the procedure of applying the non-linear activation into stages.
The first will be simply to assume the combination of inputs, weights, and bias are stored in a continuous variable \emph{collector} register, $\textsc{c}$.
That is, we will assume that we have prepared $\ket{\bm{x}}_\textsc{i} \ket{\bm{w},b}_\textsc{w} \ket{0}_\textsc{c} \mapsto \ket{\bm{x}}_\textsc{i} \ket{\bm{w},b}_\textsc{w} \ket{z}_\textsc{c}$, where $z = \bm{w} \cdot \bm{x} + b$.
The aim is to take the value stored in the collector register, and approximate the computation $\ket{z}_\textsc{c} \ket{0}_\textsc{a} \mapsto \ket{z}_\textsc{c} \ket{\sigma(z)}_\textsc{a}$, where $\textsc{a}$ is the label for the (continuous) activation register.

\subsubsection{Hybrid CV-DV Neurons}

For the current case of hybrid discrete-continuous variable neurons, we will also make use of an intermediate discrete variable \emph{filter} register, $\textsc{f}$, which will be taken to be a collection of $N$ qubits.
The purpose of this intermediate filter is to determine the sign of the value of the collector.
This will allow for an implementation of a step function (as an approximation to the sigmoid function) as well as ReLU.

The first stage of this version of the neuron will be to perform phase estimation on the collector using the filter subsystems as the pointer system.
We will use the notation of Section~\ref{sec:bkgd} and write a simulated position operator $\hat{\Phi}_{d,\textsc{f}}$ for the filter system, where $d = 2^N$.
The spectrum of $\hat{\Phi}_{d,\textsc{f}}$ should be taken to encompass the range of expected values of the collector variable $z$.
For convenience, we will assume that the range is symmetric about the origin, and will denote the maximum value by $R$.
Thus, the discrete variable system aims to provide a simulation of the collector variable $z$ on the interval $[a,b] = [-R,R]$.
The phase estimation step can then be written as
\begin{widetext}
\begin{equation}
\begin{split}
  \omega_d^{-\hat{z}_\textsc{c} \hat{\Pi}_{d,\textsc{f}}} \ket{z}_\textsc{c} \ket{0}_\textsc{f} &= \ket{z}_\textsc{c} \otimes \sum_{k \in \mathds{Z}_d} \Delta \left( z \left( \frac{d-1}{2R} \right) - k \right) \ket{k}_\textsc{f} \\
  &= \ket{z}_\textsc{c} \otimes \sum_{x_1,\dots,x_N=0}^1 \Delta \left( (2^N-1) \frac{z}{2R} - \sum_{n=1}^N x_n 2^{n-1} \right) \ket{x_1,\dots,x_N}_\textsc{f}.
\end{split}
\end{equation}
\end{widetext}
Note that although we only wish to determine the sign of $z$, thus only the value of the most significant qubit, $x_N$, the use of additional qubits aids in suppressing the probability of error (as discussed in Section~\ref{sec:bkgd}).

Now, we can proceed to implement the non-linear activation by conditioning on the value of the most significant qubit after the phase estimation step.
For example, the sigmoid function can be approximated with a step function by acting the unitary
\begin{equation}
  e^{-\frac{i}{2} (1 - \hat{Z}_{2,\textsc{f}}^{(N)}) \hat{\Pi}_\textsc{a}} = \ket{1}\!\bra{1}_\textsc{f}^{(N)} \otimes e^{-i \hat{\Pi}_\textsc{a}} + \ket{0}\!\bra{0}_\textsc{f}^{(N)} \otimes \hat{I}_\textsc{a},
\end{equation}
where $\frac{1}{2} (1 - \hat{Z}_{2,\textsc{f}}^{(N)}) = \ket{1}\!\bra{1}_\textsc{f}^{(N)}$ is the projector onto the value 1 of the most significant qubit in the filter register.
Therefore, if the value of this register is 1, which corresponds to $z > 0$, the activation register is shifted to a value of 1 (otherwise it retains its original value of 0).

The case of ReLU can be approximated similarly with the unitary
\begin{equation}
  e^{-\frac{i}{2} \hat{z}_\textsc{c} (1 - \hat{Z}_{2,\textsc{f}}^{(N)}) \hat{\Pi}_\textsc{a}} = \ket{1}\!\bra{1}_\textsc{f}^{(N)} \otimes e^{-i \hat{z}_\textsc{c} \hat{\Pi}_\textsc{a}} + \ket{0}\!\bra{0}_\textsc{f}^{(N)} \otimes \hat{I}_{\textsc{c} \textsc{a}}.
\end{equation}
Here, we see that if $z > 0$, the unitary which is implemented is effectively an adder $\ket{z}_\textsc{c} \ket{0}_\textsc{a} \mapsto \ket{z}_\textsc{c} \ket{z}_\textsc{a}$, otherwise it is just the identity.

\subsubsection{CV-only}

The case of a fully continuous variable implementation of the quantum neurons does not involve a filter register, but a single unitary applied to the collector and activation registers,
\begin{equation}
  e^{-i P(\hat{z}_\textsc{c}) \hat{\Pi}_\textsc{a}} : \ket{z}_\textsc{c} \ket{0}_\textsc{a} \mapsto \ket{z}_\textsc{c} \ket{P(z)}_\textsc{a},
\end{equation}
where $P$ is some polynomial.\cite{marshall2015repeat}
The idea here is to choose a polynomial to approximate the desired activation function on a particular interval $[a,b]$. 

Suppose we wish to approximate the activation function with a polynomial of fixed degree $N$, so that $P(z) = \sum_{n=0}^N c_n z^n$.
One possibility is to truncate a Taylor series of the activation function $\sigma$ (if it exists) to order $N$.
Another possibility would be to choose the coefficients $\{ c_n \}_{n=0}^N$ to minimize the distance between this polynomial and the desired activation function $\sigma$ in some norm.
For example, one could choose a weighted $L_2$ norm on the interval $[a,b]$ and minimize the mean-squared error
\begin{equation}
  MSE = \frac{1}{2} \int_a^b w(z) dz \hspace{1mm} | \sigma(z) - P(z) |^2.
\end{equation}
The weight function, $w(z)$, can be used to demand more accuracy on particular regions of the interval which are of interest.
The minimum is achieved by choosing coefficients which solve the matrix equation:
\begin{equation}
  \sum_{m=0}^N \left( \int_a^b w(z) dz \hspace{1mm} z^{n+m} \right) c_m = \left( \int_a^b w(z) dz \hspace{1mm} z^n \sigma(z) \right).
\end{equation}
Solving for these coefficients amounts to inverting the Hankel matrix with elements $T_{nm} := \int_a^b w(z)dz \hspace{1mm} z^{n+m}$, where $n,m = 0,\dots,N$, and applying the inverse to the right-hand side of the equation.
For a sigmoid function, $\sigma_\beta(z) = 1/(1+e^{-\beta z})$, and uniform weight function $w(z) = 1$, the right-hand side involves calculating a collection of incomplete Fermi-Dirac integrals.
However, if we approximate the sigmoid function with a step function or in the case where we are using ReLU as the activation function, then evaluation of the elements of the right-hand side is trivial (assuming a simple weight function).
One can also straightforwardly use this technique to build a polynomial approximation to other non-linear activation functions, provided one can calculate (or approximate) the integrals on the right-hand side of the above equation.

\section{Quantum Parametric Circuit Learning} \label{sec:qdata_algs}

Quantum Deep Learning of quantum data will generally consist of having to learn a certain quantum map.
As all quantum operations can be seen as unitaries in a dilated Hilbert space (possibly along with a standard basis measurement), learning a certain quantum map will often reduce to optimizing over a space of candidate unitaries in order to minimize some loss function of the output.
In general, a certain unitary transformation over a large Hilbert space of multiple registers can be decomposed into a composition of unitaries which are unitary on smaller sets of registers.
Each unitary can be seen as a form of generalized rotation in the Hilbert space of its registers.
It is then natural to consider parametrized ansatze constructed by composition of such generalized rotations, each with a given ``direction'' (Hamiltonian generator) and a certain angle. We call \emph{parametric quantum circuits} such hypothesis classes of unitaries composed of multiple unitaries with are each parametrized by real numbers. The key to learning is then to leverage efficient optimization strategies to search over the space of possible parameters in order to minimize some loss function.

The traditional approach to the optimization of these parameters has been a classical-quantum hybrid approach. In this case the circuit for a certain set of parameter values would be executed, and the expectation value of the loss function for a given set of parameters would be estimated. Then, by querying the expectation value for multiple values of the parameters for multiple runs, one could use a classical optimizer to find a suitable set of parameters which minimize the loss to a satisfying degree. For example, a finite-difference gradient method \cite{farhi2018classification,chen2018universal} is often used, but this approach necessitates $\mathcal{O}(N)$ (where $N$ is the number of parameters) runs to obtain enough expectation values of the loss for various values of the parameters in order to estimate the gradient. 

Instead of using a hybrid quantum-classical method based on estimation of multiple expectation values of the loss function for the optimization of quantum parametric circuits, we can harness the Backwards Quantum Propagation of Phase Errors (Baqprop) principle to descend the optimization landscape more efficiently. Given a way to quantum coherently evaluate the exponential of the loss function of a potential candidate solution, one will be able to use either Momentum Measurement Gradient Descent or Quantum Dynamical Descent to optimize over the set of possible circuits.

In this section, we will explore various use cases of parametric quantum circuits, explain in greater detail how to query exponential loss functions for various cases, and explore how the update rule derived in previous sections specializes in these various cases.

\subsection{Parametric Ansatze \& Error Backpropagation}\label{sec:par}

Before we talk about applications of parametric circuits to various problems and how to adapt the Quantum Feedforward and Baqprop procedure to each application, let us briefly review parametric circuits in a formal manner, and provide an overview of how error signals backpropagate through the circuit during uncomputation.

\subsubsection{From Classically- to Quantumly-Parametrized Ansatze}\label{sec:qpar_ans}

Let us first consider a generic \textit{classically} parametrized circuit ansatz. Consider a set of indices for the parameteric operations, partitioned into the indices for each layer, $\mathcal{I} = \cup_{\ell=1}^\mathscr{L}\mathcal{I}_{\ell}$, we can write the parametric unitary as
\begin{equation}\label{eq:class_par}
    \hat{U}(\bm{\Phi}) = \prod_{\ell=1}^\mathscr{L}\hat{U}^{(\ell)}(\bm{\Phi}^{(\ell)}),
\end{equation}
where $\hat{U}^{(\ell)}(\bm{\Phi}^{(\ell)})$ is the multi-parameter unitary corresponding to the $\ell^\text{th}$ layer, which can itself be composed of multiple parametric unitaries $\{\hat{U}_{j_\ell}(\Phi_{j_\ell})\}_{j_\ell}$ as follows,
\begin{equation}
    \hat{U}^{(\ell)}(\bm{\Phi}^{(\ell)})\equiv \bigotimes_{j_\ell\in \mathcal{I}_\ell} \hat{U}_{j_\ell }(\Phi_{j_\ell}).
\end{equation}

Now, suppose we wish to optimize some loss operator $\hat{L}_j$, for the above parametric unitary applied onto an initial state $\ket{\xi_j}$ the typical approach to optimizing this is to compute the expectation value of the loss operator for this feedforwarded state
\begin{equation}
    \braket{\hat{L}_j}_{\bm{\Phi}} := \bra{\xi_j}\hat{U}^\dagger(\bm{\Phi})\hat{L}_j\hat{U}(\bm{\Phi})\ket{\xi_j}_{\bm{\Phi}}.
\end{equation}
A classical optimizer is then tasked to find the set of parameters which minimize the cost function which in general can be the su of multiple loss operators, i.e. $\text{argmin}_{\bm{\Phi}}(\sum_{j\in\mathcal{B}_k}\braket{\hat{L}_j}_{\bm{\Phi}})$. In general, for a quantum-classical optimization procedure, multiple expectation values of the loss operators will need to be estimated. For example, one may perform finite-difference gradient descent by estimating derivatives of each loss at a time
\begin{equation}
    \left.\partial_{\Phi_k} \braket{\hat{L}_j}_{\bm{\Phi}}\right|_{\bm{\Phi}^\ast}  \approx \tfrac{1}{\epsilon }\left(\braket{\hat{L}_j}_{\bm{\Phi}^\ast+\bm{\delta}_k}-\braket{\hat{L}_j}_{\bm{\Phi}^\ast}\right)
\end{equation}
where $(\bm{\delta}_k)_j = \epsilon \delta_{jk}$, $\epsilon\ll1$. For an $N$-parameter ansatz and $M$ terms in the loss function, in order to estimate the gradient, this requires $\mathcal{O}(M\cdot N)$ expectation value estimations, which in some cases must each taken in separate feedforward runs.

Instead of classically parametrizing the circuits, as we have covered extensively in this paper, we can use quantum parameters in order to leverage either MoMGrad or QDD. As we know from section \ref{sec:opt}, gradients can then be estimated via MoMGrad in $\mathcal{O}(M)$ feedforward and Baqprop queries, which we then gave techniques to fully parallelize this gradient acquisition over the minibatch (sec. \ref{sec:misc}) with only $\mathcal{O}(\log M)$ added depth over the single-replica QFB operation.

To convert a classically parameterized a circuit of the form \eqref{eq:class_par} to a quantumly-parametrized circuit, we convert $\hat{U}(\bm{\Phi})\mapsto \hat{U}(\bm{\hat{\Phi}})$ where
\begin{equation}
      \hat{U}(\bm{\hat{\Phi}}) := \prod_{\ell=1}^\mathscr{L}\hat{U}^{(\ell)}(\bm{\hat{\Phi}}^{(\ell)}),\quad 
  \hat{U}^{(\ell)}(\bm{\hat{\Phi}}^{(\ell)})\equiv \bigotimes_{j_\ell\in \mathcal{I}_\ell} \hat{U}_{j_\ell }(\hat{\Phi}_{j_\ell}),
\end{equation}
and each unitary is converted to a continuously-controlled unitary with a quantum parameter register, \begin{equation}
  \hat{U}_{j_\ell }(\hat{\Phi}_{j_\ell})= \sum_{\Phi_{j_\ell}}\ket{\Phi_{j_\ell}}\!\bra{\Phi_{j_\ell}}\otimes  \hat{U}_{j_\ell }(\Phi_{j_\ell}).
\end{equation}
Now assuming each unitary is generated by a certain Hamiltonian, i.e., $ \hat{U}_{j_\ell }(\Phi_{j_\ell})= e^{-i\Phi_{j_\ell}\hat{h}_{j_\ell }}$ then the above becomes \begin{equation}
  \hat{U}_{j_\ell }(\hat{\Phi}_{j_\ell})=   e^{-i\hat{\Phi}_{j_\ell}\otimes \hat{h}_{j_\ell }},
\end{equation}
which we see is an exponential with a generator $\hat{\Phi}_{j_\ell}\otimes \hat{h}_{j_\ell }$. 

Let us examine how to synthesize such an exponential into basic gates. For a given index $j\in\mathcal{I}$, suppose we know a way to synthesize the classically parametrized gate $e^{-i\Phi_{j} \hat{h}_{j}}$ (with $\Phi_j$ as the classical parameter) into a product of non-parametric unitaries and of one or multiple parametric unitaries of the form  $\{e^{-i\beta_{k_j} \Phi_j \hat{Z}}\}_{k_j}$ where all $\beta_{k_j}\in\mathds{R}$. We can then convert this synthesis of the classically parametric gate into a synthesis for its respective quantum-parametric analogue by converting all the parametric exponentials of $\Phi_j\hat{Z}$ into quantum-parametric exponentials of $\hat{\Phi}_j\otimes \hat{Z}$, i.e.

\begin{equation}
\{e^{-i\beta_{k_j} \Phi_j \hat{Z}}\}_{k_j}\mapsto \{e^{-i\beta_{k_j} \hat{\Phi}_j \otimes\hat{Z}}\}_{k_j}.\end{equation} Each quantum-parametric exponential  $e^{-i\beta_{k_j} \hat{\Phi}_j \hat{Z}}$ is essentially like a single-qubit observable phase estimation unitary; as discussed in section \ref{sec:QPE}, it can be
can be broken up into $\lceil\log(d)\rceil$ exponentials of $\hat{Z}\otimes \hat{Z}$ where $d$ is the effective qudit dimension of the parameter register. 

Often, the generators of these exponentials are chosen to be simple (e.g. $n$-local Paulis $\mathcal{P}_n$ \cite{gottesman2010introduction}), hence as an explicit example, we can consider a case where $\hat{h}_{j}$ is a Pauli operator on $n$ qubits. For any  $\hat{h}\in\mathcal{P}_n$, there exists a $\hat{V}\in\mathcal{C}_n$, where $\mathcal{C}_n$ is the $n$-qubit Clifford group \cite{gottesman2010introduction}, 
such that $\hat{h} = \hat{V}^\dagger\hat{Z}^{(r)}\hat{V}$, where $\hat{Z}^{(r)}$ is the Pauli $Z$ on a register of choice, which we label as having an index $r$. To decompose Clifford group operator $\hat{V}$ into basic Clifford gates there are multiple known algorithms for this synthesis \cite{kliuchnikov2016practical} and Clifford gates are very efficiently implementable on error-corrected quantum computers \cite{barends2014superconducting}. For such an operator $\hat{h}$, a parametric exponential of the form 
\begin{equation}
    e^{-i\Phi \hat{h}} =    e^{-i\Phi\hat{V}^\dagger  \hat{Z}^{(r)}\hat{V}}= \hat{V}^\dagger e^{-i\Phi \hat{Z}^{(r)}}\hat{V}
\end{equation}
thus to convert this parametric exponential into a quantum-parametric exponential, we need to apply
\begin{equation}
 e^{-i\hat{\Phi} \otimes \hat{h}} =    \hat{V}^\dagger e^{-i\hat{\Phi} \otimes \hat{Z}^{(r)}}\hat{V}
\end{equation}
the quantum-phase-estimation-like exponential in the middle can then be broken down into $\lceil\log(d)\rceil$ exponentials of Paulis between the qubits of the parameter register and that of the $r$ register. As a side note, for analog quantum computers, for parameter registers which are physical qumodes, the quantum-parametric exponential $e^{-i\hat{\Phi} \otimes \hat{Z}}$ can be implemented using an interaction Hamiltonian of the form
\begin{equation}\label{eq:interact}
    \hat{H}_{\text{int}} = \lambda \hat{\Phi}\otimes \hat{Z}
\end{equation}
where $\hat{\Phi}$ is a quadrature of a qumode and $\lambda$ some coupling strength. Such an interaction should be feasible to implement in various quantum computing implementations of today \cite{shore1993jaynes}.

Now that we have seen how to execute quantum-parametric unitaries, let us recall that in order to optimize the parameters $\bm{\Phi}$ such as to minimize some loss operator on the output $\hat{L}_j$, we can execute the quantum feedforward and quantum phase error backwards propagation (QFB) procedure with quantum-parametric circuits, and leverage techniques from section \ref{sec:opt} for optimization. Recall that the QFB consists of applying the quantum-parametric feedforward operation, an exponential of the loss function, followed by the uncomputation of the feedforward, 
\begin{equation}
  e^{-i \eta \hat{L}(\bm{\hat{\Phi}}) }=  \hat{U}(\hat{\bm{\Phi}})^\dagger e^{-i \eta \hat{L}_j} \hat{U}(\hat{\bm{\Phi}}).
\end{equation}
Recall \eqref{eq:momm_update} that for an input state $\ket{\xi_j}$, to first order in $\eta$, the momenta of the parameters get kicked by the gradient
\begin{equation}
\begin{split}
  \bm{\hat{\Pi}} \quad \mapsto \quad & e^{i \eta \mathcal{L}(\bm{\hat{\Phi}})} \bm{\hat{\Pi}} e^{-i \eta \mathcal{L}(\bm{\hat{\Phi}})} + \mathcal{O}(\eta^2) \\
  &= \bm{\hat{\Pi}} - \eta \frac{\partial \mathcal{L} (\bm{\hat{\Phi}})}{\partial \bm{\hat{\Phi}}} + \mathcal{O}(\eta^2).
\end{split}
\end{equation}
where the effective loss function is given by
\begin{equation}
  \mathcal{L}(\bm{\hat{\Phi}}) := \bra{\xi} \hat{L}(\bm{\hat{\Phi}}) \ket{\xi}.
\end{equation}
We can then leverage this momentum shift to optimize the parameters via MoMGrad or QDD (see sec. \ref{sec:opt}). Notice that \emph{all components} of the momentum get kicked, but each component of the parameters comes into contact with the compute at a different time during both the feedforward and Baqprop phases. In order to understand how exactly the error signal backpropagates and influences the various parameter's momenta during the uncomputation, we can further examine how the parameters get kicked in a layerwise fashion, which we do now below.  

\subsubsection{Quantum Parametric Circuit Error Backpropagation}

In Section~\ref{sec:deep_baq}, we elaborated upon the mechanism through which the QFB circuit propagates the errors, layer-by-layer, for quantum-coherent neural networks through the recursive formulas \eqref{eqn:activation_backprop}, \eqref{eqn:weight_backprop}, and \eqref{eqn:bias_backprop}.
Here, we will briefly discuss a layerwise analysis of the quantum phase error backpropagation for layered quantum parametric circuits.
Of course, since we are using a very general ansatz for the quantum parametric circuits, we cannot repeat the analysis in the same level of detail as for the quantum-coherent neural networks.

Consider once again a parametrized circuit decomposed into layers:
 \begin{equation}
      \hat{U}(\bm{\hat{\Phi}}) := \prod_{\ell=1}^\mathscr{L}\hat{U}^{(\ell)}(\bm{\hat{\Phi}}^{(\ell)}),\quad 
  \hat{U}^{(\ell)}(\bm{\hat{\Phi}}^{(\ell)})\equiv \bigotimes_{j_\ell\in \mathcal{I}_\ell} \hat{U}_{j_\ell }(\hat{\Phi}_{j_\ell}).
\end{equation}
where $\hat{\bm{\Phi}}= \{\bm{\hat{\Phi}}^{(\ell)}\}_{\ell=1}^{\mathscr{L}}$ is the operator vector of all the parameters, and $\bm{\hat{\Phi}}^{(\ell)} = \{\Phi_{j_\ell}\}_{ j\ell \in \mathcal{I}_\ell}$ is that of the parameters for a single layer.

For convenience, let us write the circuits of operations before and after the layer $k$ as:
\spliteq{
   &\hat{U}^{(<k)}(\bm{\hat{\Phi}}^{(<k)}):=\prod_{\ell=1}^{k} \hat{U}^{(\ell)}(\bm{\hat{\Phi}}^{(\ell)}),\\ &\hat{U}^{(>k)} (\bm{\hat{\Phi}}^{(>k)}):= \prod_{\ell=k}^{\mathscr{L}} \hat{U}^{(\ell)}(\bm{\hat{\Phi}}^{(\ell)}).
}
Recall the QFB circuit for the entire circuit is:
\begin{equation}
  \hat{U}^\dagger(\bm{\hat{\Phi}}) e^{-i \eta \hat{L}} \hat{U}(\bm{\hat{\Phi}}) 
\end{equation}
Now suppose we would like to focus on a certain layer $\ell$ in the QFB circuit above, we could group the feedforward, phase kick and uncomputation operations for layers beyond layer $\ell$ as
\begin{equation}
\hat{U}^{(>\ell)\dagger}(\bm{\hat{\Phi}}^{(>\ell)}) e^{-i \eta \hat{L}} \hat{U}^{(>\ell)}(\bm{\hat{\Phi}}^{(>\ell)})   := e^{-i \eta \hat{L}(\bm{\hat{\Phi}}^{(>\ell)})}
\end{equation}
we see that this is just a loss exponential with respect to a different loss operator
\begin{equation}
\hat{L}(\bm{\hat{\Phi}}^{(>\ell)}):= \hat{U}^{(>\ell)\dagger} (\bm{\hat{\Phi}}^{(>\ell)}) \hat{L}\hat{U}^{(>\ell)}(\bm{\hat{\Phi}}^{(>\ell)})
\end{equation}
which is effectively a backpropagated loss operator in the Heisenberg picture. Similarly, we can group the operations for layers below the layer $\ell$, combined with the backpropagated loss exponential from above, the whole QFB circuit can then be seen as 
\begin{equation}\label{eq:l_focus}
\begin{split}
  &\hat{U}^{(<\ell)\dagger}(\bm{\hat{\Phi}}^{(<\ell)})\hat{U}^{(\ell)\dagger}(\bm{\hat{\Phi}}^{(\ell)}) \\
  &\qquad \qquad \qquad \times e^{-i \eta \hat{L}(\bm{\hat{\Phi}}^{(>\ell)})}  \hat{U}^{(\ell)}(\bm{\hat{\Phi}}^{(\ell)})   \hat{U}^{(<\ell)}(\bm{\hat{\Phi}}^{(<\ell)})
\end{split}
\end{equation} which is effectively like a single-layer QFB, with the backpropagated loss, and a modified input state being the fedforward input state. 

In a sense, we can view the above picture as a quantum form of \emph{automatic differentiation} \cite{rall1981automatic}. In classical automatic differentiation, for a composite of functions composed in layers, in order to compute the gradient of the output with respect to a certain parameter, the gradient of layers beyond that of the parameter are computed layerwise starting from the output. Using both this backpropagated gradient and the value of the fedforward input up to the layer of the given parameter of interest, one can compute the gradient for the said parameter. We can try to examine how automatic differentiation is naturally executed in each branch of the multi-parameter wavefunction by the Quantum Phase Error Backpropagation.

Let us label eigenstates of all parameters other than those of layer ${\ell}$ as $\ket{\bm{\Phi}^{(\neg \ell)}} := \bigotimes_{j\neq \ell} \ket{\bm{\Phi}^{(j)}}$ where $\bm{\hat{\Phi}}^{(\neg \ell)} = \{\bm{\hat{\Phi}}^{(\ell)}\}_{j\neq \ell}$ are the correponding parameter operators. Furthermore, consider the very initial input state to the whole QFB circuit to be $\ket{\xi}$, and let us define the conditional feedforwarded state up to layer $\ell$ as
\begin{equation}
    \ket{\xi_{\bm{\Phi}^{(< \ell)}}}: = \hat{U}^{(<\ell)}(\bm{\Phi}^{(< \ell)})\ket{\xi}.
\end{equation}

Suppose we consider each branch of the wavefunction of parameters of layers other than $\ell$, e.g., each term in $\sum_{\bm{\Phi}^{(\neg\ell)}}\psi_{\bm{\Phi}^{(\neg\ell)}}\ket{\bm{\Phi}^{(\neg \ell)}}$ then conditioning each branch of this wavefunction, we get an effective phase kick on the parameters of layer $\ell$. Equation \eqref{eq:l_focus} becomes the following conditional effective phase on the parameters of layer $\ell$,
\begin{equation}
   \sum_{\bm{\Phi}^{(\neg\ell)}}\ket{\bm{\Phi}^{(\neg \ell)}}\!\bra{\bm{\Phi}^{(\neg \ell)}} \otimes \mathcal{L}_{\bm{\Phi}^{(\neg \ell)}}(\bm{\hat{\Phi}}^{(\ell)}) 
\end{equation}
where
\begin{equation}
\begin{split}
  &\mathcal{L}^{(\ell)}\!(\bm{\hat{\Phi}}^{(\ell)}\!) \\
  &\quad \!=\!\bra{\xi_{\bm{\Phi}^{(< \ell)}}\!}\!\hat{U}^{(\ell)\dagger}(\bm{\hat{\Phi}}^{(\ell)})  e^{-i \eta \hat{L}(\!\bm{\Phi}^{(>\ell)}\!) }  \hat{U}^{(\ell)}(\bm{\hat{\Phi}}^{(\ell)})    \!\ket{\xi_{\bm{\Phi}^{(< \ell)}}\!}\!.
\end{split}
\end{equation}
We see above that for each case (branch of the parameter values wavefunction for value  $\bm{\Phi}^{(< \ell)}$) there is an incoming state to layer $\ell$  ($\ket{\xi_{\bm{\Phi}^{(< \ell)}}}$) and there is a backpropagated phase kick operator ($e^{-i \eta \hat{L}(\!\bm{\Phi}^{(>\ell)}\!) }  $). This is similar in vein to classical automatic differentiation, but this happens in every branch of the wavefunction in parallel. In a sense, it is automated automatic differentiation. Because each parameter can take gradients of the loss conditioned on previous and later layers' quantum states, all parameters' momenta can thus get nudged simultaneously by the gradient of the conditional loss in each branch of the wavefunction. This allows for single-sweep gradient estimation of all parameters, in contrast to some other techniques for parametric circuits which require each derivative to be computed one at a time \cite{dallaire2018quantum}.

Now, recall from section \ref{sec:qnn}, in the case of the quantum neural networks, we rewrote analysed how the backpropagating error signal is carried between parameter registers by the compute registers. In the case of coherent neural networks, the phase kick corresponding to the error signal would kick the activation's momenta (which in turn kicks the momenta of the weights and biases).
Here, to see which operator is getting kicked, one would need to examine more concretely the conjugate of the generator of the unitaries in each layer of the circuit. Performing such an analysis could shed some light as to what makes a good choice parametric circuit ansatz such as to avoid the vanishing gradient problem of most currently known ansatze \cite{mcclean2018barren}.
 We leave this remaining analysis for future work.

\subsection{Quantum State Exponentiation}
\label{sec:qse}

In this subsection, we will delve into greater detail into the ways to enact a certain set of loss function exponentials for quantum data. In previous section \ref{sec:opt}, we showed how we could harness the MoMGrad and QDD optimization procedures given access to a phase kick (complex exponential) of a loss function operator for which we would like to minimize the error. For many applications of quantum parametric circuit learning, it will be useful to create the phase kick for a loss function which will provide a notion of distance between the output state and the target state, and in our case this notion of metric will be induced by some form of inner product between states.

\subsubsection{Single state exponentiation}\label{sec:sse}
Well-known in quantum information is the notion of \textit{fidelity} between quantum states. For pure quantum states, the fidelity $F$ between states $\ket{\psi}$ and $\ket{\phi}$ is simply the magnitude of the inner product  $F(\phi,\psi)= |\braket{\phi|\psi}|$. Note that clearly fidelity itself is not a metric, but one can create a proper metric on the space of states by considering the sine distance $\mathcal{S}$ \cite{nielsen2002quantum}, which is related to the fidelity by the equality $\mathcal{S} = \sqrt{1-F}$. In order to perform gradient ascent on the fidelity in the case of pure state learning (which we will treat in-depth in the next subsection) we will need to be able to exponentiate states, i.e., perform 
$e^{-i\ket{\psi}\!\bra{\psi}}$ given multiple copies of $\ket{\psi}$ in memory.

That is, given a set of $n$ copies of pure states $\ket{\psi}^{\otimes n}$ held in memory, we would like to execute the unitary $e^{-i\eta \ket{\psi}\!\bra{\psi}}$ to a certain precision by consuming some of these copies. More generally, for a set of mixed states $\hat{\rho}^{\otimes n}$ held in memory, we would like to be able to enact the unitary $e^{-i\eta \hat{\rho}}$ on our target state. As we will see in \ref{sec:qpsl}, the exponential of mixed states will induce a gradient ascent on the Hilbert-Schmidt inner product rather than the fidelity.

This task is referred to as quantum state exponentiation 
(QSE) \cite{bromley2018batched}.  The original protocol to perform quantum state exponentiation was first formulated by Lloyd, Mohseni, and Rebentrost \cite{lloyd2014quantum}. This approach was recently proven to be optimal for the Quantum State Exponentiation task \cite{kimmel2017hamiltonian}. For the target state $\hat{\sigma}$ and a copies of mixed states in data, the original QSE protocol applies the map 
\begin{equation}
    \hat{\sigma} \otimes \hat{\rho}^{n} \mapsto e^{-i\hat{\rho} \eta} \hat{\sigma} e^{i\hat{\rho} \eta}  = \text{Ad}[e^{-i\hat{\rho} \eta}](\hat{\sigma}) 
\end{equation}
up to an error accuracy $\epsilon$ in the diamond norm, by using $n\sim \mathcal{O}(\eta^2/\epsilon)$ steps, each consuming a copy of $\hat{\rho}$.

\begin{figure}[h!]
 \begin{center}
\includegraphics[width=0.9\columnwidth]{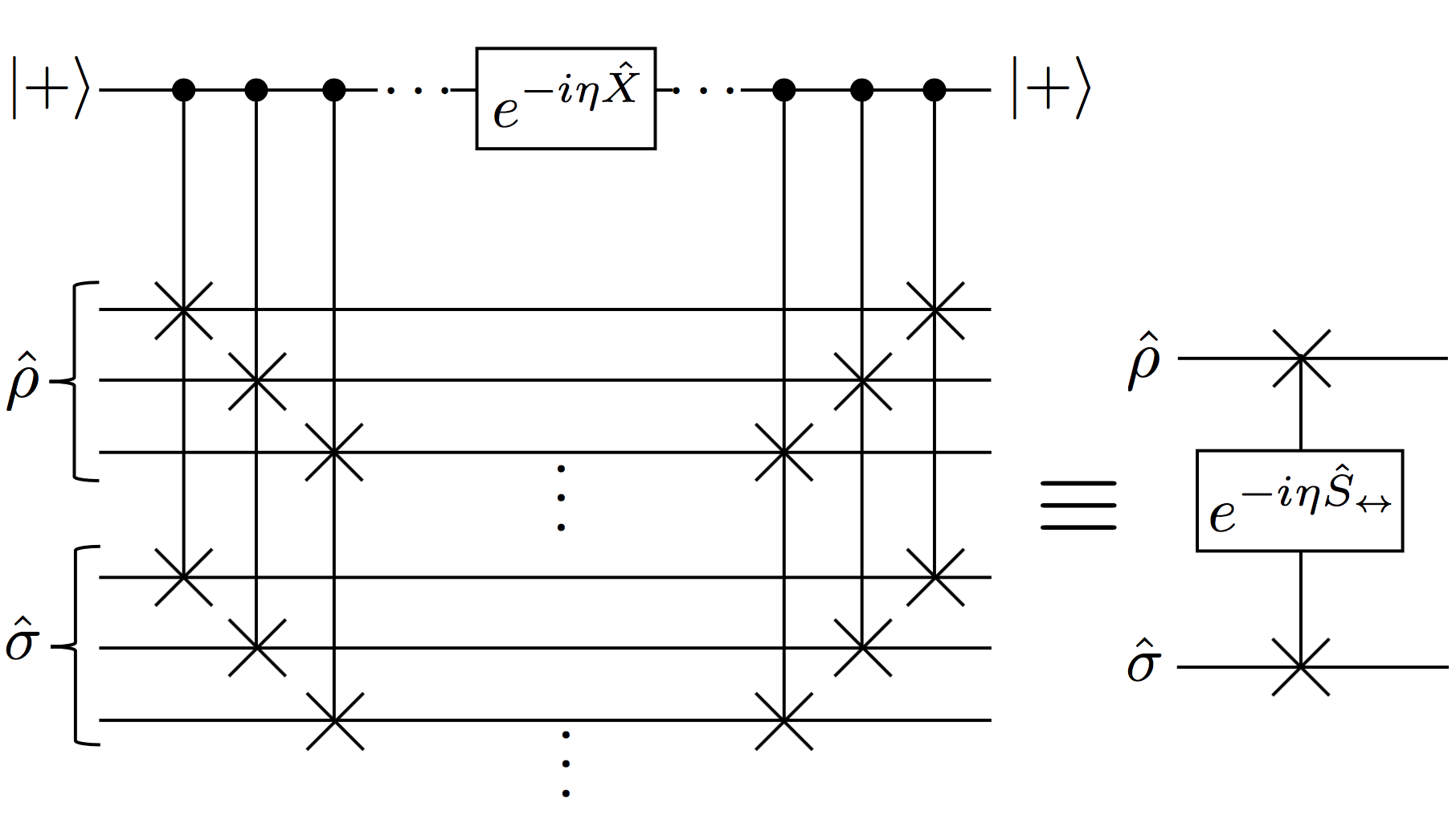}
\caption{Given two states $\rho$ and $\sigma$ each supported on $n$ qubits, one can perform an ancilla-assisted exponential swap $ e^{-i\eta S_{\leftrightarrow}}$ as above \cite{lloyd2014quantum}. Using an ancilla qubit initially in the $\ket{+}$ state, by applying individual controlled-swaps sequentially, with the ancilla as control and corresponding qubit registers of $\hat{\rho}$ and $\hat{\rho}$ as targets, then applying an exponential $e^{-i\eta\hat{X}}$ on the ancilla, and later undoing the control-swap sequence, as the reader can readily check, the ancilla is left unchanged, and an effective unitary exponential of swap between sets of registers $ e^{-i\eta S_{\leftrightarrow}}$ is thus applied. } \label{fig:exp_swap}
\end{center}
\end{figure}

More explictly, this quantum state exponentiation approach consists of approximating within $\epsilon$ diamond norm error the final target state
\begin{equation}
    \text{Ad}[e^{-i\hat{\rho} \eta}](\hat{\sigma}) =  \hat{\sigma} -i[\hat{\rho},\hat{\sigma}] \eta -\tfrac{1}{2!}[\hat{\rho},[\hat{\rho},\hat{\sigma}]]\eta^2 + \ldots
\end{equation}  with $n$ steps each consisting of partial-swapping of a copy of $\hat{\rho}$ onto the target $\hat{\sigma}$ using an exponential swap operation $e^{-i \delta \hat{S}} $, where $\delta = \epsilon/\eta$, for a total of $n\sim \mathcal{O}(\eta^2/\epsilon)$ steps/copies. In Figure \ref{fig:exp_swap}, we provide further detail as to the implementation of exponential swaps via more standard gates.

\begin{figure}[h!]
 \begin{center}
\includegraphics[width=0.8\columnwidth]{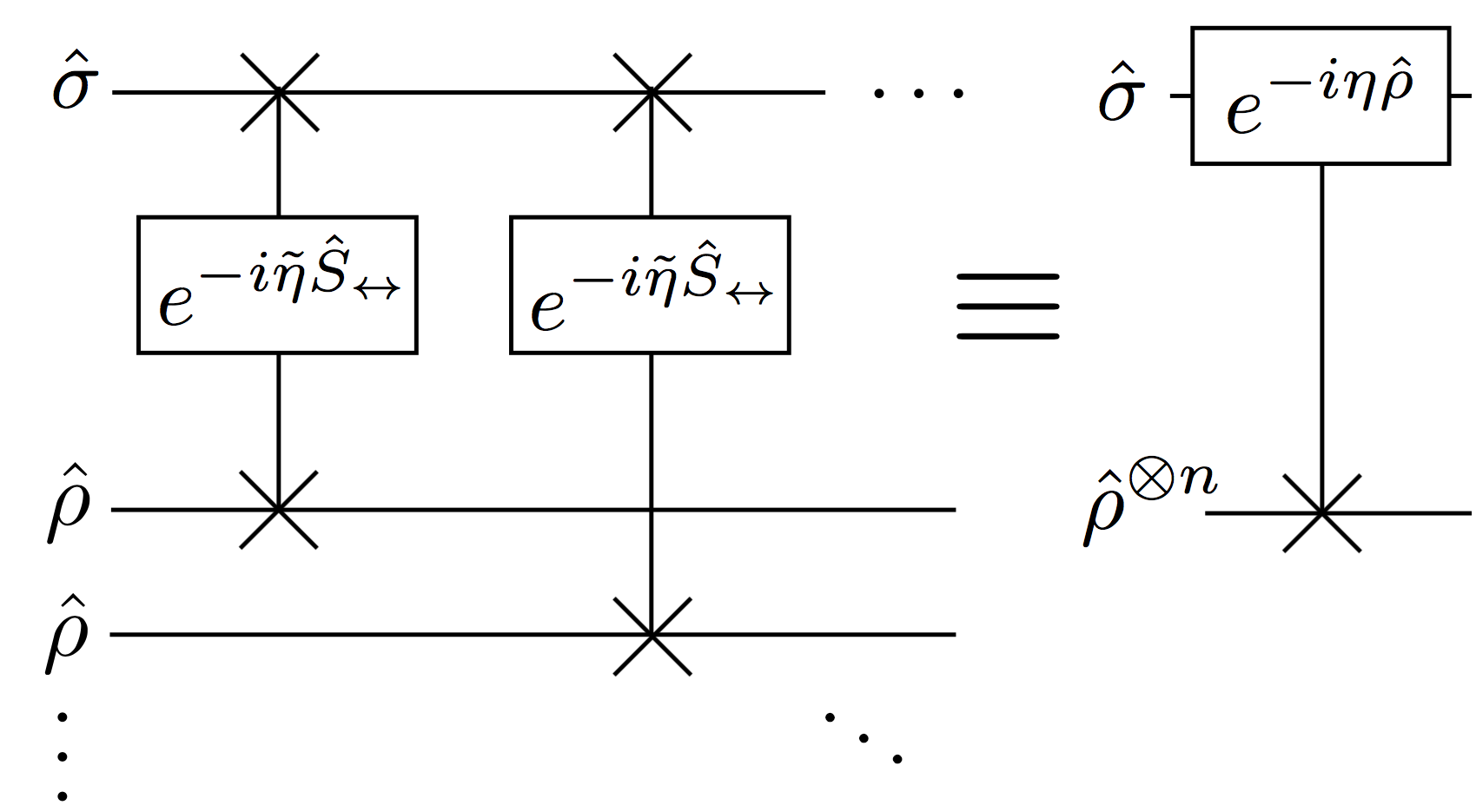}
\caption{Given $n$ copies of a state $\hat{\sigma}$, we can batch exponential swaps (pictured in figure \ref{fig:exp_swap}) with the target state $\hat{\rho}$, for angles $\tilde{\eta} = \epsilon/\eta$ to induce a phase $e^{-i\eta \hat{\sigma}}$ on $\hat{\rho}$ up to an error $\mathcal{O}(\epsilon)$ for a number of copies scaling as $n\sim \epsilon/\eta$. } \label{fig:batch_exp}
\end{center}
\end{figure}

If we look at the effective operation acted upon the  on the target register, we have
\spliteq{
    \text{tr}_2\left(e^{-i\delta S} (\hat{\sigma}\otimes \hat{\rho})e^{i\delta S} \right) &=   \hat{\sigma} -i[\hat{\rho},\hat{\sigma}] \delta +\mathcal{O}(\delta^2 )\\
    &=\text{Ad}[e^{-i\hat{\rho} \delta}](\hat{\sigma})+\mathcal{O}(\delta^2 ),}
thus, by repeating this process $n$ times, we get 
\begin{equation}
    \text{Ad}[e^{-i\hat{\rho} n\delta}](\hat{\sigma}) +\mathcal{O}(n\delta^2) \approx \text{Ad}[e^{-i\hat{\rho} \eta}](\hat{\sigma}) + \mathcal{O}(\epsilon)
\end{equation}
for $n\sim \mathcal{O}(\eta^2/\epsilon)$ and $\delta = \epsilon/\eta$. This is represented in Figure \ref{fig:batch_exp}.

\subsubsection{Sequential Exponential Batching}
There are multiple ways to perform this exponentiation of the mixed state, one of which is to perform a serially batched state exponentiation \cite{bromley2018batched}. Assuming our mixed state is a classical mixture of pure states of the form
\begin{equation}
    \hat{\rho} = \tfrac{1}{|\mathcal{M}|}\sum_{j\in\mathcal{M}} \ket{\psi_j}\!\bra{\psi_j}
\end{equation}
then, from the Baker-Campbell-Hausdorff lemma, notice we can split the exponential of the mixed state into a sort of Trotterization \cite{bromley2018batched} 
of exponentials, sequentially exponentiating each state one at a time and performing multiple sweeps over the data set. Suppose we perform $N$ sweeps over the dataset, let $\tilde{\eta} \equiv \eta/|\mathcal{M}|$, we can then approximate the exponential as
\spliteq{
    e^{-i\eta \hat{\rho}}   &=e^{-i\tilde{\eta}\sum_{j\in\mathcal{M}}\ket{\psi_j}\!\bra{\psi_j}}\\&= \prod_{n=1}^N \Big(\!\!\prod_{j\in \mathcal{M}}\!e^{-i\frac{\tilde{\eta}}{N} \ket{\psi_j}\!\bra{\psi_j}}\Big) + \mathcal{O}(\tfrac{|\mathcal{M}|\tilde{\eta}^2}{N})
}
where the error is of order $\mathcal{O}(\tfrac{|\mathcal{M}|\tilde{\eta}^2}{N}) = \mathcal{O}(\tfrac{\eta^2}{|\mathcal{M}| N})$ in the diamond norm \cite{bromley2018batched}. We can call this sequential mini-batching of quantum state exponentiation. Trivially, a similar bound can be derived for a decomposition of the mixed state into other mixed states, e.g. if $\hat{\rho} = |\mathcal{M}|^{-1}\sum_{j\in\mathcal{M}} \hat{\rho}_j$ we can then batch the state exponential as
\spliteq{
    e^{-i\eta \hat{\rho}}   &=e^{-i\tilde{\eta}\sum_{j\in\mathcal{M}}\hat{\rho}_j}= \prod_{n=1}^N \Big(\!\!\prod_{j\in \mathcal{M}}\!e^{-i\frac{\tilde{\eta}}{N} \hat{\rho}_j}\Big) + \mathcal{O}(\tfrac{|\mathcal{M}|\tilde{\eta}^2}{N}).
}
Since we will be considering both mixtures of mixed states and pure state as input the above techniques are an important option.
Note this batching is used for the data loading, which is different from batching phase kicks on the parameters Quantum Feedforward and Baqprop iterations as discussed in \ref{sec:misc}.

\subsubsection{QRAM Batching}

Another option to create the mixed state is to use a Quantum Random Access Memory (QRAM). Although using a QRAM is not essential, the QRAM will create a mixture of various states, thus effectively preparing a mixed state. Given a set of states $\{\ket{\psi_j}\}_{j\in\mathcal{M}}$, using a QRAM with a uniform superposition over addresses in the index set $\mathcal{M}$, we can prepare a state

\begin{equation}
    \tfrac{1}{\sqrt{|\mathcal{M}|}}\sum_{j\in \mathcal{M}} \ket{j}_{\textsc{a}}\mapsto \tfrac{1}{\sqrt{|\mathcal{M}|}}\sum_{j\in \mathcal{M}} \ket{j}_{\textsc{a}} \ket{\psi_j}_{\textsc{d}}
\end{equation}
where $A$ is the quantum address index and $D$ is the data register, using a tree-like network of Fredkin gates, of depth $\mathcal{O}(\log|\mathcal{M}|)$ \cite{giovannetti2008quantum}. The reduced state of the data register with the address traced out is the desired mixed state
\begin{equation}
   \hat{\rho}_{\textsc{d}}=  \tfrac{1}{|\mathcal{M}|}\sum_{j\in \mathcal{M}} \ket{\psi_j}\!\bra{\psi_j}_{\textsc{d}}.
\end{equation}
Through multiple queries of the QRAM, multiple copies of the mixed state $\hat{\rho}_{\textsc{d}}$ can be obtained, then using the same single-state exponentiation techniques as described above in \ref{sec:sse}, for a number $n\sim \epsilon/\eta$ copies, we can enact the exponential $e^{-i\eta \hat{\rho}_\textsc{d}}$ within an error $\epsilon$.

Note that apart from requiring a lower depth, there is no clear advantage of using a QRAM batching over sequential batching. Once again, this batching of state exponentiation is only for data lodaing, one can also use a QRAM for minibatching of the descent of the wavefunction in the parameter landscape, as discussed in subsection \ref{sec:qramm}. Then again, there does not seem to be a necessity for QRAM in that scenario either.

\subsection{Quantum State Learning}\label{sec:qstate_learn}

Quantum state learning can be seen as the quantum analogue of unsupervised learning. In classical ML, given samples from a certain distribution, using neural network anstaze such as Restricted Boltzmann machines, autoencoders, or Generative Adversarial Networks, one learns a way to sample for the underlying distribution of the data. The statistics of the classical probability distribution are replicated by learning a map which can take as input simple (often taken to be uncorrelated) random variables, called the latent variables, and transforms their joint distribution into an approximation of the data's underlying distribution.

In quantum mechanics, instead of strictly classical probability distributions, there are wavefunctions, and classical distributions of wavefunctions. These are known as \textit{pure states} and \textit{mixed states} respectively. Similarly to the classical case, we can learn a way to map simple distributions, such as tensor products of pure states, or tensor products of mixed states, to the quantum distribution which underlies the data.

We begin by learning how to generate pure states, given many copies of the same state from data. Following this, we will cover a way to recover mixed states, given copies of the mixed state or access to pure state samples from the distribution.

\subsubsection{Quantum Pure State Learning}\label{sec:qpsl}

The pure state learning task is the following: given $n$ copies of an unknown state, $\ket{\psi}$, we would like to learn a circuit decomposition which prepares the state, $\ket{\tilde{\psi}}$, with a high fidelity to the desired state $\ket{\psi}$.
One can achieve this by employing the framework of this paper of optimizing over a family of parametrized circuits, $\hat{U}(\bm{\hat{\Phi}})$, which are applied to an initial resource state, $\ket{\psi_0}$.
This resource state, for example, could be the computational null state, $\bigotimes_j \ket{0}_j$, of a collection of qubits.
Depending on the complexity of the pure state to be learned, it may be advantageous to exploit any available prior knowledge to begin in a state which is closer to the target state.

\begin{figure}[h!]
\includegraphics[width=0.65\columnwidth]{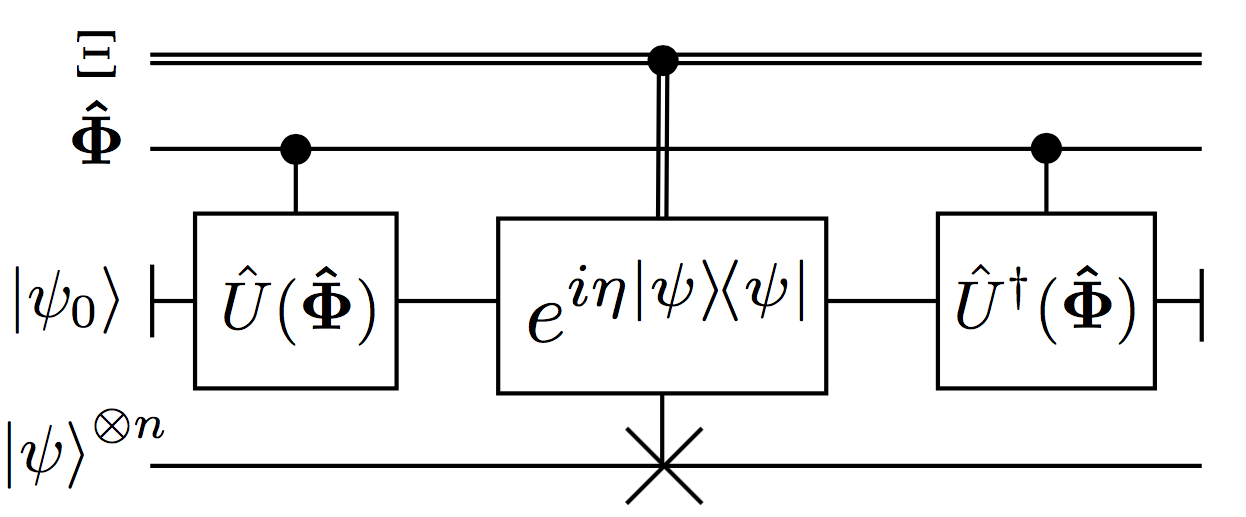}
\caption{The QFB circuit for quantum pure state learning. This consists of a feedforward unitary, $\hat{U}(\bm{\hat{\Phi}})$, controlled by quantum parameters $\bm{\hat{\Phi}}$, a phase kick achieved through state exponentiation of $n$ copies of $\ket{\psi}$ in the lower registers (and classically-controlled by the hyper-parameter $\eta$), followed by the uncomputation $\hat{U}^\dagger(\bm{\hat{\Phi}})$. The initial state on the computational registers is an input resource pure state $\ket{\psi_0}$.
} \label{fig:q_pure_learn}
\end{figure}

Now we will explain how this task can be solved using the Quantum Feedforward and Phase-Kick Backpropagation (QFB) algorithm in conjunction with either Momentum Measurement Gradient Descent (MoMGrad) or Quantum Dynamical Descent (QDD).
Recall that a single run of QFB entails an application of the parametrized unitary, $\hat{U}(\bm{\hat{\Phi}})$, on the input state, $\ket{\psi_0}$, followed by the exponentiated loss function, $e^{-i \eta \hat{L}}$, and the uncompute, $\hat{U}^\dagger(\bm{\hat{\Phi}})$.
In the present task of pure state learning, the loss function will be $\hat{L} = - \ket{\psi}\! \bra{\psi}$.
Exponentiation of this loss function can be achieved, using multiple copies of $\ket{\psi}$, through the methods described in Section~\ref{sec:qse}.
This circuit is illustrated in Figure~\ref{fig:q_pure_learn}.

For this loss function, the effective phase which generates the kick in the momenta of the parameters is:
\begin{equation}
\begin{split}
  \mathcal{L}(\bm{\hat{\Phi}}) &= \bra{\psi_0} \hat{L}(\bm{\hat{\Phi}}) \ket{\psi_0} \\
  &= - | \bra{\psi} \hat{U}(\bm{\hat{\Phi}}) \ket{\psi_0} |^2.
\end{split}
\end{equation}
Recall that above we defined $\hat{L}(\bm{\hat{\Phi}}) := \hat{U}^\dagger(\bm{\hat{\Phi}}) \hat{L} \hat{U}(\bm{\hat{\Phi}})$, i.e., the evolution of the cost function under the parametrized algorithm $\hat{U}(\bm{\hat{\Phi}})$.
Notice that the effective phase (for each value of $\bm{\Phi}$) is minus the squared fidelity between the output of the parametrized circuit on the input resource state with the desired state.
Since the momenta are kicked according to $\bm{\hat{\Pi}} \mapsto \bm{\hat{\Pi}} - \eta \partial \mathcal{L}(\bm{\hat{\Phi}}) / \partial \bm{\hat{\Phi}} + \mathcal{O}(\eta^2)$, we see that the use of QFB along with MoMGrad or QDD performs gradient ascent on the squared fidelity of the output of the parametrized circuit with the state we wish to learn.

\subsubsection{Quantum Mixed State Learning}\label{sec:qmsl}

The task of mixed state learning is similar to the case of pure states: given $n$ copies of an unknown state, $\hat{\rho} \in \mathcal{B}(\mathcal{H})$, one would like to learn a parametrized circuit which prepares a state close to $\hat{\rho}$.
The methods presented here will use the notion of proximity induced by the Hilbert-Schmidt inner product on $\mathcal{B}(\mathcal{H})$.

The parametrized circuit, $\hat{U}(\bm{\hat{\Phi}})$, will act on a pure initial resource state, $\ket{\psi_0}$, on a larger Hilbert space, $\tilde{\mathcal{H}}$, of sufficient size to be capable of containing the purification of the state to be learned.
We will then identify a subsystem of $\tilde{\mathcal{H}}$ as the Hilbert space $\mathcal{H}$, so that we can decompose $\tilde{\mathcal{H}} = \mathcal{H} \otimes \mathcal{H}^c$.
The goal is for the reduced state on $\mathcal{H}$, after applying the parametrized circuit to the input state, to approximate $\hat{\rho}$.
Let us denote this reduced state as
\begin{equation}
  \hat{\tilde{\rho}}(\bm{\hat{\Phi}}) := \tr_{\mathcal{H}^c} [ \hat{U}(\bm{\hat{\Phi}}) \ket{\psi_0}\!\bra{\psi_0} \hat{U}^\dagger (\bm{\hat{\Phi}}) ].
\end{equation}
For example, if the state $\hat{\rho}$ is a mixed state on $N$ qubits, then one can take the extended Hilbert space to be a space containing $2N$ qubits.
Then the goal is to create a mixed state on a subset of $N$ qubits which approximates $\hat{\rho}$.

\begin{figure}[h!]
\includegraphics[width=0.6\columnwidth]{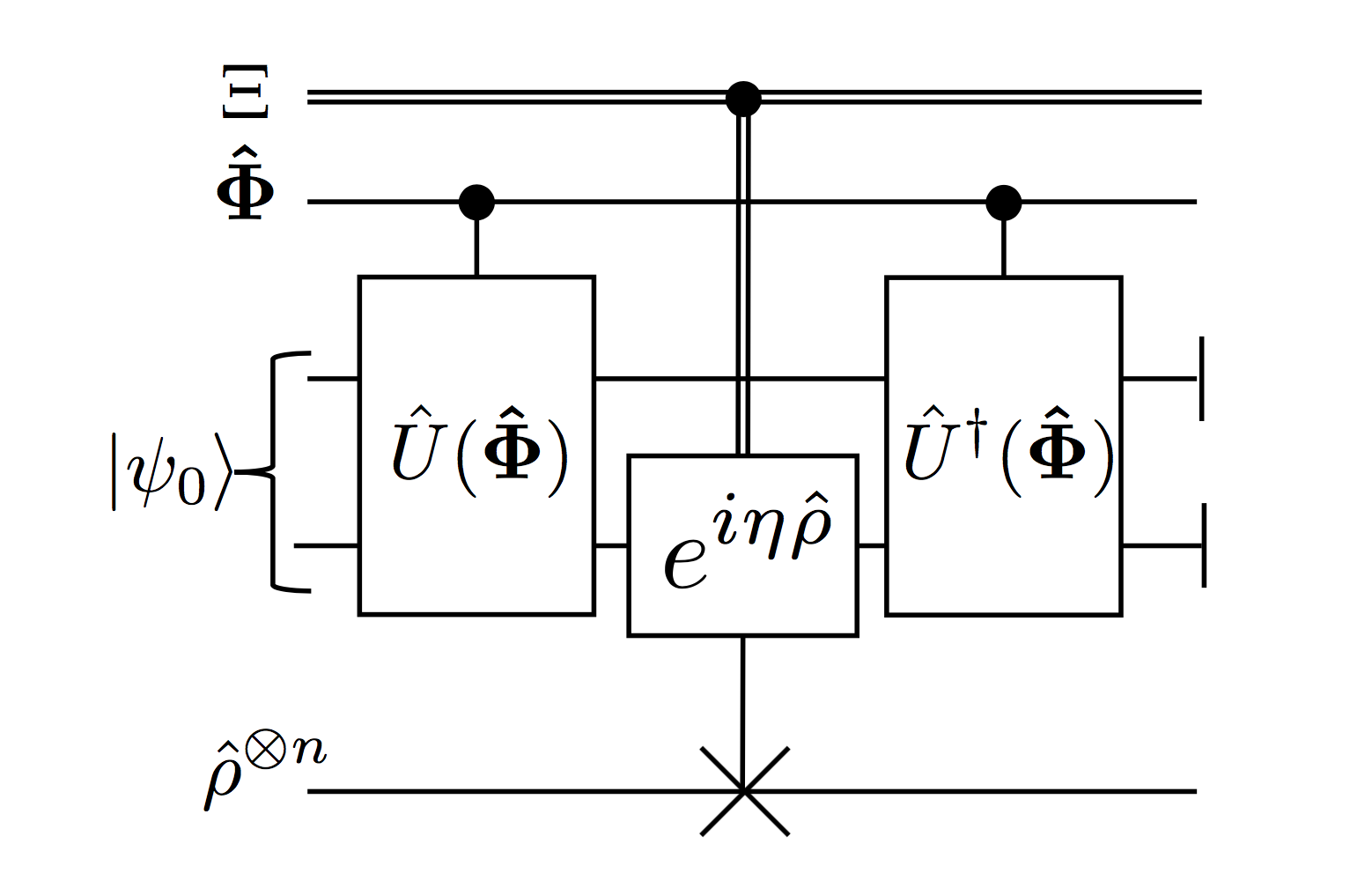}
\caption{The QFB circuit for quantum mixed state learning. Again, the phase kick is achieved through quantum state exponentiation using $n$ copies of the mixed state $\hat{\rho}$ (lower registers). The phase kick gate is also classically-controlled by the hyper-parameter $\eta$. The feedforward and uncomputation unitaries act on a dilated Hilbert space with initial pure resource state $\ket{\psi_0}$. The task of mixed state learning is for the feedforward unitary to prepare the desired mixed state on a subset of these registers (those upon which the phase kick acts).
} \label{fig:q_mixed_learn}
\end{figure}

The loss function will be $\hat{L} = - \hat{\rho} \otimes \hat{I}_{\mathcal{H}^c}$ acting on $\tilde{\mathcal{H}} = \mathcal{H} \otimes \mathcal{H}^c$.
As before, exponentiation of this loss function can be achieved using the methods of Section~\ref{sec:qse} given multiple copies of the state.
It is straightforward to show that the effective phase will be minus the Hilbert-Schmidt inner product between the desired state and the reduced state on $\mathcal{H}$ after applying the parametrized circuit on the input:
\begin{equation}
\begin{split}
  \mathcal{L}(\bm{\hat{\Phi}}) &= - \bra{\psi_0} \hat{U}^\dagger(\bm{\hat{\Phi}}) \hat{\rho} \otimes \hat{I}_{\mathcal{H}^c} \hat{U}(\bm{\hat{\Phi}}) \ket{\psi_0} \\
  &= - \tr_{\mathcal{H}} [ \hat{\rho} \tr_{\mathcal{H}^c} [ \hat{U}(\bm{\hat{\Phi}}) \ket{\psi_0}\!\bra{\psi_0} \hat{U}^\dagger(\bm{\hat{\Phi}}) ] ] \\
  &= - \tr_{\mathcal{H}} [ \hat{\rho} \hat{\tilde{\rho}}(\bm{\hat{\Phi}}) ].
\end{split}
\end{equation}
Therefore, the training algorithm will perform gradient ascent on this inner product.
The circuit for this procedure is illustrated in Figure~\ref{fig:q_mixed_learn}.

\subsection{Quantum Unitary \& Channel Learning}

\subsubsection{Supervised Unitary Learning}\label{sec:qunitary_sup_learn}

One means of learning a unitary operator, $\hat{V}$, is via samples of input/output pairs, $\{ ( \ket{\psi_j^\textsc{i}}, \ket{\psi_j^{\textsc{o}}} ) \}_j$.
Ideally, these pairs are such that $\ket{\psi_j^\textsc{o}} = \hat{V} \ket{\psi_j^\textsc{i}}$ for all $j$.
However, it is possible that the source of these samples is noisy, in which case one may need to assume some of the data states are not pure and hence be represented as mixed states related through a channel.
Such a situation will be subsumed by the following subsection where we describe the process for supervised channel learning.
In that context, one can use a unitary ansatz for the channel mapping between mixed states.
For this section, we will focus on the more particular case where the data states are pure, and we want to learn a unitary which approximates the ideal unitary, $\hat{V}$.

For each input/output data pair, indexed by $j$, the input to the parametrized algorithm, $\hat{U}(\bm{\hat{\Phi}})$, is $\ket{\psi_j^\textsc{i}}$.
The loss function will be $\hat{L}_j = - \ket{\psi_j^\textsc{o}}\!\bra{\psi_j^\textsc{o}}$, which, as opposed to state learning, will be different for every data pair $j$.
Again, this loss function can be implemented as a phase using state exponentiation, given multiple copies of the state.
Using these, the effective phase on the parameters for the data pair $j$ will be:
\begin{equation}
  \mathcal{L}_j(\bm{\hat{\Phi}}) = - | \bra{\psi_j^\textsc{o}} \hat{U}(\bm{\hat{\Phi}}) \ket{\psi_j^\textsc{i}} |^2,
\end{equation}
i.e., the negative squared fidelity between the output of the parametrized algorithm (upon input $\ket{\psi_j^\textsc{i}}$) and the desired output, $\ket{\psi_j^\textsc{o}}$.
This is quite similar to the phase obtained for pure state learning, but here the input and loss functions are different for every kick of the momenta.

\begin{figure}[h!]
\includegraphics[width=0.7\columnwidth]{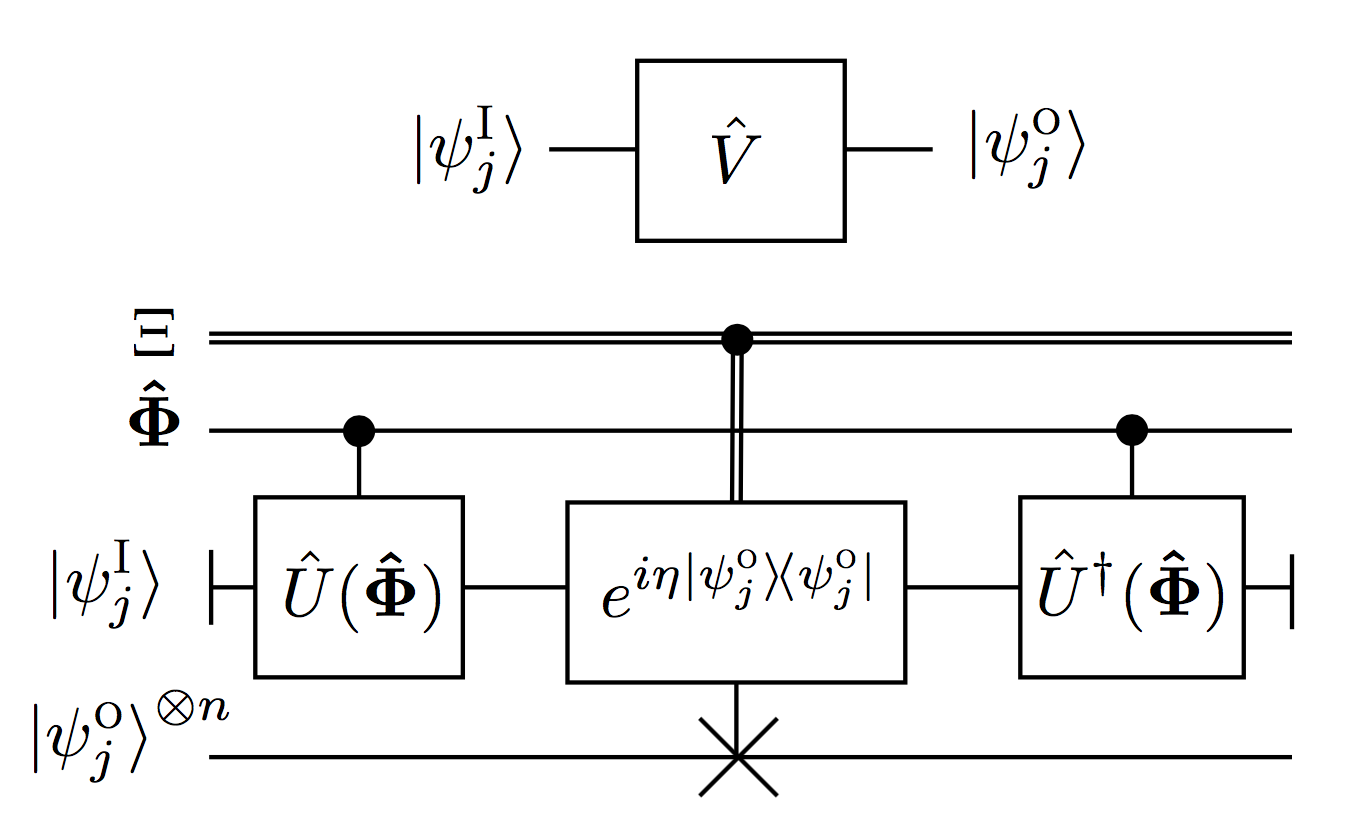}
\caption{QFB circuit for supervised unitary learning. The data points $(\ket{\psi_j^\textsc{i}},\ket{\psi_j^\textsc{o}})$ are ideally generated by some unitary operator $\hat{V}$, which we wish to approximate with $\hat{U}(\bm{\hat{\Phi}})$. This is achieved by using $\ket{\psi_j^\textsc{i}}$ as an input to the QFB circuit, and the corresponding projector onto $\ket{\psi_j^\textsc{o}}$ is used as a loss function via state exponentiation.
} \label{fig:q_sup_U_learn}
\end{figure}

This setup is illustrated in Figure~\ref{fig:q_sup_U_learn}.

\subsubsection{Supervised Channel Learning}\label{sec:qchannel_sup_learn}

Supervised learning of a quantum channel, $\Lambda$, requires input/output pairs, $\{ ( \hat{\rho}_j^\textsc{i}, \hat{\rho}_j^\textsc{o} ) \}_j$, which will generally be mixed states acting on a Hilbert space $\mathcal{H}$.
Ideally, these pairs satisfy $\hat{\rho}_j^\textsc{o} = \Lambda( \hat{\rho}_j^\textsc{i} )$, but of course there may be noise in the dataset.

In a similar fashion to mixed state learning, we will employ a parametrized unitary, $\hat{U}(\bm{\hat{\Phi}})$, acting on an extended Hilbert space $\tilde{\mathcal{H}} := \mathcal{H} \otimes \mathcal{H}^c$.
We will then train this algorithm so that, when restricted to $\mathcal{H}$, the algorithm approximates the channel $\Lambda$.
Explicitly, for each data pair $j$, we will have the parametrized unitary, $\hat{U}(\bm{\hat{\Phi}})$, act on $\hat{\rho}_j^\textsc{i}$ and an initial resource state $\ket{\psi_0} \in \mathcal{H}^c$.
Tracing out $\mathcal{H}^c$ after the unitary gives a quantum-parametrized channel:
\begin{equation}
\label{eq:param_channel}
  \Lambda(\bm{\hat{\Phi}}) : \hat{\rho}_j^\textsc{i} \mapsto \tr_{\mathcal{H}^c} [ \hat{U}(\bm{\hat{\Phi}}) \hat{\rho}_j^\textsc{i} \otimes \ket{\psi_0}\!\bra{\psi_0} \hat{U}^\dagger(\bm{\hat{\Phi}}) ].
\end{equation}
We will also denote the output of this channel, for input $\hat{\rho}$, as $\Lambda(\bm{\hat{\Phi}})[\hat{\rho}]$. 
The goal is to parametrize the channel so that, for each $j$, this output is close to $\hat{\rho}_j^\textsc{o}$.

To this end, we will take the loss operator to be $\hat{L}_j = - \hat{\rho}_j^\textsc{o} \otimes \hat{I}_{\mathcal{H}^c}$, similar to the case of mixed state learning.
The effective phase we obtain is:
\begin{equation}
\begin{split}
  \mathcal{L}_j(\bm{\hat{\Phi}}) &= \tr_{\tilde{\mathcal{H}}} [ \hat{L} \hat{U}(\bm{\hat{\Phi}}) \hat{\rho}_j^\textsc{i} \otimes \ket{\psi_0}\!\bra{\psi_0} \hat{U}^\dagger(\bm{\hat{\Phi}}) ] \\
  &= - \tr_{\mathcal{H}} [ \hat{\rho}_j^\textsc{o} \Lambda(\bm{\hat{\Phi}})[\hat{\rho}_j^\textsc{i}] ],
\end{split}
\end{equation}
which is negative the Hilbert-Schmidt inner product between the output of the parametrized channel (upon input $\hat{\rho}_j^\textsc{i}$) and the desired output state, $\hat{\rho}_j^\textsc{o}$.

\begin{figure}[h!]
\includegraphics[width=0.6\columnwidth]{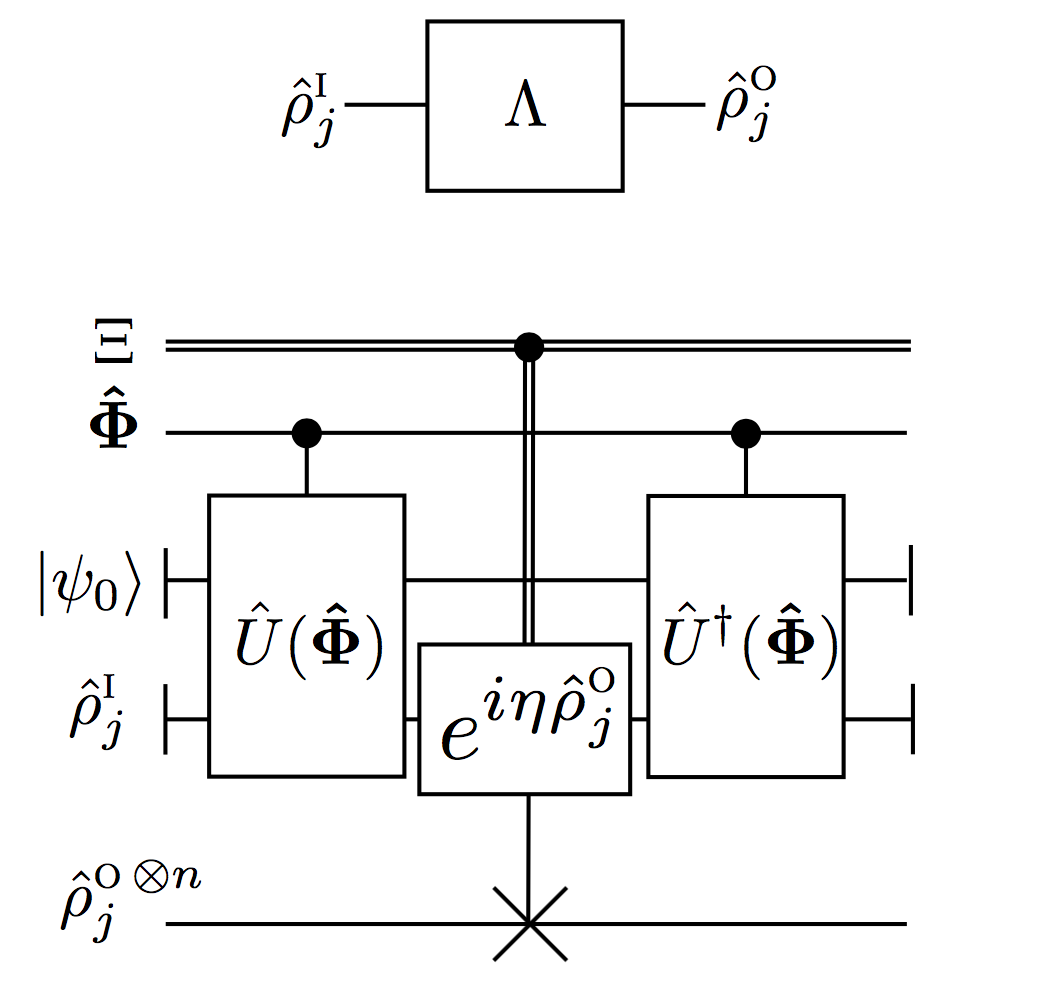}
\caption{QFB circuit for supervised channel learning. Similar to unitary learning, the data points $(\hat{\rho}_j^\textsc{i},\hat{\rho}_j^\textsc{o})$ ideally correspond to the input and output of a quantum channel $\Lambda$. The goal of supervised channel learning is to find a unitary on a dilated Hilbert space, such that the desired channel is approximated when this unitary is restricted to a subset of the input and output registers.
} \label{fig:q_sup_chan_learn}
\end{figure}

The QFB circuit for this task is illustrated in Figure~\ref{fig:q_sup_chan_learn}.

\subsubsection{Unsupervised Unitary Learning}\label{sec:qunitary_unsup_learn}

Another situation in which the methods presented herein can be used to learn a unitary, $\hat{V}$, is when one is given an oracle for $\hat{V}$ which can be queried, rather than a set of input/output data pairs.
The basic idea is to turn the problem into that of state learning on the Choi state of the unitary.
The oracle for $\hat{V}$ will be used to create the desired Choi state in order to use it as a loss function.
This technique can also be used to learn the Choi state of a channel (next section), but first we will describe the special case of learning a unitary.

To generate the appropriate loss function, we employ the unitary oracle mapping $\hat{V} : \mathcal{H} \to \mathcal{H}$.
First, let us denote $\ket{\phi^+} := \tfrac{1}{\sqrt{\dim \mathcal{H}}} \sum_j \ket{jj}$ as a maximally entangled state on $\mathcal{H}^{\otimes 2}$, equivalent to the identity map in the Choi-Jamiolkowski picture.
The loss function will be the Choi state, obtained by acting the oracle on one of the two subsystems of this maximally entangled state:
\begin{equation}
  \hat{L} = - (\hat{I}_\mathcal{H} \otimes \hat{V}) \ket{\phi^+}\!\bra{\phi^+} (\hat{I}_\mathcal{H} \otimes \hat{V}^\dagger) =: - \hat{\sigma}_{\hat{V}}.
\end{equation}
Exponentiation of this state to obtain a phase operator will require multiple queries to the oracle.

The parametrized algorithm, $\hat{U}(\bm{\hat{\Phi}})$, will similarly be applied to one of the two subsystems of $\ket{\phi^+} \in \mathcal{H}^{\otimes 2}$ as an input state, i.e., $(\hat{I}_\mathcal{H} \otimes \hat{U}(\bm{\hat{\Phi}})) \ket{\phi^+}$.
Then the above loss function will be applied as a phase, yielding an effective phase on the parameters:
\begin{equation}
\begin{split}
  \mathcal{L}(\bm{\hat{\Phi}}) &= - | \bra{\phi^+} (\hat{I}_\mathcal{H} \otimes \hat{V}^\dagger) (\hat{I}_\mathcal{H} \otimes \hat{U}(\bm{\hat{\Phi}})) \ket{\phi^+} |^2 \\
  &= - \tr_{\mathcal{H}^{\otimes 2}} [ \hat{\sigma}_{\hat{V}} \hat{\sigma}_{\hat{U}(\bm{\hat{\Phi}})} ] \\
  &= - | \tr_{\mathcal{H}} [ \hat{V}^\dagger \hat{U}(\bm{\hat{\Phi}}) ] |^2.
\end{split}
\end{equation}
Notice we have defined, analogous to $\hat{\sigma}_{\hat{V}}$,
\begin{equation}
  \hat{\sigma}_{\hat{U}(\bm{\hat{\Phi}})} := (\hat{I}_\mathcal{H} \otimes \hat{U}(\bm{\hat{\Phi}})) \ket{\phi^+}\!\bra{\phi^+} (\hat{I}_\mathcal{H} \otimes \hat{U}^\dagger(\bm{\hat{\Phi}})).
\end{equation}
This effective phase can be seen in terms of either the Hilbert-Schmidt inner product on the Choi states of the two unitaries or the square of the Hilbert-Schmidt inner product of the parametrized unitary and the desired unitary.

\begin{figure}[h!]
\includegraphics[width=0.7\columnwidth]{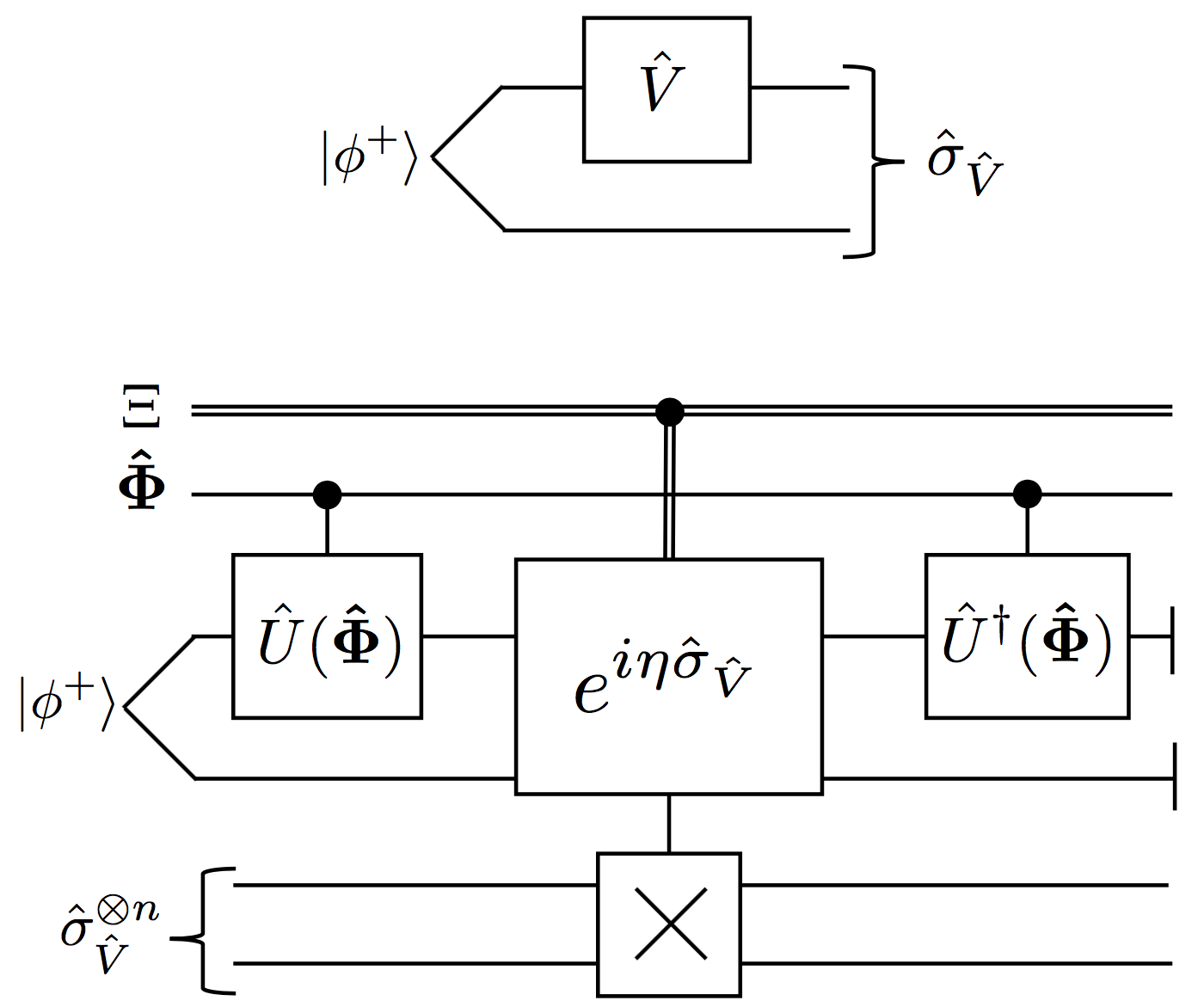}
\caption{QFB circuit for unsupervised unitary learning. Using an oracle for $\hat{V}$, one can prepare multiple copies of the Choi state of the unitary for the purposes of state exponentiation. The QFB circuit first involves creating the Choi state of the parametrized ansatz, $\hat{U}(\bm{\hat{\Phi}})$, as the feedforward, then using the desired Choi state as a loss function before the uncomputation.
} \label{fig:q_unsup_U_learn}
\end{figure}

The setup for this task is illustrated in Figure~\ref{fig:q_unsup_U_learn}.

\subsubsection{Unsupervised Channel Learning}\label{sec:qchannel_unsup_learn}

Unsupervised channel learning will be very similar to unsupervised unitary learning, with the addition of using a parametrized unitary, $\hat{U}(\bm{\hat{\Phi}})$, acting on an extended space $\tilde{\mathcal{H}} = \mathcal{H} \otimes \mathcal{H}^c$ as in mixed state learning and supervised channel learning.
Here, we assume access to an oracle for a quantum channel, $\Lambda$, and the task is to use this to learn a set of parameters for the unitary, $\hat{U}(\bm{\hat{\Phi}})$, acting on $\tilde{\mathcal{H}}$, so that the parametrized channel, $\Lambda(\bm{\hat{\Phi}}) : \mathcal{B}(\mathcal{H}) \to \mathcal{B}(\mathcal{H})$, given by
\begin{equation}
  \Lambda(\bm{\hat{\Phi}}) : \hat{\rho} \mapsto \tr_{\mathcal{H}^c} [ \hat{U}(\bm{\hat{\Phi}}) \hat{\rho} \otimes \ket{\psi_0}\!\bra{\psi_0}_{\mathcal{H}^c} \hat{U}^\dagger(\bm{\hat{\Phi}}) ],
\end{equation}
approximates $\Lambda$.
Note that $\ket{\psi_0} \in \mathcal{H}^c$ is some resource state, as described in the mixed state learning section.

As in the previous section, we will use the oracle $\Lambda$ and a maximally entangled state, $\ket{\phi^+}$, to generate a loss function which will be the Choi state of $\Lambda$:
\begin{equation}
  \hat{L} = - (\mathcal{I} \otimes \Lambda)( \ket{\phi^+}\!\bra{\phi^+} ) =: - \hat{\sigma}_\Lambda.
\end{equation}
The procedure, then, is to apply $(\hat{I}_\mathcal{H} \otimes \hat{U}(\bm{\hat{\Phi}}))$ to the input state $\ket{\phi^+}\ket{\psi_0} \in \mathcal{H}^{\otimes 2} \otimes \mathcal{H}^c$, apply $e^{-i \eta \hat{L}}$, followed by the uncompute.
After tracing over everything except the parameter registers, we obtain an effective phase:
\spliteq{
  \mathcal{L}(\bm{\hat{\Phi}}) &= \tr_{\mathcal{H}^{\otimes 2} \otimes \mathcal{H}^c} \big[ \hat{L}( \hat{I}_\mathcal{H} \otimes \hat{U}(\bm{\hat{\Phi}}) ) ( \ket{\phi^+}\!\bra{\phi^+}_{\mathcal{H} \otimes \mathcal{H}} \\&\qquad\qquad\qquad  \otimes \ket{\psi_0}\!\bra{\psi_0}_{\mathcal{H}^c} ) ( \hat{I}_\mathcal{H} \otimes \hat{U}^\dagger(\bm{\hat{\Phi}}) ) \big] \\
  &= - \tr_{\mathcal{H} \otimes \mathcal{H}} [ \hat{\sigma}_\Lambda \hat{\sigma}_{\Lambda(\bm{\hat{\Phi}})} ],
}
where the Choi state of the parametrized channel is:
\begin{equation}
  \hat{\sigma}_{\Lambda(\bm{\hat{\Phi}})} := (\mathcal{I} \otimes \Lambda(\bm{\hat{\Phi}}))( \ket{\phi^+}\! \bra{\phi^+} ).
\end{equation}
Hence, the effective phase is the Hilbert-Schmidt inner product between the Choi state of the parametrized channel with that of the desired channel.

\begin{figure}[h!]
\includegraphics[width=0.55\columnwidth]{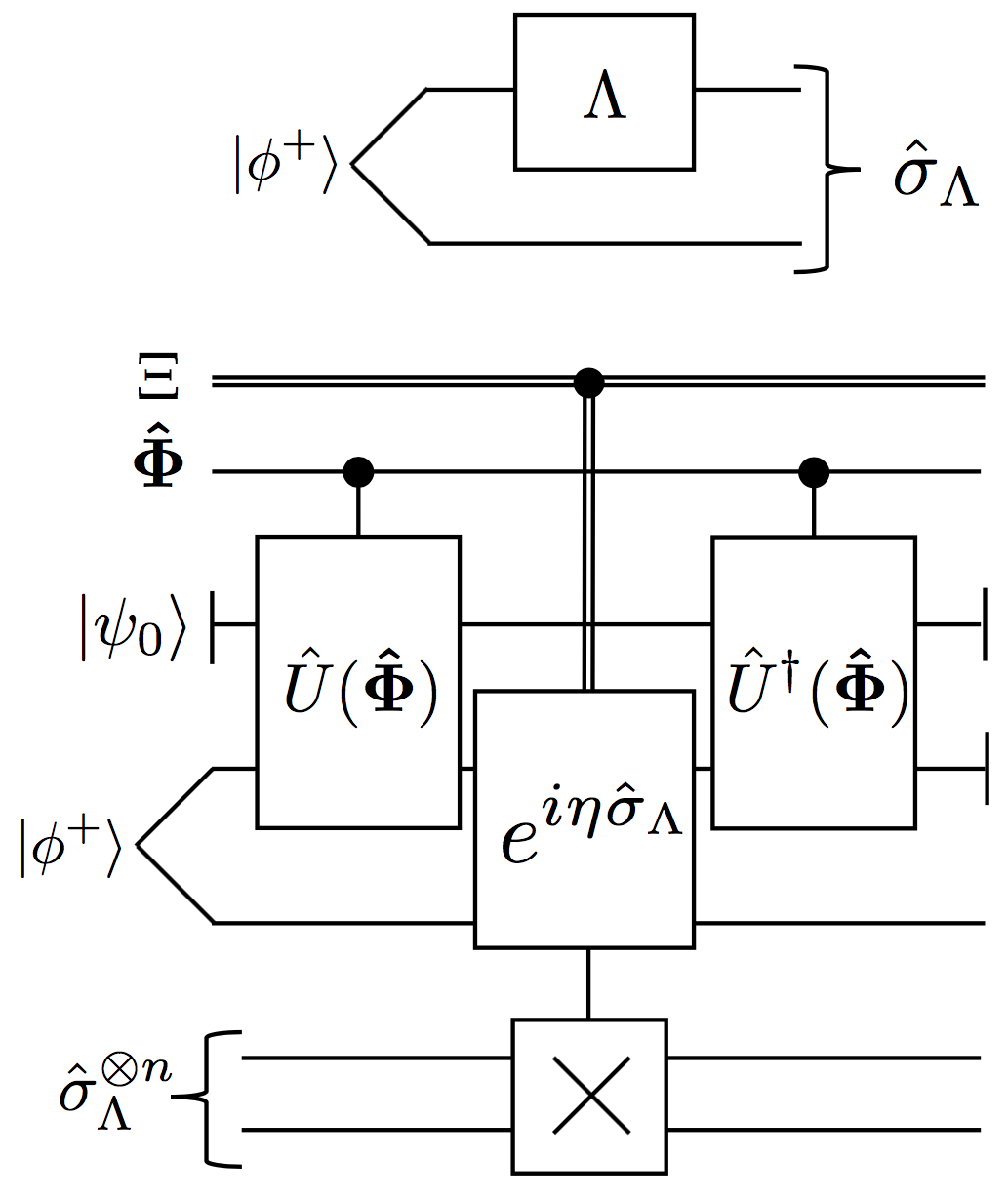}
\caption{QFB circuit for unsupervised channel learning. Given an oracle for the channel $\Lambda$, one can create multiple copies of the Choi state of the channel, $\hat{\sigma}_\Lambda$, for state exponentiation in the phase kick step of QFB. The ansatz for the channel consists of a parametrized unitary on a dilated Hilbert space. The goal is for this unitary to approximate the channel $\Lambda$ on a subset of the input and output registers. The feedforward step of QFB involves creating the Choi state of the parametrized channel on this subset of registers. The phase kick applies the Choi state of the desired channel as a loss function on the output of the feedforward. Of course this is followed by uncomputation. The input to the parametrized unitary on the dilation of the input space of the channel is some initial resource state $\ket{\psi_0}$.
} \label{fig:q_unsup_chan_learn}
\end{figure}

The setup for unsupervised channel learning is illustrated in Figure~\ref{fig:q_unsup_chan_learn}.

\subsection{Quantum Classification/Regression/Measurement Learning}\label{sec:meas_learn}

\subsubsection{Overview}
Classification is the task of associating collections of objects with some set of discrete labels.
Regression is essentially a similar task, but where the labels are continuous.
The present discussion will apply to both cases of discrete and continuous labels, hence we will not restrict the discussion to either case and simply denote the labels by a parameter $\alpha$.

Here, we will describe how one can train a quantum algorithm to assign labels to quantum states, using a set of training examples.
Let us denote the set of labelled example quantum states by $\{ \hat{\rho}_j^{\alpha_j} \}_j \subset \mathcal{B}(\mathcal{H})$, with $\alpha_j$ denoting the label for example of index $j$.
The set of labels will be denoted $\mathcal{A}$.
The goal of the classification/regression task is to build a measurement scheme so that, upon input of a state, the measurement outcome corresponds to the appropriate label.
Therefore, this task could also be called measurement learning.

Ideally, the labels for the example states are exactly characterized by a POVM with effects $\{ \hat{E}_\alpha \}_\alpha$, so that $\tr( \hat{E}_\alpha \hat{\rho}_j^{\alpha_j} ) = \delta_\alpha^{\alpha_j}$.
Thus, we wish to design a set of quantum-parametrized effects $\{ \hat{E}_\alpha(\bm{\hat{\Phi}}) \}_\alpha$ to approximate this assignment of labels.

Note that if the example states are joint eigenstates of some collection of observables, then the problem is essentially classical since we would simply be assigning labels to elements of the configuration space.
Also, for the cases where all of the example states are pure, one may imagine attempting to build a measurement scheme by using a unitary to map to a fixed basis of \textit{label states}, $\ket{\alpha}$, which, upon measurement in this basis, would provide a label $\alpha$.
This problem would correspond to learning a PVM.
However, it is clear that this task is simply providing exact labels to some basis of the Hilbert space, which again is essentially a classical labelling task.
Of course, the classical task of learning a PVM will be included here as a special case, but here we will focus on the more general case of learning a POVM.

Naimark's dilation theorem reduces the problem of learning a POVM to learning a unitary and a projective measurement on an extended space.
The projective measurements can be the projectors onto the label states, $\ket{\alpha}$, and the unitary will be a parametrized algorithm, $\hat{U}(\bm{\hat{\Phi}})$, acting on an extended Hilbert space $\tilde{\mathcal{H}} = \mathcal{H} \otimes \mathcal{H}^c$, with an initial resource state $\ket{\psi_0}$ in $\mathcal{H}^c$.

For the input $\hat{\rho}_j^{\alpha_j}$ on $\mathcal{H}$ to the parametrized algorithm, a possible choice of loss operator is $\hat{L}_j = - \hat{I}_\mathcal{H} \otimes \ket{\alpha_j}\!\bra{\alpha_j}_{\mathcal{H}^c}$. In the next subsection we discuss various loss function options which share the same optimum.
The corresponding effective phase on the parameters is:
\begin{equation}
  \mathcal{L}_j(\bm{\hat{\Phi}}) = - \tr_\mathcal{H} ( \hat{E}_{\alpha_j}(\bm{\hat{\Phi}}) \hat{\rho}_j^{\alpha_j} ),
\end{equation}
where the parametrized effects, $\hat{E}_\alpha(\bm{\hat{\Phi}}) : \mathcal{H} \to \mathcal{H}$, are
\begin{equation}
  \hat{E}_\alpha(\bm{\hat{\Phi}}) := \bra{\psi_0} \hat{U}^\dagger(\bm{\hat{\Phi}}) \ket{\alpha}\! \bra{\alpha} \hat{U}(\bm{\hat{\Phi}}) \ket{\psi_0}.
\end{equation}
Note that this is similar to the result obtained for supervised channel learning.
The difference is that, here, the loss function penalizes the incorrect label states on $\mathcal{H}^c$, rather than the incorrect output states on $\mathcal{H}$.

\begin{figure}[h!]
\includegraphics[width=0.55\columnwidth]{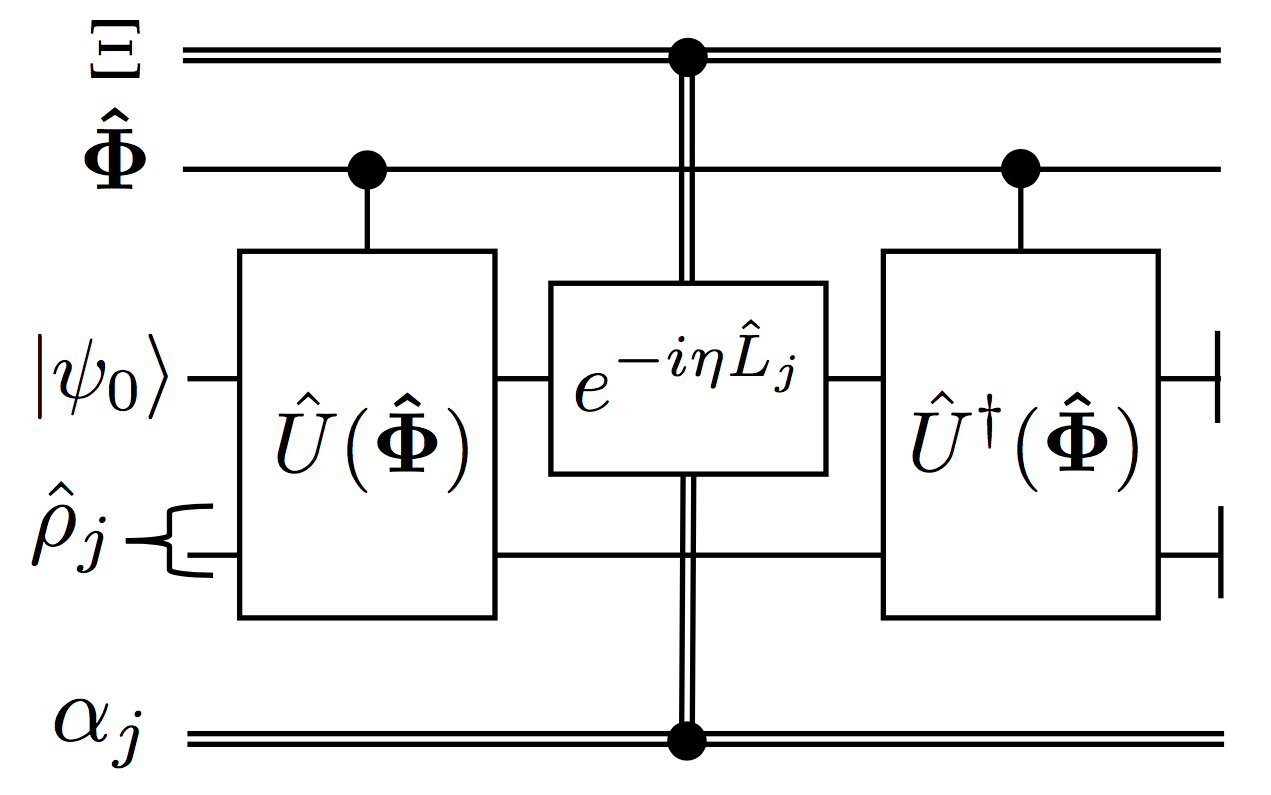}
\caption{QFB circuit for quantum measurement learning. The parametrized ansatz for a POVM consists of a parametrized unitary acting on a dilated Hilbert space followed by a standard basis measurement. The loss function is a projector onto the corresponding basis state assigned to a particular label. Naimark's theorem ensures that this measurement can be performed solely on the registers extending the original Hilbert space of the states to be classified.
} \label{fig:Quantum_classifier}
\end{figure}

The setup for this task is illustrated in Figure~\ref{fig:Quantum_classifier}.

\subsubsection{Output Encodings \& Implementation Options}

There exists multiple ways of encoding the output, especially when considering both continuous-variable and discrete labels. As such, there exists multiple options for our choice of loss function, which all reward having the correct label but may penalize incorrect labels differently. It is worth discussing these options as some have varying implementation and compilation overheads.

In all cases, if we denote the output label registers as $A$, the negative projector onto the correct label state $\hat{L}_j = - \ket{\alpha_j}\!\bra{\alpha_j}_{A}$ gives us a valid loss function which clearly has minimal value when the correct label is assigned in each case. Let us denote the label state $n
$-qubit projector
\begin{equation}
    \ket{\alpha_j}_A=\bigotimes_{k=1}^n\ket{\alpha^{(k)}_{j}}_{A_k}
\end{equation}
in which $\alpha_j$ is an $n$-bit string of bit values $\alpha^{(k)}_{j}$ for the $k^\text{th}$ bit of the $j^\text{th}$ label, and in which we denote the quantum registers of each qubit in this label as $A_k$. To implement an exponential of this loss function, since the label states $\ket{\alpha_j}_A$ are computational basis states, for an $n
$-qubit projector we can implement the exponential of this projector by applying $\hat{U}_{\bm{\alpha_j}}e^{i\eta \ket{\bm{1}}\!\bra{\bm{1}}}\hat{U}^\dagger_{\bm{\alpha_j}}$ where $\hat{U}_{\bm{\alpha_j}} \equiv \bigotimes_k \hat{X}_k^{\neg \alpha^{(k)}_j}$ is the product of bit flips corresponding to the bitwise-negated label bit string. To implement the exponential of the multi-$\ket{1}$ state, we can use an additional ancilla \textit{work} register $W$, onto which we apply a C$^{n}$-NOT (i.e., and n-qubit control generalized Toffoli gate, which itself can be broken down into a linear number of Toffolis \cite{nielsen2002quantum}), then apply an exponential of $\hat{Z}$ on the work register, and undo the multi-control-Toffoli, that is  $\text{C}^{n}\text{-NOT}_{\textsc{aw}}e^{i\eta \hat{Z}_\textsc{w}}\text{C}^{n}\text{-NOT}_{\textsc{aw}}\ket{0}_{\textsc{w}} = e^{i\eta \ket{\bm{1}}\!\bra{\bm{1}}}\,\!\!_{\textsc{a}}\otimes\ket{0}_{\textsc{w}} $. 

Hence, we have described how to enact the exponential of any given bit-string represented label. Now, the cost function is a quite sparse in the Hilbert space of the possible bit string states for the label. It might then be advantageous in some cases to have a cost function whose representation in the computational basis has a greater support (larger rank) than a single-state projector, such as to nudge the optimization of parameters even when the output of the network for a given set of parameters is wrong with high probability. To do so, we can construct a Hamiltonian whose ground state coincides with that of the correct label. For example, for the one-hot encoded label $\ket{\alpha_j}= \bigotimes_{k\in\mathcal{A}}\ket{\delta_{jk}}_{k}$, we can use the following Hamiltonian for which it is the ground state:
\begin{equation}
    \hat{L}_j \equiv -\sum_{k\in\mathcal{A}} (-1)^{\delta_{jk}}\hat{Z}_k.
\end{equation}
One of the draws of this approach is that this loss function exponential is easy to synthesize using a product of individual qubit exponentials;
\begin{equation}
    e^{-i\eta    \hat{L}_j } = \bigotimes_{k\in\mathcal{A}} e^{i\eta (-1)^{\delta_{jk}}\hat{Z}_k}
\end{equation}
which is much easier than synthesizing a single-state projector. Additionally, the rank of this Hamiltonian is the same as that of dimension of the label space, i.e., of $|\mathcal{A}|$, this can, in turn, provide a better kickback on the parameters being optimized, especially for the sectors of the wavefunction which have minimal overlap with the correct label.

In the case of continuous-label classification (i.e., regression), we can imagine having each class label be a tensor product of multiple qudit-computational basis states. That is, $
    \ket{\alpha_j} = \bigotimes_{k\in \mathcal{A}} \ket{\alpha_j^{(k)}}_{A_k}
$
where each component $\alpha_j^{(k)}$ is a $d$-ary number, and the states of each label subregister are qudit states; $\ket{\alpha_j^{(k)}}_{\textsc{a}_k} =\bigotimes_{j\in\mathcal{A}} \hat{X}^{\alpha_j^{(k)}}\ket{0}_{\textsc{a}_k}$ where $\hat{X}^{\alpha_j^{(k)}}$ are qudit shifts. 

In terms of cost function, one option is to use once again the negative projector on the joint label eigenstate, i.e., $\hat{L}_j = - \ket{\alpha_j}\!\bra{\alpha_j}_{A}$. To apply an exponential of this projector, we can apply $\hat{U}_{\bm{\alpha_j}}e^{i\eta \ket{\bm{0}}\!\bra{\bm{0}}}\hat{U}^\dagger_{\bm{\alpha_j}}$ where $\hat{U}_{\bm{\alpha_j}} \equiv \bigotimes_k \hat{X}^{ \alpha^{(k)}_j}$. To apply an exponential of the joint null state, one could consider using the same trick as outlined above, using multi-controlled Toffolis, but now multi-qudit-controlled generalize Toffolis. If each qudit is made of qubits, one can shift the state of each qudit from $\ket{0}$ to whichever state has all qubits be in their $\ket{1}$ state, then use a C$^{N}$-NOT gate with an ancilla work qubit as before, where $N = n\lceil\log_2 d\rceil$ is the total number of qubits. This can be achieved in $\mathcal{O}(N)$ gates.

Another possible choice of loss function is the mean-squared loss, where we consider the loss as 
\begin{equation}
    \hat{L}_j = -\sum_{k\in \mathcal{A}} (\hat{\phi}_{A_k}- \alpha_k)^2
\end{equation}
where $\hat{\phi}_k$ is the simulated position operator of the qudit register, similar to the $\hat{\Phi}_j$ operators of the parameters. Note that the state $\ket{\alpha_j}_A$ is the ground state of this loss Hamiltonian, hence optimizing the above will also result in the correct label being output, and there is a less sparse error signal since the rank of this loss function spans the whole space of possible labels, rather than being rank 1 in the case of the projector.

In terms of implementation of the exponential loss,  $(\hat{\phi}_{A_k}- \alpha_k)^2 = \hat{X}^{ \alpha^{(k)}_j} \hat{\phi}_{A_k}^2\hat{X}^{ \alpha^{(k)\dagger }_j}$, hence a simple way to enact the exponential loss is by applying
\begin{equation}
     e^{-i\eta \hat{L}_j} = \hat{U}_{\bm{\alpha_j}} e^{i\eta\hat{\phi}_{A_k}^2}\hat{U}^\dagger_{\bm{\alpha_j}},
\end{equation}
which is similar to the weight decay exponentials described in section \ref{sec:opt}, and can be synthesized into a circuit of depth $\mathcal{O}(\lceil\log_2 d\rceil^2)$. 

\subsection{Quantum Code Learning}

In this section we consider how to automate the learning of quantum codes for compression and error correction. In both cases, there exists a skew subspace $\mathcal{H}_G$ of the input Hilbert space $\mathcal{H}$ which we would like to isolate into a subset of registers. This nonlocal subspace could be the subspace where most of the input space has its support, or the logical subspace of a quantum error correcting code. Finding the \textit{code} (transformation) which concentrates this non-locally encoded subspace onto a subset of registers will be the task we will automate with quantum learning. To find a good transformation of the input space, we can optimize over a family of parametric quantum circuits, this can be achieved by imposing cost functions which either maximize the fidelity of reconstruction (after encoding and decoding), or minimize the information leakage to the environment, or maximize the fidelity of the state in the logical subspace. We will briefly introduce the information theoretic task in each case, and outline how to evaluate the cost function and execute the Quantum Feedforward and Phase Backpropagation (QFB) procedure in each case. This leaves all options discussed in section \ref{sec:opt} for optimization over the space of parameters open.

\subsubsection{Quantum Autoencoders: Compression Code Learning}\label{sec:q_auto_learn}

We first consider regular quantum autoencoders and later consider the more specialized case of denoising quantum autoencoders.

The information theoretic task automated by autoencoders is that of compression of a source's signal, also known as the quantum source coding task \cite{wilde2013quantum} in quantum Shannon theory. Consider a quantum source to be akin to a sender of a quantum messages, where the sender picks from a set of possible quantum states to send through, and the variable representing the decision to send a specific state is modelled by a classical random variable.

More specifically, we can consider having a classical random variable $X$ with a probability distribution $ p(X=x)\equiv p_x$. This classical random variable is an index for a certain \textit{alphabet} of states, which we can consider to be either a set of pure states $\{\ket{\psi_j}\}_{j\in X}$ or mixed states $\{\hat{\rho}_j\}_{j\in X}$. Each incoming message can be represented as a classical mixture of states in the alphabet, with the classical probability distribution being that of the alphabet index, $\hat{\rho} = \sum_{j\in X}p_j\ket{\psi_j}\!\bra{\psi_j}$ or $\hat{\rho} = \sum_{j\in X}p_j\,\hat{\rho}_j$.

In general, this message will be send using an alphabet made of states of multiple registers, whether these be qubits, qudits, or qumodes.
The goal of compression is to map these states to a space of fewer qubits/qudits, while retaining sufficient information so that they can be recovered with high probability.

The theoretical optimum \textit{rate} (i.e., number of qubits per message) at which we can encode our messages without loss (considering the asymptotic limit of sending many messages), is given by the Von Neumann entropy of the mixed state
\begin{equation}
    S(\hat{\rho}) = -\text{tr}\left[\hat{\rho}\log(\hat{\rho})\right] = -\!\!\!\!\!\sum_{\lambda\in\text{spec}(\hat{\rho})}\!\!\!\!\! \lambda \log(\lambda),
\end{equation}
and the scheme which achieves this optimal rate is called Schumacher's quantum data compression protocol.

In the following, we will outline how one can train a parametrized unitary as an encoder to perform this compression task. Note that for the case of the regular autoencoder, we consider the source to be \textit{noiseless}, i.e., the messages are brought to the network as is. When we consider the denoising autoencoder later in this section we will consider adding noise to the input. Thus for the regular autoencoder the information theoretic task is akin to \textit{noiseless} Shannon compression, except that in general there is no guarantee to reach the theoretical entropy limit, nor to be completely lossless.

The inputs to the quantum autoencoder will run through the collection of states in the alphabet, on the Hilbert space $\mathcal{H}$.
For simplicity, we will denote a general input state to the autoencoder as $\hat{\rho}$.
The autoencoder will consist of a parametrized unitary, $\hat{U}(\bm{\hat{\Phi}})$, acting on $\mathcal{H}$.
We will factorize the Hilbert space at the output of the unitary into $\mathcal{H} = \mathcal{H}_{\textsc{g}} \otimes \mathcal{H}_{\textsc{a}}$, where $\mathcal{H}_{\textsc{g}}$ is the sector containing the compressed representation of the input state and $\mathcal{H}_{\textsc{a}}$ corresponds to \emph{trash registers}.

Before we can discuss appropriate loss functions, first we must determine a means of characterizing the success of an encoder.
One means of characterizing the success of the encoder is by measuring the fidelity between the state at the input of the encoder with that of a decoding of the compressed state.
A decoding scheme would be to input the compressed state along with a reference state into $\hat{U}^\dagger(\bm{\hat{\Phi}})$.
Explicitly, let us introduce a new register, $\mathcal{H}_{\textsc{a}'}$, to denote the source of the reference state used during the decoding.
Then, an encoding followed by a decoding involves applying $\hat{U}(\bm{\hat{\Phi}})$ to $\hat{\rho}_{\textsc{g} \textsc{a}}$ (where the subscripts $\textsc{g}$, $\textsc{a}$ have been introduced to specify the appropriate subsystems), then applying a swap, $S_{\textsc{a} \textsc{a}'}$, between $\mathcal{H}_{\textsc{a}}$ and a reference state $\ket{\psi_0} \in \mathcal{H}_{\textsc{a}'}$, followed by $\hat{U}^\dagger(\bm{\hat{\Phi}})$ acting on $\mathcal{H}_{\textsc{g} \textsc{a}}$.
Then we can write the decompressed state as:
\begin{equation}
\begin{split}
  &\hat{\tilde{\rho}}_{\textsc{g} \textsc{a}} = \\
  &\quad\!\! \tr_{\textsc{a}'} [ \hat{U}^\dagger(\bm{\hat{\Phi}}) S_{\textsc{a} \textsc{a}'} \hat{U}(\bm{\hat{\Phi}}) \hat{\rho}_{\textsc{g} \textsc{a}} \otimes \ket{\psi_0}\!\bra{\psi_0}_{\textsc{a}'} \hat{U}^\dagger(\bm{\hat{\Phi}}) S_{\textsc{a} \textsc{a}'} \hat{U}(\bm{\hat{\Phi}}) ].
\end{split}
\end{equation}
Then the success of the compression is quantified by the fidelity between $\hat{\rho}_{\textsc{g} \textsc{a}}$ and $\hat{\tilde{\rho}}_{\textsc{g} \textsc{a}}$: $F(\hat{\rho}_{\textsc{g} \textsc{a}},\hat{\tilde{\rho}}_{\textsc{g} \textsc{a}})$.

An alternative, and for our purposes more convenient, means to quantify the quality of the encoding begins with the observation that any information lost during the compression will manifest itself as entropy in the trash register after the encoding.
Therefore, we can train the algorithm to minimize the entropy in the trash register.

\begin{figure}[h!]
 \begin{center}
\includegraphics[width=0.8\columnwidth]{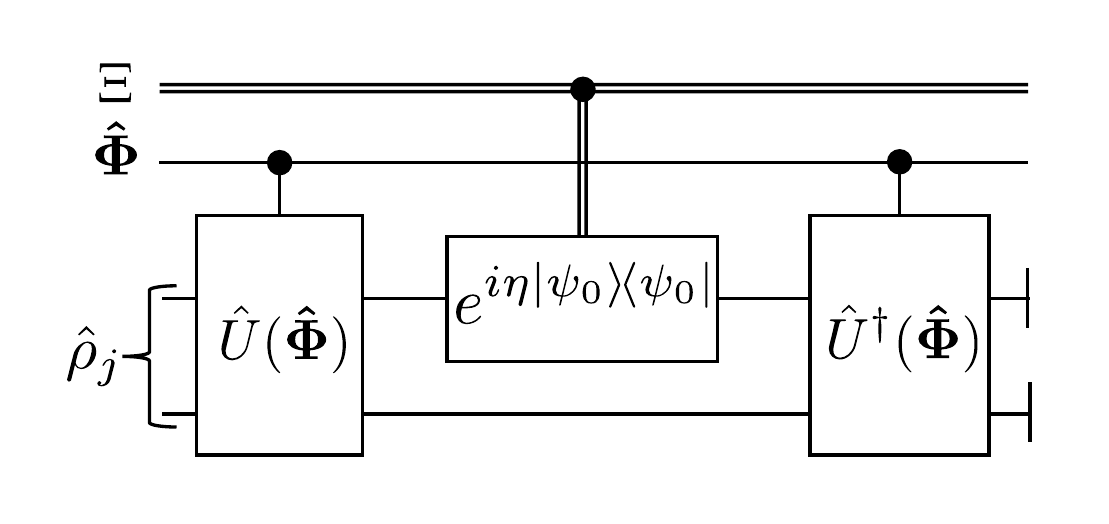}
\caption{QFB circuit for the regular autoencoder. In the feedforward step, the parametrized unitary acts on the input state to be compressed. The loss function is a projector onto a pure resource state on the trash registers at the output of the encoding.
} \label{fig:reg_autoenc}
\end{center}
\end{figure}

\begin{figure}[h!]
 \begin{center}
\includegraphics[width=0.8\columnwidth]{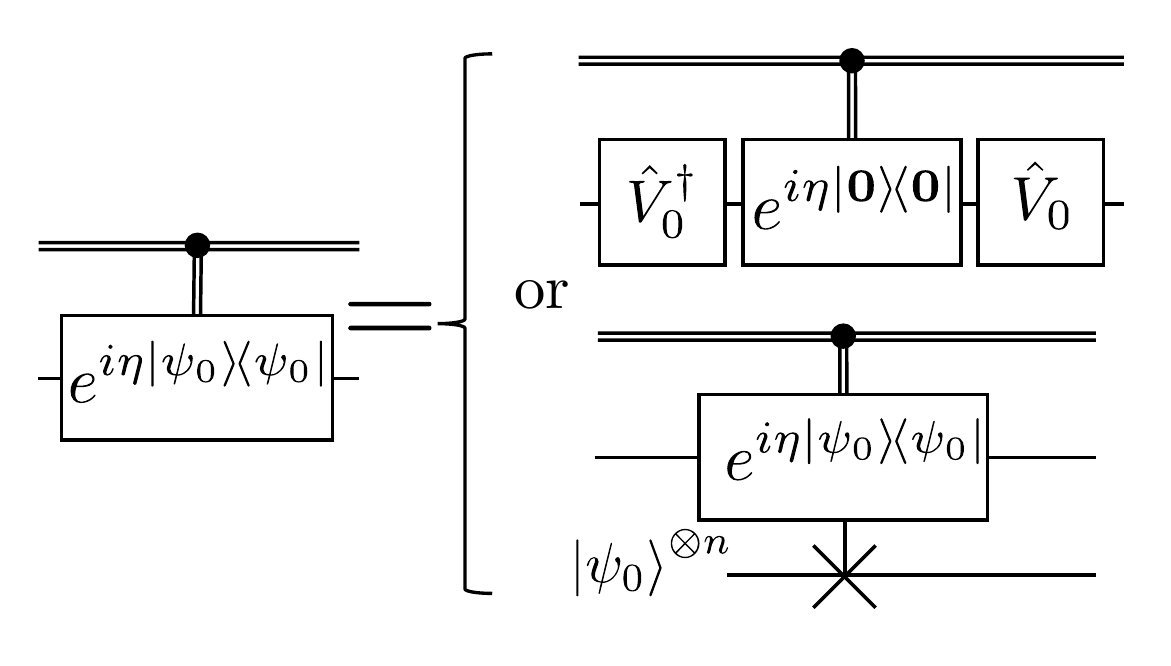}
\caption{Reference state exponentiation options. The loss function in the QFB circuit is a projector onto a pure resource state $\ket{\psi_0}$. The exponentiated projector can be constructed either through a unitary transformation of the exponentiated null projector (top), or through state exponentiation using multiple copies of the resource state (bottom). 
} \label{fig:ref_state_exp}
\end{center}
\end{figure}

Here, we will describe two different loss functions that may be used for training.
The first is based on maximizing the fidelity of the trash register with the reference state $\ket{\psi_0} \in \mathcal{H}_{\textsc{a}}$.
Therefore, in order to enact this, we will use the projector onto the pure resource state, $\ket{\psi_0} \in \mathcal{H}_{\textsc{a}}$, as the loss function: $\hat{L} = - \hat{I}_{\textsc{g}} \otimes \ket{\psi_0}\!\bra{\psi_0}_{\textsc{a}}$.
Of course, any other pure state would suffice, but such a state can be related to $\ket{\psi_0}$ via a unitary operator which can be absorbed into $\hat{U}(\bm{\hat{\Phi}})$.
Illustrations of this setup are provided in Figures~\ref{fig:reg_autoenc} and \ref{fig:ref_state_exp}.

The effective phase we obtain for this loss function is:
\begin{equation}
\begin{split}
  \mathcal{L}(\bm{\hat{\Phi}}) &= - \tr_\mathcal{H}[ \ket{\psi_0}\!\bra{\psi_0}_{\textsc{a}} \hat{U}(\bm{\hat{\Phi}}) \hat{\rho}_{\textsc{g} \textsc{a}} \hat{U}^\dagger(\bm{\hat{\Phi}}) ] \\
  &= - F \left( \tr_{\textsc{g}} [ \hat{U}(\bm{\hat{\Phi}}) \hat{\rho}_{\textsc{g} \textsc{a}} \hat{U}^\dagger(\bm{\hat{\Phi}}) ], \ket{\psi_0}\!\bra{\psi_0}_{\textsc{a}} \right).
\end{split}
\end{equation}
(Note that throughout this section, for convenience of notation, it should be understood that these fidelity functions remain operator-valued since we have not traced over the Hilbert space of the parameters.)
Thus, as desired, the effective phase is the negative fidelity between the state on the trash registers at the output of the algorithm and the pure reference state.

One can relate this fidelity to the fidelity of reconstruction in the following manner:
\begin{widetext}

\begin{equation}
\begin{split}
  F( \hat{\rho}_{\textsc{g} \textsc{a}}, \hat{\tilde{\rho}}_{\textsc{g} \textsc{a}} ) &= F \left( \hat{\rho}_{\textsc{g} \textsc{a}} , \tr_{\textsc{a}'} [ \hat{U}^\dagger(\bm{\hat{\Phi}}) S_{\textsc{a} \textsc{a}'} \hat{U}(\bm{\hat{\Phi}}) \hat{\rho}_{\textsc{g} \textsc{a}} \otimes \ket{\psi_0}\!\bra{\psi_0}_{\textsc{a}'} \hat{U}^\dagger(\bm{\hat{\Phi}}) S_{\textsc{a} \textsc{a}'} \hat{U}(\bm{\hat{\Phi}}) ] \right) \\
  &= F \left( \hat{U}(\bm{\hat{\Phi}}) \hat{\rho}_{\textsc{g} \textsc{a}} \hat{U}^\dagger(\bm{\hat{\Phi}}) , \tr_{\textsc{a}'} [  S_{\textsc{a} \textsc{a}'} \hat{U}(\bm{\hat{\Phi}}) \hat{\rho}_{\textsc{g} \textsc{a}} \otimes \ket{\psi_0}\!\bra{\psi_0}_{\textsc{a}'} \hat{U}^\dagger(\bm{\hat{\Phi}}) S_{\textsc{a} \textsc{a}'} ] \right)\\
  &\leq F \left( \tr_\textsc{g} [ \hat{U}(\bm{\hat{\Phi}}) \hat{\rho}_{\textsc{g} \textsc{a}} \hat{U}^\dagger(\bm{\hat{\Phi}}) ] , \tr_{\textsc{g} \textsc{a}'} [  S_{\textsc{a} \textsc{a}'} \hat{U}(\bm{\hat{\Phi}}) \hat{\rho}_{\textsc{g} \textsc{a}} \otimes \ket{\psi_0}\!\bra{\psi_0}_{\textsc{a}'} \hat{U}^\dagger(\bm{\hat{\Phi}}) S_{\textsc{a} \textsc{a}'} ] \right)\\
  &= F \left( \tr_\textsc{g} [ \hat{U}(\bm{\hat{\Phi}}) \hat{\rho}_{\textsc{g} \textsc{a}} \hat{U}^\dagger(\bm{\hat{\Phi}}) ] , \ket{\psi_0}\!\bra{\psi_0}_{\textsc{a}} \right).
\end{split}
\end{equation}

\end{widetext}
Note that in the first step of the above calculation, we can pull the unitary out of the partial trace since $\hat{U}(\bm{\hat{\Phi}})$ does not act on $\mathcal{H}_{\textsc{a}'}$.
Then we use the unitary invariance property of the fidelity, i.e., $F(\hat{\rho},\hat{U}^\dagger\hat{\sigma}\hat{U}) = F(\hat{U}\hat{\rho}\hat{U}^\dagger,\hat{\sigma})$.
In the second step we use the monotonicity of the fidelity under trace preserving operations (e.g., partial trace): $F(\hat{\rho},\hat{\sigma}) \leq F(\Lambda(\hat{\rho}),\Lambda(\hat{\sigma}))$.
In the last step we used the simple fact that: $\tr_{\textsc{g} \textsc{a}'} [ S_{\textsc{a} \textsc{a}'} \hat{\rho}_{\textsc{g} \textsc{a}} \otimes \ket{\psi_0}\!\bra{\psi_0}_{\textsc{a}'} S_{\textsc{a} \textsc{a}'} ] = \ket{\psi_0}\!\bra{\psi_0}_\textsc{a}$.

Hence, we see that the algorithm will train to maximize an upper bound to the reconstruction fidelity.
Although maximizing an upper bound does not guarantee that we are maximizing the reconstruction fidelity, maximizing the fidelity of the trash state relative to a pure reference state will indirectly enforce a maximization of purity of the trash state.
If we consider the entire compression procedure as a channel, i.e., the composition of enacting the unitary, swapping out the trash state for a fresh copy, and acting the reverse of the encoder, then enforcing the purity of the trash state will enforce a null entropy leakage to the environment. The coherent information of this channel \cite{wilde2013quantum} 
will be the maximum over isometric extensions of the input state of the difference between the entropy of the output of the channel minus the entropy of the environment:
\begin{equation}
    I_c = \max_{\psi}[S(B)-S(E)],
\end{equation}
where the maximization over $\psi$ denotes maximization over isometric extensions of the input.
Thus minimizing the entropy leakage to the environment will necessarily increase our coherent mutual information of our channel. 

As a proxy for this entropy, we can use the purity as an alternative loss function for training the quantum autoencoder, as it is operationally easier to implement as a cost function.
However, note that one must use the state of the trash register at the output of the encoding along with the compressed state in order to later decompress the state.
By simply maximizing the purity of the trash registers, and not training the register to map to a particular state (as before), we will not be able to decompress unless we also perform state learning on this trash state.
Thus we see this means of performing the compression task involves splitting the problem into two tasks: encoding and state learning.
In some cases, this may prove to be advantageous instead of enforcing a particular ancilla state on $\mathcal{H}_\textsc{a}$ and training a possibly more complicated encoder, $\hat{U}(\bm{\hat{\Phi}})$.
Here, we will proceed to describe the training of the encoder, and one can use the methods of Section~\ref{sec:qmsl} to learn the state of the ancilla.

In order to accomplish the training using the purity as a loss function, one must run two copies of the algorithm in parallel, but which can be trained simultaneously by "tenting" the weights.
We will show that one can obtain the purity of the trash state as an effective phase by using a swap gate, $\hat{L} = - S_{\textsc{a} \textsc{a}'}$, as a loss function.
A means of exponentiating this loss function was described in Section~\ref{sec:qse}.

\begin{figure}[h!]
 \begin{center}
\includegraphics[width=0.65\columnwidth]{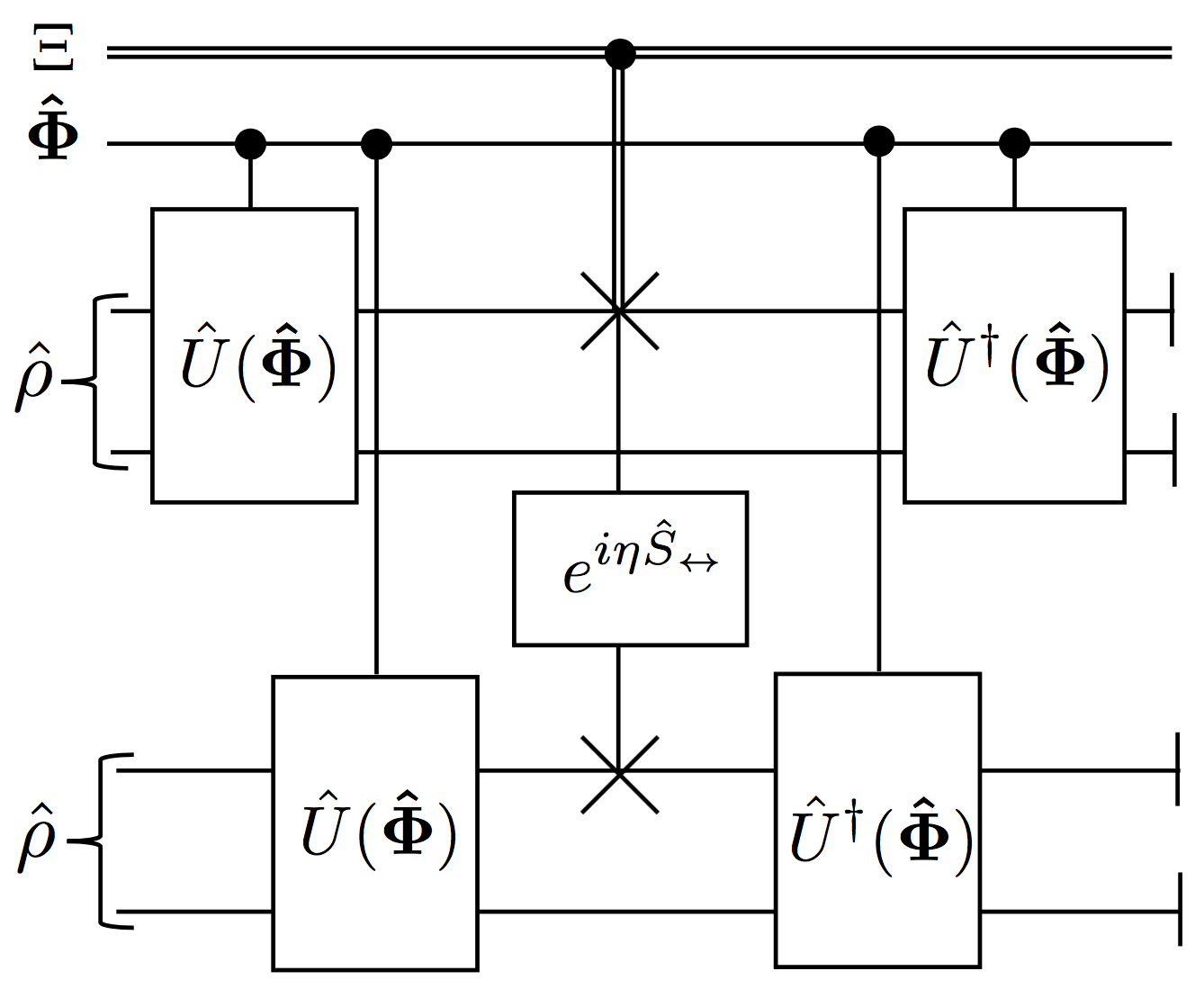}
\caption{QFB circuit for the purity-based autoencoder training. We run two copies of the feedforward parametrized unitary, $\hat{U}(\bm{\hat{\Phi}})$, in parallel acting on two copies of the input state. Both parallel unitaries are controlled by the same parameter registers. The phase kick is an exponentiated swap operator applied to the trash registers of the two instances. After uncomputation, the effective phase on the parameters is the (negative) purity of the trash registers.
}
 \label{fig:switcharoo_autoenc}
\end{center}
\end{figure}

Let us denote the state after the parametrized unitary by $\hat{\rho}_{\textsc{g} \textsc{a}}(\bm{\hat{\Phi}}) := \hat{U}(\bm{\hat{\Phi}}) \hat{\rho}_{\textsc{g} \textsc{a}} \hat{U}^\dagger(\bm{\hat{\Phi}})$ and the trash state after the compression as $\hat{\rho}_{\textsc{a}}(\bm{\hat{\Phi}}) := \tr_{\textsc{g}} \hat{\rho}_{\textsc{g} \textsc{a}}(\bm{\hat{\Phi}})$ (and similar for $\textsc{g}'$ and $\textsc{a}'$).
The effective phase we obtain for the parameters is:
\begin{equation}
\begin{split}
  \mathcal{L}(\bm{\hat{\Phi}}) &= - \tr_{\textsc{g} \textsc{a} \textsc{g}' \textsc{a}'} [ S_{\textsc{a} \textsc{a}'} \hat{\rho}_{\textsc{g} \textsc{a}}(\bm{\hat{\Phi}}) \otimes \hat{\rho}_{\textsc{g}' \textsc{a}'}(\bm{\hat{\Phi}}) ] \\
  &= - \tr_{\textsc{a} \textsc{a}'} [ S_{\textsc{a} \textsc{a}'} \hat{\rho}_{\textsc{a}}(\bm{\hat{\Phi}}) \otimes \hat{\rho}_{\textsc{a}'}(\bm{\hat{\Phi}}) ] \\
  &= - \tr_{\textsc{a}} [ \hat{\rho}_{\textsc{a}}(\bm{\hat{\Phi}})^2 ],
\end{split}
\end{equation}
i.e., the purity of the trash state at the output of the autoencoder.
The QFB circuit for this version of the autoencoder is shown in Figure~\ref{fig:switcharoo_autoenc}

\subsubsection{Denoising Quantum Autoencoder}\label{sec:q_noiseauto_learn}

In this section, we will examine a task similar to the previous.
The difference is that we will assume that the state we wish to compress has first gone through a some noise channel.
The goal here will be to train an autoencoder to not only compress, but also denoise the state after having passed through the noise channel.
Ideally, the network will learn how to filter noise into the trash registers, creating a more robust autoencoder.

The algorithm will be trained as follows.
The feedforward will consist of applying both the encoding and decoding maps.
Ideally, we would like this encoding/decoding process to recover the initial state $\ket{\psi_\textsc{i}}$, i.e., after compression and denoising it should recover the state at the input of the noise channel.
Therefore, we will apply the projector onto this state as a loss function at the output of the \emph{decoder} (as opposed to the previous case), in order to penalize the algorithm if it does not output the correct state.
As before, to employ QFB, the uncompute will consist of the inverse of the feedforward.
In this case, this inverse is comprised of the encoding, swap with the ancilla register, and decoding.

More concretely, let us assume that we have multiple copies of the input states, $\ket{\psi_\textsc{i}}$, available for the training, as well as an oracle for the noise channel $\mathcal{N}$.
The first step for the training is to send a copy of $\ket{\psi_\textsc{i}}$ through the noise channel to obtain $\hat{\rho} := \mathcal{N}( \ket{\psi_\textsc{i}}\!\bra{\psi_\textsc{i}} )$.
We then proceed analogously to the previous autoencoder.
We apply a parametrized ansatz for the encoder, $\hat{U}(\bm{\hat{\Phi}})$, acting on $\mathcal{H} = \mathcal{H}_\textsc{g} \otimes \mathcal{H}_\textsc{a}$ (where $\textsc{g}$ indexes the Hilbert space containing the compressed state and $\textsc{a}$ the trash register).
This is followed by performing a swap operation, $S_{\textsc{a} \textsc{a}'}$, between the trash register and a pure resource state $\ket{\psi_0} \in \mathcal{H}_{\textsc{a}'}$.
Now, as opposed to the previous case, we first apply the decoding map, $\hat{U}(\bm{\hat{\Phi}})$, before applying the loss function $\hat{L} = - \ket{\psi_\textsc{i}}\!\bra{\psi_\textsc{i}}_{\textsc{g} \textsc{a}}$ on the output.
The effective phase we obtain on the parameters is:
\begin{widetext}
\begin{equation}
\begin{split}
  \mathcal{L}(\bm{\hat{\Phi}}) &= - \tr_{\textsc{g} \textsc{a} \textsc{a}'} [ \ket{\psi_\textsc{i}}\!\bra{\psi_\textsc{i}}_{\textsc{g} \textsc{a}} \hat{U}^\dagger(\bm{\hat{\Phi}}) S_{\textsc{a} \textsc{a}'} \hat{U}(\bm{\hat{\Phi}}) \mathcal{N}( \ket{\psi_\textsc{i}}\!\bra{\psi_\textsc{i}}_{\textsc{g} \textsc{a}} ) \otimes \ket{\psi_0}\!\bra{\psi_0}_{\textsc{a}'} \hat{U}^\dagger(\bm{\hat{\Phi}}) S_{\textsc{a} \textsc{a}'} \hat{U}(\bm{\hat{\Phi}}) ] \\
  &= - \tr_{\textsc{g} \textsc{a}} [ \hat{\rho}_{\textsc{g} \textsc{a}}(\bm{\hat{\Phi}}) \hat{\tilde{\rho}}_\textsc{g}(\bm{\hat{\Phi}}) \otimes \ket{\psi_0}\!\bra{\psi_0}_\textsc{a} ] \\
  &= - \tr_\textsc{g} [ \bra{\psi_0} \hat{\rho}_{\textsc{g} \textsc{a}}(\bm{\hat{\Phi}}) \ket{\psi_0}_\textsc{a} \hat{\tilde{\rho}}_\textsc{g}(\bm{\hat{\Phi}}) ],
\end{split}
\end{equation}
\end{widetext}
where we have denoted the noisy input state after the encoding as $\hat{\tilde{\rho}}_{\textsc{g} \textsc{a}}(\bm{\hat{\Phi}}) := \hat{U}(\bm{\hat{\Phi}}) \mathcal{N}( \ket{\psi_\textsc{i}}\!\bra{\psi_\textsc{i}}_{\textsc{g} \textsc{a}} ) \hat{U}^\dagger(\bm{\hat{\Phi}})$, its partial trace on $\textsc{a}$ as $\hat{\tilde{\rho}}_\textsc{g}(\bm{\hat{\Phi}}) := \tr_\textsc{a} [ \hat{\tilde{\rho}}_{\textsc{g} \textsc{a}}(\bm{\hat{\Phi}}) ]$, and the noiseless pure state, $\ket{\psi_\textsc{i}}$, after the encoding as $\hat{\rho}_{\textsc{g} \textsc{a}}(\bm{\hat{\Phi}}) := \hat{U}(\bm{\hat{\Phi}}) \ket{\psi_\textsc{i}}\!\bra{\psi_\textsc{i}}_{\textsc{g} \textsc{a}} \hat{U}^\dagger(\bm{\hat{\Phi}})$.
Then we see that the effective phase is the Hilbert-Schmidt inner product between the encoded noiseless input state with the compressed state reduced to the $\textsc{g}$ register along with the pure resource state on the trash register.

\begin{figure}[h!]
 \begin{center}
\includegraphics[width=0.8\columnwidth]{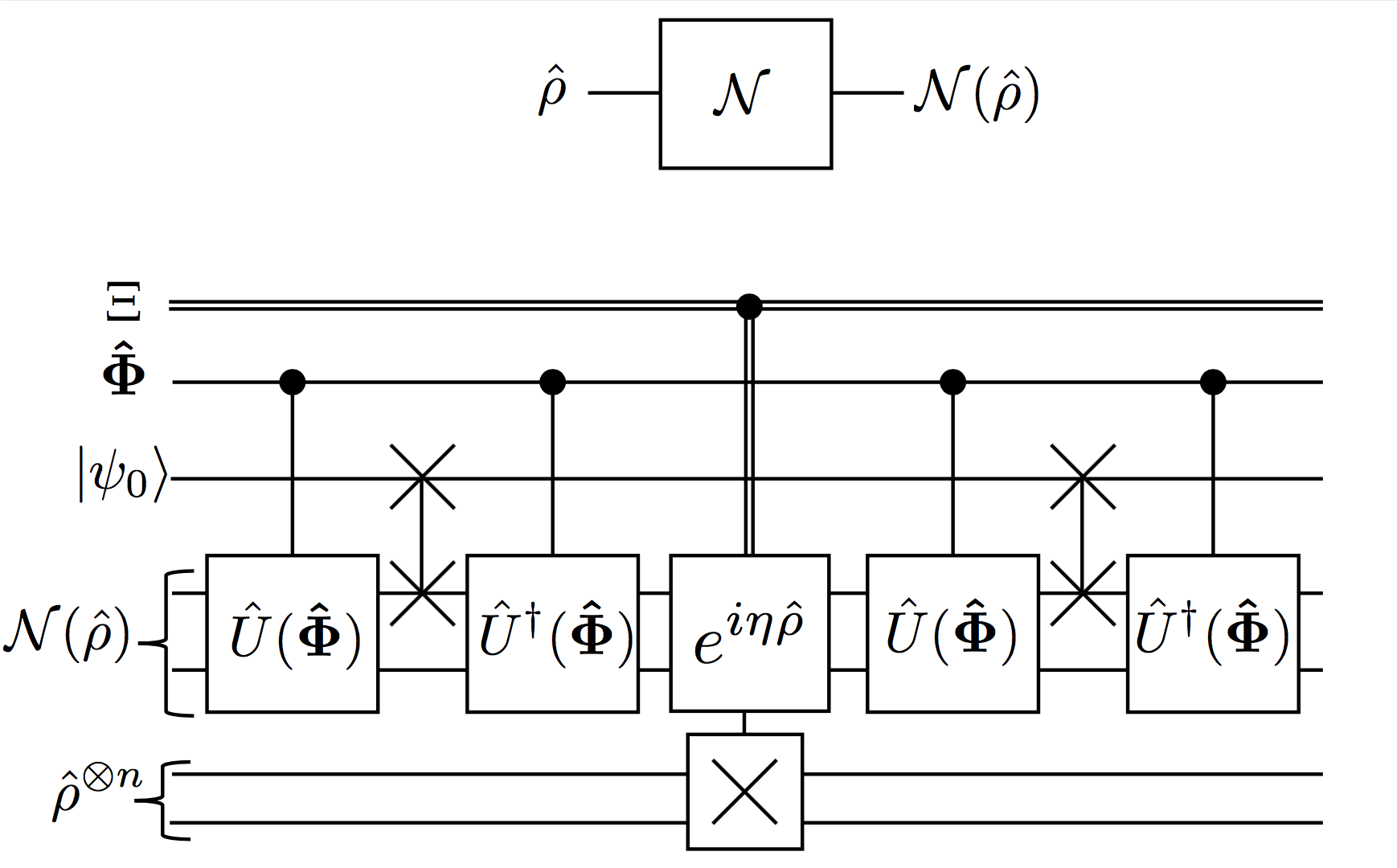}
\caption{Denoising autoencoder. The input states to the denoising autoencoder are created from feeding the desired state through an oracle to the noise channel. The feedforward of the QFB circuit consists of the entire encoding, swapping out of the trash registers, and decoding circuit. The loss function is the input state to the noise channel, whose exponentiation requires multiple copies.
}
\label{fig:denoise_auto}
\end{center}
\end{figure}

An illustration of this procedure is provided in Figure~\ref{fig:denoise_auto}.

\subsubsection{Quantum Error Correcting Code Learning}\label{sec:qecc_learn}

The final parametric coding task we will examine is that of quantum channel coding.
The scheme considered here will first involve a parametrized encoding unitary, $\hat{U}(\bm{\hat{\Phi}})$, which acts on the logical sector of the Hilbert space, $\mathcal{H}_\textsc{l}$, as well as the syndrome Hilbert space, $\mathcal{H}_\textsc{s}$.
The output of this encoding then passes through a noise channel, $\mathcal{N}$.
We then apply a parametrized recovery map, $\hat{W}(\bm{\hat{\Phi}})$, to the output of the channel along with a set of refresh qubits $\ket{\bm{0}} \in \mathcal{H}_\textsc{r}$.
This is followed by a decoding map consisting of the inverse of the encoding unitary, $\hat{U}^\dagger(\bm{\hat{\Phi}})$.
The goal is to simultaneously train the parametrized encoding and recovery maps to counteract the noise channel.
Hence, we want to train these unitaries so that this full channel is the identity channel on the logical sector.

Let us denote the full channel by:
\begin{equation}
  \Lambda_{\textsc{l} \textsc{s} \textsc{r}}(\bm{\hat{\Phi}}) := \tr_{\textsc{s} \textsc{r}} \circ \Ad[\hat{U}^\dagger_{\textsc{l} \textsc{s}}(\bm{\hat{\Phi}}) \hat{W}_{\textsc{l} \textsc{s} \textsc{r}}(\bm{\hat{\Phi}})] \circ \mathcal{N}_{\textsc{l} \textsc{s}} \circ \Ad[\hat{U}_{\textsc{l} \textsc{s}}(\bm{\hat{\Phi}})],
\end{equation}
where, for convenience, we have included subscripts to denote the Hilbert space factors that each operator acts on.
Note that we will use the symbol $\bm{\hat{\Phi}}$ to encompass the parameters of both the encoder and the recovery map, although generically these maps would not share any parameters.
Thus, we could decompose $\bm{\hat{\Phi}} = \bm{\hat{\Phi}}_U \oplus \bm{\hat{\Phi}}_W$, where $\bm{\hat{\Phi}}_U$ are the parameters for $\hat{U}$ and $\bm{\hat{\Phi}}_W$ are the parameters for $\hat{W}$.

Note that this task is essentially the same as the channel learning task we have described before.
As before, we will describe two means of training the channel.
The first is analogous to supervised channel learning, where the channel is trained on a set of logical input states.
The second is similar to unsupervised channel learning.

In order to perform training, we must have access to an oracle or an implementation of the noise channel, $\mathcal{N}$ (e.g., this channel could be learned using the channel learning techniques described above).
Furthermore, to apply the uncomputation of the channel, we must dilate the noise map with an auxiliary Hilbert space $\mathcal{H}_\textsc{p}$ to a unitary operator.
In the following we will not need to refer to this unitary operator explicitly, but we will assume access to the dilation for QFB.

\textbf{Supervised QEC Learning.}\cite{johnson2017qvector}
For supervised learning, one can simply input different logical states of $\mathcal{H}_\textsc{l}$ into the channel $\Lambda(\bm{\hat{\Phi}})$, and train the parameters to learn the identity map.
In order to simplify the implementation of the loss function (as we shall see below), it will be more convenient to describe the generation of these logical states as acting unitaries, $\hat{V}_\textsc{l}$, on some logical reference state $\ket{\bm{0}} \in \mathcal{H}_\textsc{l}$.
For example, if $\mathcal{H}_\textsc{l} = (\mathds{C}^2)^{\otimes k}$, then we could choose $\ket{\bm{0}} = \ket{0}^{\otimes k}$.
These unitaries can be chosen from a set which forms a unitary 2-design, in order to provide a uniform set of sample data for the input in the logical space.

Without loss of generality, let us also denote the input states on $\mathcal{H}_\textsc{s}$ and $\mathcal{H}_\textsc{r}$ each as $\ket{\bm{0}}$.

The feedforward for the training algorithm involves acting $\hat{V}_\textsc{l}$ on the logical input reference state, applying the channel $\Lambda_{\textsc{l} \textsc{s} \textsc{r}}$, and then uncomputing the logical operator $\hat{V}_\textsc{l}$.
Since we want to train the channel to learn the identity on the logical sector, one should choose the loss function to be (negative) projector onto the logical input reference state: $\hat{L} = - \ket{\bm{0}}\!\bra{\bm{0}}_\textsc{l}$.
With this, one can see that the corresponding effective phase is the fidelity between the logical state after the channel and the input logical state:
\begin{equation}
\begin{split}
  \mathcal{L}_V(\bm{\hat{\Phi}}) &= - \tr_\textsc{l} [ \ket{\bm{0}}\!\bra{\bm{0}}_{\textsc{l}} \hat{V}^\dagger_\textsc{l} \Lambda_{\textsc{l} \textsc{s} \textsc{r}}(\bm{\hat{\Phi}})[ \hat{V}_\textsc{l} \ket{\bm{0}}\!\bra{\bm{0}}_{\textsc{l} \textsc{s} \textsc{r}} \hat{V}^\dagger_\textsc{l} ] \hat{V}_\textsc{l} ] \\
  &= - F(\Lambda_{\textsc{l} \textsc{s} \textsc{r}}(\bm{\hat{\Phi}})[ \hat{V}_\textsc{l} \ket{\bm{0}}\!\bra{\bm{0}}_{\textsc{l} \textsc{s} \textsc{r}} \hat{V}^\dagger_\textsc{l} ], \hat{V}_\textsc{l} \ket{\bm{0}}\!\bra{\bm{0}}_{\textsc{l}} \hat{V}^\dagger_\textsc{l} )
\end{split}
\end{equation}
where $\ket{\bm{0}}_{\textsc{l} \textsc{s} \textsc{r}} \equiv \ket{\bm{0}}_\textsc{l} \otimes \ket{\bm{0}}_\textsc{s} \otimes \ket{\bm{0}}_\textsc{r}$. 
As discussed in \ref{sec:meas_learn}, acting an exponential of a projector can be slightly costly to synthesize into gates. A good alternating to the loss function $\hat{L} = - \ket{\bm{0}}\!\bra{\bm{0}}_\textsc{l}$ is to use 
$\hat{L} \equiv -\sum_{k\in\textsc{l}} \hat{Z}_k$. This loss function shares the same ground state as the projector but is an operator of higher rank hence provides a richer error signal for the quantum parameters when the output is far from correct. Synthesizing exponentials of such a loss function is more straightforward;
\begin{equation}
    e^{-i\eta    \hat{L}_j } = \bigotimes_{k\in\textsc{l}} e^{-i\eta\hat{Z}_k}
\end{equation}
which is much easier than synthesizing a single-state projector.

Now, since we would like to learn the identity channel for all $\hat{V}_\textsc{l}$, we could draw these from a unitary 2-design $\{ \hat{V}_j \}_{j=1}^N$.
If this were done in parallel (or in a minibatch) for each element in the 2-design, then if the effective phases were combined, we would obtain an effective phase,
\begin{equation}
  \bar{\mathcal{L}}(\bm{\hat{\Phi}}) := \frac{1}{N} \sum_{j=1}^N \mathcal{L}_{V_j}(\bm{\hat{\Phi}})
\end{equation}
which would correspond to the (negative) average code fidelity.

\textbf{Unsupervised QEC Learning.}
The unsupervised version of learning channel codes is more straightforward.
Instead of generating various logical states at the input using unitaries $\hat{V}_\textsc{l}$, here we act the channel $\Lambda(\bm{\hat{\Phi}})$ on one of the subsystems of a maximally entangled state $\ket{\phi^+} \in \mathcal{H}_{\textsc{l}'} \otimes \mathcal{H}_\textsc{l}$, where $\mathcal{H}_{\textsc{l}'} \cong \mathcal{H}_\textsc{l}$ is an auxiliary copy of the logical space.

Since we want to train the algorithm to learn the identity map on the logical sector, the loss function should be $\hat{L} = - \ket{\phi^+}\!\bra{\phi^+}_{\textsc{l} \textsc{l}'}$.
One can use state exponentiation or some other means to prepare the exponential of this state.
The effective phase we obtain for this process is:
\begin{equation}
\begin{split}
  &\mathcal{L}(\bm{\hat{\Phi}}) \\
  &\quad\!\!= - \tr_{\textsc{l}' \textsc{l}} [ \ket{\phi^+}\!\bra{\phi^+}_{\textsc{l}' \textsc{l}} \Lambda_{\textsc{l} \textsc{s} \textsc{r}}(\bm{\hat{\Phi}})[ \ket{\phi^+}\!\bra{\phi^+}_{\textsc{l}' \textsc{l}} \otimes \ket{\bm{0}}\!\bra{\bm{0}}_{\textsc{s} \textsc{r}} ] ] \\
  &\quad\!\!= - \tr_{\textsc{l} \textsc{l}'} [ \hat{\sigma}_\mathcal{I} \hat{\sigma}_{\Lambda(\bm{\hat{\Phi}})} ],
\end{split}
\end{equation}
where we have denoted $\hat{\sigma}_{\Lambda(\bm{\hat{\Phi}})} := \Lambda_{\textsc{l} \textsc{s} \textsc{r}}(\bm{\hat{\Phi}})[ \ket{\phi^+}\!\bra{\phi^+}_{\textsc{l}' \textsc{l}} \otimes \ket{\bm{0}}\!\bra{\bm{0}}_{\textsc{s} \textsc{r}} ]$ as the Choi state of the channel from $\textsc{l}$ to $\textsc{l}$, and $\hat{\sigma}_\mathcal{I} = \ket{\phi^+}\! \bra{\phi^+}$ as the Choi state of the identity channel.
If we write the effective phase in this manner, we see that, similar to unsupervised channel learning, it is the Hilbert-Schmidt inner product between these two Choi states.

\subsection{Generative Adversarial Quantum Circuits}\label{sec:gaq}

\subsubsection{Classical Generative Adversarial Networks Review}

Generative adversarial networks \cite{2014arXiv1406.2661G} are a class of networks used in classical machine learning which allows for the generation of new samples from a given dataset, hence the name \textit{generative}, by pitting two sub-networks against each other in an adversarial game.
One of these sub-networks is dubbed the \textit{generator}, while the other is dubbed the \textit{discriminator}.
The goal of the generator is to mimic samples from the given dataset, while the discriminator attempts to discern which datapoints came from the generator and which came from the dataset.
By progressively training both the discriminator and the generator, the networks can converge to a Nash equilibrium, where the generator is particularly good at generating convincing samples mimicking the data, and the discriminator particularly good at filtering out unconvincing samples. 

Let us briefly review how to train these classical networks.
The classical approach is to first sample a set of random noise variables independently from some simple probability distribution.
Typical choices for this probability distribution would be a Bernoulli distribution for discrete random variables, or a Gaussian distribution for continuous random variables.
These samples, $\{\bm{r}_j\}_j$, are used as random seeds for the generator.
That is, they are the input to the generator network, $G$, which outputs a candidate datapoint, $G(\bm{r}_j)$, which is supposed to mimic a datapoint from the actual dataset, $\{ \bm{x}_j \}_j$.
More precisely, the generator is trained so that the distribution of outputs, $\{ G(\bm{r}_j) \}_j$, matches the distribution of the dataset $\{ \bm{x}_j \}_j$.
Before considering how the generator should be trained, let us discuss how to train the discriminator.

The discriminator network, in the most simple case, is taken to be a binary classifier (a network with a single bit as output).
To train the discriminator network, $D$, first we sample a single Bernoulli random variable (coin flip), denoted $l_j \in \{0,1\}$.
Based on the value of $l_j$ we feed in to the discriminator either a point from the real dataset, $\bm{x}_j$, or a fake datapoint from the generator, $G(\bm{r}_j)$.
This datapoint (real or fake) is fed forward through the discriminator, and the loss at the classifier's output is some function which is \emph{minimized} when the datapoint is correctly classified as originating from either the real dataset or the generator.
There is some flexibility in the choice of this loss function, and the gradient can be backpropagated through the discriminator network.
(At this stage only the parameters of the discriminator network are trained, and the generator network parameters are held fixed.)
This process is repeated for a few data points (each time flipping a coin to decide whether the input is from the real or fake datasets), and then perform minibatch gradient descent on the classifier.

After a few iterations of gradient descent on the discriminator's parameters, we can begin to train the generator network.
Training the generator network involves connecting the output of the generator to the input of the discriminator, and then \emph{maximizing} the error of the discriminator by performing gradient \emph{ascent} on the parameters of the generator network (while keeping the parameters of the discriminator network fixed).

To summarize, we can consider the random bit $l_j$ during the discriminator training to be the ground truth label for the real or fake datapoint.
If we denote the output of the discriminator by $o_j$, then we can frame the problem as the discriminator trying to enforce correlation, $l_j \oplus o_j = 0$, while the generator tries to enforce anti-correlation, $l_j \oplus o_j = 1$ (where $\oplus$ is the binary exclusive-or/modulo 2 addition).
Phrasing the training in this manner will help formulate the quantum version of the problem, which we will now consider.

\subsubsection{Generative Adversarial Quantum Circuits}

Now that we have reviewed how to train a typical classical Generative Adversarial Network (GAN), we can describe how to make a quantum parametric circuit equivalent of these GANs, which we call Generative Adversarial Quantum circuits (GAQs). 

Similar to GANs, GAQs can be used to generate samples from a certain distribution.
Since we are considering quantum data, GAQs will be used to replicate samples from a distribution of quantum states.
The datasets we consider can be a mixture of pure states or mixed states:
\begin{equation}
    \hat{\rho}= \sum_{j} p_j \ket{\psi^{\textsc{i}}_j}\!\bra{\psi^{\textsc{i}}_j} \quad \text{ or } \quad \hat{\rho}= \sum_{j} p_j\hat{\rho}_j.
\end{equation}
The goal of the generator will be to mimic states that are part of this distribution, while the discriminator network will attempt to discern the real quantum states from the generated ones. 

The generator network will be a parametric quantum circuit which takes in some quantum randomness as input, and outputs candidate quantum states to mimic samples from the data distribution.
We will denote this generator's parametric circuit as $\hat{G}(\bm{\hat{\Phi}}_{\textsc{g}})$, where  $\bm{\hat{\Phi}}_{\textsc{g}}$ are the parameters for the generator.
The randomness is provided in the form of a state, $\ket{r}_{\textsc{ge}}$, where G and E are the Hilbert spaces of the input to the generator, and the \textit{environment}, which is simply the purification of this input.
Since there generally can be entanglement across the G-E bipartition, the input to the network will generally be a mixed state, $\rho_{r}\in\mathcal{B}(\mathcal{H}_{\textsc{g}})$.
We consider the preparation unitary for the purified state, $\ket{r}_{\textsc{ge}} = \hat{U}_r(\bm{\theta})\ket{\bm{0}}_{\textsc{ge}}$, to be dependent on a set of preparation hyper-parameters, $\bm{\theta}$.
Hence, we can append these to our preparation hyper-parameters $\bm{\Theta}$ (which also includes our pointer state preparation hyper-parameters).
Thus, for a given set of parameters, $\bm{\Phi}_{\textsc{g}}$, the purified output mixed state of the generator is given by $\hat{G}(\bm{\Phi}_{\textsc{g}})\ket{r}_{\textsc{ge}}$.
Tracing this over the environment E gives us the mixed state of samples generated by the generator, $\hat{G}(\bm{\Phi}_{\textsc{g}})\hat{\rho}_r\hat{G}^\dagger(\bm{\Phi}_{\textsc{g}})$, which we will feed to the discriminator network.

The discriminator is simply a binary quantum classifier, as treated in Subsection~\ref{sec:meas_learn}, with a parametric circuit $\hat{D}(\bm{\hat{\Phi}}_{\textsc{d}})$ and corresponding parameters $\bm{\Phi}_{\textsc{d}}$.
We write the standard basis Pauli operator of the output register to be $\hat{Z}_o$.
As a first version of the GAQ, for training the discriminator, we can consider having the the ground truth label for the iteration $j$ to be a classical random bit $l_j$.
We thus sample a random Bernoulli distribution to determine $l_j$.
If $l_j = 1$, we perform the QFB on the discriminator by feeding it a datapoint (quantum state) sampled from the dataset $\{\hat{\rho}_j\}_{j}$.
In the case of $l_j=0$, we feed the state output by the generator to perform QFB.
In both cases the loss function is,
\begin{equation}
    \hat{L}_j^{(\textsc{d})} = (-1)^{l_j+1} \hat{Z}_o.
\end{equation}
Training the parameters to minimize this loss will move to positively correlate the output of the discriminator with the ground truth label $l_j$.
The effective phase we get (on average) for the discriminator parameters (assuming an unbiased coin flipped for the truth label $l_j$) is
\spliteq{
    \mathcal{L}_j(\bm{\hat{\Phi}}_{\textsc{d}}) =&\tfrac{1}{2} \text{tr}[\hat{D}^\dagger(\bm{\hat{\Phi}}_{\textsc{d}})\hat{Z}_{o}\hat{D}(\bm{\hat{\Phi}}_{\textsc{d}})\hat{\rho}_j]\\ &-\tfrac{1}{2}\text{tr}[\hat{D}^\dagger(\bm{\hat{\Phi}}_{\textsc{d}})\hat{Z}_{o}\hat{D}(\bm{\hat{\Phi}}_{\textsc{d}})\hat{G}(\bm{\Phi}_{\textsc{g}})\hat{\rho}_r\hat{G}^\dagger(\bm{\Phi}_{\textsc{g}})].
 }
(Note that the traces in all of the formulas in this subsection are understood to be taken over everything \emph{except} the parameter Hilbert spaces.) 
The parameter optimization here is only for the discriminator parameters, $\bm{\hat{\Phi}}_{\textsc{d}}$, which are quantum and are optimized quantum dynamically.
The generator parameters, $\bm{\Phi}_{\textsc{g}}$, can be kept kept classical (or equivalently in an eigenstate of  $\bm{\hat{\Phi}}_{\textsc{g}}$) in the case of MoMGrad, or can be kept fixed (no kinetic pulse) in the case of QDD
For concreteness, if we were to train a few iterations using QDD, the unitary to be applied would be 
 \begin{equation}\label{eq:qdderp}
    \hat{U}_{\textsc{qdd}} = \prod_{j} e^{-i\gamma_j\bm{\hat{\Pi}}_{\textsc{d}}^2}e^{-i\eta_j\mathcal{L}_j(\bm{\hat{\Phi}}_{\textsc{d}})},
\end{equation}
where $\bm{\hat{\Pi}}_{\textsc{d}}$ is the vector of canonical conjugate operators of $\bm{\hat{\Phi}}_{\textsc{d}}$.
 
To train the generator network, we connect the generator directly into the discriminator network, i.e., the feedforward unitary becomes 
\begin{equation}
    \hat{U}(\bm{\hat{\Phi}}) = \hat{D}(\bm{\hat{\Phi}}_{\textsc{d}})\hat{G}(\bm{\hat{\Phi}}_{\textsc{g}}),
\end{equation}
which acts upon the input resource quantum random state $\ket{r}_{\textsc{ge}}$.
The parameters of the adversary network (discriminator) are fixed, i.e., we can consider the parameters registers, $\bm{\hat{\Phi}}_{\textsc{d}}$, to be classical or to be in an eigenstate of parameter values.
We can perform the Quantum Feedforward and Phase Kick Baqprop (QFB) procedure on this joint network with the loss function 
\begin{equation}
     \hat{L}_j^{(\textsc{g})} = \hat{Z}_o,
\end{equation}
which, when minimized via either of the quantum parameter descent techniques, drives the generator's parameters to fool the discriminator.
That is, the generator's weights will be optimized so that, for current discriminator's parameters, there is an increased chance for the discriminator to output $Z_o = 1$ when fed the output of the generator.
The effective phase induced on the generator's parameters via the QFB procedure with this loss function is given by
\spliteq{
 \mathcal{L}_j(\bm{\hat{\Phi}}_{\textsc{g}}) =\text{tr}[\hat{G}^\dagger(\bm{\hat{\Phi}}_{\textsc{g}})\hat{D}^\dagger(\bm{\Phi}_{\textsc{d}})\hat{Z}_{o}\hat{D}(\bm{\Phi}_{\textsc{d}})\hat{G}(\bm{\hat{\Phi}}_{\textsc{g}})\hat{\rho}_r].
}

Thus, by alternating the training of discriminator network and the generative network, both networks should reach an adversarial equilibirum \cite{2014arXiv1406.2661G}, and near this equilibrium the generator should be able to provide good candidate states to mimic the quantum data distribution.

\begin{figure}[h!]
 \begin{center}
\includegraphics[width=0.666\columnwidth]{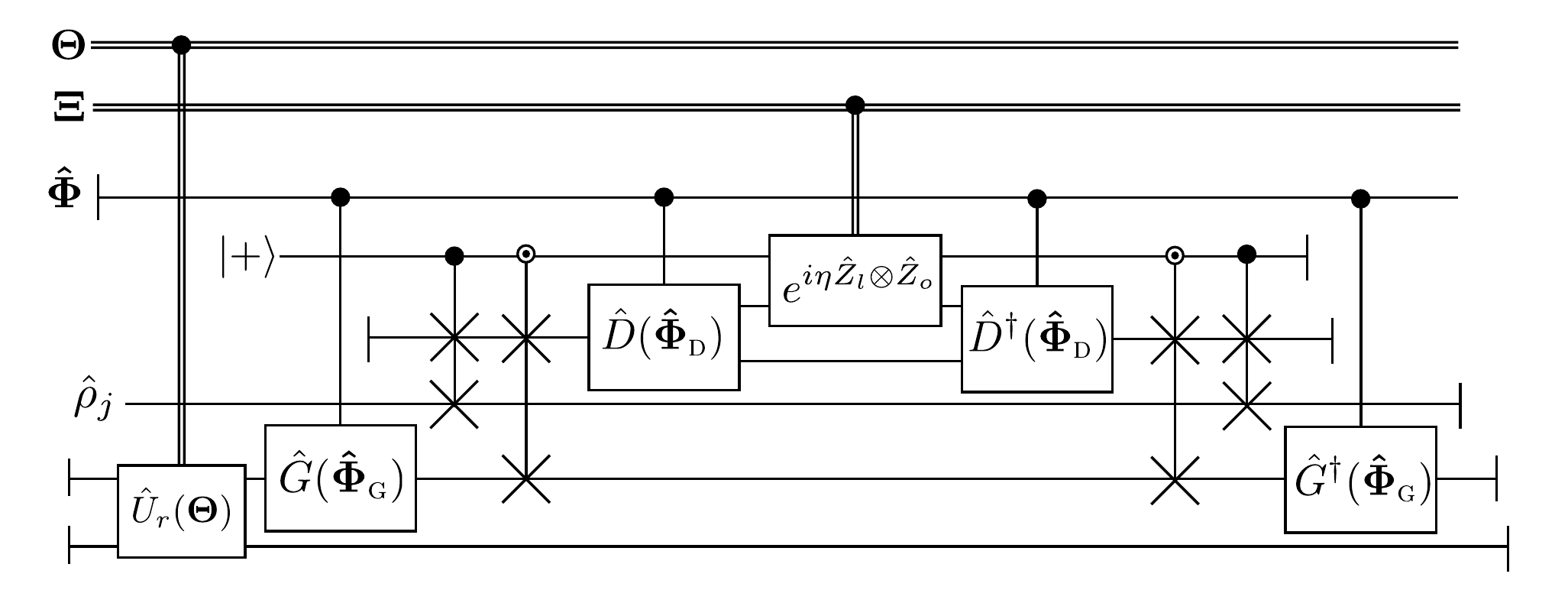}
\caption{QFB procedure for the fully coherent Generative Adversarial Quantum circuit. The input to the parameter-controlled discriminator, $\hat{D}(\bm{\hat{\Phi}}_\textsc{d})$, is swapped in from either the real data set, with sample $\hat{\rho}_j$, or from the output of the generator network, $\hat{G}(\bm{\hat{\Phi}}_\textsc{g})$. The determination of which sample to swap in is controlled by a label qubit in the state $\ket{+}$. The sample $\hat{\rho}_j$ is connected to the discriminator input via a controlled-SWAP (CSWAP) gate with the label qubit as the control. The output of the generator network is connected to the discriminator input via a negated CSWAP gate (the negation is depicted by a white circle in the control register). The phase kick is applied with the loss function acting on the output of the discriminator and the label qubit. The phase kick is followed by an uncomputation of the entire feedforward circuit as prescribed by the QFB procedure.
} \label{fig:GAQ}
\end{center}
\end{figure}

An option to simplify the number of steps involved in the training algorithm and to train both networks simultaneously is to use a qubit for the ground truth label.
With this, we will be able to make the entire algorithm fully coherent, and we can use the same loss function for both networks, except that the discriminator will be trained to \emph{descend} the loss function landscape while the generator will be trained to \emph{ascend}.
This setup, which we will now proceed to describe, is illustrated in Figure~\ref{fig:GAQ}.

We first replace the Bernoulli random variable $l_j$, representing the ground truth label for iteration $j$, with a qubit beginning in a state of uniform superposition of two label values, i.e., $\ket{+}_l = \tfrac{1}{\sqrt{2}} ( \ket{0}_l + \ket{1}_l )$.

Now, we keep the generator and quantum randomness seed the same, but the input to the discriminator will be swapped in based on the computational value of the label qubit.
That is, we use the label qubit as the control for both a Fredkin (controlled-SWAP) gate and a negated Fredkin gate (i.e., a Fredkin gate conjugated by qubit flips $\hat{X}_l$), as depicted in Figure~\ref{fig:GAQ}.
The first of these will be used to swap in a sample from the real data set, $\hat{\rho}_j$, in the branch of the superposition of the label qubit corresponding to $\ket{1}_l$.
The second will swap in the output of the generator network (which, along with the quantum randomness seed will remain the same as in the non-coherent version of the network).
With this setup, we can phase kick the entire network at the output of the discriminator network with the loss function,
\begin{equation}
    \hat{L}_j = -\hat{Z}_l\otimes \hat{Z}_o,
\end{equation}
where $\hat{Z}_l$ is the Pauli-Z operator for the label qubit, and $\hat{Z}_o$ is the Pauli-Z operator for the output classifier of the discriminator circuit.
Thus, this cost function is minimized when the output of the discriminator is positively correlated with the ground truth label.
Hence, the training of the discriminator will aim to minimize this loss, whereas the training of the generator will aim to maximize it.

First, let us consider the effective phase, which can be shown to recover the formula from before,
\spliteq{
    \mathcal{L}_j(\bm{\hat{\Phi}}) =&\tfrac{1}{2} \text{tr}[\hat{D}^\dagger(\bm{\hat{\Phi}}_{\textsc{d}})\hat{Z}_{o}\hat{D}(\bm{\hat{\Phi}}_{\textsc{d}})\hat{\rho}_j]\\ &-\tfrac{1}{2}\text{tr}[\hat{D}^\dagger(\bm{\hat{\Phi}}_{\textsc{d}})\hat{Z}_{o}\hat{D}(\bm{\hat{\Phi}}_{\textsc{d}})\hat{G}(\bm{\hat{\Phi}}_{\textsc{g}})\hat{\rho}_r\hat{G}^\dagger(\bm{\hat{\Phi}}_{\textsc{g}})].
 }
For training the parameters of the discriminator, one can employ either MoMGrad or QDD as in other situations we have considered.
For the parameters of the generator, since we want to \emph{ascend} the average landscape of this effective phase, we can act a squared Fourier transform on the parameters of the generator, denoted $\bm{\hat{F}}_\textsc{g}^2$, before and after the kinetic term of QDD, or before the measurement of the parameters in the case of MoMGrad.
The squared Fourier transform acts effectively as a NOT gate on the parameter registers, and hence will act to update the parameters in the opposite direction as we have seen before.
Concretely, if we were to perform QDD for this network, we would enact 
 \begin{equation}\label{eq:qdderpp}
    \hat{U}_{\textsc{qdd}} = \prod_{j\in \mathcal{M}} \bm{\hat{F}}^{\dagger 2}_{\textsc{g}}e^{-i\gamma_j\bm{\hat{\Pi}}^2}\bm{\hat{F}}^2_{\textsc{g}}e^{-i\eta_j\mathcal{L}_j(\bm{\hat{\Phi}})},
\end{equation}
where $\bm{\hat{\Pi}}^2 = \bm{\hat{\Pi}}^2_{\textsc{g}}\otimes\bm{\hat{\Pi}}^2_{\textsc{d}}$ is the kinetic term for all registers, while $\bm{\hat{F}}^2_{\textsc{g}}$ is the squared Fourier transform in each of the generator's registers.
Thus, negating the phase kick effectively forces the generator network to ascend the cost function rather than descend, which will drive the generator network to anti-correlate the output of the discriminator with the ground truth.
Note that, in practice, although we used the same kicking and kinetic rates for both the generator's and discrimnator's parameters in \eqref{eq:qdderpp}, it might be best to use different rates for both networks as attaining the adversarial equilibrium may require some hyper-parameter fine-tuning.

\onecolumngrid

\vspace*{25px}

\pagebreak
\twocolumngrid

\subsection{Parametric Hamiltonian Optimization}\label{sec:hammer}

Parametric Hamiltonian optimization algorithms consist of a broad class of algorithms \cite{farhi2014quantum,peruzzo2014variational} where the goal is to optimize over a parametrized hypothesis class of states in order to minimize the expectation value of a certain Hamiltonian, $\hat{H}$.
That is, if we denote the parametrized class of states as $\ket{\psi(\bm{\Phi})}$, then we want to find
\begin{equation}
 \operatornamewithlimits{argmin}_{\bm{\Phi}}  \bra{\psi(\bm{\Phi})}\hat{H} \ket{\psi(\bm{\Phi})}.
\end{equation}

Such algorithms includes the Variational Quantum Eigensolver (VQE) \cite{peruzzo2014variational}, which is used to find approximate eigenstates of non-commuting Hamiltonians in chemistry, the Quantum Approximate Optimization Algorithm (QAOA) \cite{farhi2014quantum}, which is used for quantum-enhanced optimization, as well as other parametric circuit ansatze like the Deep Multiscale Entanglement Renormalization Ansatz (DMERA) \cite{kim2017robust}, which is a hierarchically-structured ansatz which allows for sampling statistics of local observables in strongly-correlated quantum systems.

Such an optimization problem fits very naturally within the framework introduced in this paper. In our case, we consider the optimization over the hypothesis class of states as the task of optimizing of a class of quantum parametric circuits acting upon a reference state: $\ket{\psi(\bm{\Phi})} = \hat{U}(\bm{\Phi})\ket{\psi_0}$.
Then we can simply use the Hamiltonian as the loss function we wish to minimize:
\begin{equation}
    \hat{L} = \hat{H}.
\end{equation}
The main challenge with the implementation of a general Hamiltonian as a loss function is to construct its exponentiation, i.e., enacting the operator
\begin{equation}
    e^{-i\eta \hat{L}}= e^{-i\eta \hat{H}}.
\end{equation}
For a Hamiltonian which is a sum of various terms of index $\mathcal{X}$, 
\begin{equation}
    \hat{H} = \sum_{j\in \mathcal{X}} \hat{H}_{j},
\end{equation}
the task of exponentiating such a Hamiltonian is the same as that of quantum simulation of the time evolution generated by this Hamiltonian \cite{lloyd1996universal}.
This is a task for which there is much literature, as it is a core concept of quantum computing \cite{nielsen2002quantum}.
There exist many techniques to approximate such an exponential, and for a given desired operator norm error, the overhead will depend on the locality and operator norms of the Hamiltonian terms \cite{lloyd1996universal}.
A theoretically simple approach is the Suzuki-Trotter method, which is a divide-and-conquer method where each term is exponentiated independently,
\begin{equation}\label{eq:STformula}
    e^{-i\eta \hat{H}}\approx (\prod_{j\in \mathcal{X}}    e^{-i\eta \hat{H}_j/M})^{M}.
\end{equation}
The operator norm error $\epsilon$ in this approximation \cite{lloyd1996universal,poulin2011quantum} 
is $\epsilon =\eta^2\sum_{j,k \in\mathcal{X}}\lVert[\hat{H}_j,\hat{H}_k]\rVert /2M +\mathcal{O}(\eta^3)$.
Therefore, as long as we choose $M \sim \sum_{j,k \in\mathcal{X}}\lVert[\hat{H}_j,\hat{H}_k]\rVert$, we have an error of order $\epsilon \sim\mathcal{O}(\eta^2)$.

Now, for the QFB procedure, if we begin in a reference state $\ket{\psi_0}$, apply the parametric unitary $\hat{U}(\bm{\hat{\Phi}})$, apply a quantum simulated exponential of $\hat{H}$ (with error $\epsilon$), followed by an uncomputation of the parametric unitary, then we arrive at the effective phase kick on the parameters generated by,
\spliteq{
    \mathcal{L}(\bm{\hat{\Phi}}) &= \bra{\psi_0}\hat{U}^\dagger(\bm{\hat{\Phi}})\hat{H} \hat{U}(\bm{\hat{\Phi}})\ket{\psi_0},
}
up to an error of order $\mathcal{O}(\epsilon)$.
Recall that, in general, the effective phase kick is only accurate to first order in $\eta$, i.e., it has an error of order $\mathcal{O}(\eta^2)$.
Hence a first order Suzuki-Trotter formula as in equation \eqref{eq:STformula} should suffice.

\begin{figure}[h!]
 \begin{center}
\includegraphics[width=0.6\columnwidth]{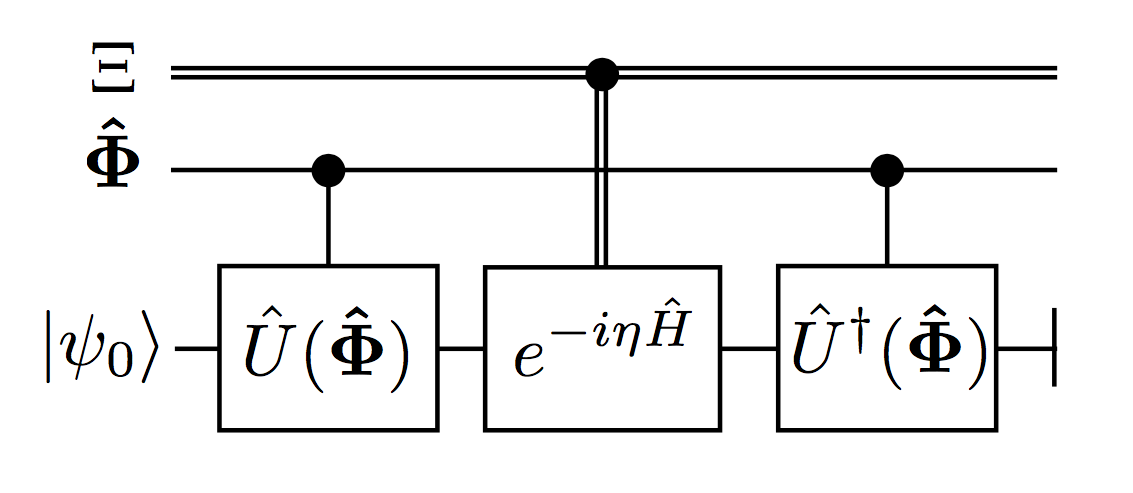}
\caption{QFB circuit for Parametric Hamiltonian Optimization in the case of a single QPU. The circuit simply consists of a feedforward of the parametric unitary acting on a reference state, followed by the simulated Hamiltonian exponentiation as a phase kick and the uncomputation.
} \label{fig:Hamopt}
\end{center}
\end{figure}

The circuit to implement the Quantum Feedforward and Baqprop (QFB) on a single QPU is simple, pictured in Figure \ref{fig:Hamopt}.
Note that the implementation of the exponential of the Hamiltonian can come with large depth overhead, thus it may be convenient to have a method with higher space overhead but with lower depth, i.e., a way to parallelize the accumulation of the gradient over the terms in the Hamiltonian.
We discuss this in the next subsubsection. 

\subsubsection{Hamiltonian-Parallelized Gradient Accumulation}

Here we discuss various methods to parallelize the accumulation of phase kicks and gradients in order to reduce circuit depth (time) overhead, at the cost of higher space overhead.
Notice that since the effective phase from above is an expectation value, it is linear and can be split up over the terms of the Hamiltonian:
\spliteq{
    \mathcal{L}(\bm{\hat{\Phi}}) &= \bra{\psi_0}\hat{U}^\dagger(\bm{\hat{\Phi}})\hat{H} \hat{U}(\bm{\hat{\Phi}})\ket{\psi_0}\\
    &=\sum_{j\in\mathcal{X}} \bra{\psi_0}\hat{U}^\dagger(\bm{\hat{\Phi}})\hat{H}_j \hat{U}(\bm{\hat{\Phi}})\ket{\psi_0}.
}
This trick is the fundamental principle behind the Quantum Expectation Estimation algorithm \cite{mcclean2016theory}, which parallelizes the expectation values of each term in the Hamiltonian over different QPUs/runs.
Recall that in our case we are looking to obtain a gradient of these effective phases (which are expectation values).
Since the gradient is a linear operator, we can accumulate the gradient of the sum by the sum of the gradients.

\begin{figure}[h!]
 \begin{center}
\includegraphics[width=1\columnwidth]{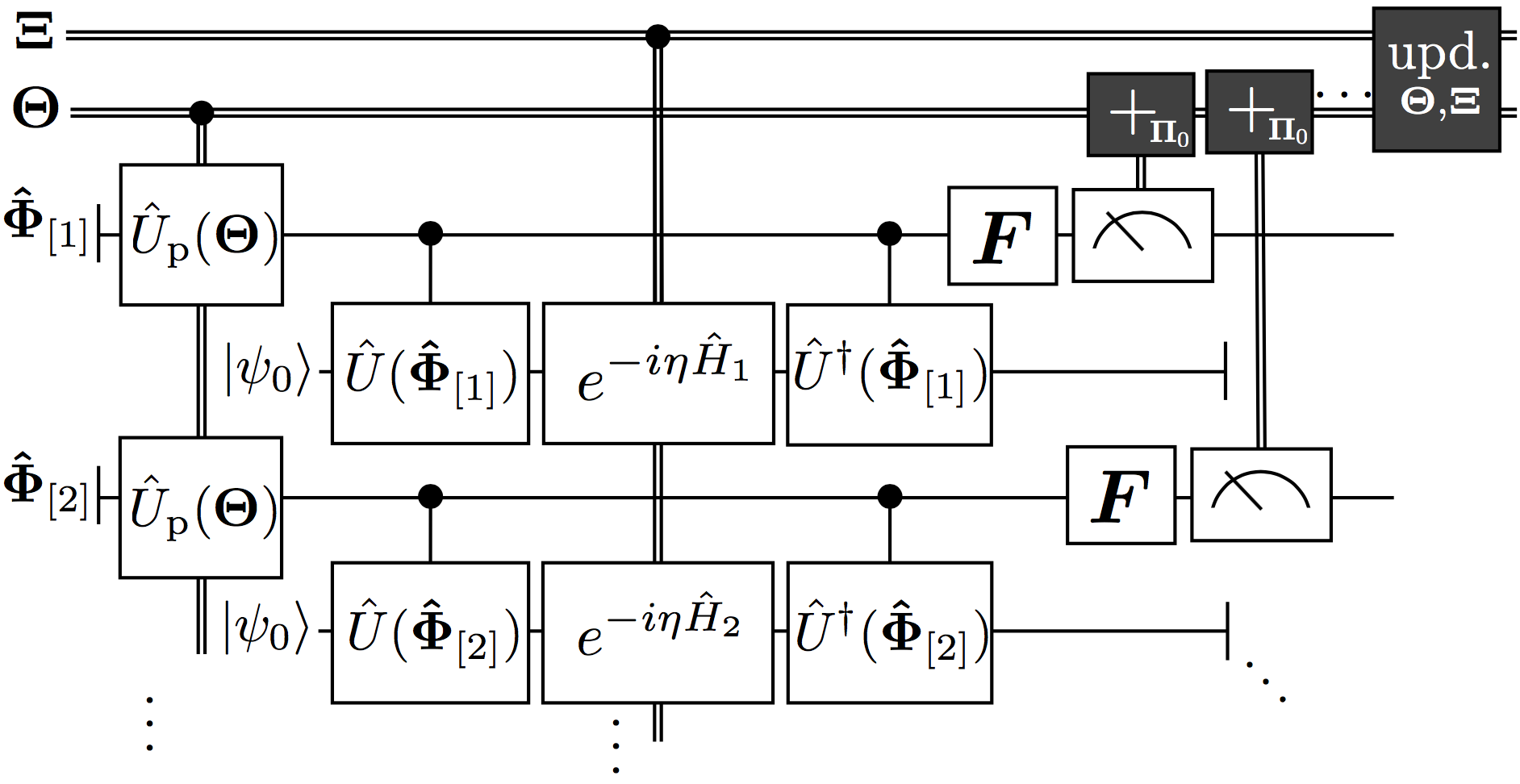}
\caption{MoMGrad + GEEP: Momentum Measurement Gradient Descent iteration via Gradient Expectation Estimation Parallelization. The parameter pointer states in each replica are prepared using a unitary $\hat{U}_\text{p}(\bm{\Theta})$. The QFB circuit is applied in each replica, with each parametrized unitary controlled by the corresponding replica parameters and the phase kick generated by the corresponding term in the Hamiltonian. The shift in the momenta of the parameters in each replica are measured, and after many runs the expectation values are classically added to obtain an averaged gradient of the total loss function.
} \label{fig:GEEP}
\end{center}
\end{figure}

Operationally, by using multiple sets of parameter registers, $\{\bm{\hat{\Phi}}_{[j]}\}_{j\in\mathcal{X} }$, and dividing up the terms in the Hamiltonian into individual loss functions over different QPUs, we can classically parallelize the accumulation of gradients, i.e., using classical addition we can sum up the gradient contribution of each term.
We call this approach Gradient Expectation Estimation Parallelization (GEEP), which is technique mostly relevant to Momentum Measurement Gradient Descent (MoMGrad), since the gradient has to be measured to be stored as classical information.
Mathematically, by acting a QFB with a loss $\hat{L}_{[j]} = \hat{H}_j$ on each replica, we get the following effective QFB phase on replica $j$:
\begin{equation}\label{eq:loss_hams}
\begin{split}
  \mathcal{L}_j(\bm{\hat{\Phi}}_{[j]}) &= \bra{\psi_0} \hat{L}_j(\bm{\hat{\Phi}}_{[j]}) \ket{\psi_0} \\
  &= \bra{\psi_0}\hat{U}^\dagger(\bm{\hat{\Phi}}_{[j]})\hat{H}_j \hat{U}(\bm{\hat{\Phi}}_{[j]})\ket{\psi_0}.
\end{split}
\end{equation}
Thus, to first order in $\eta$, the effective phase kick is $e^{-i \eta \mathcal{L}_j(\bm{\hat{\Phi}}_{[j]})}$ on each of the parameter sets.
The corresponding shift in momenta of each set of parameters is
\begin{equation}
\begin{split}
  \bm{\hat{\Pi}}_{[j]} \quad \mapsto \quad & e^{i \eta \mathcal{L}_j(\bm{\hat{\Phi}}_{[j]})} \bm{\hat{\Pi}}_{[j]} e^{-i \eta \mathcal{L}_j(\bm{\hat{\Phi}}_{[j]})} + \mathcal{O}(\eta^2) \\
  &= \bm{\hat{\Pi}}_{[j]} - \eta \frac{\partial \mathcal{L}_j (\bm{\hat{\Phi}}_{[j]})}{\partial \bm{\hat{\Phi}}_{[j]}} + \mathcal{O}(\eta^2).
\end{split}
\end{equation}
Therefore, by preparing identical momentum pointer states (cenetered at zero momentum) in each of the parameter registers of the replicas, i.e., $\ket{\Psi_0}^{\otimes |\mathcal{X}|}$ for some pointer state $\ket{\Psi_0}$, and by classically summing up the expectation values of the momenta in each replica, we have 
\spliteq{
    \sum_{j\in\mathcal{X}}\braket{ \bm{\hat{\Pi}}_{[j]} }_{[j]}
    &= - \eta \sum_{j\in\mathcal{X}} \,_{[j]}\!\bra{\Psi_0}  \frac{\partial \mathcal{L}_j (\bm{\hat{\Phi}}_{[j]})}{\partial \bm{\hat{\Phi}}_{[j]}}\ket{\Psi_0}_{[j]} + \mathcal{O}(\eta^2)\\
 &= - \eta  \bra{\Psi_0} \sum_{j\in\mathcal{X}} \frac{\partial \mathcal{L}_j (\bm{\hat{\Phi}})}{\partial \bm{\hat{\Phi}}}\ket{\Psi_0}+ \mathcal{O}(\eta^2)\\
  &= - \eta  \bra{\Psi_0} \frac{\partial ( \textstyle\sum_{j\in\mathcal{X}}\mathcal{L}_j (\bm{\hat{\Phi}}))}{\partial \bm{\hat{\Phi}}}\ket{\Psi_0}+ \mathcal{O}(\eta^2) \\  &= - \eta  \bra{\Psi_0} \frac{\partial \mathcal{L} (\bm{\hat{\Phi}})}{\partial \bm{\hat{\Phi}}}\ket{\Psi_0}+ \mathcal{O}(\eta^2).
}
Thus we see that by classically adding up the expectation values of the momenta in each replica, we get the gradient of the total loss function as if it were applied on a single replica.
We present the quantum-classical circuit for this GEEP procedure with MoMGrad in Figure~\ref{fig:GEEP}.

To apply a similar parallelization which is applicable to Quantum Dynamical Descent, the accumulation of momenta must be done coherently.
For this purpose, we can use the technique of Coherent Accumulation of Momenta Parallelization (CAMP) introduced in Section~\ref{sec:CAMP}.
The point is that we can consider the different terms in the Hamiltonian to be analogous to datapoints in a batch, whose loss functions are exponentiated and coherently accumulated to attain a total loss function comprised of the sum of losses of each term.
Once again denoting the loss function for each Hamiltonian term as $\hat{L}_{[j]} = \hat{H}_j$, and the associated effective phase for each replica as in \eqref{eq:loss_hams}, we can apply the following unitary for parallelized Hamiltonian Quantum Dynamical Descent:
\spliteq{\label{eq:PQDD2}
      \hat{U}_{\textsc{pqdd}} &= \prod_{k} e^{-i\gamma_k\bm{\hat{\Pi}}_{[0]}^2}U^\dagger_{\textsc{tent}}\Big(\bigotimes_{j}e^{-i\eta_k \mathcal{L}_j(\bm{\hat{\Phi}}_{[j]})}\Big)\hat{U}_{\textsc{tent}}\\
      &= \prod_{k} e^{-i\gamma_k\bm{\hat{\Pi}}_{[0]}^2} \hat{U}_{\textsc{camp},k},
}
where $k$ is an index for the iterations.
Recall that the TENT unitary is simply a multi-target adder gate, as defined in equation \eqref{eq:tent}.
Also note that in the above equation \eqref{eq:PQDD2}, the phase kicking rate is $\eta_k$ in each replica whereas previously in \eqref{eq:PQDD} it is normalized by the minibatch size.
Finally, this unitary is applied on an initial state where the parameter server (replica of index 0, with parameters $\bm{\hat{\Phi}}_{[0]}$) is in a pointer state of choice and the replicas are initialized in a null-parameter eigenstate, i.e.,
\begin{equation}
   \ket{\Psi_0}_{[0]}\bigotimes_{j}\ket{\bm{0}}_{[j]}.
\end{equation}

\begin{figure}[h!]
 \begin{center}
\includegraphics[width=1\columnwidth]{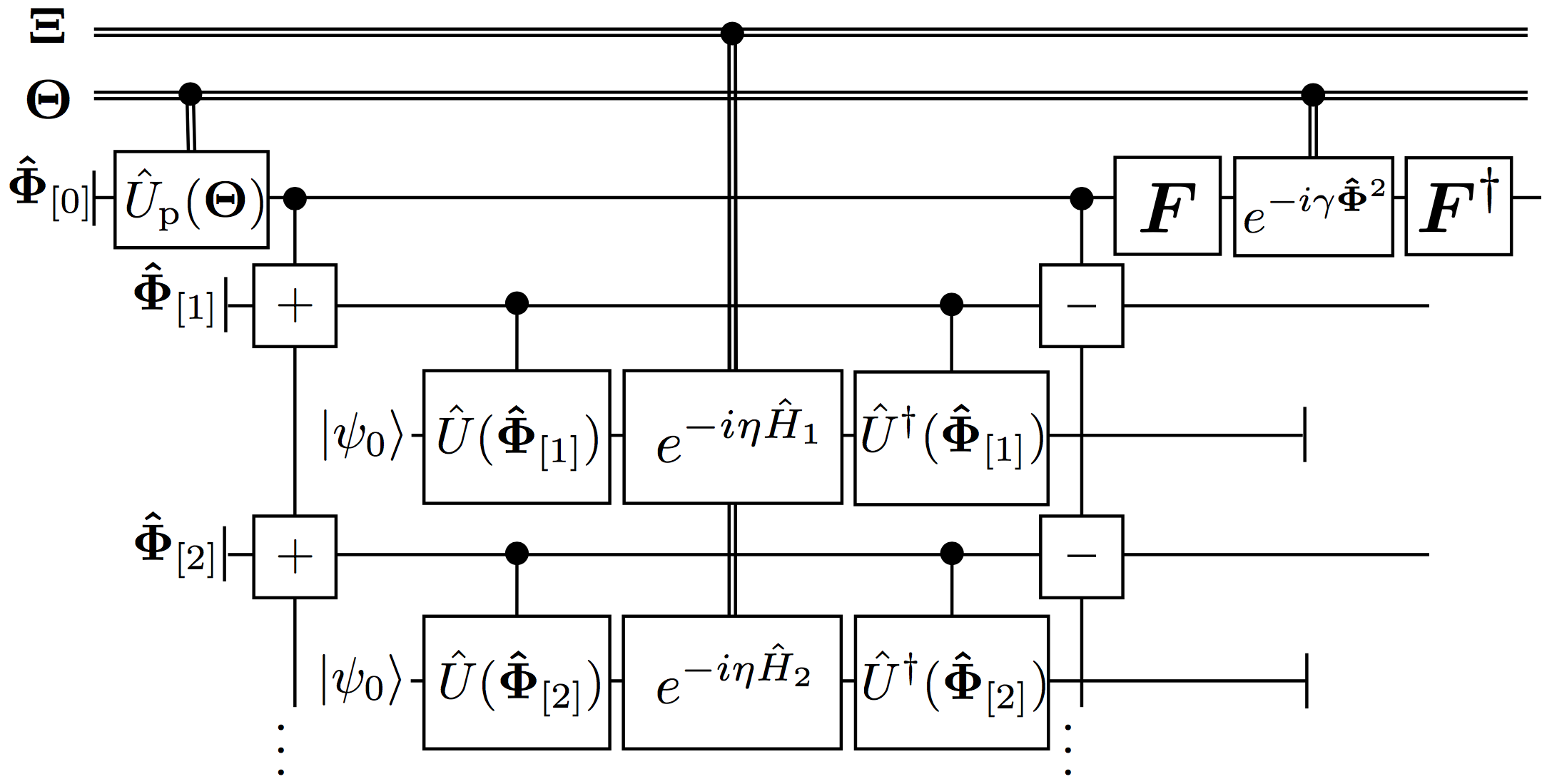}
\caption{QDD + CAMP for Parametric Hamiltonian Optimization. The parameter server is initialized to a pointer state using $\hat{U}_\text{p}(\bm{\Theta})$, and this is distributed to the replicas using the TENT operation. The QFB circuit is applied to each replica, with each parametrized unitary controlled by the corresponding replica parameters and the phase kick generated by the corresponding term in the Hamiltonian. The inverse TENT operation is applied after the QFB circuits to accumulate the phase kicks in the parameter server as a phase kick according to the total loss function. The standard QDD kinetic pulse is then applied to the parameter server at the end of the iteration.
} \label{fig:QDDCAMP}
\end{center}
\end{figure}

We represent the circuit for an iteration of Quantum Dynamical Descent with Coherent Accumulation of Momenta Parallelization for the Hamiltonian Optimization task in Figure~\ref{fig:QDDCAMP}.

\subsection{Hybrid Quantum Neural-Circuit Networks}

The method for regression/classification using quantum parametric circuits, outlined in Subsection~\ref{sec:meas_learn}, is effectively a method for learning a quantum-to-classical map.
One could then imagine having a classical neural network taking in this signal to perform some further processing.
More generally, one may wish to perform further processing on the outcomes of the measurement of some observable at the output of a general parametrized quantum circuit.
Further, the methods we describe here will, in principle, also apply to a general parametrized classical circuit, although we will focus on neural networks for concreteness.
The challenge examined here is to efficiently train both the quantum and classical parts of the hybrid network in unison.

In this section we focus on methods to backpropagate error signals through such a quantum-classical interface, i.e., how to train networks which are hybrids of quantum parametric circuits connected to classical neural networks.
We consider two cases, first is to embed the classical neural network into a quantum computation, i.e., both parts for the network are trained on a Quantum Processing Unit (QPU). 

In the second case, we have the classical neural network being trained on a Classical Processing Unit (CPU), which is connected to a QPU on which the Quantum Parametric Circuit is being trained.
We propose two methods for simultaneous quantum-classical training, in both cases we propose a way to backpropagate the error signal through the quantum-classical boundary.

The first of these latter methods depends only on expectation values of the observables of the quantum output registers, which are used as input activations for the neural network.
Using classical feedforward and backpropagation of gradients, we can approximate the error signal as a linear potential centered around this expectation value, and enact a linear relative phase kick on the quantum system (momentum kick) to convert this approximate error signal back to quantum.

The second method follows a similar philosophy, but allows for a more non-trivial error signal tomography, hence a higher-order approximation to the error signal.
The approach relies on sampling various measurement results from the  output of the parametric circuit and feeding these through the classical neural network.
For each sample point, a gradient is obtained through classical feedforward and backpropagation.
For sample points that are relatively close to each other (low variance of output from quantum regression net), a higher order interpolation of the effective backpropagated cost function can be obtained. 
This can then be applied as a higher-order phase kick on the quantum network, which can then be leveraged by the usual quantum phase kick backpropagation method for MoMGrad.

\subsubsection{Fully Coherent Hybrid Networks}

To begin, let us examine the case where both the quantum-parametrized circuit (QPC) and the classical neural network (NN) are trained on a QPU.
This setup simply involves connecting one of the neural nets from Section~\ref{sec:qnn} to the output of the QPC.
The presence of both the QPC and NN on the quantum chip allows one to use QFB in a straightforward manner.
Of course, once the QPC and NN have been trained on the QPU, there is an option to do inference with the NN on a CPU.

Although, in essence, using QFB in this situation is similar to before, it will be worth describing explicitly in order to compare to the subsequent cases.
Let us write $\hat{U}_\textsc{qpc}(\bm{\hat{\Phi}})$ as the quantum-parametric circuit, where $\bm{\hat{\Phi}}$ are the parameters of the circuit.
We will also write $\tilde{y} = f(\bm{\theta},\bm{x})$ as the prediction at the output of the classical parametric circuit (i.e., the NN) with parameters $\bm{\theta}$ and input $\bm{x}$.
Once embedded in a quantum chip, the input $\bm{\hat{x}}$ and parameters $\bm{\hat{\theta}}$ are quantum.
As in Section~\ref{sec:qnn}, the circuit for the feedforward in the NN is $\hat{U}_{\textsc{ff}}(\bm{\hat{\theta}}) = e^{-i f(\bm{\hat{\theta}},\bm{\hat{x}}) \hat{p}_{\tilde{y}}}$, where $\hat{\tilde{y}}$ denotes the output register of the network (prediction) and $\hat{p}_{\tilde{y}}$ its conjugate momentum.
Recall also that the QFB circuit for the NN is $e^{-i \eta \hat{L}( \hat{\tilde{y}} + f(\bm{\hat{\theta}},\bm{\hat{x}}), y )}$.
The full QFB for the QPC and NN involves the feedforward of the QPC< followed in turn by the feedforward of the NN, phase kick for the output of the NN, backpropagation for the NN, and finally backpropagation for the QPC.
We will find it illustrative to absorb the middle three steps as simply the QFB circuit for the NN alone:
\begin{equation}
  \hat{U}_\textsc{qpc}^\dagger(\bm{\hat{\Phi}}) e^{-i \eta \hat{L}( \hat{\tilde{y}} + f(\bm{\hat{\theta}},\bm{\hat{x}}), y )} \hat{U}_\textsc{qpc}(\bm{\hat{\Phi}}).
\end{equation}
In this way, we can abstract away the entire NN, so that insofar as the QPC is concerned, the QFB of the NN appears as a phase kick on the momentum of the observable $\bm{\hat{x}}$ at the output of the QPC.

In the following, we will discuss cases where the NN is implemented on a CPU, so that $\bm{\theta}$ are classical NN parameters.
The tasks that remain are to find an appropriate replacement for the input to the classical network, $\bm{x}$, determined from the output of the quantum-parametric circuit, as well as a means of employing the notion of backpropagation at the quantum-classical interface.

\subsubsection{Hybrid Quantum-Classical Networks}\label{sec:hybrid}

To obtain a classical number from the output of the QPC, one has to perform a measurement.
In the case of classification or regression, this could correspond to a generalized measurement.
However, a generalized measurement can always be seen as a projective measurement on a larger system.
Thus, without loss of generality, we will define an observable $\bm{\hat{z}}$ at the output of the QPC as an operator whose spectrum consists of some encoding of the measurement outcomes.

After declaring such an observable, we need to decide how to use outcomes of measurements of $\bm{\hat{z}}$ to feed into the classical circuit.
That is, if we again write the output of the classical circuit as $f(\bm{\theta},\bm{x})$, then we will describe some choices of maps from measurements of $\bm{\hat{z}}$ to values of $\bm{x}$.
For example, the first map we will examine is the expectation value: $\bm{x} = \langle \bm{\hat{z}} \rangle$.
The second map we will discuss will be able to accommodate some variance in the variable $\bm{\hat{z}}$.
However, simultaneously training a quantum circuit combined with a classical network will typically only work well if the uncertainty in the QPC parameters, $\bm{\hat{\Phi}}$, is low (i.e., when their distributions are highly concentrated close to their expectation values).

Once we have chosen such a map, we can feedforward the input through the network, and perform classical backpropagation to obtain $\partial f(\bm{\theta},\bm{x}) / \partial \bm{x}$.
In the previous case, where the QPC and NN were both placed on the QPU, we saw that QFB involved feeding forward the QPC, and applying the phase kick
\begin{equation}
  e^{-i \eta \hat{L}( f(\bm{\hat{\theta}},\bm{\hat{z}}), y )},
\end{equation}
followed by uncomputing the QPC.
Note that in this formula we have removed $\hat{\tilde{y}}$, since we will assume that the register for $\hat{\tilde{y}}$ is initialized to zero and none of the other circuit elements act on the output register of the NN embedded in the QPU.
Below we will discuss analogues of the QFB for the NN constructed from the gradients, $\partial f(\bm{\theta},\bm{x}) / \partial \bm{x}$, obtained form the classical backpropagation in order to obtain a means of propagating the error from the classical network as a phase kick on the output of the quantum circuit.

\textbf{First-order method.}
The simplest means of mapping the observable $\bm{\hat{z}}$ at the output of the QPC to a classical input is to assign $x = \langle \bm{\hat{z}} \rangle := \tr [ \hat{U}^\dagger(\bm{\hat{\Phi}}) \bm{\hat{z}} \hat{U}(\bm{\hat{\Phi}}) \hat{\rho}_0 ]$.
Note the trace is taken over the computational Hilbert space as well as the Hilbert space of the parameters $\bm{\hat{\Phi}}$.
In practice, this expectation value is obtained from measuring $\bm{\hat{z}}$ over multiple runs of the QPC.

With this assignment to $\bm{x}$, one can feedforward the input to obtain the output of the network, $f(\bm{\theta},\bm{x})$, and backpropagate the loss function to obtain a gradient of the loss function with respect to the input: $\left. [\partial L( f(\bm{\theta},\bm{x}), y) / \partial \bm{x}] \right|_{\bm{x} = \langle \bm{\hat{z}} \rangle}$.
This gradient can be used to approximate the QPU version of the phase when the variance of $\bm{\hat{z}}$ is small, since we can then write
\begin{equation}
\begin{split}
  L(f(\bm{\theta},\bm{\hat{z}}),y) &\approx L(f(\bm{\theta},\langle \bm{\hat{z}} \rangle),y) \\
  & \qquad + ( \bm{\hat{z}} - \langle \bm{\hat{z}} \rangle ) \cdot \left. \frac{\partial L(f(\bm{\theta},\bm{x}),y) }{ \partial \bm{x} } \right|_{\bm{x} = \langle \bm{\hat{z}} \rangle}.
\end{split}
\end{equation}
Note that when exponentiated, the c-number terms in this expression simply give global phases to the wavefunction.
Therefore, the phase kick we should apply at the output of the QPC to backpropagate the error of the classical network consists of a linear phase shift.
In summary, once we have the backpropagation of the classical network to the input, we can write the QFB for the quantum-parametric circuit as
\begin{equation}
  \hat{U}_\textsc{qpc}^\dagger(\bm{\hat{\Phi}}) e^{-i \eta \bm{\hat{z}} \cdot \left[ \left. \partial L(f(\bm{\theta},\bm{x}),y) / \partial \bm{x}  \right|_{\bm{x} = \langle \bm{\hat{z}} \rangle} \right]} \hat{U}_\textsc{qpc}(\bm{\hat{\Phi}}).
\end{equation}
As the gradients for the classical network have already been backpropagated, the classical part of the network can simply be trained using these classically backpropagated gradients, using gradient descent or any other choice classical gradient-based based optimizer \cite{kingma2014adam,zeiler2012adadelta}. This method requires relatively low-depth circuits, and only depends on easily-measured expectation values. Algorithms of low-depth which depend on simple expectation values have shown to be sufficiently robust to noise for successful implementation on near-term devices \cite{zeng2017quantum}, as such, we expect that this algorithm should be implementable on near-term devices.

\begin{figure}[h!]
 \begin{center}
\includegraphics[width=1.00\columnwidth]{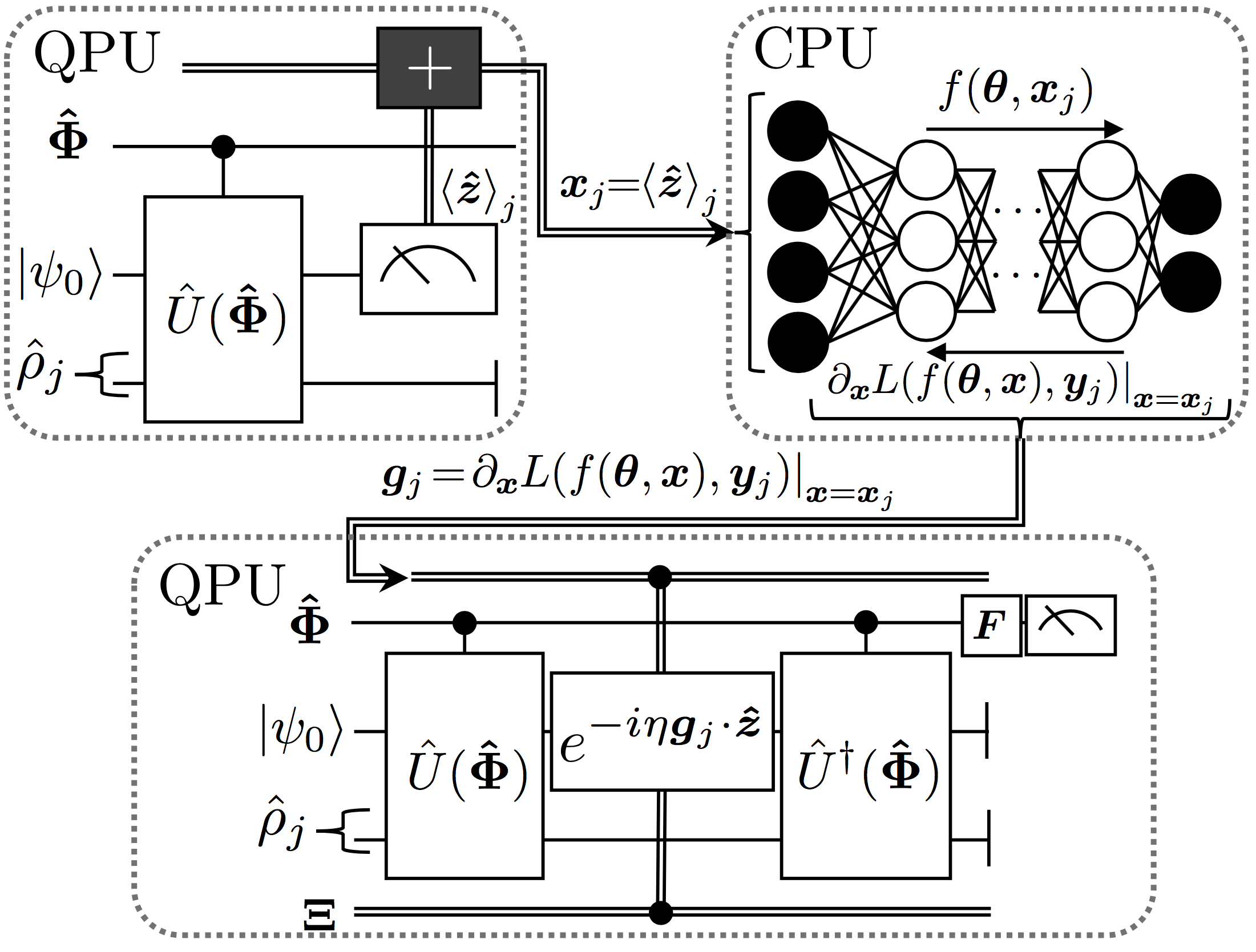}
\caption{Concurrent training of a hybrid quantum-classical network using a first-order method. The upper-left diagram shows the feedforward of a parametric quantum classifier upon input state $\hat{\rho}_j$ and auxiliary reference state $\ket{\psi_0}$. Measurements are performed for multiple runs of the feedforward (on input $\hat{\rho}_j$) to obtain the expectation value $\braket{\bm{\hat{z}}}_j$. This expectation value is fed into the classical neural network (top-right), where classical feedforward and backpropagation is performed to obtain the gradient of the loss function at the output of the classical network with respect to the input, $\bm{g}_j := \partial_{\bm{x}} L ( f(\bm{\theta},\bm{x}), \bm{y}_j ) |_{\bm{x} = \bm{x}_j}$. The gradient, $\bm{g}_j$, is then used to employ a phase kick, $\exp(-i\eta \bm{g}_j \cdot \bm{\hat{z}})$, in the MoMGrad procedure for the parametric quantum circuit on the QPU (bottom).
} \label{fig:hybrid_net}
\end{center}
\end{figure}

An illustration of this first-order method for training hybrid quantum-classical networks is shown in Figure~\ref{fig:hybrid_net}.

\textbf{Higher-order method.}
Instead of inputting the expectation value $\langle \bm{\hat{z}} \rangle$ into the classical network, here we will input a sample for the outcome of a measurement of $\bm{\hat{z}}$, i.e., we draw a sample point, $\bm{z}^\ast$, from the distribution $p(\bm{z}) = \tr[ \ket{\bm{z}}\!\bra{\bm{z}} \hat{U}(\bm{\hat{\Phi}}) \hat{\rho}_0 \hat{U}^\dagger(\bm{\hat{\Phi}}) ]$.
Now, if we perform backpropagation on the classical network back to the input, we obtain a gradient:
\begin{equation}
  g( \bm{z}^\ast ) := \left. \frac{ \partial L( f(\bm{\theta},\bm{x}), y) }{ \partial \bm{x} } \right|_{\bm{x} = \bm{z}^\ast }.
\end{equation}

We can repeat this for multiple samples, $\{ \bm{z}_i^\ast \}_{i=1}^N$, in order to collect multiple gradients, $\{ g(\bm{z}_i^\ast) \}_{i=1}^N$.
Now the idea is to use this collection of gradients in an interpolation scheme to obtain an approximation to the quantum phase kick,
\begin{equation}
  G( \bm{\hat{z}}, \{ g(\bm{z}_i^\ast) \}_{i=1}^N ) \approx L(f(\bm{\theta},\bm{\hat{z}}),y).
\end{equation}
For example, if the sample points happen to be near one another, one could try to reconstruct a second-order Taylor approximation to the function $L$.
Otherwise, if the sample points are too far apart, one could use a different interpolation scheme.
After making a choice for the function $G$, one can write the QFB circuit for the QPC as:
\begin{equation}
  \hat{U}_\textsc{qpc}^\dagger(\bm{\hat{\Phi}}) e^{-i \eta G( \bm{\hat{z}}, \{ g(\bm{z}_i^\ast) \}_{i=1}^N )} \hat{U}_\textsc{qpc}(\bm{\hat{\Phi}}).
\end{equation}

\section{Numerical Experiments}\label{sec:num}

In this section we demonstrate the capabilities of the heuristics proposed in sections \ref{sec:opt} by implementing these methods to optimize various quantum neural networks and quantum parametric circuits. We compare the performance of Quantum Dynamical Descent (QDD) versus Momentum Measurement Gradient Descent (MoMGrad). We begin with the training of a classical deep neural networks on a quantum computer to demonstrate how the algorithm performs for classical computation embedded in quantum computation. Following this, we show how QDD and MoMGrad can leveraged to enhance quantum Hamiltonian optimization algorithms, we use the Quantum Alternating Operator Ansatz parametric circuit as our example. To show how the heuristics deal with loss operators that are not Hamiltonian-based, but rather state-based, we show how one can perform gradient ascent on the fidelity in order to learn a parametric circuit approximating a unitary. Finally, to show how Quantum Phase Backpropagation interfaces with classical backprop, we demonstrate the training of a hybrid network on a simulated quantum computer running a quantum parametric circuit connected to a classical neural network running on a classical processor.

All the experiments featured in this section were classical numerical simulations of quantum computation, which were on the Rigetti Forest Quantum Virtual Machine, with code written in PyQuil \cite{Forest}.

\subsection{Quantum Neural Deep Learning}

In this subsection we train a classical deep neural network y embedding it into a quantum computation in order to leverage MoMGrad and QDD. 

In order to demonstrate the capabilities of the quantum descent algorithms of section \ref{sec:opt} to train a \textit{deep} neural network, we chose a problem which is a non-linearly separable classification task. Due to being one of the most elementary canonical counter-example to the learning capabilities of single-layer networks, we chose the task of learning the exclusive-or (XOR) Boolean function using a 2-layered perceptron network. 
Learning a set of optimal parameters which minimize the loss for this classification task counts as \textit{deep} learning, since it requires a neural network of depth at least 2.

\subsubsection{Application \& Methods}

Recall the XOR function (denoted $\oplus$) takes binary pairs of bits and maps them $\{b_1,b_2\}\in\mathds{Z}_2\times \mathds{Z}_2$ and maps them to a single binary value corresponding to their addition modulo 2; $b_1 \oplus b_2 \equiv (b_1 +b_2) (\text{mod}\ 2)$. 

\begin{figure}[h!]
 \begin{center}
\includegraphics[width=0.5\columnwidth]{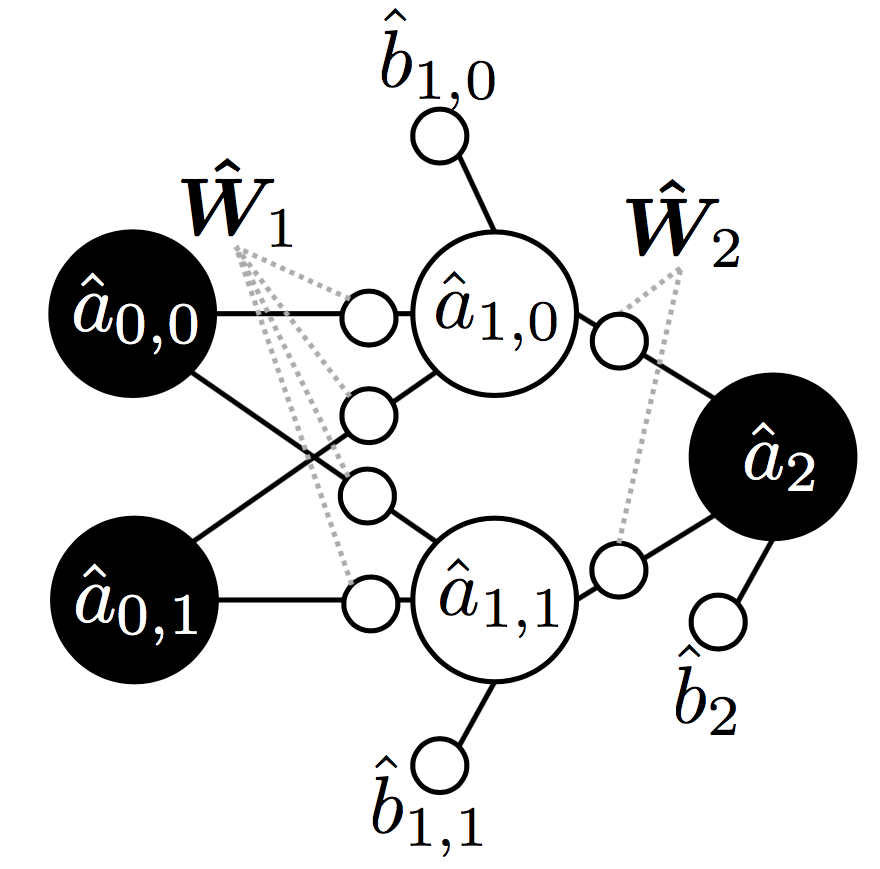}
\caption{Neural network used for learning the XOR function. Input neurons and output layers' neurons are black, hidden layer neurons are white, while the quantum weights and biases are represented with white dots. Both the hidden layer and output layer have biases.
} \label{fig:nn_xor_diag}
\end{center}
\end{figure}

In order to learn this function, we then have to use a neural network which has 2 input units and 1 output unit. The particular network chosen for the implementation in this paper is pictured in figure \ref{fig:nn_xor_diag}. This network has an input layer, a single hidden layer, and one neuron constituting the output layer. 

In order to encode this network on the quantum simulator, we use finite-dimensional qudits for each neuron, weight and bias. In terms of notation, we denote the qudit standard basis position operators of the neurons as $\bm{\hat{a}}_{\ell,j_\ell}$ for the $j_{\ell}^\text{th}$ neuron of the $\ell^\text{th}$ layer, $\bm{\hat{W}}_{\ell}$ is the matrix of operators corresponding to the weight parameters for the $\ell^\text{th}$ layer, and $\bm{\hat{b}}_{\ell}$ is the vector of operators corresponding to the bias parameters for the $\ell^\text{th}$ layer's neurons.

\begin{figure}[h!]
 \begin{center}
\includegraphics[width=1\columnwidth]{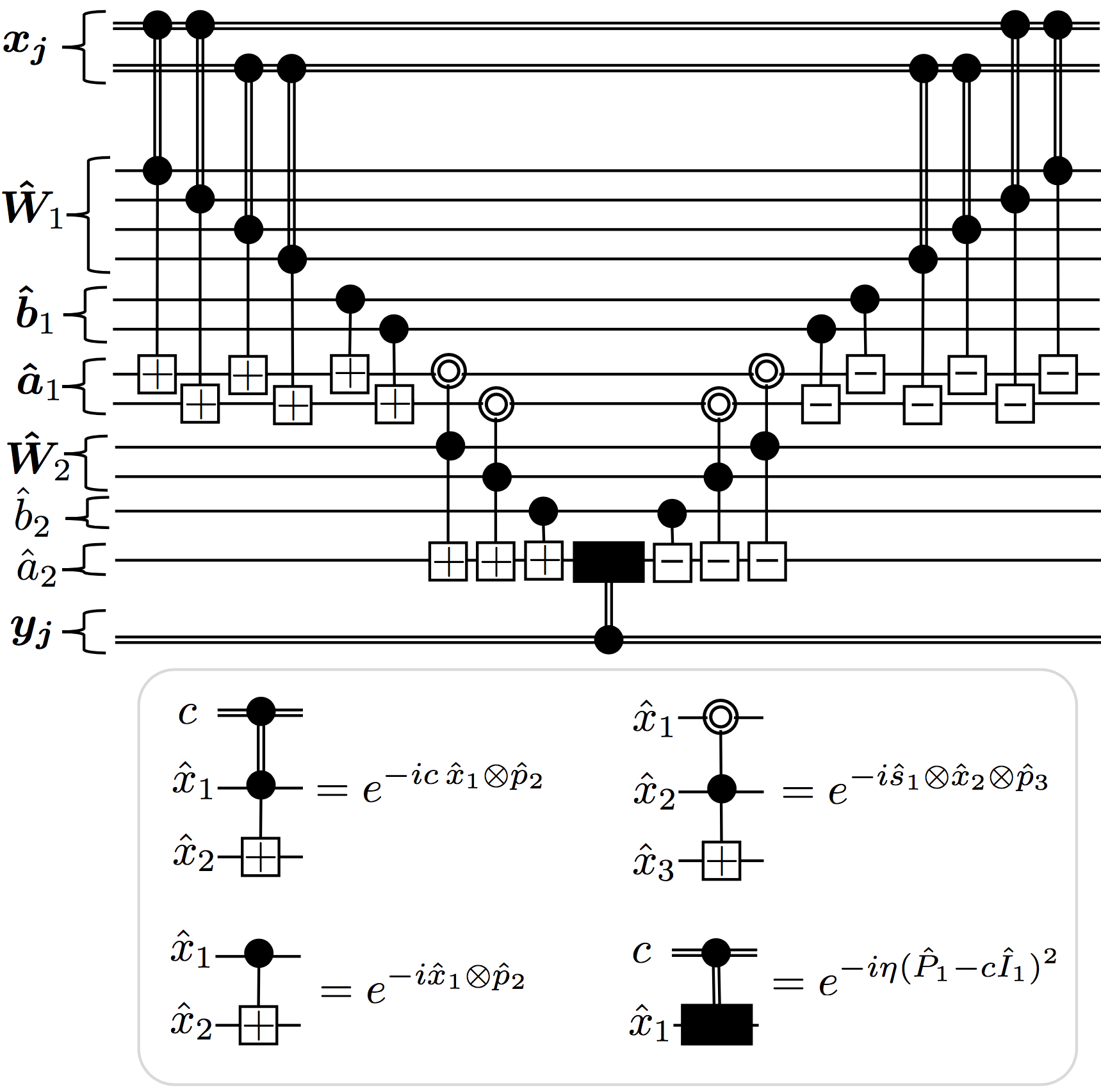}
\caption{Quantum Feedforward and Baqprop circuit for the Neural Network tasked with learning the XOR function. Refer to figure \ref{fig:nn_xor_diag} for a schematic of the neural network architecture. The data are input-output pairs $\{\bm{x}_j,y_j\}$ where $\bm{x}_j\in \mathds{Z}_2^2$ and $y_j = (x_{j,0}\oplus  x_{j,1})\in\mathds{Z}_2$. The solid lines are qudits (simulated qumodes), while classical registers are classical bits. The legend for the diagram is boxed below the circuit, in which $c\in\mathds{Z}_2$ represents an arbitrary bit, $\hat{P}\equiv \sum_{x\geq0}\ket{x}\!\bra{x}$ is the projector onto the qudit's positive-position states, and $\hat{s}\equiv \hat{P}\hat{x}\hat{P}$ is the position operator projected onto the positive states; akin to a RELU operator. Note the controlled-shifts with a $[-]$ are the Hermitian conjugate of their respective counterpart with a $[+]$, since they serve to uncompute the feedforward orperations. The loss function is the squared difference between the desired bit value and the truth value whether the output activation's is positive of not; we thus read out any output activation of positive value as 1 and any of negative value as 0.
} \label{fig:nn_xor_circ}
\end{center}
\end{figure}

For simplicity, we use a simulated Rectified Linear Unit (RELU) activation function, as it is standard in modern classical deep learning. Instead of using separate qudit registers for the input acumulation of the neuron and the activation value of this neuron's input, we perform the activation \textit{in-situ}, by using a modified position operator projected onto its positive values as the generator of shifts (see figure \ref{fig:nn_xor_circ}).

Let us describe more explicitly the circuit that was applied in order to train the network from figure \ref{fig:nn_xor_diag}. The QFB circuit which was applied is represented in figure \ref{fig:nn_xor_circ}. The input and output data registers were kept as classical controls in order to save memory space for the simulation. The parametric unitary for the first layer feedforward was
\spliteq{
\hat{U}^{(1)}(\bm{\hat{W}}_1,\bm{\hat{b}}_1)= \prod_{j,k\in\{0,1\}} e^{-i x_{m,j} \hat{W}_{1,jk}\hat{p}_{a_{1,k}}}e^{-i  \hat{b}_{1,k}\hat{p}_{a_{1,k}}}
}
where here the $\bm{x}_{m}\in \mathds{Z}_2^2$ are the possible input data points. This is simply the addition of the weight values conditioned on the input bit values, and the addition of the bias values onto the second layer's neurons. The feedforward operation for the following layer is given by
\spliteq{
\hat{U}^{(2)}(\bm{\hat{W}}_2,\bm{\hat{b}}_2)= \prod_{j\in\{0,1\}} e^{-i \hat{s}_{1,j} \hat{W}_{2,j}\hat{p}_{a_{2}}}e^{-i  \hat{b}_{2}\hat{p}_{a_{2}}}
}
where $\hat{s}$ is a neuron's activation value, which is the quadrature value projected onto the positive values;
\begin{equation}
    \hat{s}_{1,j} \equiv  \hat{P}\hat{a}_{1,j}\hat{P},\quad \hat{P}\equiv \sum_{a\geq0}\ket{a}\!\bra{a}
\end{equation}
which is an operator which assigns the ReLU eigenvalue to the neuron's input; effectively
\begin{equation}
    \hat{s}_{1,j} = \sigma(\hat{a}_{1,j}), \quad \sigma(x) = \begin{cases}x,& x>0\\
    0,&x<0\end{cases}.
\end{equation}
To synthesize this operation, an ancilla qubit would normally be necessary, the above option was implemented in order to reduce the effective dimension of the Hilbert space and reduce memory overhead during simulation.

Now, after the feedforward of both layers has been applied, we apply a phase kick according to the following cost function, 
\begin{equation}
    (\hat{P}_{a_2}-y_j\hat{I}_{a_2})^2 = \hat{P}_{a_2}+y_j\hat{I}_{a_2}-2y_j\hat{P}_{a_2}
\end{equation}

where $\bm{y}_j\in\mathds{Z}_2$ is the classical data bit desired output. The above cost function forces the output activation to be positive to indicate a value $1$ versus being nonpositive for the output $0$; in other words the above loss foces the network to encode the XOR value of the inputs into the eigenvalue of the obersvable $\hat{P}$, which is the projector onto the positive value qudit states of the output. One could consider the $\hat{P}$ as a step function activation operator for the output. For the full Quantum Feedforward and Baqprop circuit that was applied, see figure \ref{fig:nn_xor_circ}.

\subsubsection{Implementation \& Results}

In this section we present neural network training results from leveraging Momentum Measurement Gradient Descent (MoMGrad) and Quantum Dynamical Descent (QDD) to train the neural network from figure \ref{fig:nn_xor_diag} to learn the classical XOR function. We use the Quantum Feedforward and Baqprop circuit presented in figure \ref{fig:nn_xor_circ} in order to query the effective phase on the parameters for the cost function. The parameters, neurons, and bias registers were all chosen to be qudits of dimension 7 in the simulation. The parameters to be optimized via MoMGrad/QDD are the the weights and biases; in the notation of section \ref{sec:opt}, $\bm{\hat{\Phi}}= \{\bm{\hat{W}},\bm{\hat{b}}\}$.

\begin{figure}[h!]
 \begin{center}
\includegraphics[width=1\columnwidth]{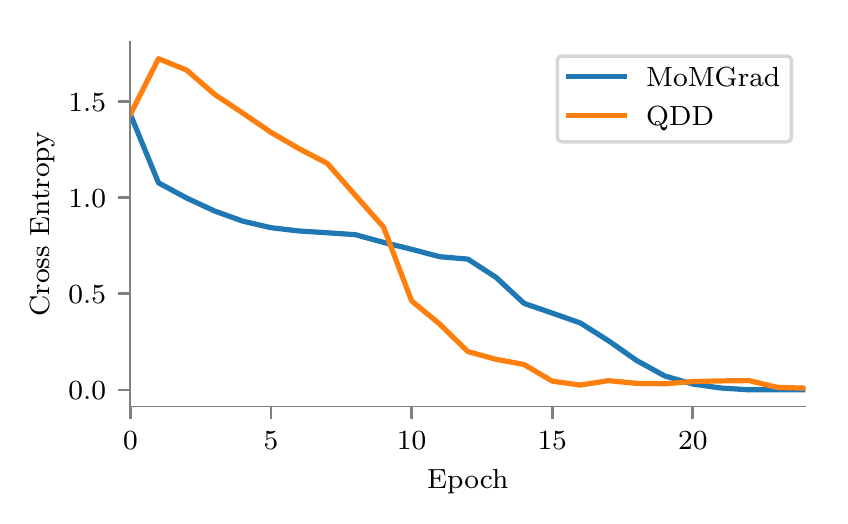}
\caption{Plot of the cross entropy between the neural networks output and the XOR of the input, at various iterations. The above is the average loss for 3 separate runs at each iteration index, for both training via MoMGrad and QDD.
} \label{fig:xor_learning}
\end{center}
\end{figure}

In figure \ref{fig:xor_learning}, we show the cross-entropy (Kullback-Leibler divergence \cite{Goodfellow-et-al-2016}
) between the desired output bit value and the value obtained through the feedforward. We consider any output of postive eigenvalue of the output's position quadrature as a $1$, and any nonpositive value as $0$ (effectively like a step function activation).

In both the QDD and MoMGrad cases, we begin with Gaussian wavefunctions for the quantum parameters $\bm{\hat{\Phi}}$. 
In terms of hyper-parameters, the initial means of the Gaussian wavefunctions; the components of $\bm{\Phi}_0$, were sampled randomly from a classical Gaussian distribution of mean 0 and standard deviation 0.5, while all the momenta hyper-parameters $\bm{\Pi}_0$ were initialized at $0$.

In the case of training via QDD, the initial standard deviation of the Gaussian wavefunction was chosen to be $\Sigma_0 = 1 $ for all parameters in the case of QDD, the kicking rate was kept at a constant $\eta_j=0.5\ \forall j$ and the kinetic rate for iteration $j$ was adjusted as $\gamma_j=0.5 -0.1\lfloor j/5 \rfloor $.

For the MoMGrad case this standard deviation of the Gaussian pointer state was adjusted at each iteration as $\Sigma_{0}^{(j)} = 0.95^{j} $ for the $j^\text{th}$ iteration, the kicking rate held constant at $\eta_j =0.5$ and kinetic rate held at $\gamma_j=1$ for all iterations. 

\begin{figure}[h!]
\includegraphics[width=1.0\columnwidth]{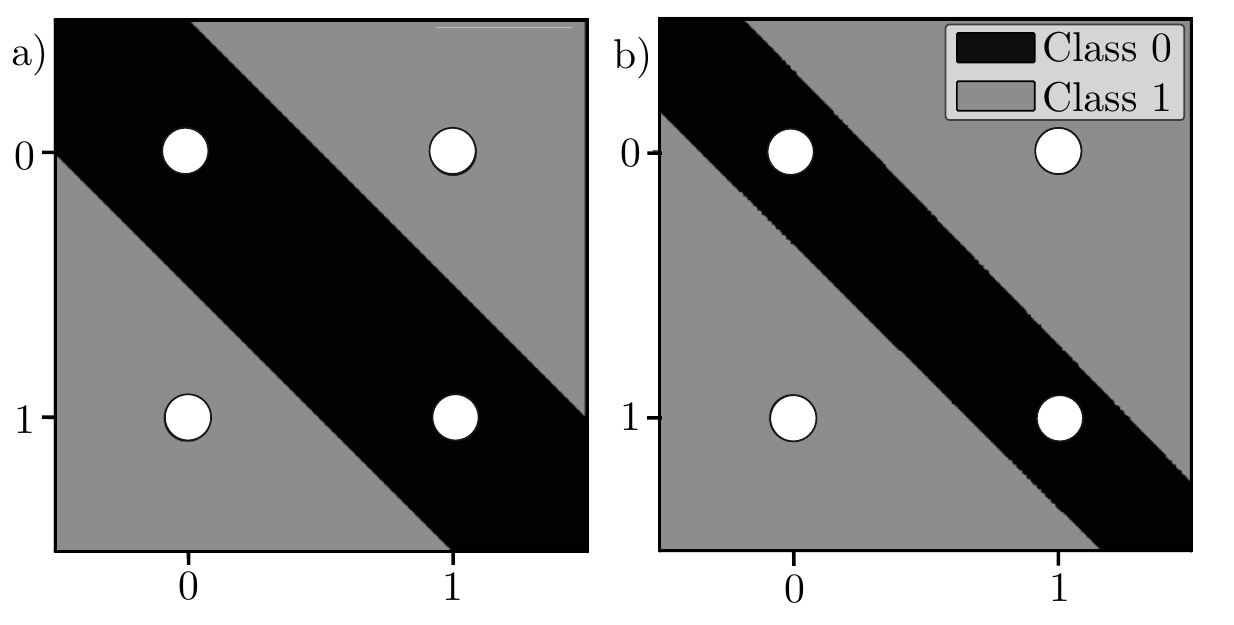}
\caption{XOR Quantum neural network learning decision boundary obtained from numerical quantum simulations, left (a) is via MoMGrad, right (b) is via QDD. The decision boundary was obtained by feeding a continuum of values in the input qudits, and observing the value of the output. For a positive-valued output, the corresponding decision is $1$, whereas a nonpositive output is considered as $0$. We see that both the QDD and MoMGrad correctly classified the output of the XOR.
} \label{fig:decision_XOR}
\end{figure}

In figure \ref{fig:decision_XOR} we show the decision boundary when considering a continuous input as opposed to simply binary. For a given continuously valued input in the range $[-0.5,1.5]\times [-0.5,1.5]$, we show the domain where the output is postitive (hence would be decided corresponding to an output 1), versus where the output is negative. We see that there is a striped domain characteristic of the non-linear separability of this domain. The QDD seems to have a tighter interval around the two points of XOR value 0. Due to our choice of cost function and due to feeding only binary data points, there was no incentive for the network to find an optimal hyperplane separating the inputs into the 0 and 1 classes. For the desired domain the neural network was trained for, binary inputs and output, the network performs the correct classification.

\subsection{Quantum Parametric Hamiltonian Optimization}

As mentioned in section \ref{sec:hammer}, there exists multiple possible applications of Parametric Hamiltonian Optimization we could chose to implement.
We choose to focus our numerical experiments on the QAOA, since we have established in section \ref{sec:meta} that the meta-learning problem is technically a QAOA-class problem. Given the large overheads of simulation of many parameters on a classical computer, testing the meta-learning directly for an interesting problem size would be intractable, hence by simply showing that our quantum-enhanced parameter optimization methods work for an instance of QAOA, we can thereby verify that it would work for a meta-learning problem.

In terms of specific QAOA implementation, we look at the canonical application of QAOA, i.e., applied to the optimization problem corresponding to finding the Maximum Cut (Max-Cut) of a graph \cite{farhi2014quantum}. We briefly review this application below, before showing our results for enhancing this optimization algorithm using MoMGrad and QDD.

\subsubsection{Application \& Methods}
Consider a graph $\mathcal{G}= \{\mathcal{V},\mathcal{E}\}$ of vertices $\mathcal{V}$ and edges $\mathcal{E}$. A \textit{cut set} $C\subseteq \mathcal{E}$ is defined as a set of edges which partitions the set of vertices in two. The maximum cut is the largest such subset of edges. We can consider the following Hamiltonian:
\begin{equation}\label{eq:max_Ham}
        \hat{H}_{\textsc{c}} = \sum_{\{j,k\}\in\mathcal{E}} \tfrac{1}{2}(\hat{I}- \hat{Z}_j \hat{Z}_k)
\end{equation}
where each vertex in $j\in\mathcal{V}$ is assigned a qubit, with $\ket{0}_j$ or $\ket{1}_j$ representing whether a given vertex is in partition $0$ or partition $1$. Each edge $\{j,k\}\in\mathcal{E}$ is associated a coupling of the form $\tfrac{1}{2}(\hat{I}- \hat{Z}_j \hat{Z}_k)$, which is an operator of eigenvalue 1 if the both vertices of the edge are of different partitions, or of eigenvalue 0 if they are in the same partition. 

Thus, finding the computational basis state $\ket{\bm{b}}$ which is the maximal eigenvalue eigenstate of the Hamiltonian \eqref{eq:max_Ham}, would be equivalent to finding the bitstring $\bm{b}$ of partition labels for each vertex $\bm{b} =\{b_j\}_{j\in\mathcal{V}}$, $b_j\in\mathds{Z}_2 \forall j$ which represents the Max-Cut set. 

In order to find this optimal state, we can apply the Quantum Approximate Optimization Algorithm, with $\hat{H}_\textsc{c}$ from \eqref{eq:max_Ham} as the cost Hamiltonian and
\begin{equation}\label{eq:mix_Ham}
    \hat{H}_{\textsc{m}} = \sum_{j\in\mathcal{V}} \hat{X}_j
\end{equation}
as the mixer Hamiltonian. The parametric circuit to be applied for the QAOA is then given by 
\begin{equation}\label{eq:KWAWAH}
    \hat{U}(\bm{\hat{\Phi}}) = \prod_{j=1}^P e^{-i\hat{\Phi}_{2j}\hat{H}_{\textsc{m}}}e^{-i\hat{\Phi}_{2j-1}\hat{H}_{\textsc{c}}}
\end{equation}
where $P$ is the number of alternating exponential steps, and the loss function to be minimized is $\hat{L} = -\hat{H}_{\textsc{c}}$. We can use MomGrad or QDD to optimize this parametric circuit in order to minimize the loss function (maximize the Hamiltonian). The canonical choice of initial state onto which one applies the above parametric circuit is the superposition of all bitsrings, \begin{equation}
    \ket{\psi_0} \equiv  \bigotimes_{j\in\mathcal{V}}\ket{+}.
\end{equation}
In general, after applying the QAOA parametric circuit for some choice of parameters deemed sufficiently optimal $\bm{\Phi}$, the final state should have a certain probablity of being in the Max-Cut state, or at least a probability of having states with a cut size close to this Max-Cut.

\begin{figure}[h!]
 \begin{center}
\includegraphics[width=1\columnwidth]{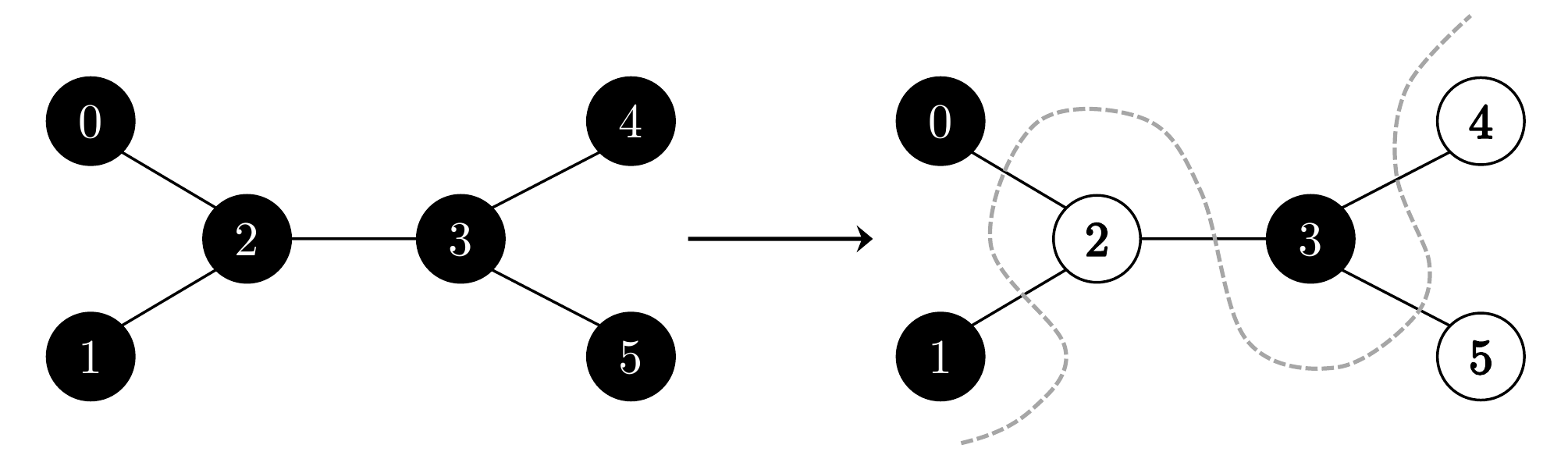}
\caption{Graph $\mathcal{G}= \{\mathcal{V},\mathcal{E}\}$ of $|\mathcal{V}|=6$ vertices and $|\mathcal{E}|=5$ edges for which we would like to leverage the QAOA in order to find the maximum cut. The maximum cut is represented on the right, with the partition index being represented by the vertex coloring, with either black (0) or white (1). Note the Max-Cut set has cardinality 5 hence any cut set $C$ for this graph has $|C|\leq 5$.
} \label{fig:max-cut-graph}
\end{center}
\end{figure}

For our particular implementation of QAOA, we apply it to find the Max-Cut of the graph depicted in figure \ref{fig:max-cut-graph}, which has a maximum cut of size $5$. In figure \ref{fig:qaoa_learning}, we plot the probability of measuring a state which has a cut size of 4 or more, for the expected parameters at various iterations of the optimization. We see that the probability of obtaining a near-optimal cut becomes high ($\text{Pr}(|C|\geq 4) \gtrapprox 0.8$ where $C$ is the eigenvalue of $\hat{H}_{\textsc{c}}$ for the measured bit string) as the training progresses, a sign that the approximate optimization is working. For this particular implementation, we chose a circuit with $P=2$, hence with only 4 parameters to be optimized, which we depict in figure \ref{fig:par_circ_QAOA}.

\subsubsection{Implementation \& Results}
\begin{figure}[h!]
 \begin{center}
\includegraphics[width=1\columnwidth]{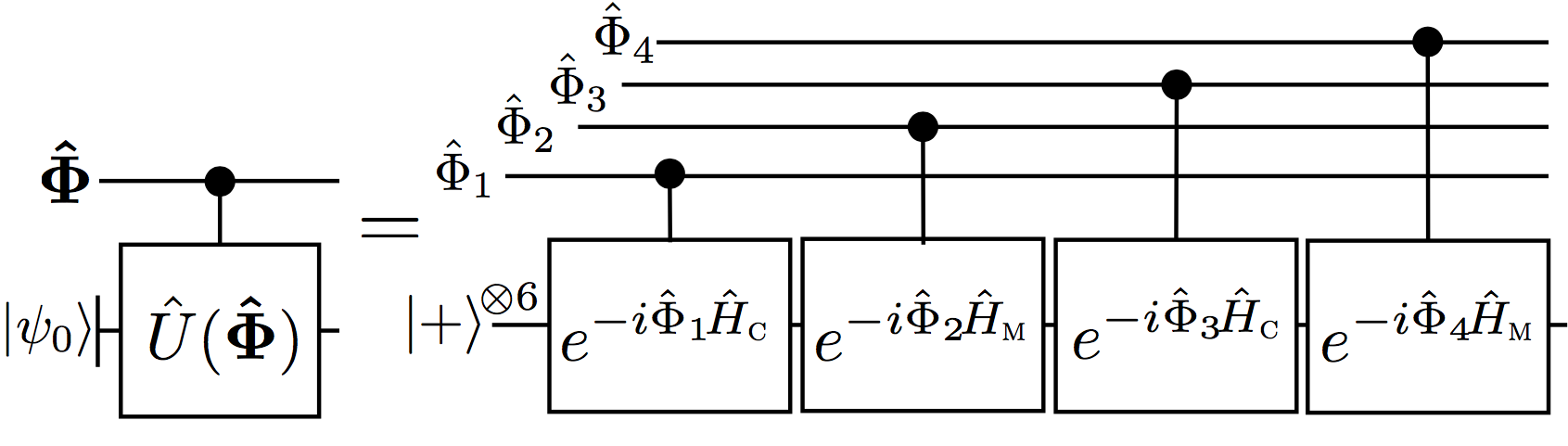}
\caption{QAOA $P=2$ parametric circuit from equation \ref{eq:KWAWAH} which was optimized for the results displayed in figure \ref{fig:qaoa_learning}. The cost and mixer Hamiltonians are those from equations \eqref{eq:max_Ham} and \eqref{eq:mix_Ham} for the graph depicted in figure \ref{fig:max-cut-graph}.
} \label{fig:par_circ_QAOA} 
\end{center}
\end{figure}

In this subsection, we present training results for the optimization of the QAOA parametric circuit using both MoMGrad and QDD. This parametric circuit consists of a $P=2$ QAOA ansatz (depicted in figure \ref{fig:par_circ_QAOA}), with cost Hamiltonian from \eqref{eq:max_Ham} and mixer Hamiltonian from \eqref{eq:mix_Ham} for the graph depicted in figure \ref{fig:max-cut-graph}.

\begin{figure}[h!]
 \begin{center}
\includegraphics[width=1\columnwidth]{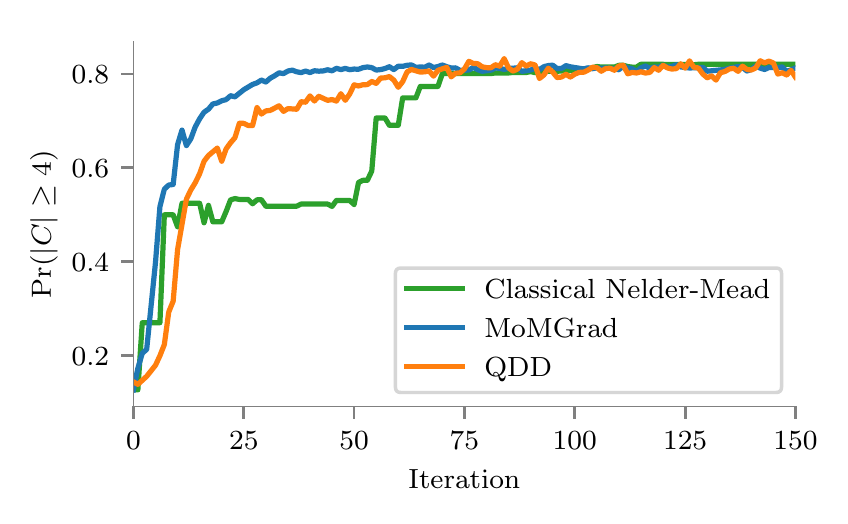}
\caption{Training results for optimizing the QAOA circuit from figure \ref{fig:par_circ_QAOA} via MoMGrad and QDD. Displayed is the probability of measuring a bitstring which corresponds to a cut set $C$ of size $4\leq C\leq 5$; near the optimum of $5$ which is the Max Cut set size. Additionally plotted above for comparison is a Nelder-Mead optimized QAOA, which converges slower than QDD and MoMGrad. We see that all 3 optimizers converge to a probability $\text{Pr}(|C|\geq 4) \gtrapprox 0.8$.
} \label{fig:qaoa_learning}
\end{center}
\end{figure}

In figure \ref{fig:qaoa_learning}, we represent the probability of obtaining a near-optimal cut, over the training iterations, for MoMGrad, QDD, and a quantum-classical Nelder-Mead method \cite{lagarias1998convergence} (for comparison). For this implementation, the parameters were simulated qudits of dimension $d=7$. For the hyper-parameters, the kicking rate for both QAOA and MoMGrad cases was kept at $\eta_j = 0.35\ \forall j$. The kinetic rate for QDD and MoMGrad were updated as $\gamma_j = 0.98^j/4$. The initial wavefunction for both QDD and MoMGrad was a Gaussian of $\Sigma_0=1$ for all parameters, with a mean $\bm{\Phi}_0$ with each component sampled from an independent classical Gaussian distribution of standard deviation 0.5 and mean 0. For MoMGrad, the subsequent standard deviation $\Sigma^{(j)}_{0} = 0.98^j$ for all components for the $j^\text{th}$ iteration.

\subsection{Quantum Unitary Learning}


In this subsection we demonstrate the implementation of quantum supervised unitary learning, see section \ref{sec:qunitary_sup_learn} for more details on this task. 

\subsubsection{Application \& Methods}
The task of supervised unitary learning, as described in section \ref{sec:qunitary_sup_learn}, is the following: given a set of input-output pairs $\{\ket{\psi_j^{\textsc{i}}},\ket{\psi_j^{\textsc{o}}}\}$ which are related by a unitary mapping $\ket{\psi_j^{\textsc{o}}} = \hat{V}\ket{\psi_j^{\textsc{i}}}$,  find a parameteric unitary ansatz $\hat{U}(\bm{\Phi})$ and sufficiently optimal parameters $\bm{\Phi}^\ast$ such that $\ket{\psi_j^{\textsc{o}}} \approx\hat{U}(\bm{\Phi}^\ast) \ket{\psi_j^{\textsc{i}}}$ so as to generally approximate the unitary $\hat{U}(\bm{\Phi}^\ast)\approx \hat{V}$, which should hold ideally for input-output pairs which lie outside the given dataset.

\begin{figure}[h!]
 \begin{center}
\includegraphics[width=1\columnwidth]{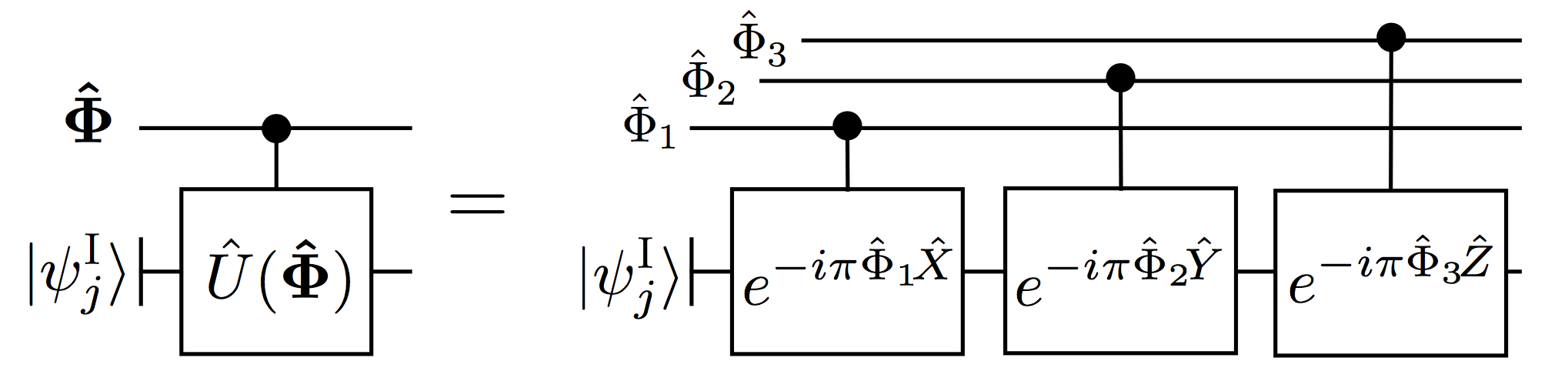}
\caption{Parametric circuit ansatz applied in this implementation of supervised unitary learning. The rotations are about the $x$, $y$ and $z$ axes of the Bloch sphere, in that order.
} \label{fig:1qb-parcirc}
\end{center}
\end{figure}

For the implementation in this paper, we consider a fairly simple case of learning a random single-qubit unitary. Using a uniform measure on the unit sphere, we can sample random points on the Bloch sphere and generate uniformly random single-qubit pure states. The input states $\ket{\psi_j}$ are thus generated by sampling from the Bloch sphere. As for the unitary $\hat{V}$ to be learned, we first sample a random state on the Bloch sphere, call it $\ket{\varphi_V}$. Then we define the unitary $\hat{V}$ to be learned as the unitary such that $\hat{V}^\dagger \ket{\varphi_V} =\ket{0}$, where $\ket{0}$ is the computational basis null state of the qubit. The parametric ansatz we use is represented in figure \ref{fig:1qb-parcirc}, it is a sequence of parametric rotations about the $x$, $y$ and $z$ axes of the Bloch sphere, in that order.

\subsubsection{Implementation \& Results}

Here we describe the details of the implementation of the learning of random single-qubit unitaries via the use of both QDD and MomGrad, for the parametric ansatz presented in figure \ref{fig:1qb-parcirc}. The qudit dimension of the simulated quantum parameters was $d=7$.

Note that in order to apply the phase kick according the output state projector loss function, $e^{i\eta\ket{\psi^{\textsc{o}}_j}\!\bra{\psi^{\textsc{o}}_j}}$, we implement the exponential of these states directly in the numerics rather than with the quantum state exponentiation tricks described in \ref{sec:qse}. This was done to minimize the classical memory overhead of simulation.

The phase kicks were applied in a sequential minibatches (see sec. \ref{sec:batch}) of size 10. 

\begin{figure}[h!]
 \begin{center}
\includegraphics[width=1\columnwidth]{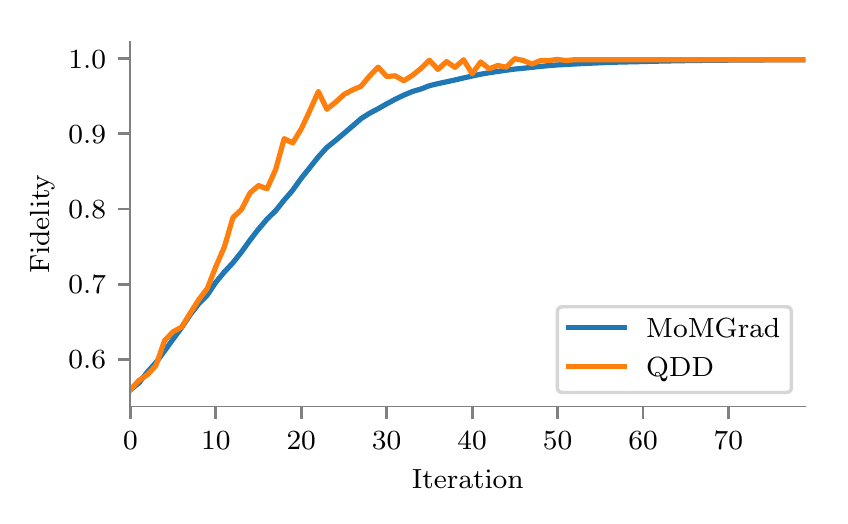}
\caption{
Training results for single qubit random unitary learning problem. Shown is the average fidelity between the states generated by applying the true unitary and the current best estimate to the parametric unitary, averaged over 5 runs. Each run had a different random unitary to be learned. Both QDD and MoMGrad converge to $\gtrapprox 99.75\%$ fidelity. 
} \label{fig:1QB_learning}
\end{center}
\end{figure}

Now, for the hyperparameters chosen for the training. For both the QDD and MoMGrad training, the kicking rates were kept at $\eta=0.2$ for all iterations. For both QDD and MoMGrad, the initial Gaussian wavefunction over parameters was chosen to have standard deviation $\Sigma_0=0.9$ for all parameters, and the initial means $\bm{Phi}_0$ which were sampled from a normal distribution of null mean and standard deviation 0.5. For QDD the kinetic rate for iteration $j$ was $0.2 \cdot 0.98^{j}$. Featured in  \ref{fig:1QB_learning} are the results of the average fidelity throughout the training, averaged over 5 different optimization runs with different random unitaries to be learned in each case.

\subsection{Hybrid Neural-Circuit Learning}\label{sec:num_hybrid}

In this section we numerically implement the training of a hybrid quantum-classical neural-circuit hybrid network, as described in section \ref{sec:hybrid}. That is, we consider having a quantum parametric circuit whose output is connected to a classical neural network.
\subsubsection{Application \& Methods}
In this particular implementation, we consider asking the hybrid network to learn to readout the momentum eigenvalue of a input momentum eigenstate, i.e., a computational basis state in the canonical dual (Fourier) basis. This can be seen as a hybrid quantum state classification task. For this particular implementation, we look at learning the quantum Fourier transform on 3 qubits.

Mathematically, we prepare a set of states $\{\ket{\psi_j} = \hat{F}\ket{j}_{012}\}_{j\in\mathds{Z}_8}$ where $\ket{j} = \bigotimes_{k=0}^{2}\ket{j_k}$ is the binary representation of the computational basis state of eigenvalue $j = \sum_{k=0}^2 j_k 2^k$, $\hat{F}$ is the 3-qubit Quantum Fourier transform.
The network is fed quantum states along with their corresponding desired label $\{\ket{\Psi_j}, j\}$. The network is tasked to learn how to correctly classify these states according to their label. 
The task is then effectively to learn decode the eigenvalue of the operator $\hat{F}^\dagger \hat{J} \hat{F}$, where
\begin{equation}
    \hat{J} = \sum_{j\in\mathds{Z}_8} j\ket{j}\!\bra{j} =  \sum_{k=0}^2 2^{k-1}(\hat{I}_k-\hat{Z}_k).
\end{equation}

We decompose this task of decoding the spectrum of this operator in such a way to force cooperation between the classical processing unit and the quantum processing unit in order to obtain correct classification. The learning is hybrid, as the quantum parametric circuit has to learn the inverse Fourier transform $F^\dagger$ gate decomposition, while the classical network learns the correct weighted combination of the readouts from the different qubit registers. Both networks must be optimized in a joint fashion in order for the composite quantum-classical mapping to guess the correct scalar corresponding to the eigenvalue of $\hat{F}^\dagger \hat{J} \hat{F}$ for each possible input state. 

\begin{figure}[h!]
 \begin{center}
\includegraphics[width=1\columnwidth]{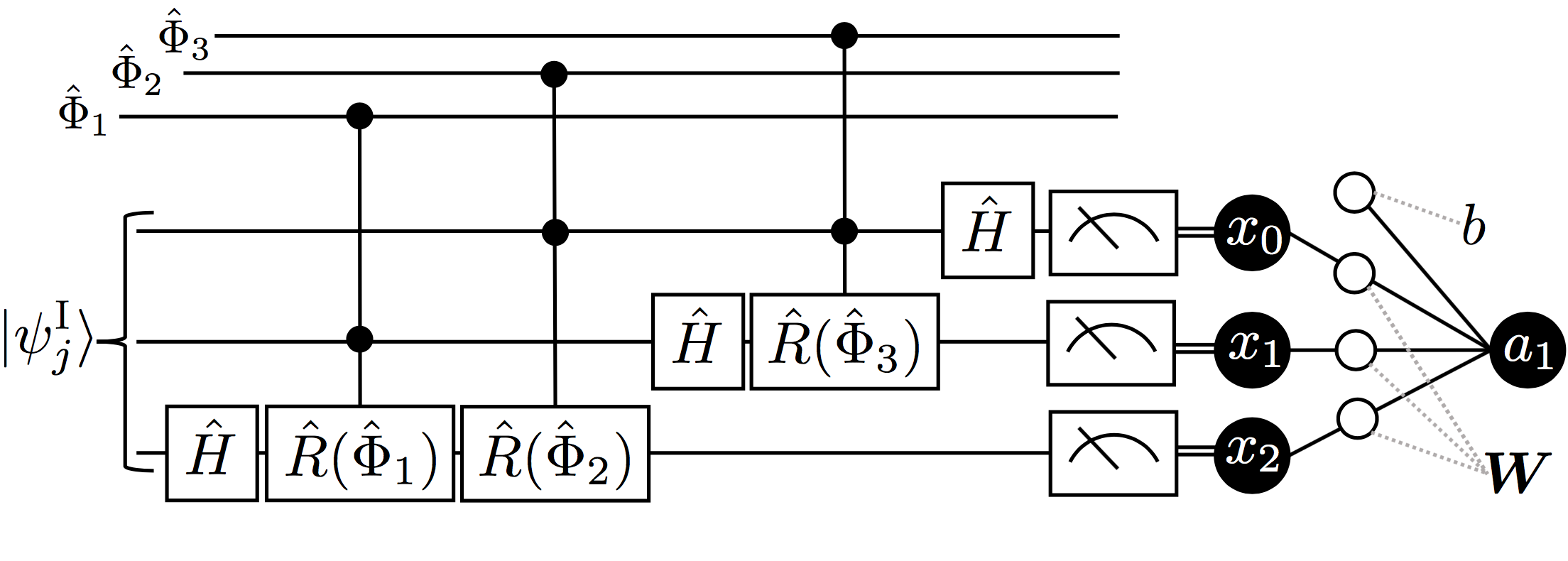}
\caption{Quantum Parametric Circuit Hybridized with a classical Neural Network to learn the Quantum Fourier Transform. Here each parametric rotation is of the form $\hat{R}(\Phi) = \ket{0}\!\bra{0}+ e^{i\Phi\pi/4}\ket{1}\!\bra{1}$. We use quantum-parametric versions of these rotations, $\hat{R}(\hat{\Phi}_j)$, in order to perform quantum-enhanced optimization of the latter via MoMGrad. The neural network connected to the output of the parametric circuit is a single neuron with rectified linear unit (ReLu) activation.
} \label{fig:hybrid}
\end{center}
\end{figure}

The classical network must learn the affine transformation which converts vectors of expectation values as $\bm{z} = \{\braket{\hat{Z}}_0,\braket{\hat{Z}}_1,\braket{\hat{Z}}_2\} \mapsto \sum_{k=0}^2 2^{k-1}(1-\braket{\hat{Z}_k}) \equiv y$.
Meanwhile the quantum parametric circuit must learn the canonical decomposition of the 3-qubit inverse QFT.
To restrict the number of quantum parameters needed to simulate the learning of the inverse Fourier transform, we only parametrize the controlled-$R_z$ rotations of this decomposition; the parametric circuit ansatz for this is represented in figure \ref{fig:hybrid}. The neural network is a single-layer of input activations with one neuron with RELU activation as output.

\subsubsection{Implementation \& Results}

Using a numerical simulation of the QPU-CPU interaction, we simulate the implementation of the first-order hybrid quantum-classical MoMGrad described in section \ref{sec:hybrid}. The qudit dimension of the simulated quantum parameters was $d=7$ once again.

\begin{figure}[h!]
 \begin{center}
\includegraphics[width=1\columnwidth]{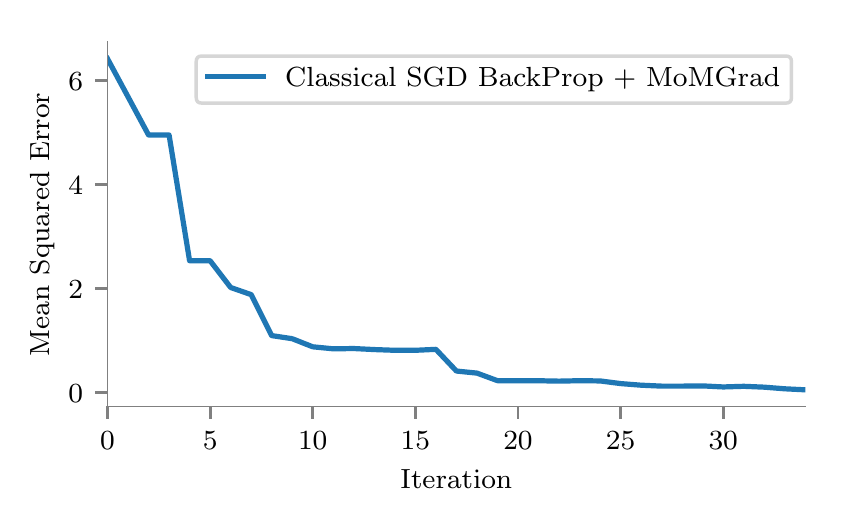}
\caption{
Training results for parametric circuit and classical neural network hybrid learning, for the network featured in figure \ref{fig:hybrid}. Shown is the mean squared error between the neural network output and the true label. The training was executed using the hybrid gradient descent technique described in section \ref{sec:hybrid}. Towards the end of training the Mean Squared Error (average squared distance between labels averaged over the dataset) converges to $\lessapprox 0.12$, thus indicating successful training.
} \label{fig:Hybrid_learning}
\end{center}
\end{figure}

We use a hybrid network stochastic gradient descent, where an iteration of gradient descent is performed for each state and label combination $\{\ket{\Psi_{j_k}},j_k\}_k$. The loss function to be optimized was the mean squared error: for a network prediction at the output of value $\tilde{y}_k$, and a desired label value $j_k$, the loss function is given by
\begin{equation}
    L(\tilde{y}_k,j_k):= (\tilde{y}_k -   j_k)^2 .
\end{equation}
The gradient of such a loss function is straightforward to obtain. The optimization procedure is that which is described in section \ref{sec:hybrid}. The results of the hybrid training are presented in figure \ref{fig:Hybrid_learning}.

Let us now describe the set of hyperparameters chosen to generate the results featured in figure \ref{fig:Hybrid_learning}. The learning rate for the classical network and the kicking rate for the parametric circuit QFB were both kept at $\eta=0.15$ throughout the training. The quantum parameters' initial wavefunction was a Gaussian of $\Sigma_0=0.65$ for all parameters, with a mean $\bm{\Phi}_0$ whose components were each sampled from independent classical Gaussian distributions of standard deviation 0.5 and mean 0. For the MoMGrad pointer states of further iterations, the subsequent standard deviations were $\Sigma^{(j)}_{0} = 0.65 \cdot 0.98^j$ in all components for the $j^\text{th}$ iteration.

\section{Discussion \& Outlook}
In this section we discuss potential implementations, implications, and possible future extensions of this work.

\subsubsection{Near-term considerations}

We begin with a discussion of potential near-term implementations.
Parametric circuits have been shown to be successfully implementable on NISQ (Noisy Intermediate Scale Quantum) devices \cite{preskill2018quantum}.
Due to the variational nature of parametric circuits optimization algorithms, in the presence of noise, the parametric transformations can adjust in order to partially counter the effects of noise.
As there is currently no standard criterion quantifying how robust a certain algorithm is to noise, and since execution performance can vary greatly depending both on the device and the algorithm, most approaches have resorted to empirically checking performance on a case-wise basis.
The common conception is that algorithms with low-depth quantum circuits using a quantum-classical optimization loop which relies only on expectation values of simple observables tend to be somewhat robust to noise. Thus, it is difficult to predict whether a certain optimizer and ansatz will perform well under various noise conditions but in this section we shall speculate as to which protocols have the best chance of being implementable in the near-term.

From the optimizers presented in this paper, although QDD has the potential for non-trivial tunneling in the optimization landscape, MoMGrad is the protocol with the best chance of execution on near-term devices due to its lower circuit depth requirements. In the case of MoMGrad, for a low-depth circuit ansatz, assuming having quantum parameters does not increase depth of execution of a parametric gate, the quantum feedforward and Baqprop should generally also be a low-depth circuit. The MoMGrad circuit includes twice the depth of the original ansatz plus an added depth due the exponential of the loss function. For simple loss functions, such as is the case for quantum classifiers for example, the exponential of the loss adds very minimal depth, while for loss operators with non-commuting terms (e.g. for Hamiltonian optimization), one can leverage the Gradient Expectation Estimation technique from section \ref{sec:hammer} to split up the gradient over multiple runs for the various terms. 

As for the efficient execution of quantum-parametric gates, there are a few options that could be tractable while adding minimal depth relative to a classically parametrized circuit ansatz. The most elementary form of a quantum parameter register would be using a single qubit instead of a qudit. As mentioned in section \ref{sec:discrete}, one can use a qubit to estimate the phase kickback induced by Baqprop, analogous to single-qubit phase estimation, at the cost of having to execute multiple runs in order to estimate the gradient to a sufficiently high precision. Generally, one could use perhaps only a few qubits (e.g. on the order of 2 or 3) to form a qudit of potentially sufficient dimension for multiple applications. For such low numbers of qubits, the Quantum Fourier transform is quite low depth, hence the gradient readout should be relatively robust to noise. Our numerical experiments in section \ref{sec:num} showed a good performance with only 7-dimensional qudits (achievable with 3 qubits). On the other hand since these were classical simulations of quantum computation, the expectation values could be extracted directly from the simulator, whereas a real quantum computation would necessitate multiple runs. Thus using a small-dimensional qudit (using a few qubits) for the parameters may be sufficient for some applications, but in general one would expect the performance to decay for many parameters since the readout of the gradient value is stochastic and could be greatly influenced by noise during the execution of the Quantum Fourier transform.
Using language from section \ref{sec:bkgd}, not only can there be underflow error (phase kick too small to be well detected) but there can also be overflow error, where the gradient phase kick exceeds the range of the qudit or qubit.

A possible alternative to qudits and qubits for implementation of the quantum parameters would be to use a continuous variable (CV) quantum mode (qumode) for each parameter. Note that the formalism and derivations throughout this paper were compatible with both simulated (qudit) qumodes and physical qumodes. Most current implementations of quantum computing, whether it be via superconducting qubits or trapped ions, have ways to build and control quantum harmonic oscillators on-chip \cite{poyatos1996quantum,hofheinz2009synthesizing}. Using CV modes as quantum parameters would require the ability to prepare squeezed states, the ability to perform measurements of the position/momentum quadratures, and for the execution of qubit-based circuit ansatze, the qumodes would need to be able to couple to qubits via an interaction of the form depicted in equation \eqref{eq:interact}. A potential advantage of using a physical qumode for readout is its robustness to small perturbations in its phase or position. For contrast, a small error on one qubit of in a multi-qubit Quantum Fourier transform can lead to a significant change in the readout value of the qudit position, whereas a small nudge of the qumode leads to a small change in readout value. Thus one could expect that the readout of the gradient values would be more robust to noise using analog qumodes. The effective phase estimation capacity of the qumode will then be determined by its degree of squeezing \cite{liu2016power,verdon2017probing}.

As for the implementation of the quantum-coherent classical neural networks from section \ref{sec:qnn}, both small-dimensional qudits or qumodes could work for the neurons in the near-term, the same arguments from above concerning the quantum parameters apply. One problem that may arise chaining many low-dimensional qudits' controlled-displacements (feedforward operations) is that any sort of under/overflow errors could add up exponentially with the depth. On the other hand, the current trend in classical machine learning has been to employ low-precision arithmetic \cite{gupta2015deep} for deep learning, which would suggest that not all classical deep learning algorithms necessitate high-precision floating points for effective operation and training. As such, one may consider few-qubit precision quantum-coherent neurons potentially sufficient in precision. One could even potentially consider the noise induced from the qudit imprecision as a form of regularization during both the feedforward an Baqprop phases. For further details on the influence of qudit imprecision on the feedforward operation, and for more details on potential physical CV implementations, see section \ref{sec:zoo}.

Let us now consider which applications from section \ref{sec:qdata_algs} have the best chance of being near-term implementable. As mentioned above, apart from the overheads of using quantum parameters to execute the feedforward of the parametric circuit, a key component to determining whether or not a certain Quantum Feedforward and Baqprop circuit is implementable in the near-term is the circuit depth required for the execution of the the exponential of the loss function. For any quantum data application which requires quantum state exponentiation, one could consider the near-term implementation of such an algorithm as unlikely, mainly due to the large overheads of have multiple Fredkin gates, and of batching quantum state exponentials sequentially. On the other hand, quantum classification, including measurement learning and quantum regression, have fairly simple cost functions which can be exponentiated as simple exponentials of standard basis observables. Another set of networks with a chance of near-term implementation are the Quantum-classical Hybrid neural-circuit hybrids, which as one may recall from section \ref{sec:hybrid}, can be built from parametric circuits for quantum classifiers/regression. One may imagine that having some additional classical neural processing after a quantum parametric circuit may reduce the need for depth of the quantum circuit to attain the same transformation or accomplish a given learning task many cases. Additionally, the feedforward step only relies on simple expectation values of simple observables, hence it should be robust to noise according to our criterion mentioned above. Finally, Hamiltonian Optimization, which includes the Variational Quantum Eigensolver and the Quantum Approximate Optimization Algorithm, should be implementable on near-term hardware with Baqprop. For Hamiltonians that are a sum of commuting terms, the Quantum Feedforward and Baqprop approach is straightforward, and for non-commuting terms in the Hamiltonian, one can use the Gradient Expectation Estimation technique (GEEP, sec. \ref{sec:hammer}), which allows for parallelization of the gradient accumulation over multiple runs. 

\subsubsection{Further-term considerations}

We now proceed to considering potential interesting applications in the further-term, as well as future work.

In the long-term, with the advent of large-scale error-corrected fault-tolerant quantum computers, the possibility of training large-scale classical neural networks on quantum computers, as presented in section \ref{sec:qnn}, will become tractable. At that moment, one may want to consider training neural networks with the Quantum Dynamical Descent (QDD) approach. QDD may be more powerful than simple gradient descent in some instances, due to being an effective Quantum Approximate Optimization of the parameters. Furthermore, with a large scale error-corrected quantum computer one could test the training of quantum neural networks using Quantum Meta-Learning (sec. \ref{sec:meta}), for either the optimization of hyper-parameters or network architecture to improve generalization error. If one were to apply the Meta-QDD protocol to either of these meta-learning applications, which would consist of a quantum dynamical simulation of descent (with possible tunneling) in the space of possible network architectures or hyper-parameters, one could imagine the distribution over such hyper-parameters difficult to simulate. Again this is would be due the known difficulty of simulating samples from a QAOA \cite{farhi2016quantum}. Empirical testing of possible advantages of Quantum Meta-Learning via its large-scale implementation could yield potentially interesting results. 

As a side-note, although we only treated how to quantize and train classical feedforward networks, the optimization methods featured in this paper could potentially be used to train classical Boltzmann machines. In recent work, it was shown that one could train Quantum Boltzmann machines using QAOA-type quantum parametric circuits \cite{verdon2017quantum}. The QAOA was used to approximately sample from various Gibbs distributions of networks, such sampling is a necessary step to perform (classical) gradient descent of the network's weights. Thus, using techniques developed in this paper, one could potentially consider enhancing the optimization of the parametric QAOA circuit via MoMGrad or QDD, such as to leverage Baqprop to accelerate the Gibbs sampling at each gradient descent. If one were to go further and also consider the Boltzmann machine's weights as quantum parameters along with the corresponding parametric circuit's parameters, one could then potentially us a meta-QDD optimization loop to quantumly the optimize Boltzmann machine weights, similar to that featured in section \ref{sec:meta}. We leave further details of this approach for future work.

Another interesting avenue of future exploration is the possibility of performing quantum deep learning in a massively quantum-parallelized fashion across a quantum network. Very recently, the first experimental demonstration of quantum state transfer between quantum computing chips was successfully implemented \cite{kurpiers2018deterministic}. Eventually, with Quantum Error Corrected state transfers, parallelization of algorithms across multiple quantum chips will become a feasibility. As we showed in section \ref{sec:misc}, various parallelization and regularizaton protocols such as the Coherent Accumulation of Momenta Parallelization protocol (CAMP, sec. \ref{sec:CAMP}) and the Meta-networked Swarm Optimization (MISO, sec. \ref{sec:miso}) can take advantage of a quantum network of quantum processing units to improve the precision and time requirements of training networks. In the particular case of CAMP, there is a square root speedup to get the expectation value of the gradient over a minibatch within a certain precision as compared to classical parallelization.
In modern classical deep learning, parallelization is key to training large-scale neural networks in a feasible time frame \cite{dean2012large}, 
 it is thus to be expected that once quantum algorithms can reach a certain scale, parallelization becomes indispensable, just as it is in the classical case.

Now, let us mention some avenues for further possible mathematical analyses which could be conducted. A first one is to provide a more detailed analysis of the resource overheads of synthesizing gate sequences for the various protocols studied in this paper. As we established multiple connections with quantum simulation theory, perhaps tools from this subfield
could be ported over to quantum deep learning. In terms of the further analysis of the effective physics of the parameters for QDD, one could view the stochastic QFB phase kicks as a repeated interaction with an environment (in this case the compute registers). One could then perform an analysis of the effective open system dynamics and disspation terms at higher-orders of the kicking rate.

Finally, now that we have added multiple optimization techniques to the repertoire of quantum deep learning tools, the key to making quantum deep learning feasible for large-scale quantum parametric circuits will require new quantum parametric ansatze. As pointed out by McClean et al. \cite{mcclean2018barren}, most current quantum parametric ansatze relying on random circuits of qubits have vanishing gradients. Similar to the problem of vanishing gradients in classical machine learning, current parametric ansatz have exponentially vanishing gradients in the number of degrees of freedom. Further study into the mechanism behind the obtention of gradients of parametric circuits is necessary in order to allow for the design of new ansatze which could solve this vanishing gradient problem. In section \ref{sec:deep_baq}, we explicitly detailed the mechanism for the backward quantum propagation of phase errors through the quantum-coherent neural networks. One could then potentially extend the analysis presented in section \ref{sec:qdata_algs} for the layerwise backpropagation of the gradient signal in general quantum parametric circuits such as to provide the same level of detail as to how the gradient (phase kick) signal travels through the compute registers in order to influence the parameters. By choosing specific ansatze, one could examine the generators of each parametric circuit element, and possibly replicate the level of detail of the analysis from \ref{sec:deep_baq} for this specific parametric ansatz. 
Such an analysis would have the potential to shed new light on the vanishing gradient problem and point towards solutions. We leave an analysis of this kind for general parametric quantum circuit ansatze for future work.

\section{Conclusion}

The goal of this paper was to establish a bridge between the theories of classical and quantum deep learning, such as to allow for the exchange of tools and the gain of new insights in both fields. In alignment with this goal, we took inspiration from classical deep learning techniques to create numerous new methods for the quantum-enhanced optimization of quantum parametric networks on a quantum computer. Furthermore, we explored various ways classical deep learning can leverage quantum computation for optimization, and how classical and quantum deep learning optimization strategies can directly interface with one another.

More specifically, we introduced a unified approach to the optimization of quantum parametric circuits and classical neural networks on a quantum computer, based on a universal error backpropagation principle for quantum parametric networks. We then further extended these quantum optimization methods with a compatible set of tools for parallelization, regularization, and meta-learning.
Furthermore, we detailed how to leverage these optimization strategies for the effective training of any classical feedforward neural network on a quantum computer, as well as for numerous quantum parametric circuit applications. We numerically tested both core optimization algorithms on multiple such applications, empirically demonstrating their effectiveness. Finally, we introduced a way to merge classical and quantum backpropagation between classical and quantum computers, opening up the possibility for the field of truly hybrid quantum-classical deep learning.

We hope that the work presented in this paper will bolster further work exploring this nascent field of Quantum Deep Learning.

\section{Acknowledgements}
Quantum circuit simulations featured in this paper were executed on the Rigetti Forest Quantum Virtual Machine, with code written in PyQuil \cite{Forest}.
The authors would like to thank Rigetti Computing and its team for providing computing infrastructures and continued support for Forest. The authors would also like to thank Steve Weiss and the Information Technology team at the IQC for providing additional computing infrastructures and IT support for this project. The authors would like to thank Atmn Patel for useful discussions, as well as Achim Kempf for the support. GV and JP acknowledge funding from NSERC. 

\bibliographystyle{apsrev4-1}
\bibliography{library}

\end{document}